\documentclass{aa}

\usepackage[sort]{natbib}
\usepackage[utf8]{inputenc}
\usepackage{amsmath}
\usepackage[amssymb]{SIunits}
\usepackage{listings}
\usepackage{xspace}
\usepackage{graphicx}
\usepackage{pict2e}
\usepackage{supertabular}
\usepackage{tabularx}
\usepackage{lscape}
\usepackage{longtable}
\usepackage{multirow}
\usepackage[x11names, rgb]{xcolor}
\usepackage[utf8]{inputenc}
\usepackage{tikz}
\usetikzlibrary{arrows,shapes,positioning}
\newcommand{\ab}{\ensuremath{\text{ab}}\xspace}
\newcommand{\weff}[1][]{\ensuremath{\lambda_\text{eff}^{#1}}\xspace}
\newcommand{\wshift}[1][]{\ensuremath{\delta\lambda_\text{eff}^{#1}}\xspace}

\newcommand{\bdtruc}{BD~+17~4708\xspace}
\newcommand{\bdref}{{bd17}}

\newcommand{\Fixed}[1]{{#1}}
\newcommand{\madu}[1][m]{\ensuremath{#1_{\text{ADU}}}\xspace}

\newcommand{\cref}[1][]{\ensuremath{{c_{\text{ref}#1}}}\xspace}
\newcommand{\umag}[1][m]{\ensuremath{#1_{|\rpos}}\xspace}

\makeatletter
\newcommand*{\cmag}{\@ifstar\cmagsynt\cmagmes}
\newcommand{\cmagmes}[1][m]{\ensuremath{#1_{|\spos}}\xspace}
\newcommand{\cmagsynt}[1][m]{\ensuremath{\tilde{#1}_{|\spos}}\xspace}
\makeatother
\newcommand{\smag}[1][m]{\ensuremath{#1_{\text{2.5}}}\xspace}
\newcommand{\dzp}[1][]{\ensuremath{{\delta Z_{#1}}}\xspace}
\newcommand{\zp}[1][]{\ensuremath{{Z_{#1}}}\xspace}
\newcommand{\dk}[1][]{\ensuremath{\delta k_{#1}}\xspace}

\newcommand{\rpos}{{\ensuremath{\vec{x}_0}}\xspace}
\newcommand{\arcdeg}{{\ensuremath{^\circ}}\xspace}
\newcommand{\spos}{\ensuremath{\vec{x}}\xspace}
\newcommand{\x}{\ensuremath{\vec{x}}\xspace}
\newcommand{\band}[1]{\ensuremath{#1_M}\xspace}

\newcommand{\LtoM}{\ensuremath{f_{\rm L}}\xspace}
\DeclareMathOperator{\logdec}{\log_{10}}

\DeclareMathOperator{\cov}{cov}
\DeclareMathOperator{\diag}{diag}

\graphicspath{{fig/}}

\newcommand{\igraph}[2][]{\includegraphics[width=#1\linewidth]{#2}}

\newcommand{\megacam}{MegaCam\xspace}

\newcommand{\mapc}{MAPC\xspace}

\makeatletter
\let\jnl@style=\rm
\def\ref@jnl#1{{\jnl@style#1}}
\def\aj{\ref@jnl{AJ}}                   
\def\araa{\ref@jnl{ARA\&A}}             
\def\apj{\ref@jnl{ApJ}}                 
\def\apjl{\ref@jnl{ApJ}}                
\def\apjs{\ref@jnl{ApJS}}               
\def\ao{\ref@jnl{Appl.~Opt.}}           
\def\apss{\ref@jnl{Ap\&SS}}             
\def\aap{\ref@jnl{A\&A}}                
\def\aapr{\ref@jnl{A\&A~Rev.}}          
\def\aaps{\ref@jnl{A\&AS}}              
\def\azh{\ref@jnl{AZh}}                 
\def\baas{\ref@jnl{BAAS}}               
\def\jrasc{\ref@jnl{JRASC}}             
\def\memras{\ref@jnl{MmRAS}}            
\def\mnras{\ref@jnl{MNRAS}}             
\def\pra{\ref@jnl{Phys.~Rev.~A}}        
\def\prb{\ref@jnl{Phys.~Rev.~B}}        
\def\prc{\ref@jnl{Phys.~Rev.~C}}        
\def\prd{\ref@jnl{Phys.~Rev.~D}}        
\def\pre{\ref@jnl{Phys.~Rev.~E}}        
\def\prl{\ref@jnl{Phys.~Rev.~Lett.}}    
\def\pasp{\ref@jnl{PASP}}               
\def\pasj{\ref@jnl{PASJ}}               
\def\qjras{\ref@jnl{QJRAS}}             
\def\skytel{\ref@jnl{S\&T}}             
\def\solphys{\ref@jnl{Sol.~Phys.}}      
\def\sovast{\ref@jnl{Soviet~Ast.}}      
\def\ssr{\ref@jnl{Space~Sci.~Rev.}}     
\def\zap{\ref@jnl{ZAp}}                 
\def\nat{\ref@jnl{Nature}}              
\def\iaucirc{\ref@jnl{IAU~Circ.}}       
\def\aplett{\ref@jnl{Astrophys.~Lett.}} 
\def\apspr{\ref@jnl{Astrophys.~Space~Phys.~Res.}}
\def\bain{\ref@jnl{Bull.~Astron.~Inst.~Netherlands}} 
\def\fcp{\ref@jnl{Fund.~Cosmic~Phys.}}  
\def\gca{\ref@jnl{Geochim.~Cosmochim.~Acta}}   
\def\grl{\ref@jnl{Geophys.~Res.~Lett.}} 
\def\jcp{\ref@jnl{J.~Chem.~Phys.}}      
\def\jgr{\ref@jnl{J.~Geophys.~Res.}}    
\def\jqsrt{\ref@jnl{J.~Quant.~Spec.~Radiat.~Transf.}}
\def\memsai{\ref@jnl{Mem.~Soc.~Astron.~Italiana}}
\def\nphysa{\ref@jnl{Nucl.~Phys.~A}}   
\def\physrep{\ref@jnl{Phys.~Rep.}}   
\def\physscr{\ref@jnl{Phys.~Scr}}   
\def\planss{\ref@jnl{Planet.~Space~Sci.}}   
\def\procspie{\ref@jnl{Proc.~SPIE}}   

\makeatother
\usepackage[pagebackref,colorlinks,citecolor=blue,urlcolor=blue,linkcolor=blue]{hyperref}

\title{Improved Photometric Calibration of the SNLS and the SDSS Supernova Surveys\thanks{ Based on
observations obtained with MegaPrime/MegaCam, a joint project of CFHT
and CEA/DAPNIA, at the Canada-France-Hawaii Telescope (CFHT) which is
operated by the National Research Council (NRC) of Canada, the
Institut National des Sciences de l'Univers of the Centre National de
la Recherche Scientifique (CNRS) of France, and the University of
Hawaii. This work is based in part on data products available at the
Canadian Astronomy Data Centre as part of the Canada-France-Hawaii
Telescope Legacy Survey, a collaborative project of NRC and CNRS.}~\thanks{Tables 22 and E1 -- E10 are  
available in electronic form at the CDS via anonymous ftp to {\tt cdsarc.u-strasbg.fr (130.79.128.5)}
or via {\tt http://cdsweb.u-strasbg.fr/cgi-bin/qcat?J/A+A/}}}
\author{M.~Betoule \inst{1,2}
  \and J.~Marriner \inst{3} 
  \and N.~Regnault \inst{1}
  \and J.-C.~Cuillandre\inst{4}
  \and P.~Astier\inst{1}
  \and J.~Guy\inst{1}
  \and C.~Balland\inst{1} 
  \and P.~El~Hage\inst{1}
  \and D.~Hardin\inst{1}
  \and R.~Kessler\inst{5,6}
  \and L.~Le~Guillou\inst{1}
  \and J.~Mosher\inst{7}
  \and R.~Pain\inst{1}
  \and P.-F.~Rocci\inst{1}
  \and M.~Sako\inst{7}
  \and K.~Schahmaneche\inst{1}
}

\institute{
  LPNHE, CNRS-IN2P3 and Université Paris 6 \& 7, 4 place Jussieu, Paris Cedex 05, France\\
  \email{betoule@lpnhe.in2p3.fr}
  \and  PCCP, 10 rue Alice Domont et L\'eonie Duquet, Paris Cedex 13, France
  \and Center for Particle Astrophysics, Fermi National Accelerator Laboratory, P.O. Box 500, Batavia, IL 60510, USA
  \and Canada-France-Hawaii Telescope Corp., Kamuela, HI 96743, USA
  \and Department of Astronomy and Astrophysics, University of Chicago, 5640 South Ellis Avenue, Chicago, IL 60637
  \and Kavli Institute for Cosmological Physics, University of Chicago, 5640 South Ellis Avenue Chicago, IL 60637
  \and Department of Physics and Astronomy, University of Pennsylvania, 209 South 33rd Street Philadelphia, PA 19104-6396.
}

\titlerunning{Photometric Calibration of the SNLS and the SDSS.}
\authorrunning{M.~Betoule et al.}

\date{Preprint online version: Accepted 15 December 2012 }

\usepackage{layouts}
\usepackage{txfonts} 

\begin{document}

\abstract{We present a combined photometric calibration of the
  Supernova Legacy Survey (SNLS) and the SDSS supernova survey, which
  results from a joint effort of the SDSS and the SNLS
  collaborations.} {Our primary motivation is to eventually sharpen
  cosmological constraints derived from type Ia supernova measurements
  by improving the accuracy of the photometric calibration. We deliver
  fluxes calibrated to the HST spectrophotometric star network for
  large sets of tertiary stars \Fixed{that cover} the science fields
  of both surveys \Fixed{in all photometric bands.}  We also
  cross-calibrate directly the two \Fixed{surveys} and demonstrate
  their consistency.}  {For \Fixed{each survey} the flat-fielding is
  revised \Fixed{based} on the analysis of dithered star observations.
  The calibration transfer from the HST spectrophotometric standard
  stars to the multi-epoch tertiary standard star catalogs in the
  science fields follows three different paths: observations of
  primary standard stars with the SDSS PT telescope; observations of
  Landolt secondary standard stars with \Fixed{SNLS MegaCam
    instrument at CFHT}; and direct observation of faint HST standard stars
  with MegaCam.  In addition, the tertiary stars for the two surveys
  are cross-calibrated using dedicated MegaCam observations \Fixed{of
    stripe 82}.  This \Fixed{overlap enables} the comparison of
  \Fixed{these} three calibration paths and \Fixed{justifies using}
  their combination to improve the calibration accuracy.}{
  \Fixed{Flat-field corrections have improved the uniformity of each
    survey as demonstrated by the comparison of photometry in
    overlapping fields: the rms of the difference between the two
    surveys is 3~mmag in $gri$, 4~mmag in $z$ and 8~mmag in $u$.}
  We \Fixed{also} find a remarkable agreement
  (better than 1\%) between the SDSS and \Fixed{the SNLS calibration
    in $griz$}. The cross-calibration and the introduction of direct
  calibration observations bring redundancy and strengthen the
  confidence in the resulting calibration. We conclude that the
  surveys are calibrated to the HST with a precision of about 0.4\% in
  $griz$. This \Fixed{precision} 
  is comparable to the external uncertainty affecting the
  color of the HST primary standard stars.}  {} 

\maketitle
\section{Introduction}
\label{sec:intro}

Substantial efforts have been spent in the last few years to improve
the accuracy of the calibration of large photometric surveys
(e.g. \citealt[][for SDSS]{ivezi_sloan_2007,2008ApJ...674.1217P}
\citealt[for SNLS]{R09} \citealt[for
Pan-STARRS]{2012arXiv1201.2208S}), establishing large catalogs of
stars with broadband magnitudes known with an accuracy of 1 or 2\%.
The scientific goals pursued cover a broad domain from stellar physics
to cosmology.  \Fixed{One of the main drivers for precision
  photometry} is the quest for precise cosmological constraints from
the Hubble diagram of type Ia supernovae.

In spite of the \Fixed{vast improvements in photometric calibration
  over the last decade}, \Fixed{calibration uncertainties} remain the
dominant source of systematic error limiting the precision of
cosmological constraints obtained from \Fixed{more than 400
  well-measured} type-Ia supernova \Fixed{(SN~Ia) light curves} (see
\emph{e.g.} the discussion of systematics in
\citealt{2011ApJS..192....1C}).  As an illustration, the accuracy of
the measurement of the dark energy equation of state parameter $w$
provided in \cite{2011ApJ...737..102S} is about 8\% while it would
reach 5.7\% with a perfect photometric calibration. The calibration
accuracy is likely to remain a serious issue in upcoming surveys
expecting an order of magnitude increase in the number of supernovae.

The photometric component \Fixed{of the SNLS} was conducted at the
Canada France Hawaii Telescope (CFHT) using the wide field
MegaPrime/MegaCam camera. During its five years of \Fixed{survey
  operations} (2003-2008), about $1000$ \Fixed{multi-band SN~Ia light
  curves were discovered} in the redshift range $0.2<z<1$, $500$ of
which were spectroscopically identified. The first three years of the
data sample are \Fixed{analyzed and published} in
\citet{2010A&A...523A...7G,2011ApJS..192....1C,2011ApJ...737..102S}.
The SDSS-II SN Survey \Fixed{was} one of the three components of the
SDSS-II project.  During three three-month seasons of \Fixed{survey
  operations} (Fall 2005-2008), the SDSS SN survey discovered
multi-band light curves for 500 spectroscopically confirmed SNe Ia in
the redshift range $0.01<z<0.45$. The processing and analysis of the
first year sample are given in
\cite{holtzman_sloan_2008,kessler_first-year_2009}.

This work is part of the SDSS+SNLS joint analysis. It presents the
collaborative effort made to improve the accuracy of photometric
calibration. Previous calibrations for the two surveys are described
in \citet{holtzman_sloan_2008} and \citet{R09} (hereafter R09)
respectively. This paper supersedes both by providing a common and
consistent calibration for the two surveys. 
\Fixed{We} exploit dedicated \Fixed{CFHT}
observations that complement the usual calibration data. 
The complementary data consists \Fixed{of} 
direct observations of primary spectrophotometric standard stars 
and of SDSS science fields.

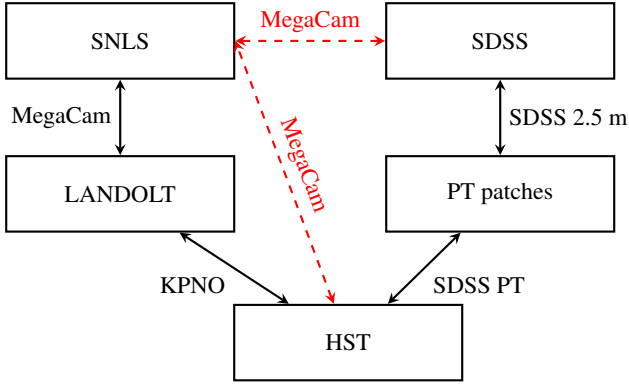
\begin{figure}
\centering
\setlength{\unitlength}{1cm}
  \begin{tikzpicture}
  [align=center,thick,meas/.style={rectangle,draw=black,minimum height=1cm,minimum width=3cm},new/.style={stealth-stealth, dashed,above,sloped,midway,red},old/.style={stealth-stealth}, auto]
\node  [meas] (snls) at (0,4) {SNLS};
\node  [meas] (land) [below=of snls] {LANDOLT};
\node  [meas] (sdss) at (5,4) {SDSS};
\node  [meas] (pt) [below=of sdss] {PT patches};
\node  [meas] (hst) at (3,0) {HST}
edge [new] node {MegaCam} (snls.east) ;
\draw [old] (snls) to node [swap] {MegaCam} (land) ;
\draw [old] (land) to node [swap] {KPNO} (hst);
\draw [old] (sdss) to node {SDSS 2.5~m} (pt);
\draw [old] (pt) to node  {SDSS PT} (hst);
\draw [new] (snls) to node {MegaCam} (sdss);
\end{tikzpicture}

\caption{Schematic view of calibration data \Fixed{used} in the
  present work.  \Fixed{Each box represents a} set of standard stars
  established in \Fixed{that} photometric system.  \Fixed{Each arrow
    shows} the available cross-calibration data \Fixed{corresponding}
  to the measurement of flux ratios between \Fixed{the} two different
  sets of standards.  The two new sets of \Fixed{SNLS} measurements
  introduced in this paper are \Fixed{indicated} with red dashed
  lines.}
\label{fig:schema}
\end{figure}

The advantage of \Fixed{these complementary} observations is
threefold: \Fixed{1)} they provide a direct cross-calibration of the
tertiary standard stars used to calibrate the supernova measurement in
both surveys, \Fixed{2)} they enable an accurate consistency check of
the photometry between both instruments, \Fixed{and 3)} they provide
redundant paths, subject to different systematics, to anchor the
common cross-calibrated set of tertiary standards on the flux scale
defined by the HST white dwarfs
\citep{bohlin_spectrophotometric_2001}.  The various calibration paths
to the HST system are illustrated \Fixed{in} Fig.~\ref{fig:schema},
\Fixed{where the red dashed lines show the paths enabled by the
  complementary CFHT observations.}

Since the publication of the SNLS calibration in R09, \Fixed{we use
  additional calibration data including a factor of two more data to
  map the photometric response and data to calibrate the replacement
  $i$ band filter that was installed in 2007.}
In addition, the consistency \Fixed{checks} introduced in this work
\Fixed{showed subtle problems} in the flat-fielding of both surveys
that required corrections.  As a consequence, this paper also
describes a reanalysis of the original calibration data for the two
surveys.

The paper is organized as follows. An overview of the subject is given
in section~\ref{sec:rationale}. The determination of the MegaCam
flat-fielding solution, the \Fixed{determination} of a consistent set
of tertiary standards in the SNLS fields, and the calibration of the
SNLS are detailed in sections~\ref{sec:instrument-model},
\ref{sec:tert-catal-constr} and \ref{sec:megac-absol-calibr}.
Section~\ref{sec:SDSS} and \ref{sec:sdss-pt-calibration}
\Fixed{describe} a redetermination of the photometric calibration of
the SDSS supernova survey. Section~\ref{sec:sdsssnls-direct-inte}
describes the analysis of the dedicated cross-calibration data that
anchor the two surveys \Fixed{to} the same flux scale.  We gather all
the relevant data to provide a combined calibration solution to both
\Fixed{surveys} in section~\ref{sec:concl}.  \Fixed{Finally, in
  section~\ref{sec:calibration-products}} we summarize the properties
of the calibration products and discuss the consistency of the
available calibration data.  We conclude by identifying survey and
detector design limitations and present perspectives for future
improvements.

\section{Overview}
\label{sec:rationale}

\subsection{Definitions}
\label{sec:definitions}

The purpose of photometric calibration is to relate the
\Fixed{instrumental fluxes} measured in the image pixels to the
physical fluxes of the observed objects.  \Fixed{Instrumental
  magnitudes are defined as $\madu = -2.5\logdec \madu[\phi]$.}
\Fixed{The calibration} is commonly divided in two steps. 
The first step consists in the compensation of the variation of the
effective instrument throughput (in space and time) to deliver 
a homogeneous and consistent set of measurements. The second step
consists in the delivery of a model relating the homogenized
quantities to the physical flux of the astrophysical objects.

The results \Fixed{of these two steps are}
calibrated broadband magnitudes  \Fixed{($m$)}
of objects in a photometric system whose interpretation in 
\Fixed{terms} of physical fluxes is given by
\begin{equation}
  \label{eq:}
  m = -2.5\logdec\frac{\int_\lambda \lambda T(\lambda) S(\lambda) d\lambda}{\int_\lambda \lambda T(\lambda) S_\text{ref}(\lambda) d\lambda} ~~.
\end{equation}
$S(\lambda)$ is the spectral energy density (SED) of the object above
the earth's atmosphere in units of ${\rm erg s^{-1} cm^{-2}
  {\angstrom}^{-1}}$, $T(\lambda)$ is the overall instrument
throughput (including the atmosphere) \Fixed{that defines the
  \emph{natural} photometric system}, and $S_\text{ref}$ is a
reference flux scale to which the object flux is compared.
\Fixed{$T(\lambda)$} is expressed as the detector response to incident
photons \Fixed{and its normalization is irrelevant because it cancels
  in the ratio.}  The factor $\lambda$ is required to convert the SED
to a function proportional to the number of photons per unit
wavelength.

\Fixed{In what follows we refer to the AB magnitude system as defined in
\citet{fukugita_sloan_1996}, unless otherwise specified. Accordingly,
$S_\text{ref} = S_\ab$ where $S_\ab$ is:
\begin{equation}
S_\ab(\lambda)= 10^{-48.6/2.5} c \lambda^{-2}{\rm erg \cdot s^{-1}
  \cdot cm^{-2} \cdot {\angstrom}^{-1}}\label{eq:2}
\end{equation}
with the wavelength $\lambda$ expressed in angstrom and $c$ the speed
of light expressed in~$\rm \angstrom \cdot s^{-1}$.}

The model \Fixed{describing} the physical interpretation of the
instrumental fluxes has two components. The first is the determination
of the effective passband corresponding to the instrument measurements
up to a normalization factor. The second is the determination of the
flux scale of the homogeneous measurements in each passband. The
latter is obtained from \Fixed{a} comparison to a flux reference,
\Fixed{which we choose to be} the HST-CALSPEC stellar library
(see the CALSPEC
website\footnote{\url{http://www.stsci.edu/hst/observatory/cdbs/calspec.html}}
and references therein).

Our primary goal is the calibration of the SNLS and the SDSS supernova
surveys. The main output of this work consists \Fixed{of} calibrated AB
natural magnitudes for large sets of stars selected in the science
fields. 
\Fixed{These} stars constitute in-situ photometric references directly
usable to calibrate supernova flux measurements. They will be referred
to as the tertiary standards, while in the same terminology, the
spectrophotometric standards established by the HST constitute the
primary standards, and stars used in the calibration transfer from the
HST to the science fields are referred to as secondary standards.

We also deliver an estimate of uncertainties related to this
calibration that can be propagated to the cosmological
result. Accurately measuring the flux ratios between different
photometric bands is of paramount importance for cosmological
constraints, as this enables the comparison of supernova luminosity at
different redshifts.  \Fixed{Since} cosmological constraints are
insensitive to a global calibration offset on the flux scale, we
concentrate on the accuracy of the relative calibration between the
different photometric \Fixed{bands in} the two surveys.  We do not
\Fixed{include} the uncertainty \Fixed{on} the global flux scale of
the calibration reference (see \emph{eg.}
\citealt{bohlin_hubble_2004}).

\subsection{Description of Instruments and Surveys}
\label{sec:survey-description}

The SNLS covers the four low extinction fields of the CFHT Legacy
Survey \emph{Deep} component (hereafter D1 to D4). They were
repeatedly imaged in the 4 photometric bands
\band{g}\band{r}\band{i}\band{z} every 3--4 nights during dark
astronomical time\footnote{MegaPrime shares the rest of the lunation
  with other instruments. A ``run'' covers at least a 14 days period
  around the new Moon.}. Photometry in the \band u band is also
available for those fields but was not part of the supernova survey.
The original \band{i} filter was broken in July 2007 and replaced by a
slightly different \Fixed{$i_M$} filter in October of the same year.
We will refer to the photometric band defined by the replacement
filter as \band{i2}. The MegaPrime instrument is mounted on the CFHT
prime focus and hosts the MegaCam camera, a mosaic of 36 e2v
$2048\times4612$ CCDs, that covers a field of $0.96\times0.94$ deg$^2$
\citep{MegacamPaper}. Raw MegaCam images are processed by the Elixir
pipeline \citep{2004PASP..116..449M} that handles bias subtraction,
flat-fielding and defringing in \band i and \band z bands.

The SDSS-II Supernova Survey primary instrument is the SDSS CCD camera
\citep{1998AJ....116.3040G} mounted on a dedicated 2.5-m telescope
\citep{2006AJ....131.2332G} at Apache Point Observatory (APO), New
Mexico. The focal plane hosts a CCD array organized as six columns of
5 CCDs. Each CCD of the column \Fixed{records} images in one of five
broad optical bands: $ugriz$ \citep{fukugita_sloan_1996}. The camera
was used in time-delay-and-integrate (or drift scan) mode, which
provides efficient sky coverage.  The Supernova Survey scanned at the
normal (sidereal) SDSS survey rate, which yielded 55 second integrated
exposures in each passband. The survey covers a 300 square-degree
region ($2.5\degree$ wide over $8$ hours in right ascension) that was
repeatedly observed (every fourth day in average) over the course of
three three-month seasons.  The region is centered on the celestial
equator and \Fixed{is} referred to as ``Stripe 82''.

Many aspects of the analysis are complicated by two instrumental
effects: spatial variation of the instrument passband
\Fixed{response}, and stray light \Fixed{which adversely affects}
common flat-fielding methods.  \Fixed{Below we describe these two
  issues.}

\subsubsection{Varying passbands}
\label{sec:effective-passbands}
It was realized in R09 that \Fixed{spatial variations of the passband
  response resulted in} non-negligible color terms \Fixed{between}
photometric measurements obtained at different positions on the
MegaCam focal plane. The size of the color terms between the center
and the edge of the focal plane is about 0.02 (depending on the band)
for main sequence stars indexed by their $g-i$ color (\emph{cf.}
Sect.~\ref{sec:instrument-model} and Fig.~\ref{fig:synthdk}).  The
major contribution to this variation is related to the manufacturing
process of the large MegaCam interference filters
($30\times30$~cm$^2$). The variation was found to follow an almost
perfect radial pattern, \Fixed{with the} filters being redder at the
center of the focal plane than at their edge. The typical
\Fixed{variation} of the filter \Fixed{transmission's} mean wavelength
amounts to a few nanometers.  Other expected contributors to
variations of the effective passband are differences of quantum
efficiency between CCDs (they were found small enough to be
neglected), and variations in time of the spectral shape of the
atmospheric extinction (that average out over multi-epoch measurements
and will not be considered).

The measurement of SDSS effective passbands are described in
\citet{2010arXiv1002.3701D}. Small differences (at most 2~nm
variations of the effective wavelength against the average
transmission for $ugri$, 4~nm in $z$ band) were found between the
individual filters mounted on the different columns of the
camera. The SDSS calibration strategy is such that this issue
can generally be neglected in the calibration transfer.

Varying passbands bring two complications to the analysis of
photometric measurements. First, the flux interpretation of a
broadband magnitude (Eq.~\ref{eq:}) depends on the position of the
measurement on the focal plane of the instrument.  \Fixed{Defining
  \spos as the focal plane position and \cmag[T] as the passpand
  response at \spos,} magnitudes in the natural (and position
dependent) photometric system \Fixed{are denoted by}
\begin{equation}
   \label{eq:24}
   \cmag = -2.5\logdec\frac{\int_\lambda \lambda \cmag[T](\lambda) S(\lambda) d\lambda}{\int_\lambda \lambda \cmag[T](\lambda) S_\ab(\lambda) d\lambda}\,.
\end{equation}

\Fixed{
Second, the variation of response depends on the object SED
and therefore it is not possible to ensure the uniformity of 
the instrument response for objects with an arbitrary SED. }
There is thus a degree of freedom in the choice of the spectrum 
for which the response of the instrument is made uniform 
(refer to Sect.~~\ref{sec:phot-flat-field} for further discussion).

\subsubsection{Stray light in the wide field corrector}
\label{sec:flat-fielding}

In addition to the problem of the passbands varying with the position
on the focal plane, there are other position dependent optical effects
that complicate the determination of the instrument photometric
response. Parasitic reflections between filters and optical elements
of the wide-field corrector are the most \Fixed{problematic} because
they cannot be modeled and subtracted easily.  \Fixed{These effects}
produce stray light that pollutes the images obtained by observation
of uniform illumination sources such as twilight sky, dome screen, or
night sky background.  \Fixed{These} techniques thus cannot be used
directly to produce the instrument flat-fields.\footnote{Another
  effect that must be accounted for when computing the photometric
  response from images of uniform illumination is the distortion of
  the plate-scale (see Fig.~4 in R09).  \Fixed{This distortion can} be
  easily computed from the astrometry.  Other effects can also arise
  from the variation of the PSF across the focal plane. When not
  properly accounted for by the photometry method, the effective
  fraction of the flux captured by the photometry can vary. This makes
  the effective throughput spatially variable in a way that depends on
  the photometry method.}

\subsection{Current state of the SNLS and SDSS calibration}

\subsubsection{Absolute calibration transfer}
\label{sec:calibration-transfer}

The original calibration strategies for both the SDSS
\citep{2002AJ....123..485S} and the SNLS
\citep{astier_supernova_2006,R09} photometric surveys follow roughly
the same scheme. As neither survey 
\Fixed{telescope}
could directly observe the primary
spectrophotometric standard available at that time, they relied on
third party observations, to which they anchored via a set of
secondary standard stars.

In the SNLS, the secondary standard stars \Fixed{are those established
  in} \cite{landolt_ubvri_1992}. The SDSS secondary star network was
built using a dedicated instrument called the photometric telescope
(PT) \citep{2006AN....327..821T}. The PT was used to determine a
nightly photometric solution and to deliver calibrated secondary
standards.  These secondary standards are distributed in square
patches of \Fixed{0.36~sq-deg throughout} the SDSS survey area.  The
\Fixed{supernova survey area (Stripe 82)} is covered \Fixed{with} an
average of one \Fixed{PT-calibrated patch} every 3 degrees in right
ascension.

These two-step calibration strategies have limitations.  First, they
accumulate the photometric and systematic uncertainties of two
measurements.  Second, \Fixed{and more importantly,} the secondary
standard stars provide only broadband photometric references rather
than spectrophotometric \Fixed{references}.  \Fixed{Since} the
secondary \Fixed{star} photometric system is different from
\Fixed{that of} the science system \Fixed{(i.e, SNe~Ia)}, this
\Fixed{difference} ultimately limits the precision of the absolute
calibration to the accuracy of photometric color transformations
between those two systems (\emph{i.e.} to some kind of interpolation
between quantities integrated over different passbands).

The passbands in the Landolt $UBVRI$ system differ from the MegaCam
\band{u}\band{g}\band{r}\band{i}\band{z}, and the transmission
\Fixed{functions of the Landolt} filters actually used are not
\Fixed{precisely} known.  The \Fixed{resulting} color transformations
\Fixed{are difficult to model and} have significant uncertainties,
ranging between 0.3 to 2\% depending on the band.  These uncertainties
were the limiting factor in the previous release of the SNLS
calibration.

The strategy of the SDSS may appear more favorable as the PT was
specifically designed to use a filter set identical to the 2.5~m
science telescope. However, the filter sets differ significantly
primarily because of the differences in the environment (vacuum
vs. air) as explained in \citet{2010arXiv1002.3701D}.  In addition,
repeated, \textit{in situ} measurements enabled
\citet{2010arXiv1002.3701D} to detect a significant change in the
$u$-band filter \Fixed{response} after the 2001 measurement.  The PT
filter responses, however, were not measured as frequently or as
carefully, and we are again forced to rely on empirically determined
color transformations to tie the two systems together.  The problem
\Fixed{with} the determination of the color transformations is further
complicated by errors in the PT flat-fielding as discussed in section~\ref{sec:SDSS}.

\subsubsection{Survey uniformity}
\label{sec:survey-uniformity-2}

The characterization of the MegaPrime response is based on twilight
flat-fields, corrected, at large scale, for the optical effects
induced by stray light and plate-scale distortion.  This photometric
correction is \emph{measured} on dithered observations of dense
stellar fields following the scheme presented in R09.
\Fixed{Although} this measurement \Fixed{does not depend on}
assumptions about the telescope optics, we found that it is sensitive
to variations \Fixed{in} the atmospheric conditions during the
dithered observation sequences.

The determination of the instrument response presented in R09 covers
the period 2003-2006, and \Fixed{a} significant evolution of the
response \Fixed{is} attributable \Fixed{to changes in the telescope
  optical setup.}  An extension of this \Fixed{R09} study \Fixed{is}
thus required for the analysis of observations \Fixed{after} 2006, in
particular observations made with the \band {i2} replacement
filter. In addition, \Fixed{since} the determination of the
photometric corrections are affected by random variations \Fixed{in}
the atmospheric conditions, \Fixed{we expect some improvement from
  including new independent observations.}

SDSS imaging data \Fixed{was} obtained using drift scanning: each
point on the sky \Fixed{was} sampled by each CCD row so that the
effective response is averaged over all rows.  The flat-field is
therefore represented by a 1-dimensional array of values holding the
relative average response between pixel columns. Such solutions were
determined by different techniques in \citet{ivezi_sloan_2007} and
\citet{2008ApJ...674.1217P}.  The first \Fixed{technique} relies on a
combination of stellar locus analysis and comparison of the Sloan
photometry to PT secondary patches. The second \Fixed{technique}
solves for the flat-field vector by cross-correlating normal
observations, \Fixed{scanned in the right ascension direction,} with
\Fixed{infrequent} observations that scanned the sky
\Fixed{approximately along declination.}

The flat-fielding of the supernova survey currently relies on the
PT-based solution, while the cross-scan based solution is applied to
the main survey since the eighth data release (DR8). Small but
significant discrepancies between the two solutions can be found when
comparing the photometry of stars in the Stripe 82 (see
Sect.~\ref{sec:SDSS}).

\subsection{Project for this work}
\label{sec:project}

In this paper, we review, control and improve the original calibration
paths of both surveys, and we provide an important cross-check by
directly inter-calibrating the two surveys. We also describe direct
observations of \Fixed{suitably} faint primary standard stars (now
available in the CALSPEC database) with MegaCam on the CFHT, providing
an alternative path to anchor the SNLS to the HST scale.

The various calibration paths exploited in this paper are summarized
\Fixed{in} the schematic view presented in Fig.~\ref{fig:schema}.  The
two new MegaCam calibration and inter-calibration measurements (shown
by red-dashed lines \Fixed{in Fig.~\ref{fig:schema}}) \Fixed{provide}
a redundancy in the calibration of both surveys to the HST standards.
The original \Fixed{``Landolt'' and ``PT''} calibration paths involve
very large numbers of measurements accumulated over the course of the
surveys.  \Fixed{These calibrations} are essentially free from
statistical uncertainty, but, as already discussed, they are limited
\Fixed{by} the accuracy of color transformations.  On the contrary,
the direct path provides direct observation of spectrophotometric
standards in the MegaCam photometric system and \Fixed{eliminates}
this dominant systematic for SNLS.  \Fixed{The direct path, however,
  involves} a smaller number of observations, potentially affected by
systematics related to the difficulty of observing bright stars.

A general agreement between all the available measurements conducted
with different instruments and subject to different sources of
uncertainty would strongly support the conclusion that systematics are
controlled and correctly estimated.  In addition, the data gathered
for the direct cross-calibration of the SDSS and the SNLS provide
high-quality photometry in both systems for a large number of
stars. Those measurements sample a wide range of stellar types and a
large part of both focal planes. Such data can deliver two important
by-products. First, a consistency assessment of the uniformity and
quality of the photometry in both surveys. Second, a precise
determination of the color transformations between the two
systems. The latter can be compared with expectations to provide a
sensitive check of the \Fixed{passband transmissions.}

This work \Fixed{is presented} as follows. A new determination of the
MegaCam photometric response extending over the \Fixed{entire} survey
is described in Sect.~\ref{sec:instrument-model}.  It is used to build
an average catalog of photometric tertiary standards in
Sect.~\ref{sec:tert-catal-constr}. On the SDSS side, the question of
the discrepancy between the two available flat-field solutions is
investigated in Sect.~\ref{sec:SDSS}.  \Fixed{Identifying evidence}
\Fixed{of} an error in the nominal PT flat-field (appendix
\ref{sec:more-megac-flatf}), we opt for using the cross-scan solution
which is independent of the PT.  The stripe 82 tertiary catalog is
corrected accordingly.

Validation of the uniformity of the resulting tertiary catalogs is
given by the analysis of the cross-calibration data
(Sect.~\ref{sec:sdsssnls-direct-inte}, see in particular
Fig.~\ref{fig:uniformity} page \pageref{fig:uniformity}). Constraints
binding together the flux scales of both tertiary catalogs are also
derived in Sect.~\ref{sec:sdsssnls-direct-inte}.

Measurements anchoring tertiary catalogs to the HST primary standards
are described in Sect.~\ref{sec:absolute} for the path \Fixed{using}
Landolt secondary standards, in Sect.~\ref{sec:direct-hst-standard}
for the direct measurements of HST standards with MegaCam, \Fixed{and
  in} Sect.~\ref{sec:sdss-pt-calibration} for the path \Fixed{using}
the SDSS PT secondary patches.  The combination of all available data
to provide a common and improved calibration of the SDSS and SNLS
\Fixed{surveys} as well as the full estimate of the associated
uncertainties is described in Sect.~\ref{sec:concl}.  The reader
\Fixed{can skip} the details of the analysis \Fixed{and} jump directly
to Sect.~\ref{sec:calibration-products} where we discuss the
consistency between the different calibration paths (see in particular
Fig.~\ref{fig:calibcomp}) and describe the calibration products.

\section{MegaCam/MegaPrime instrument model}
\label{sec:instrument-model}

This section is dedicated to the determination of the two key
ingredients characterizing the effective model of MegaPrime
measurements: the effective \Fixed{transmission} $\cmag[T](\lambda)$,
and the photometric flat-field solution $F(\spos)$ corresponding to
our aperture photometry method.  While \Fixed{these} two subjects are
usually treated separately, they \Fixed{are correlated} in MegaPrime
due to the continuous variation of \Fixed{the} filter
\Fixed{transmission} with position.

\begin{figure}
  \centering
  \begin{tikzpicture}
  [align=center,meas/.style={rectangle,double,draw=black},emeas/.style={rectangle,double,draw=black,dashed},prod/.style={ellipse,draw=black,thick},auto,node distance=4mm]
\node  [meas] (ccd) {CCD quantum\\efficiency};
\node  [emeas] (atm) [left=of ccd] {Atmospheric transmission\\above Mauna Kea\\\citep{Buton2012}};
\node  [meas] (filt) [right=of ccd] {Filter\\transmission};
\node [prod] (passbands) [below=of ccd]  {Passbands\\model\\\emph{Sect.~\ref{sec:instrument-bandpass}}};
\node  [meas] (opt) [right=of passbands] {Optics \& mirror\\transmission};
\node [prod] (dk) [below=of passbands]  {Color\\transformations\\\emph{Sect.~\ref{sec:transl-meas}}};
\node [prod] (dzp) [below=of dk]  {Photometric\\response map\\\emph{Sect.~\ref{sec:phot-flat-field}}};
\node  [meas] (twi) [left=of dk] {Twilight\\observations};
\node  [meas] (grid) [right=of dk] {Grid\\observations};
\draw [->] (ccd) -- (passbands);
\draw [->] (atm) -- (passbands);
\draw [->] (filt) -- (passbands);
\draw [->] (opt) -- (passbands);
\draw [->] (passbands) -- (dk);
\draw [->] (dk) -- (dzp);
\draw [->] (grid) -- (dzp);
\draw [->] (twi) -- (dzp);
\end{tikzpicture}
\caption{Overview of the MegaPrime photometric response
  determination. Boxes figure the various data sets involved in the
  construction of the model (dashed for external data). Ellipses
  figure the main steps of the analysis.}
  \label{fig:overview}
\end{figure}
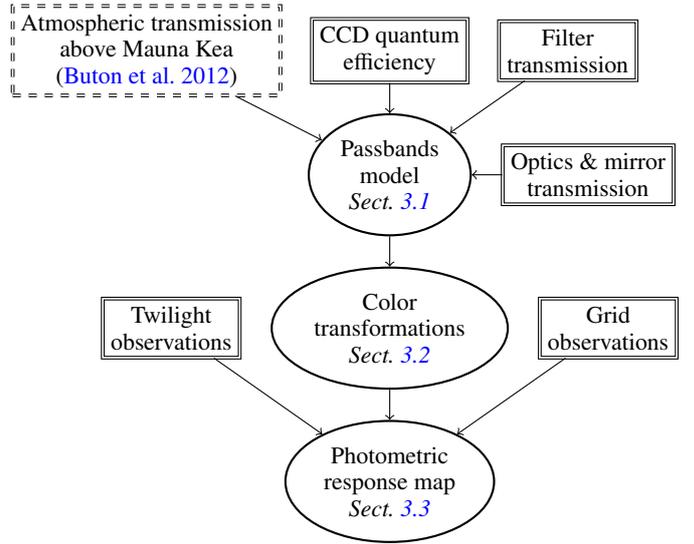

Figure~\ref{fig:overview} \Fixed{shows} an overview of the data
samples involved in the \Fixed{determination} of the instrument model,
and \Fixed{highlights} the main steps of the analysis.  In
section~\ref{sec:instrument-bandpass}, we gather independent
characterizations of the optical components to build a model of the
instrument passband \Fixed{transmissions}.  The resulting
\Fixed{transmission functions} $\cmag[T](\lambda)$ vary continuously
with the distance \Fixed{from} the focal plane center $\spos$
(hereafter \Fixed{referred to as} the radius).  In
section~\ref{sec:transl-meas} we use synthetic photometry to compute
the small color transformations relating stellar observations obtained
at different radii. We compute photometric corrections to the
instrument response maps delivered by the usual twilight observations
in section~\ref{sec:phot-flat-field}. Those corrections are obtained
from dithered observations of dense stellar fields under photometric
conditions. We use the color transformations from
Sect.~\ref{sec:transl-meas} to make sure that the resulting
photometric response maps $F(\spos)$ make the instrument response
constant for an (hypothetical) object whose SED corresponds to the AB
spectrum $S_\ab$.  \Fixed{With this choice of reference SED,} the
relation between the instrumental and AB magnitudes \Fixed{is} a
single calibration constant (the AB zero point) \Fixed{that is} valid
\Fixed{across} the entire focal plane.  The uncertainties on the
results are discussed in section~\ref{sec:instr-model-assessm}.

A comprehensive model of the MegaCam effective passband
\Fixed{transmissions} has already been released in R09. The main
modification since then is the replacement, in October 2007, of the
interference filter defining band \band{i} after its accidental
destruction on July $20^{\rm th}$ 2007.  This reanalysis was triggered
by this incident and aims mainly at providing calibration data for the
slightly different replacement filter (hereafter \band{i2}).  We also
propagate \Fixed{improvements} to other bands, mainly the benefits of
an extended data set and better understanding of the instrument
model. A summary of main differences is provided at the end of this
section.

\subsection{Instrument passbands}
\label{sec:instrument-bandpass}

\subsubsection{Passband model}
\label{sec:generic-model}

The effective transmission model for photometric band $b$ results from
the product of 5 components:
\begin{equation}
  \label{eq:11}
  \cmag[T]^b(\lambda) = T_b^f(\lambda, \x) T_c(\lambda) R_m(\lambda) T_a(\lambda) \epsilon_\text{ccd}(\lambda)\,,
\end{equation}
where $T_b^f$ is the position dependent transmission of the
interference filter, $T_c$ the transmission of the wide field
corrector, $R_m$ the primary mirror reflectivity, $T_a$ the average
atmospheric transmission at Mauna Kea (at a \Fixed{typical} airmass of
1.25) as measured by the SNFactory \citep{Buton2012} and
$\epsilon_\text{ccd}$ the measured quantum efficiency (QE) of the
MegaCam e2v CCDs. Note that $T_a$ differs quite significantly from the
preliminary version used in R09.

The measurements of quantum efficiency for individual CCDs display
slight variations between chips. They do not introduce noticeable
modification of the effective passband \Fixed{transmission} in any
photometric band \Fixed{except for} \band{g} and \band{u}.  In
\Fixed{these} two bands, we measure a maximum shift of mean wavelength
between the reddest and the bluest CCD of 1.5~nm and 2.3~nm,
respectively. The use of per chip QE curves in place of the average
one published in R09 has a negligible impact on the calibration. In
addition, they marginally improve the agreement between synthetic and
measured color terms between chips, which may indicate uncorrected
weaknesses in the individual measurements.  \Fixed{Since the per chip
  QE curves do not improve the calibration, they are not used} in this
release of the calibration and \Fixed{instead we use} the mean QE
model.\footnote{\url{http://www.cfht.hawaii.edu/Instruments/Imaging/MegaPrime/data.MegaPrime}}

\subsubsection{Laboratory measurements of filter transmission}
\label{sec:filter-transmission}

The transmissions of \Fixed{our} interference filters under normal
incidence were measured by their manufacturers at several positions.
Measurements for the original filter set, manufactured by SAGEM/REOSC,
are published in R09.  Measurements are available for each filter at
several positions: at the center of the filter, and at 23, 47, 70, 93,
117, 140, 163, 186 and 210~mm away from the center. Transmission
curves are linearly interpolated between measurements as a function of
radius to provide the continuous model \Fixed{function,}
$T_b^f(\lambda,\x)$ .

For the \band{i2} filter, measurements of the transmission provided by
the manufacturer, BARR associates, are available at \Fixed{the} center
of the filter and \Fixed{at} 20, 40, 60, 80, 100, 120, 140, 160,
180~mm away from the center.  The potential existence of leaks outside
the main passband is not excluded by these measurements that cover
only the expected wavelength range of the main passband. The 180~mm
measurements were repeated for each corner of the filter. The mean
wavelength varies along the radius from $761.2$~nm on one side of the
filter to $764.9$~nm close to the center, which makes it more uniform
than the original filter. There are indications that the filter
transmission \Fixed{function} may not follow a radial
pattern. Measurements at the 4 corners display small discrepancies,
with variations between the two extreme sides of the filter reaching
$1.7$~nm. However, without complementary measurements at intermediate
radii, it was not possible to build an alternative to a radially
variable filter model. Deviations from the radial symmetry are barely
noticeable in the observational data (\emph{cf.}
Sect.~\ref{sec:instr-model-assessm}).

The normal incidence measurements \Fixed{are} blue-shifted to account
for the average incident angle of the CFHT f/4 beam. As in R09, this
\Fixed{shift} is done using the approximate relation\Fixed{,}
\begin{equation}
  \label{eq:13}
  T_f(\lambda) \approx T_f^0\left(\frac{\lambda}{\sqrt{1-(\sin^2\theta)/n^2}}\right)\,,
\end{equation}
where $\theta$ is the incidence angle, and $n$ is an effective
\Fixed{index of refraction} for the filter.

The transmission curves of the original filters were measured again at
CFHT in 2006 at 5 angles ranging from 0 to 8$^{\circ}$ at a few
locations. This allowed us to check that the above formula applies and
\Fixed{to} determine the refractive index $n$ for each filter. 
We then integrated the transmission \Fixed{function}
over the telescope beam for each location in the focal plane.

As the replacement filter $\band{i2}$ transmission was never measured
under non-normal incidence, the exact value of its refractive index
$n$ cannot be directly determined. We used the value measured on the
original $i_M$ filter, $n=1.6$. The validity of this choice is
indirectly assessed in Sect.~\ref{sec:passband-consistency}.

Last, the comparison of the manufacturer transmission curves with the
2006 measurement at CFHT for the original filter set is provided in
\Fixed{A}ppendix~\ref{sec:megacam-band-i}. It revealed an important
discrepancy (an $8$~nm shift of the red cut-off) in \band i and
\band{r}, while \Fixed{measurements in} the other bands were found in
nearly perfect agreement.  As discussed in
Appendix~\ref{sec:megacam-band-i}, we believe that the CFHT
measurements provide a better description of the survey filters and we
\Fixed{therefore} corrected the \band i and \band r transmission
curves accordingly. The validity of this correction could not be
directly tested on the whole spatial extension of the filters.
\Fixed{An} indirect assessment using stellar locus is provided in
Sect.~\ref{sec:adeq-passb-model} and \ref{sec:passband-consistency}.

The resulting transmission curves are given in
table~\ref{tab:u_band_table}--\ref{tab:z_band_table} at several radii
away from the center. A sufficiently accurate mapping between object
coordinates and position on the filter is obtained by considering the
filter close to the focal plane and using the focal distance of the
instrument, $F=14.89$~m.

\subsection{\Fixed{Stellar measurements across the focal plane}}

While slightly variable, the MegaCam natural photometric system
remains sufficiently similar between different \Fixed{focal plane}
positions that precise transformations between stellar measurements
can be derived for a large range of the stellar population.  Passband
\Fixed{transmission} models have been used to compute synthetic values
of the color transformations between the natural system at \Fixed{a
  given radius} and the uniform system defined as the natural system
at the center of the focal plane (denoted $\spos_0$). The relations
take the form
\begin{equation}
  \label{eq:4}
  \umag \approx \cmag - \dk(\x) (\umag[c]-c^0(x))
\end{equation}
where $\umag[c]$ is the color index of the star, $c^0$ \Fixed{is} the
reference color for the linear transformation, and $\dk$ \Fixed{is}
the color term adjusted to a given stellar population.\footnote{Shall
  the selected standard $S_\text{ref}$ be the spectrum of a common
  star, one would expect by construction that $\umag[c^0]\approx
  0$. One may notice it is not strictly the case neither for a Vega or
  an AB system.}  \Fixed{The choice of $\umag[c]$ is arbitrary, and we
  choose $\umag[(g-i)]$ because it is well measured for most stars and
  it is a} good proxy of the star temperature.

The primary use of these transformations lies in the process of
determining the photometric flat-field solution (\emph{cf.}
Sect.~\ref{sec:phot-flat-field}) for MegaCam, using measurements
calibrated on Landolt stars. Since this calibration procedure delivers
magnitudes in an approximate Vega system, we determine transformations
in that system \Fixed{as described} in R09, and \Fixed{we make} use of
the primary spectrophotometric standard \bdtruc as a flux scale
reference.  More specifically, in this system the \Fixed{synthetic}
magnitudes are \Fixed{defined as}
\begin{equation}
  \cmag^V = -2.5\logdec\frac{\int_\lambda \lambda \cmag[T](\lambda) S(\lambda) d\lambda}{\int_\lambda \lambda \cmag[T](\lambda) S^V_\text{ref}(\lambda) d\lambda}\,,\label{eq:28}
\end{equation}
where \[S_\text{ref}^V(\lambda) = S_\bdref (\lambda) 10^{0.4
  m_\bdref}\,,\] with $S_\bdref$ the SED of the star
\bdtruc\footnote{The version used in this paper can be found at
  \url{ftp://ftp.stsci.edu/cdbs/current_calspec/bd_17d4708_stisnic_003.ascii}}
and $m_\bdref$ the MegaCam magnitude of this star inferred from
Landolt observations (see Sect.~\ref{sec:absolute}). As those scaling
constants ($m_\bdref$) are close to the Vega magnitudes of \bdtruc, this
system is by construction an approximate Vega system. To avoid
confusion with the AB magnitudes that we use everywhere else in this
paper, magnitudes in this system will be, when needed, distinguished
as $\cmag^V$ where the $V$ superscript stands for Vega.

\begin{figure}
  \centering
  \igraph{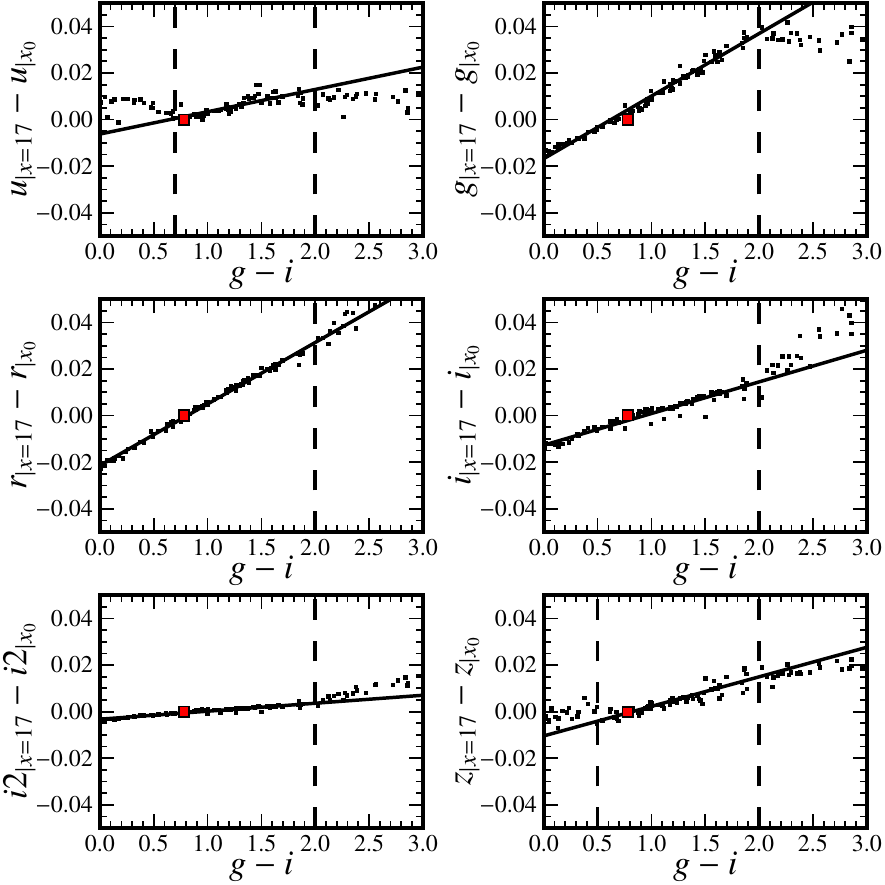}
  \caption{Synthetic color transformation between the center and the
    corner of the filters (17cm from the center) in each band. Dots
    are synthetic magnitudes computed for stellar spectra from the
    \citet{1983ApJS...52..121G} spectral library. Linear color
    transformation (solid line) are adjusted in the range delimited by
    vertical dashed lines. The color index is a Vega MegaCam
    color. The red square figures the synthesized transformation of
    the spectrophotometric standard star \bdtruc computed from its STIS
    spectrum available from the CALSPEC database. }\label{fig:synthdk}
\end{figure}

\Fixed{The relations defined by Eq.~(\ref{eq:4})} were adjusted in a
limited color range on synthetic magnitudes computed from the
\citet{1998PASP..110..863P} and \citet{1983ApJS...52..121G} stellar
libraries. As an illustration, the adjustment of synthetic
transformations between the center and the edge of the focal plane are
shown on Fig.~\ref{fig:synthdk}. The selected color range corresponds
to a region where the linear approximation is \Fixed{sufficiently}
accurate.

The uncertainty on the slope of the transformation, evaluated from the
difference obtained between the two libraries, is smaller than 2~mmag
per color unit. The dispersion around the fitted transformation is
typically smaller than 3~mmag.

We also evaluated the sensitivity of the fitted transformation to
extinction by interstellar dust. Spectra have been reddened according
to a \citet{1989ApJ...345..245C} law to account for the mean
extinction of the stellar population. The extinction vector is nearly
co-linear to the stellar locus in the selected color-range. For a
value of $E(B-V)=0.4$, the modification of the slope is typically
negligible in all bands \Fixed{except} $z$, where it reaches $3$~mmag
per color unit. We conclude that, in fields with relatively low
extinction, it is safe to translate measurements of stars lying in the
selected color range across the focal plane using the fitted
transformation.\footnote{ The parameters of the linear color
  transformation \Fixed{and their dependence on the focal plane
    position are available} in electronic form at
  \url{http://supernovae.in2p3.fr/snls_sdss/}.}

\label{sec:transl-meas}

\subsection{Determination of the photometric flat-field}
\label{sec:phot-flat-field}

A common way to measure the instrumental response is the observation
of the twilight sky, assumed to be a uniform source of light. A first
complication preventing the direct use of this technique is the
variation of filter passband \Fixed{transmission across the Megacam
  focal plane.}  This \Fixed{effect} makes the response variation
slightly dependent on the SED of the observed objects.  Consequently,
the strict uniformity of the response can only be ensured for a single
class of objects sharing the same SED shape. Our first concern here is
that the twilight images are functions of the twilight color which is
not a \Fixed{useful} reference for stellar photometry.

Putting aside this complication, the twilight images do not deliver
satisfying results \Fixed{for the photometric} uniformity of
point-like objects, \Fixed{and is} due to differences in the optical
response to isotropic and parallel illuminations (see
e.g. \citealt{1995A&AS..113..587M}).  In MegaCam, two major
contributions are identified \citep{2004PASP..116..449M}: 1)
distortions of the plate-scale and 2) parasitic reflections on the
various components of the wide field corrector. The combination of
both effects leads to differences of around 10\% between stellar
fluxes measured at the edge or center of flat-fielded images.

It is, however, expected that those purely instrumental effects are
smooth and fairly stable in time, so that the determination of the
photometric flat-field can take the form of a large-scale stable
correction to the twilight flat-field. The twilight flat-field thus
corrects for small-scales and time variations of the instrument
response, while the response at large scales is to be determined by
other means.  We will refer to this correction as the photometric
flat-field correction in what follows.

Smaller deviations (around 1\%) also arise from variations of the PSF
across the field that may not be correctly handled by the photometry
method. While small, such effects make the flat-field definition
ultimately dependent on the photometry method.  We choose to measure
fluxes in twilight flat-fielded images,\footnote{The Elixir pipeline
  can apply photometric corrections to the images. When needed, we
  suppress these corrections at the catalog level.} leading to
instrumental magnitudes written $\madu^t$, where the $t$ \Fixed{superscript}
denotes the use of twilight flat-field. 
Those magnitudes \Fixed{are related} to natural magnitudes by
\begin{equation}
  \label{eq:26}  
  \madu^t = \cmag + \dzp(x) - \zp
\end{equation}
where $\dzp(x)$ is the photometric flat-field correction, and $\zp$ is
a per-exposure global calibration offset (the zero point).  Due to the
variation of passbands the correction depends on the flux reference
chosen: roughly speaking, the color of the twilight spectrum relative
to the reference spectrum.

\subsubsection{The 'grid' observations}
\label{sec:grid-observations}

A comprehensive model of this correction could in principle be
\Fixed{built} from the knowledge of optical distortions, reflections
and of the spectral shape of the twilight. The alternative
\Fixed{strategy} that we follow in this work is to measure $\dzp$ from
a set of specifically designed observations (similarly to what was
done in R09).  To \Fixed{implement this strategy,} the CFHT has
gathered sequences of 13 dithered observations of a dense stellar
field in every passband. Since the beginning of MegaCam observations,
2 stellar fields have been observed in this manner on a yearly basis
and \Fixed{also} after each major intervention on the telescope
optics.

The characteristics of those 2 dense stellar fields are given in
table~\ref{tab:gridfields}.  Some attention has been paid to
potential effects related to Galactic dust extinction in these low
galactic latitude fields. The analysis of the average color locus of
the stellar population showed that the typical value of the dust
extinction corresponds to $E(B-V) = 0.1$ in the summer field
(hereafter grid-2), and $E(B-V)=0.4$ in the winter field (hereafter
grid-1), with noticeable variations.

\Fixed{A} summary of those observations, hereafter the "grid" observations,
as well as associated telescope events is given in
table~\ref{tab:grid_obs}. The dithering pattern followed by the
observations is described in table~\ref{tab:grid_dp}. It consists
roughly \Fixed{of} 13 exposures spaced by logarithmically increasing 
\Fixed{step sizes} in right-ascension and in declination.

\begin{table}
  \centering
  \caption{Dense stellar fields observed to model the photometric flat-field correction}
  \label{tab:gridfields}
  \begin{tabular*}{\linewidth}{@{\extracolsep{\fill}}l@{\extracolsep{\fill}}c@{\extracolsep{\fill}}c@{\extracolsep{\fill}}c}
    \hline
    \hline
    Name & RA & DEC & E(B-V)\tablefootmark{a}  \\
    &&& (mag) \\
    \hline
    Grid-1 (winter field) & 6h30m00 & 14d20m00 & $ 0.45 \pm 0.08$ \\
    Grid-2 (summer field) & 21h00m00 & 10d00m00 & $ 0.12 \pm 0.03$\\
    \hline
  \end{tabular*}
  \tablefoot{\\
    \tablefoottext{a}{Average value of the dust extinction derived from the analysis of the stellar locus displacement, assuming extinction follows \citet{1989ApJ...345..245C} law with $R_v=3.1$.}
  }
\end{table}

\subsubsection{Constraining a model of  the instrumental response with grid observations \label{sec:observations-model}}

An appropriate model for the twilight flat-fielded observations
($\madu^t$) is given by Eq.~(\ref{eq:26}).  \Fixed{The} natural
magnitudes \cmag of the grid stars are not known and have to be
determined from the observations themselves, up to a global
offset. Also, due to the variation of the effective passbands, the
natural magnitude of a star is not expected to be constant across the
focal plane.  The only tractable way to obtain a closed system is to
rely on approximate color transformation. \Fixed{Unlike what has been
  done in R09, we do not fit for color transformations and flat-field
  corrections at the same time. We choose to rely on the linear
  transformation } \Fixed{from} section~\ref{sec:transl-meas}.  For
stars lying in the validity range of those transformations,
\Fixed{Eq.}~(\ref{eq:26}) translates to the linear model already
described in R09,
\begin{equation}
  \label{eq:30}
  \madu^t - \dk(\x) [\umag[g]-\umag[i] - c^0(\x)] = \umag  + \dzp(\x) - \zp\,.
\end{equation}
Free parameters in this model are the reference instrumental
magnitudes of all stars at the center of the focal plane $\umag$, the
zero points of each exposure in the sequence $\zp$, and the shape of
$\dzp(\x)$ that has to be parameterized.  We also need the reference
$\umag[g]-\umag[i]$ colors of the stars. It is however sufficient to
have this color determined only approximately, because it appears
multiplied by $\dk$ which is typically small ($\dk \lessapprox
0.02$). The construction of an approximate catalog for the grid-field
stars, neglecting in a first pass the impact of filter changes on the
flat-fielding, is the subject of section~\ref{sec:const-grid-fields}.

An irrelevant constant can be exchanged between the 3 terms of the
right-end side of Eq.~(\ref{eq:30}).  We \Fixed{therefore} assume
$\dzp=0$ at the center of the focal plane and $\zp=0$ for the first
exposure.  However, this system remains poorly \Fixed{constrained}
\Fixed{because of} a partial degeneracy between the flat-field
correction and the variation of $\zp$ between dithered exposures.  The
degeneracy issue \Fixed{making the construction of photometric
  flat-field (or ``superflats'') for a CCD camera difficult} is
investigated in \citet{1996A&AS..118..391M}.  \Fixed{A simple
  solution} is to make the assumption that the whole sequence is
photometric, \emph{i.e.} to assume that $\zp$ is constant during the
sequence.  This assumption is hard to \Fixed{satisfy} experimentally,
even for the clearest nights, mainly \Fixed{because of} variations
\Fixed{in} the PSF between exposures inducing variations \Fixed{in}
the aperture corrections.

To \Fixed{account for} part of the aperture correction variations we
\Fixed{used} aperture photometry \Fixed{with} reasonably large radii
scaled with the image quality ($IQ$).  The $IQ$ is computed as the
average sigma of a Gaussian fit to stars on each CCD.\footnote{More
  precisely: $IQ= \sqrt[4]{\det \langle M \rangle }$, where $\langle M
  \rangle$ is the average matrix of second moments for stars on the
  CCD.} We then perform the photometry in apertures of radius $7.5
\times IQ$.

Second order variations of aperture corrections, as well as variations
of the atmospheric extinction between successive exposures are to be
expected at some level so that only a part of the \Fixed{zero point
  variation is corrected}.  Still, we rely on the assumption of
photometricity during the observation sequence to alleviate the
degeneracy problem and \Fixed{we} expect that the resulting error will
average out \Fixed{with} several determinations obtained \Fixed{from}
uncorrelated sequences.  \Fixed{This} issue is further discussed in
section~\ref{sec:instr-model-assessm}.

The exact shape of the $\dzp$ correction is unknown and \Fixed{thus}
has to be parameterized.  The scheme adopted in R09 was to subdivide
the focal plane into 1296 cells of $512\times512$ pixels on which
$\dzp$ is assumed constant \Fixed{in each cell}, leading to a
pixelized representation of the correction. Here we build a nearly
equivalent model with a smaller number of parameters from the
following considerations: 1) optical effects are expected to be a
smooth continuous function of the focal plane position, 2)
discontinuities between CCDs are introduced by the IQ-scaled aperture
photometry method \Fixed{because} different aperture radii are used to
process different CCDs on the same image, 3) variations of quantum
efficiency shape from one chip to another may introduce
discontinuities between CCDs if the twilight color is
\Fixed{significantly} different from \Fixed{that of} the standard
\Fixed{reference} and 4) occasional jumps of the amplifier gains were
noted in R09, and have to be accounted for.  As a consequence we
choose to expand $\dzp$ as the sum of a smooth function decomposed on
a uniform cubic B-Spline basis plus a discrete step function $\delta
G$ accounting for differences between amplifiers:
\begin{equation}
\dzp(\x) = \sum_{i=0}^N b_i B_{i,3}(\x) + \delta G_a\,,\label{eq:smoothcor}
\end{equation}
\Fixed{where $a$ indexes the 72 amplifiers.} We impose separately
$\delta G_{26}=0$ for the amplifier $26$ that lies close to the center
of the focal plane and $\sum_{i=0}^N b_i B_{i,3}(\x_0)=0$. The choice
of $N$, which sets the number of model parameters, is discussed in
Sect.~\ref{sec:resulting-model}.

\Fixed{Measurements} are mostly affected by photon noise, which is
approximately modeled as a Gaussian process \Fixed{whose} variance
\Fixed{is} given by the flux integrated in the aperture. Only
measurements with a signal to noise ratio exceeding 10 are kept and
\Fixed{a} weighted least-square minimization is used to estimate model
parameters.  \Fixed{Although} the \Fixed{stellar} instrumental
reference magnitudes \Fixed{contribute} a very large \Fixed{number} of
nuisance parameters, the problem remains tractable if its sparsity is
correctly taken into account.\footnote{We \Fixed{use} the sparse
  Cholesky factorization routine from the CHOLMOD library
  \citep{Chen:2008:ACS:1391989.1391995} to solve for the normal
  equation.}

\subsubsection{\Fixed{Construction} of grid field catalogs}
\label{sec:const-grid-fields}

The transformation of instrumental magnitudes to the reference system
requires the approximate knowledge of the star color in the reference
system. On a first pass, a calibrated catalog of the grid fields is
built from all the gathered observations of those fields. \Fixed{These
observations comprise} a greater number of epochs than the grid dataset
alone as those fields were routinely observed to measure the PSF of
the CFHT instrument. The latter set of observations is usually
slightly deeper than the nominal grid observations.

\begin{figure}
  \centering
  \igraph{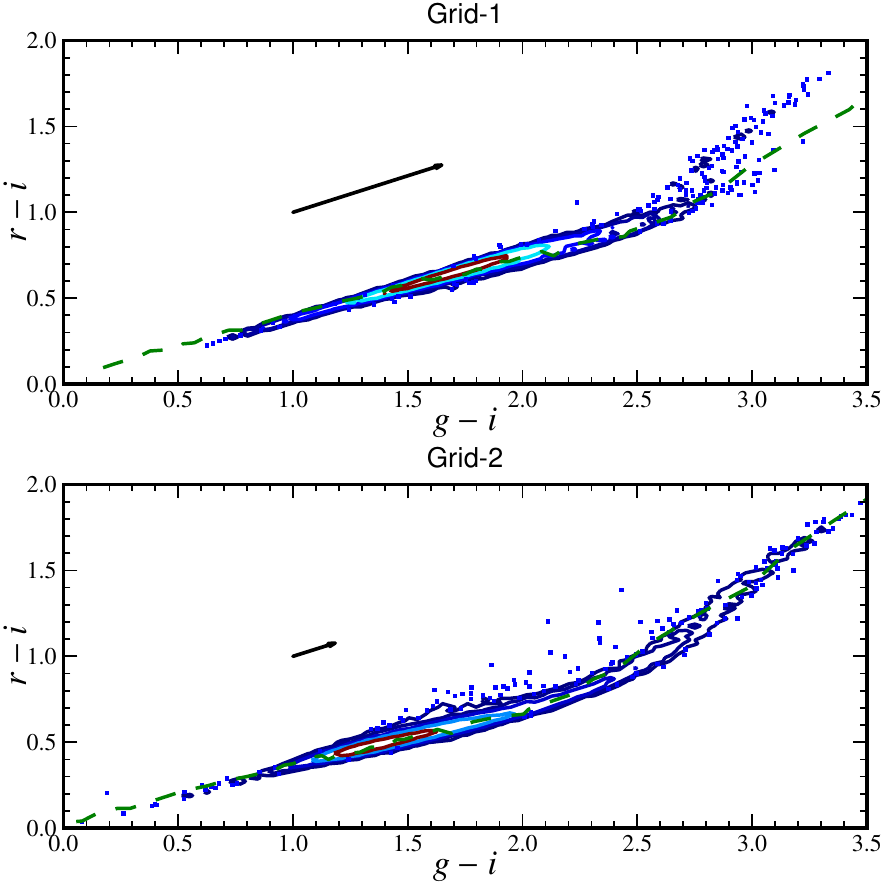}  
  \caption{Stellar population within the two grid fields. The green
    dashed line is a synthetic stellar locus computed \Fixed{from} the
    Pickles stellar library.  Spectra have been reddened to match the
    average extinction measured in the field. The black arrow
    \Fixed{shows} the corresponding extinction vector.}
  \label{fig:gridstellarpop}
\end{figure}

The flat-fielding of the observations is made using an approximate
version of the flat-field correction $\dzp^g$ ignoring the effect of
varying passbands. The simplified model \Fixed{is}
\begin{equation}
  \madu^t = \umag + \dzp^g(\x) \label{eq:14}
\end{equation}
Besides the approximate flat-fielding, the averaging procedure adopted
to build grid field catalogs is identical to the one used to establish
the tertiary star catalogs, and is extensively described in
section~\ref{sec:tert-catal-constr}. Once obtained, the average
catalogs are calibrated to the Landolt system using \Fixed{same-night}
observations of the Landolt secondary patches as described in
section~\ref{sec:absolute}.  The catalogs are cleaned \emph{a
  posteriori} to keep measurements with error below 5\% and with
reduced $\chi^2<2$ (which excludes \Fixed{some} variable stars).  The
color-color diagrams of the resulting catalogs are \Fixed{shown in}
Fig.~\ref{fig:gridstellarpop}.

Our primary interest lies in the measurement of the $\band g - \band
i$ color of the stars. This measurement is affected by the uncertainty
on the \band g and \band i calibration. From the dispersion between
the different calibration epochs, we \Fixed{estimate} that the global
uncertainty on $\band g - \band i$ \Fixed{is} below $0.008$ and
$0.004$ for the grid-1 and grid-2 catalogs, respectively. Another
error is induced by the approximate flat-field applied to the
observations and by the difference between the natural color $\cmag[g]
- \cmag[i]$ that we measure and the uniform color $\umag[g] -
\umag[i]$ that we use to index color transformations. For stars in the
color range considered the maximum difference between $\cmag[g] -
\cmag[i]$ and $\umag[g] - \umag[i]$ is expected to be 0.01.
\Fixed{Finally,} the noise on the $\band g - \band i$ color is
expected to reach at most 0.07 rms for the faintest stars in the
sample. As the \dk color term between the natural and uniform system
is at most 0.027, all error terms have a negligible impact on the
translation of star magnitudes to the reference system (at most 2~mmag
rms for the noisiest stars).  We will hence neglect the errors induced
by the color transformation in what follows.

\subsubsection{Results}
\label{sec:resulting-model}

The color transformations determined in section~\ref{sec:transl-meas}
are readily applicable to colors calibrated on the Landolt
system. Grid measurements are matched to the color catalogs described
above, and \Fixed{we use}
measurements of stars lying in the color range 
where the linear color transformation $\dk$ is valid. 
The model~(\ref{eq:30}) is then fit to the selected data.

In all bands, we have computed solutions with increasing resolution of
the spline basis. Increasing the resolution above $N=12\times12$
splines for the focal plane does not bring a significant $\chi^2$
improvement nor significant changes to the solution. We thus selected
this resolution to compute all flat-field corrections.  \Fixed{The}
systematic error on the flat-field induced by the inability of the
chosen parameterization to describe the actual shape of the
photometric correction is below 0.001~mag.

The main advantage of the parameterization adopted here is to decrease
the number of parameters needed to describe the flat-field
correction. The typical number of parameters for the selected
resolution of the spline basis \Fixed{is} $N_\text{ampli} +
N_\text{spline} = 214$, compared to 1295 in the scheme adopted in R09.
The statistical uncertainty on the reconstruction of the flat-fielding
solution is below $1$~mmag, compared to $\sim4$~mmag in the previous
R09 scheme.

We checked the sensitivity to outliers using simulations \Fixed{with}
$2\%$ \Fixed{photometric} outliers\footnote{The fraction of outliers
  introduced in the simulation is similar to what is observed in \band
  z observations (1.7\%) and significantly higher than the fraction of
  discarded measurements in the other bands.}  uniformly distributed
in the range $[-2;+2]$~mag.  \Fixed{The simulation with and without
  outliers gives the same result,} showing the efficiency of the
procedure to discard outliers.

\begin{figure*}
  \centering
  \igraph{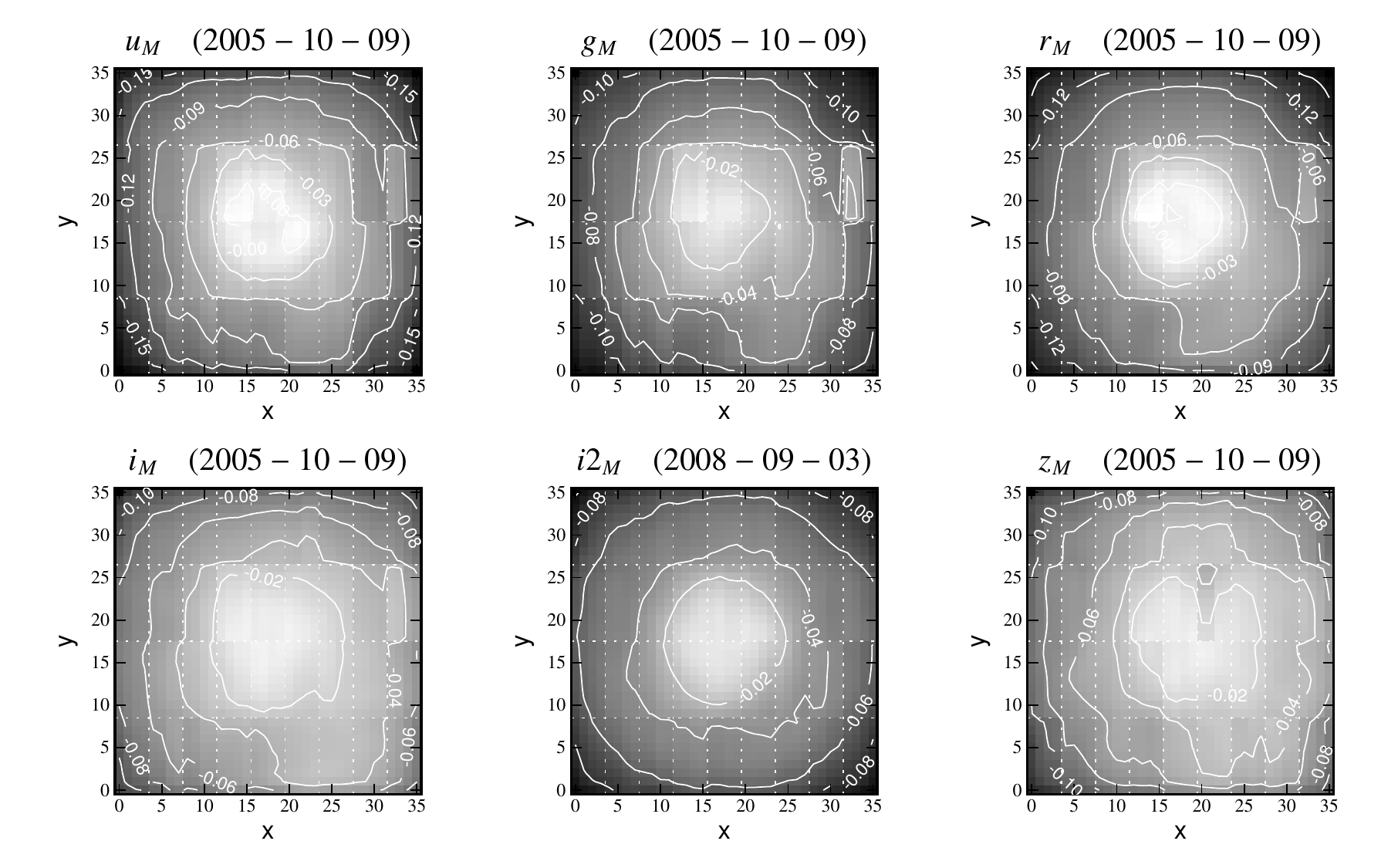}
  \caption{Maps of photometric corrections (relative to the
      twilight flat-field). Each panel displays a low resolution map
      ($36\times36$~super-pixels of $256\times256$ real pixels,
      \emph{i.e.} a resolution of about 2\arcmin) of the \dzp
      corrections. Those corrections are determined on dithered
      observations of dense stellar fields in a given photometric
      band. Observations for the corrections displayed in \band u
      \band g \band r \band i and \band z panels have been obtained
      during the same night in 2005 under remarkably stable
      atmospheric conditions. The \band{i2} panel displays the result
      of an observation taken in 2008, about 1 year after the
      replacement of the \band i filter.  CCD boundaries are figured
      by white dotted lines. }
  \label{fig:gridsolution}
\end{figure*}

As already stated, the model does not \Fixed{include} varying
photometric conditions during the observation of the grid sequence.
Unfortunately, while high quality photometric time has been dedicated
to the grid observations, very few sequences were found to exhibit
\Fixed{a} stable $IQ$. The flat-field solutions from the grid
sequences with the most stable IQ are displayed on
Fig.~\ref{fig:gridsolution}.

The error induced by variation of the photometric zero point between
exposures is expected to dominate the error on the determination of
the flat-field correction. The selection of data to minimize the
impact of zero point variations is discussed in the next section.

Finally, the flat-fielding correction that \Fixed{satisfies}
\Fixed{Eq.~}(\ref{eq:26}) depends on the flux reference of the
magnitude system.  The corrections derived \Fixed{so far} refer to the
system defined by Eq.~(\ref{eq:28}).  By construction, those
corrections make the natural magnitude of \bdtruc independent of focal
plane position. Corrections needed to ensure the same property in an
AB system are readily obtained (up to an irrelevant constant that can
be chosen to ensure $\dzp^{ab}(\x_0) = 0$) as:
\begin{equation}
  \label{eq:17}
  \dzp^{ab}(\x) = \dzp^{V}(\x) - 2.5 \logdec \frac{\int \lambda \cmag[T](\lambda) S_\ab(\lambda) d\lambda}{\int \lambda \cmag[T](\lambda) S_\text{ref}^V(\lambda) d\lambda}
\end{equation}

Once the corrections \Fixed{are} obtained, photometrically flat-fielded
instrumental magnitudes are given by:
\begin{equation}
\madu = \madu^t - \dzp^{ab}\label{eq:1}
\end{equation}

\Fixed{Since} the correction is bound to a photometry method
($IQ$-scaled here) and a photometric system (AB in the rest of this
paper), it is generally advisable to keep the image pixels scaled by
the twilight flat-field and to apply needed corrections to catalogs.

\subsection{Discussion}
\label{sec:instr-model-assessm}

We now turn to a discussion of uncertainties associated \Fixed{with}
our determination of the instrument response.  Noise does not
constitute a limiting factor in any of the measurements required to
build the response model. Instead, \Fixed{the} most significant errors
are expected to arise from imperfect verification of the assumptions
\Fixed{that} the model is built on.  Explicitly, we have made the
following hypotheses:
\begin{enumerate}
\item The transmission of the instrument as a function of wavelength
  does not evolve with time.
\item The transmission of the instrument as a function of wavelength
  depends solely on the \Fixed{focal plane} radius, 
  and this dependency is correctly mapped.
\item The photometric response can be obtained as a stable photometric
  correction to the twilight flat-fields.
\item The observations of dense stellar fields that constrain the
  photometric correction are obtained under stable photometric
  conditions.
\end{enumerate}
In what follows, we \Fixed{quantitatively} evaluate \Fixed{the
  validity of} each assumption.  We also discuss the resulting errors
on the photometric response map.

\subsubsection{Filter aging}
\label{sec:passband-variations}

The transmission model for the original Sagem/REOSC filters was based
on early measurements by the manufacturer, prior to delivery in
2002. Later measurements of these filters have been conducted at CFHT
in 2006 and display perfect agreement with the manufacturer findings
in \Fixed{the} \band{u}, \band{g} and \band{z} \Fixed{bands}.
However, quite significant discrepancies between different
measurements of the \band{r} and \band{i} filters were found,
\Fixed{and thus} a potential aging \Fixed{effect} of these filters
\Fixed{has been} investigated (\emph{cf.}
Sect.~\ref{sec:filter-transmission}).

\begin{figure}
  \centering
  \igraph{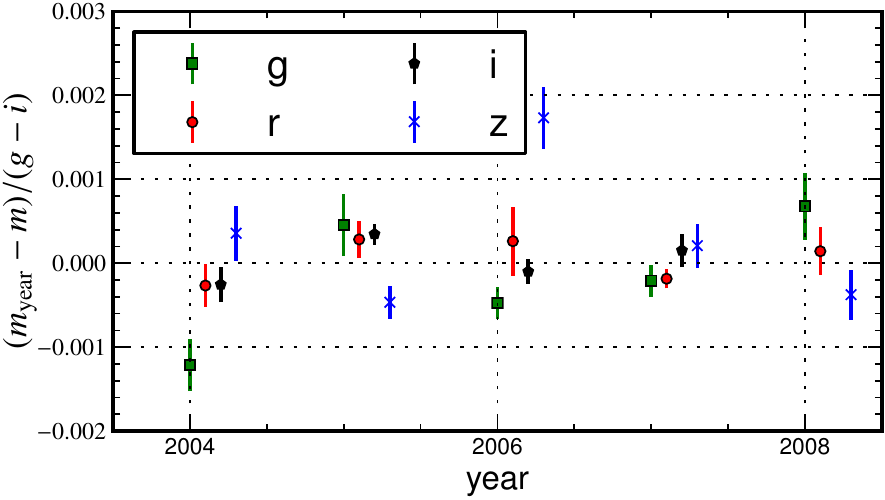}
  \caption{Color terms between tertiary catalogs averaged over
    different periods. The survey observations in the D3 field have
    been divided \Fixed{into} 5 periods covering about a year. Each data point
    is the fit of a linear color term between the catalog obtained
    from a single year and the catalog obtained from the whole
    survey. The stellar population is indexed by the $g-i$ color.}
  \label{fig:filtersvar}
\end{figure}

To investigate this effect, we formed \Fixed{independent} catalogs of
the D3 field on a yearly basis, and looked for time-dependent color
terms between the partial and the reference catalogs. The computed
color terms are displayed \Fixed{in} Fig.~\ref{fig:filtersvar}.  The
possible variation in filter response is less than $0.001$ per unit of
$\band g- \band i$ color and it can be \Fixed{safely neglected}.  This
\Fixed{test} excludes a significant continuous \Fixed{variation} of
the filters over the course of the survey.  In particular, \Fixed{this
  negligible change in} the filters cannot explain the discrepancy
between \Fixed{the} 2002 and 2006 measurements of \Fixed{the}
transmission curves; \Fixed{if these measurements corresponded to a
  real change, we would have seen} a color term of about 0.005 mag per
unit of $g-i$.  The filter measurements could both \Fixed{be} correct,
\Fixed{however,} if the filter response changed after the first
measurements were taken but before the survey began.

We conclude from this study that the passbands did not significantly
evolve during the survey and that the 2006 measurements should be
representative of the survey filters. This justifies our decision to
correct the 2002 measurements so that they now match the 2006
measurements at the few positions where the latter are available (see
appendix~\ref{sec:megacam-band-i}).

We recall, however, that this choice does not make the filter model
entirely consistent. As an example, some spatial discrepancies between
the predicted and measured color-terms for the \band r filter are
presented in the next section and are visible in
Fig.~\ref{fig:measvssynthdk}. 
This \Fixed{discrepancy} indicates that the model for this
filter is still not fully accurate, which has consequences on 
\Fixed{the} absolute calibration (\emph{cf.} Sect.~\ref{sec:concl}) 
and on the response map as discussed below.

\label{sec:survey-uniformity}

\subsubsection{Consistency of the model of passband variations}
\label{sec:adeq-passb-model}

\begin{figure}
  \centering
  \igraph{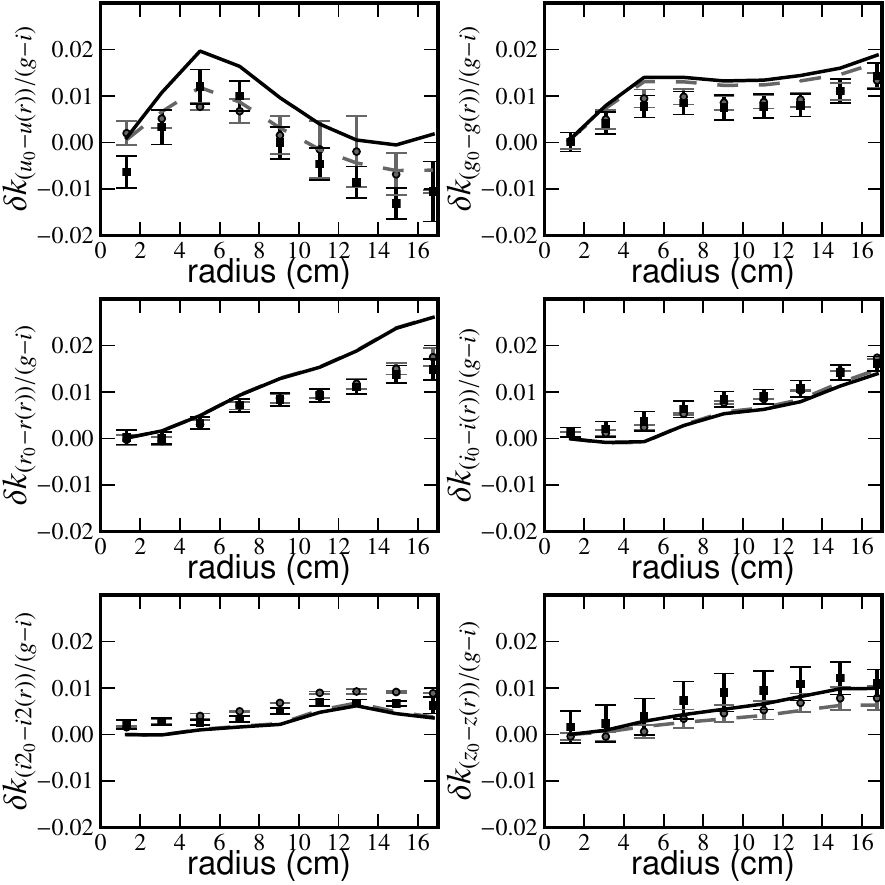}
  \caption{\Fixed{Difference} between measured and modeled color terms
    (\Fixed{relating} the natural and reference photometric system) as
    a function of radius.  \Fixed{Solid points show} the average value
    of the $\dk$ map in bins of radius. We plot separately the value
    average on all epochs of the winter field (gray dots) and summer
    field (black squares). The error bars are estimated from the
    dispersion between epochs. Data points are not independent,
    \Fixed{and} the error \Fixed{is} dominated by the uncertainty on
    the overall normalization of the curve due to the constraint
    $\dk(0)=0$. The synthetic color terms are computed on reddened
    spectra from the Pickles stellar library, to reflect the effect of
    the extinction measured in the two fields. }
  \label{fig:measvssynthdk}
\end{figure}

A first argument in favor of the adequacy of our varying passband
model is that the introduction of the color term in the model of grid
observations given by Eq.~(\ref{eq:30}) indeed improves the fit to the
grid data set without the addition of free parameters. To go one step
further and be more quantitative, one can leave $\dk$ as a free
parameter in the adjustment to the grid dataset. This was done in R09
using a coarse parameterization of the $\dk$ variation in cells of
$1536\times1024$~pixels. The $\dk$ maps have been redetermined
independently on all available grid epochs. The measured $\dk$ are
compared to synthetic ones as a function of distance to the focal
plane center in Fig.~\ref{fig:measvssynthdk}.

We have displayed the results from the two grid fields separately on
this figure to distinguish potential effects related to reddening by
dust. On can note that while the two grid fields suffer different dust
extinction, the synthetic color terms, accounting for extinction, are
almost indistinguishable.  The fitted color terms agree as well, and
there is no reason to question the results from the dusty grid-1
field.

The agreement between the model and the observations is generally
satisfying. It deteriorates a bit in band \band{r} at large radius,
where the synthetic color-term becomes the most dependent on the selected
color range. The observations also suggest that the \band{i2} filter
may be a bit more variable than measured.

\begin{figure}
  \centering
  \igraph{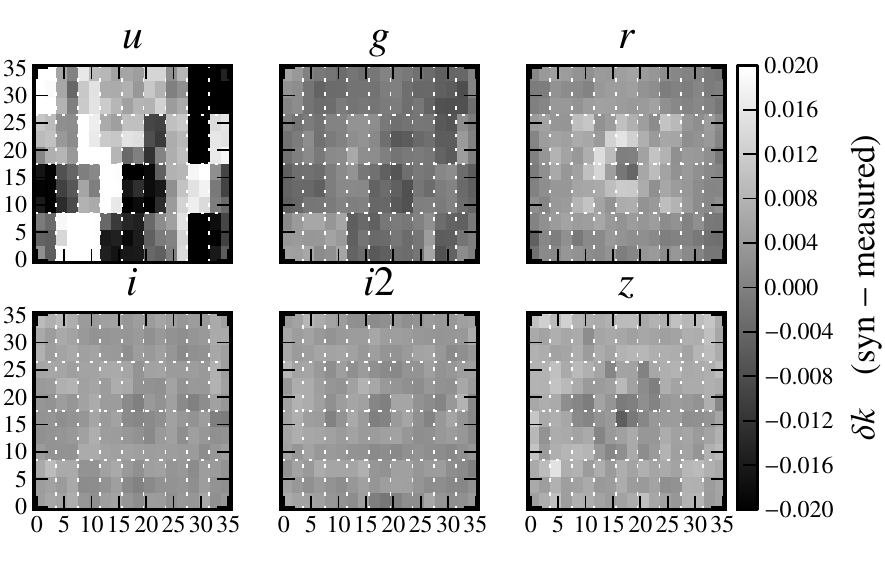}
  \caption{Maps of residuals between the predicted and measured color
    term to the reference system. The synthetic color terms are
    computed on reddened spectra from the Pickles stellar library, to
    reflect the effect of the extinction measured in field grid-1. The
    measured $\dk$ are independently adjusted on each grid sequence on
    field grid-1 and averaged over all epochs to increase the 
    signal-to-noise ratio.}
  \label{fig:dkres}
\end{figure}

Quantitatively, the standard deviations of the residuals of the
measured to the synthetic $\dk$ maps are $0.014$, $0.004$, $0.003$,
$0.003$, $0.004$, $0.002$ in band \band{u}, \band{g}, \band{r},
\band{i}, \band{i2} and \band{z} respectively. Residuals maps are
shown on Fig.~\ref{fig:dkres}. The most prominent features are the
radial deviations in band $\band{r}$ and $\band{i2}$ already noticed
on the radially averaged plot. Otherwise the residuals are consistent
with noise in the observations. The boundaries between different chips
are also visible in band $u$ and $g$, which can be explained by slight
differences in quantum efficiency curves from one CCD to
another.\footnote{Although measurements of individual chips quantum
  efficiency are available, they are known to be affected by problems
  such as ice condensation in the cryostat. Using them in the model
  does not improve the agreement with measurements and occasionally
  deteriorates it. As a consequence, we stick to the average QE
  model.}

Although statistically significant, the discrepancy between the
calculations and measurements are negligible for our purposes. We
stick to the modeled color terms as they provide some consistency with
the flux interpretation of the magnitudes, and significantly reduce
the noise of flat-field.

One can also note that the examination of the various epochs
independently exclude significant evolution of the pattern with
time. The test is however insensitive to global variations of the mean
passband with time which has been investigated above using tertiary
stars.

\subsubsection{Stability of the photometric flat-field corrections}
\label{sec:stability-flat-field}
Twilight flat-fields are assembled independently for each run of
MegaCam (every lunation). In contrast, the photometric correction is
determined about twice a year from a set of dedicated observations.
We now want to assess the hypothesis that this correction is
appropriate to other runs than the one it was derived for.

There are three main issues to consider. First, whether the
differential response between point-like and uniform illumination
(what is mapped by the photometric correction) is stable. Second,
whether the successive twilight flat-fields consistently map the
instrument response to uniform illumination. Third, whether the
variations of the instrument response occur on time scales smaller than
a \Fixed{MegaCam run}
\Fixed{Below} we qualitatively discuss each of these issues.

\paragraph{Changes in the instrumental setup :}

The origin of a difference between instrument responses to point-like
and uniform illumination is attributed to three main effects: 1) the
spatial variation of the plate-scale, 2) the pollution of twilight
images by stray light, \Fixed{and} 3) the distortion of the PSF that
\Fixed{varies the} fraction of flux recovered by the photometry.  Each
of these effects is related to the optical configuration of the
instrument and should be fairly stable.

However, several important adjustments to the telescope setup have
been made during the survey, and \Fixed{these may} have modified its
response.  We have investigated the stability of the instrument
response using the repeated observations of the SNLS deep fields.  The
compatibility between the observations conducted at different epochs
provide a stringent monitoring of changes in the survey
uniformity. The analysis is fully described in
Appendix~\ref{sec:more-megac-flatf}, and shows that telescope
modifications indeed affect the photometry. A single photometric
correction is thus not suitable to flat-field the entire survey.

We identified three telescope events with a noticeable impact on the
flat-field corrections. The first is the flip of the L3 lens in the
image corrector that produced a noticeable improvement of the
uniformity of the image quality across the focal plane.\footnote{See
  \url{http://www.cfht.hawaii.edu/News/Projects/MPIQ/}.}  The second
is the unfortunate accumulation of metal shavings on the top optical
surface below the filter mechanism until its identification and
cleaning at the end of the year 2006. The last is the accidental
breaking of the \band{i} filter in July 2007.  
\Fixed{These events} define roughly 5 epochs: 
the beginning of the survey with the original telescope setup,
the improved setup, the period of noticeable dust accumulation, 
the short period between the dust cleaning and the filter breakout, 
and the end of the survey. The exact dates of the telescope events are
given in table~\ref{tab:grid_obs}.

Grid observations were accumulated sufficiently often to account for
each of these major events.  We can \Fixed{determine} a photometric
correction suitable for each epoch of the survey.  However, this
scheme can account only for abrupt changes in the telescope setup.
The accumulation of metal dust is harder to handle \Fixed{because
  this} effect is likely to \Fixed{change slowly with time but only} a
single grid observation was taken during this period (shortly before
the cleaning when the effect was at its maximum).  That being said, we
can use the monitoring of the instrument response provided by
deep-field observations (see Appendix~\ref{sec:more-megac-flatf}) to
argue that, once corrected with the available grids, the instrument
response appears compatible across all epochs at the 5~mmag level and
does not display a significantly stronger dispersion during the period
of dust accumulation than during the other \Fixed{periods}.

\paragraph{Variations in the twilight flat-field :}

Aside from the twilight noise, which is negligible at large scales,
there is one expected contribution to systematic differences between
different runs: the twilight \Fixed{images} contain a color-dependent
term as noted in Sect.~\ref{sec:phot-flat-field}. If the effective
color of the observed twilight varies significantly from one run to
another, the photometric correction would have to be modified to
compensate the variations induced on the twilight flat-field.

The procedure of taking twilight flats is automated so we expect the
average twilight colors to be rather stable.  We checked this by
studying twilight flat-fields in the time period 04 Sep. 2007
\Fixed{through} 05 Sep. 2008 \Fixed{when} no intervention on the
telescope optics occurred.  We computed the ratio of \Fixed{each
  flat-field image obtained during this time period} to the average
flat-field image over the period. Defining an effective twilight color
as the correlation coefficient between the twilight flat and the $\dk$
maps, we have measured variations of colors between successive
flat-fields with standard deviations of $0.04$, $0.15$, $0.11$, $0.15$
and $0.45$ in units of the $\dk$ maps for
\band{u}\band{g}\band{r}\band{i} and \band{z} bands respectively.

The fraction of those variations unrelated \Fixed{to} changes in the
twilight color (but simply to actual changes in the instrument
response following roughly a similar pattern) is difficult to
estimate. Nevertheless, those \Fixed{variations} can be used to set an
upper bound on the effect: in all cases a variation of the twilight
color by the quoted amount contributes \Fixed{a} flat-field
\Fixed{error} that reaches at most 5~mmag.  The effect is small and
will be smoothed out by averaging different photometric corrections;
\Fixed{hence} it can be safely neglected.

\paragraph{Variations of the instrument response on short time scale:}

Last, variations of the amplifier gain arise on time-scales shorter
than the run and cannot be properly accounted for by the twilight
flat-field. A \Fixed{prominent example} of \Fixed{an} amplifier gain
variation is clearly visible on the 2005-10-09 grid displayed
\Fixed{in} Fig.~\ref{fig:gridsolution}.  One amplifier (the left side)
on the last CCD of the second row pops out.  This feature is present
in all bands for this epoch, \Fixed{but since} it is not reproduced on
other grid epochs \Fixed{we} interpret \Fixed{this effect} as a
variation of the gain between the acquisition of twilight and the
observation of the grid.

Such features normally contribute a small extra noise that averages
out on several lunations. \Fixed{However, when an amplifier jump alter
  a photometric flat-field correction, all the images flat-fielded
  using this correction are coherently affected.}  The jumps of the amplifier gains are
estimated to be generally small from a preliminary analysis of a
monitoring experiment of the MegaCam gains conducted with the SNDICE
experiment \citep{sndice}.\footnote{typically contained within 1\%,
  most amplifiers being much more stable.} As such deviations are
expected to \Fixed{randomly affect} different determinations of the
photometric correction, a practical way to handle this difficulty is
again to rely on the averaging of several determinations.

\paragraph{Compatibility between different epochs:}
 
From all the considerations above we conclude that the assumption of a
stable photometric correction is mostly valid, at the 5~mmag level,
over large periods during the survey. We adopt the \Fixed{subdivision
  of the survey} in five periods, as defined above, and we attribute a
range of validity to photometric corrections determined during a given
period that extends over the whole period. Remaining deviations, due
either to the instrument response evolving slowly within each period,
to variations in the twilight flat-field, or to variations of
amplifier gains are difficult to characterize with a high level of
confidence. A reliable estimate is provided by the fact that the
photometric corrections bring the different survey epochs in agreement
at the 5~mmag level.  An important point is that the remaining
deviations should be smoothed out by averaging the observations over
the 5 periods.

\subsubsection{Errors in the photometric flat-field correction}

The measurement noise contributes about 1~mmag to the flat-field
correction uncertainty and \Fixed{is therefore} negligible. The main
weakness of the grid method is the inability to adjust variations of
the photometric response during the $\sim17$ \Fixed{minute}
observation sequences.

The statistical properties of the variations of the overall zero point
during short sequences can be obtained from the main survey
observation sequences that consist \Fixed{of half-hour long visits}
split into 5-6 exposures on the same field (with a negligible
dithering).  The zero point is found to be fairly stable during clear
nights with an rms typically smaller than 5~mmag (see
Sect.~\ref{sec:catalog-constitution} for the determination of zero
points and Fig.~\ref{fig:photometricitycuts} for the distribution of
zero points variations). Most of this variation is uncorrelated from
one sequence to another. However, a small systematic \Fixed{shift} of
about $+2\pm0.5$~mmag per hour can be measured.

We used \Fixed{the measured shift and rms} as inputs to simulations in
order to evaluate the impact of such variations on the determination
of the photometric correction. We checked for the influence of a
random variation of the photometric zero point with an rms of 5~mmag
and a systematic \Fixed{shift} of 2~mmag per hour. The systematic
\Fixed{shift} results in a negligible (below $1$~mmag peak-to-peak)
gradient on the focal plane.  The random photometric zero point
\Fixed{produces} a more significant error on individual solutions
taking the form of large-scale variations with a peak-to-peak
amplitude of about 1\%. However, this random error is smoothed out as
expected when several independent determinations of the photometric
correction are averaged together.

Some epochs display stronger (and time correlated) variations of the
observing conditions than what is typically observed. We took the
conservative approach of discarding observations where conditions were
not stable. Epochs \Fixed{were not used to flat-field the survey if
  the} seeing variations \Fixed{were} greater than 50\% peak to peak,
or variations of measured aperture corrections \Fixed{were} greater
than 1.1\% peak to peak. Those epochs are likely to display zero
points variations greater than the 5~mmag rms assumed in the
simulations.  This cut removes 18, 23, 30, 28, 40 and 20\% of the
available epochs in \band{u},\band{g},\band{r},\band{i}, \band{i2} and
\band{z} band, respectively.  \Fixed{Valid} observations are displayed
in bold face in Table~\ref{tab:grid_obs}. While this choice is
somewhat arbitrary, excluding or \Fixed{keeping} the observations made
under variable conditions has a small impact on the uniformity of the
tertiary star catalogs.  When included, the catalog are modified by
about 1\% peak-to-peak (or 3~mmag rms).

\subsubsection{Internal estimate of the flat-field uncertainty}
\label{sec:uncert-flat-field}

\begin{figure*}
  \centering
  \igraph{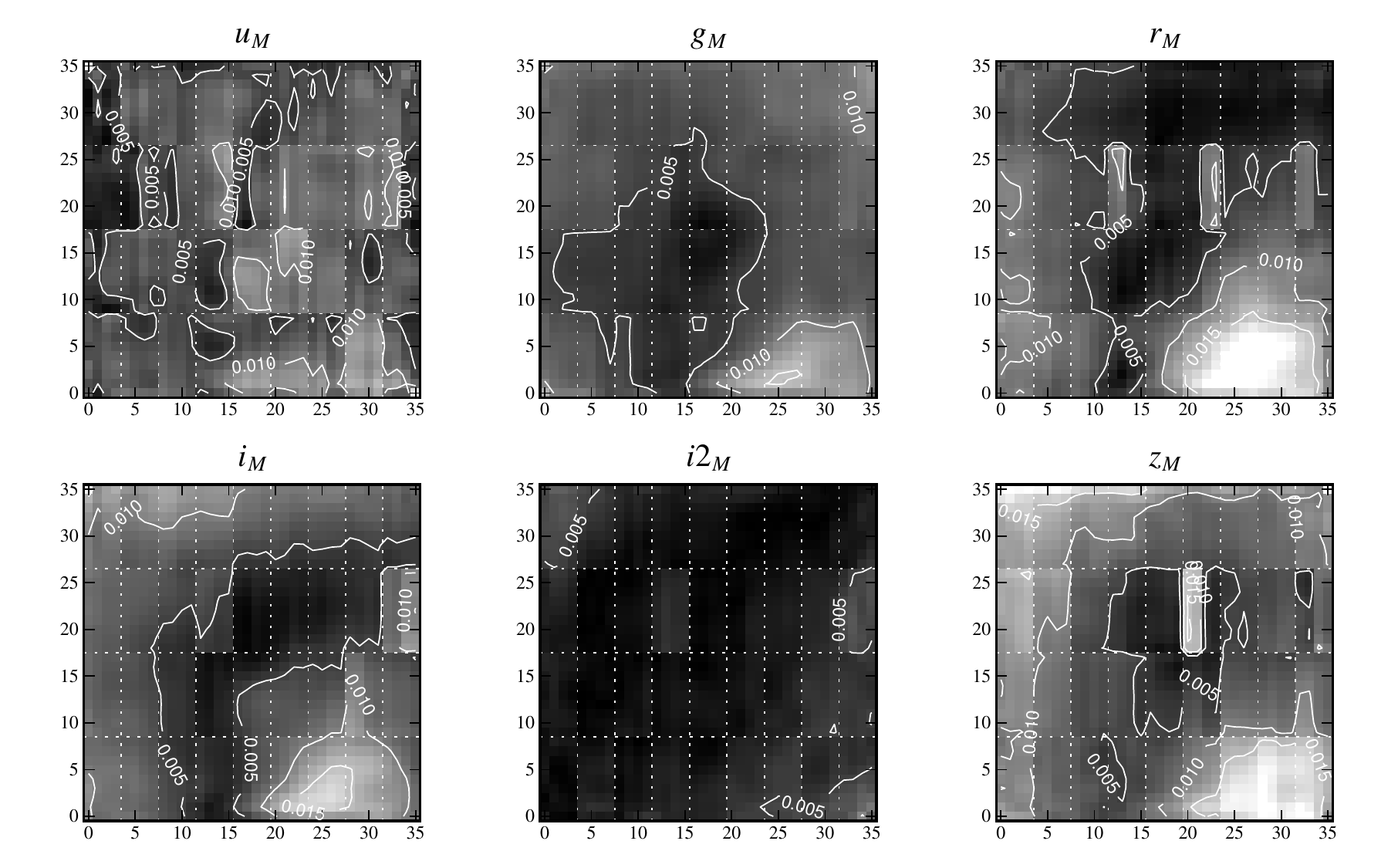}
  \caption{Variation of the photometric flat-field corrections. Each
    panel displays a low resolution map ($36\times36$~super-pixels of
    $256\times256$ real pixels, \emph{i.e.} a resolution of about
    2\arcmin) of the rms between all the determinations of the
    photometric grid corrections in a given photometric band. CCD
    boundaries are \Fixed{shown} by white dotted lines. Note that all
    photometric correction maps are normalized to be zero at the
    center of the focal plane. As a consequence the rms across epochs
    is null at the center by construction. Noisy amplifiers \Fixed{are
      easily seen} in \band r and {\band z} \Fixed{bands}.  In all
    bands (but \band {i2} that was not available), the variation of
    response induced by the 2006 deposit of metal dust is clearly
    visible at the bottom right of the focal plane.}
  \label{fig:flatvar}
\end{figure*}

The typical r.m.s scatter of the selected grid solutions across epochs
is 6, 6, 7, 7, 2 and 9~mmag in band
\band{u},\band{g},\band{r},\band{i} and \band{z}, respectively.  The
structure of the variation is visible in Fig.~\ref{fig:flatvar}.  In
all bands (but \band{i2} that was installed after the cleaning), the
accumulation of dust in the optics during \Fixed{the} year 2006,
visible at the lower right corner of the focal plane, dominates the
variation.  The pattern of noisy amplifiers also shows up on the
variation map.  \Fixed{Beyond} these recognizable features it is
difficult to distinguish between what is attributable to actual
modifications of the telescope response, and what should be considered
as part of the flat-fielding error.

As discussed above, the most reliable internal estimate of the
remaining flat-fielding error is \Fixed{determined} by the agreement
between different periods of the deep survey.  The typical rms
$\sigma_{ff}$ of the spatial discrepancies between individual
exposures are obtained in Appendix~\ref{sec:more-megac-flatf} and
\Fixed{summarized} in Table~\ref{tab:fferror}. Assuming that the
errors are smoothed out when different photometric corrections
covering the 5 periods are averaged, one concludes that a rough
estimate of the residual uncertainty on the tertiary star
flat-fielding is given by $\sigma_{ff}/\sqrt{N_g}$, where $N_g$ is the
number of independent determinations of the photometric correction. An
independent cross-check of the flat-fielding quality is obtained by
comparison with the SDSS in Sect.~\ref{sec:spatial-uniformity}; it
\Fixed{gives} similar numbers.

  \begin{table}
    \centering
    \caption{Internal consistency of the photometric flat-field.}
    \label{tab:fferror}
    \begin{tabular}{l*{6}{r}}
      \hline
      \hline
      &\band u & \band g & \band r & \band i & \band {i2} & \band z\\
      \hline
      $\sigma_{ff}$\tablefootmark{a} & 0.006 & 0.004 & 0.005 & 0.005 & 0.003 & 0.007\\
      $N_g$\tablefootmark{b} & 6 & 7 & 6 & 5 & 2 & 5\\
      \hline
      $\sigma_{ff}/\sqrt{N_g}$\tablefootmark{c} & 0.003 & 0.002 & 0.002 & 0.002 & 0.002 & 0.003\\
      \hline
    \end{tabular}
    \tablefoot{\\
      \tablefoottext{a}{Estimated as the average of the rms of the relative zero point variation on the focal plane (see appendix \ref{sec:more-megac-flatf}).}\\
      \tablefoottext{b}{Number of independent determinations of the photometric correction.}\\
      \tablefoottext{c}{This delivers a crude estimate of the rms of residual non-uniformity in the SNLS tertiary catalogs,  which are obtained by averaging over the whole survey.}
    }
  \end{table}

\subsection{Summary}
\label{sec:summary}
We have \Fixed{built} a model of the effective MegaPrime passband
corresponding to the new \band{i2} replacement filters. The \band i
and \band r models have also been significantly revised with respect
to the previous R09 release of the SNLS calibration using the more
reliable 2006 filter measurement data.

The flat-field processing has \Fixed{been} slightly \Fixed{updated}
with respect to R09.  The most significant change is the use of
modeled color terms rather than measured color terms to translate
measurements when fitting the photometric corrections.  This
\Fixed{update} is not expected to introduce significant changes in
natural magnitudes except in $\band{r}$ at the corners of the focal
plane, where the effect alters the flat-fielding by \Fixed{0.5\%}. We
also choose to make the instrument response constant for objects
having an AB \Fixed{spectrum} rather than for the primary
spectrophotometric standard \bdtruc.  This convention makes the
derivation of AB magnitudes from instrumental magnitudes more
straightforward.  Another change is a smooth parameterization of the
instrument response which is expected to provide a slight decrease
\Fixed{in} the flat-fielding noise.

The main benefits \Fixed{for flat-fielding accuracy} are related to
the extension of the data sample (except for the broken \band{i}
filter).  \Fixed{Most} sources of flat-fielding error are uncorrelated
between different epochs, \Fixed{and these} errors are thus smoothed
out with the addition of new grid observations.  \Fixed{To avoid
  biasing the average} we introduced a quality cut to \Fixed{remove}
epochs strongly affected by varying observation conditions.

An upper bound on the flat-fielding error is provided by the
variability of the photometric corrections determined on independent
data samples. It is below 1\% in all bands. Assuming the error is
uncorrelated between epochs, this also sets an upper bound of
typically 0.3 \% r.m.s on the uniformity of the resulting catalogs.
We study the consequences of flat-fielding uncertainties on
calibration \Fixed{in} Sect.~\ref{sec:flat-field-error}.

\section{SNLS tertiary star catalogs}
\label{sec:tert-catal-constr}

\subsection{Multi-epoch average photometry of tertiary stars}
\label{sec:catalog-constitution}

A large number of exposures have been accumulated on the CFHTLS deep
fields. A small dithering is generally applied to the observations for
the purpose of filling gaps between CCDs. Its typical amplitude
\Fixed{is} between 80 and 120 pixels in right ascension and 300 to 360
pixels in declination. This \Fixed{dither} is small enough that
position dependent changes in filter response can be neglected. We
thus assume that a given star is always observed through the same
effective filter.  This \Fixed{assumption} simplifies the construction
of a multi-epoch photometric catalog \Fixed{using} natural magnitudes
for the selected tertiary stars in the deep fields.

We use the stars selected as described in Sect.~11.1 of R09.  We first
\Fixed{determine} aperture flux measurements for all stars in all
\Fixed{the survey images}, \Fixed{and then apply} photometric flat-field
corrections (Sect.~\ref{sec:phot-flat-field}) to the individual
measurements according to Eq.~(\ref{eq:1}).  The survey is split
\Fixed{into} five periods, corresponding to noticeable changes in the
instrument configuration.  \Fixed{An} independent determination of the
photometric correction \Fixed{is} applied to \Fixed{each} epoch.  The
exact mapping between measurements and corrections is described in
Appendix~\ref{sec:more-megac-flatf} and illustrated in
Fig.~\ref{fig:zpperexp}.

The model for the measured flux $Y_{s,t}$ of non-variable stars after
flat-fielding is:
\begin{equation}
 Y_{s,t} = \Phi_s \alpha_t + n_{s,t}\label{eq:20}
\end{equation}
where $t$ indexes the exposure, $s$ indexes the star, $\Phi_s =
10^{-0.4 \cmag}$ is the local broadband flux of star $s$,
$\alpha_t=10^{0.4 \zp_t}$ is the photometric zero point of exposure
$t$, and $n_{s,t}$ is the measurement noise.  The overall scales of
$\Phi_s$ and $\alpha$ are not constrained from the data. Setting the
scale of the $\Phi_s$ \Fixed{to} correspond to physical flux requires
calibration data, and is the subject of
section~\ref{sec:megac-absol-calibr}.  At this stage, we are only
interested in delivering an homogeneous set of measurements, and we
will fit the model to the data with the constraint
$\langle\alpha_t\rangle = 1$.  We work in flux space rather than in
magnitude space because at low S/N, averaging logarithms causes a
bias.

We opt for a two-step, iterative method. \Fixed{We start} from a first
estimate \Fixed{of $\Phi_s = \Phi_s^{(0)}$,}
obtained as the median of instrumental flux over all epochs for each star,
\Fixed{and then} we iterate the
following steps until convergence:
\begin{enumerate}
\item $\alpha^{(i)}_{s,t} = Y_{s,t}/\Phi_{s}^{(i)}$ using the
  current estimate of $\Phi_s$, and solve for $\alpha_t$ in the linear
  model:
  \[\alpha^{(i)}_{s,t} = \alpha_t + m_{s,t}\] 
  where the measurement noise on $Y_{s,t}/\Phi_{s}^{(i)}$ 
  is modeled as \Fixed{an} independent random Gaussian variable 
   $m_{s,t}$ centered around 0 \Fixed{with} variance $\sigma^2_{s,t}$. 
\item  $\Phi_{s,t}^{(i)} = {Y_{s,t}}/{\alpha_t^{(i)}}$ and improve
  the estimate of $\Phi_s$ as the (non-weighted) mean of
  $\Phi_{s,t}^{(i)}$ across a subset of \Fixed{photometric} exposures 
\end{enumerate}
In \Fixed{this} procedure, $\pm 4\sigma_{s,t}$ outliers are flagged
and discarded.  We \Fixed{model} the relative flux uncertainty
$\sigma_{s,t}^2$ as
\begin{equation}
\sigma_{s,t}^2 \approx f_t^2 +
\frac{1}{t_{exp}G\Phi_s^{(i)}} + \frac{\gamma_t}{(\Phi_s^{(i)})^2}\,.\label{eq:31}
\end{equation}
Accurate values of $f_t^2$, and to a smaller extent, $\gamma_t$, are
difficult to derive from first principles. We thus derived robust
estimates of \Fixed{these parameters} for each exposure by
\Fixed{fitting} the rms of $\alpha_{s,t}^{(i)}$ \Fixed{computed} in
bins of $\Phi_{s}^{(i)}$.

The major contributor to the $\gamma_t$ term, which dominates the
measurement error at low fluxes, comes from the Poisson fluctuation of
the \Fixed{sky} background.  Its \Fixed{theoretically} expected value
is $\hat\gamma_t = N_\text{pix}\sigma_\text{sky}^2 G$ where
$N_\text{pix}$ is the number of pixels in the aperture
($\lesssim700$), $\sigma_\text{sky}^2$ is the variance of the
background fluctuation, and $G\sim1.6\,e^-/{\rm ADU}$ is the gain of
the readout electronics.  This \Fixed{theoretical estimate}, however,
\Fixed{is typically significantly smaller than the fitted} $\gamma_t$
value \Fixed{because it} does not account for contamination by
structured residuals in the background (PSF tails, ghost
reflections...), especially in \Fixed{the} \band i and \band z bands
where \Fixed{subtraction} residuals from \Fixed{fringing} can be
important.

The other important \Fixed{contribution to} the measurement error is
related to the flat-field error, $f_t$. The major \Fixed{components
  of} this term are the error in the flat-field estimate, the
variations of amplifier gains, and potential structures in the
atmospheric extinction (clouds).  Such errors are typically correlated
\Fixed{among} observations within the same night or the same run. This
error term dominates the measurement error at the bright end: the
error on individual measurements of star magnitudes asymptotically
reaches $1.08f_t/\alpha_t$, which typically amounts to
$6$~mmag for clear nights.

The large number of epochs over a long time period  \Fixed{makes it possible}
to control systematic errors by applying stringent \Fixed{selection
  criteria (cuts)} on the measurements entering in the averaging
process.  We applied two cuts based on the relative variations of the
zero point: \Fixed{1)} the time stability of the zero point during
each observation sequence, \Fixed{and 2)} the spatial uniformity of
the zero point in each exposure.

\begin{figure}
  \centering
  \igraph{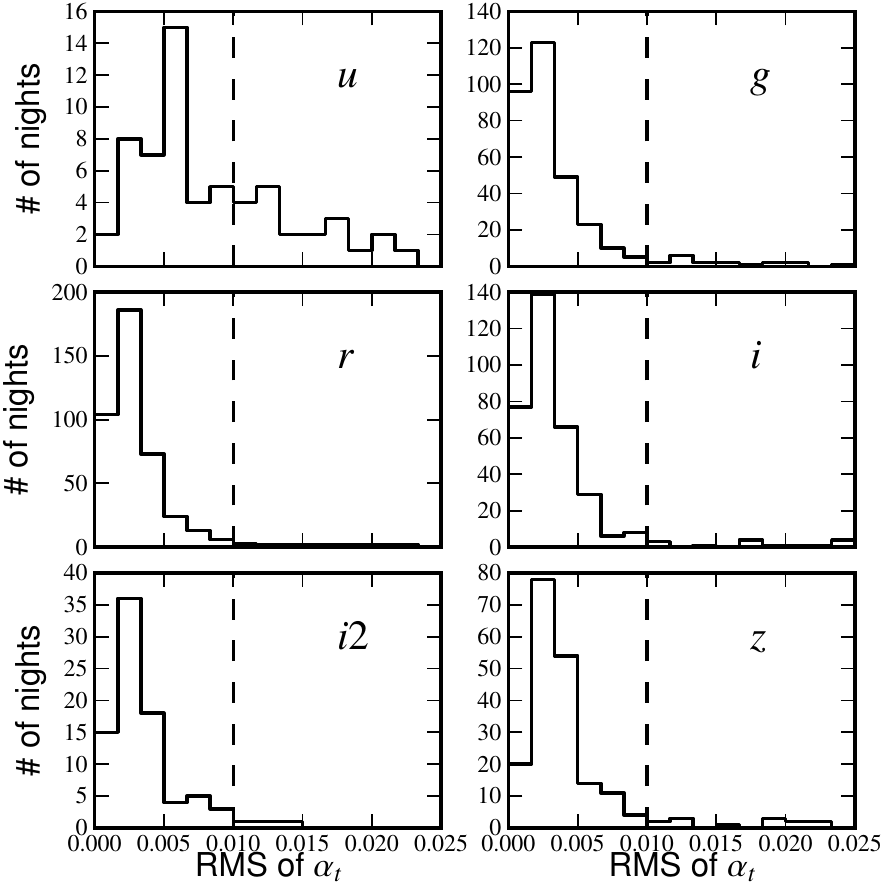}
  \caption{Top panel: Distribution of zero point variations
      during an observation sequence (typically 6 exposures of 220s in
      the same band in a row). Nights displaying a variability larger
      than 1\% are discarded from the averaging. While the plot is cut
      at 0.025, numerous sequences (about 20\%) taken under non-photometric
      conditions display a much larger variation. 
      The zero point is \Fixed{defined as} $\zp_t =-2.5\logdec \alpha_t$.}
  \label{fig:photometricitycuts}
\end{figure}

The distribution of the standard deviation of $\alpha_t$ from its mean
value for the night is displayed \Fixed{in}
Fig.~\ref{fig:photometricitycuts}. The stability of the zero point 
is generally better than 1\% for photometric sequences, with
typical values of $2-3$~mmag. We discarded 
\Fixed{observations for} nights with a relative
variation of $\alpha_t$ greater that 1\% because they might have been
affected by clouds; 
\Fixed{this requirement removes} 15 to 30 \% of the nights.

\begin{figure}
  \centering
  \igraph{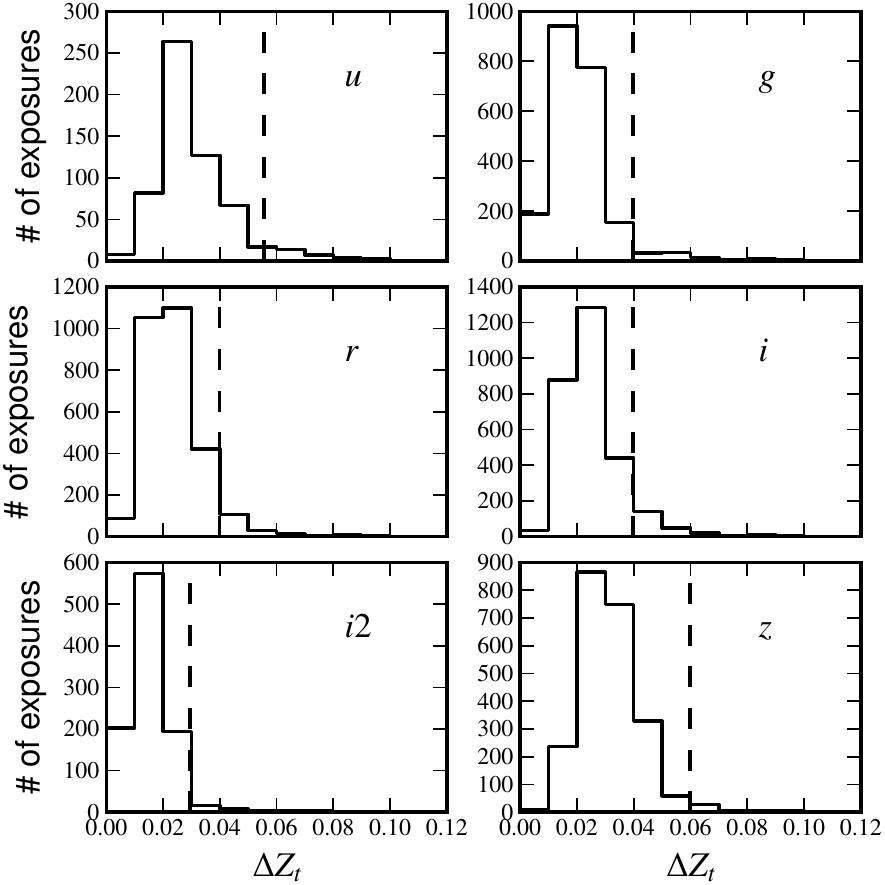}
  \caption{Distribution of the peak-to-peak zero point variation
    across the focal plane. Exposures displaying a peak-to-peak
    variation greater than 3\% (6\% in \band u and \band z) are
    discarded from the averaging. Vertical dashed lines \Fixed{show}
    the cuts applied in the averaging process. }
  \label{fig:uniformitycuts}
\end{figure}

We also compute \Fixed{the} relative variation of the zero point
$\alpha_{t,ccd}^{(i)}$ on a per CCD basis, and \Fixed{compute}
$\Delta\zp_t^{(i)} = \max_{ccd}2.5\logdec\alpha_{t,ccd}^{(i)} -
\min_{ccd} 2.5\logdec\alpha_{t,ccd}^{(i)}$.  The distribution of
$\Delta\zp_{t}$ is shown \Fixed{in} Fig.~\ref{fig:uniformitycuts}:
\Fixed{it} is generally below 4\%, with a typical value of 2\%.  This
\Fixed{result shows} that the various flat-field corrections are in
agreement at this level, \Fixed{although there are} occasional
\Fixed{outliers} related to atmospheric conditions or amplifier gain
variability.  \Fixed{These outliers are} identified and discarded by a
cut on $\Delta\zp_t$.  We discuss further the variation of
$\Delta\zp_{t}$ with time in Appendix~\ref{sec:more-megac-flatf}.

\begin{figure}
  \centering
  \igraph{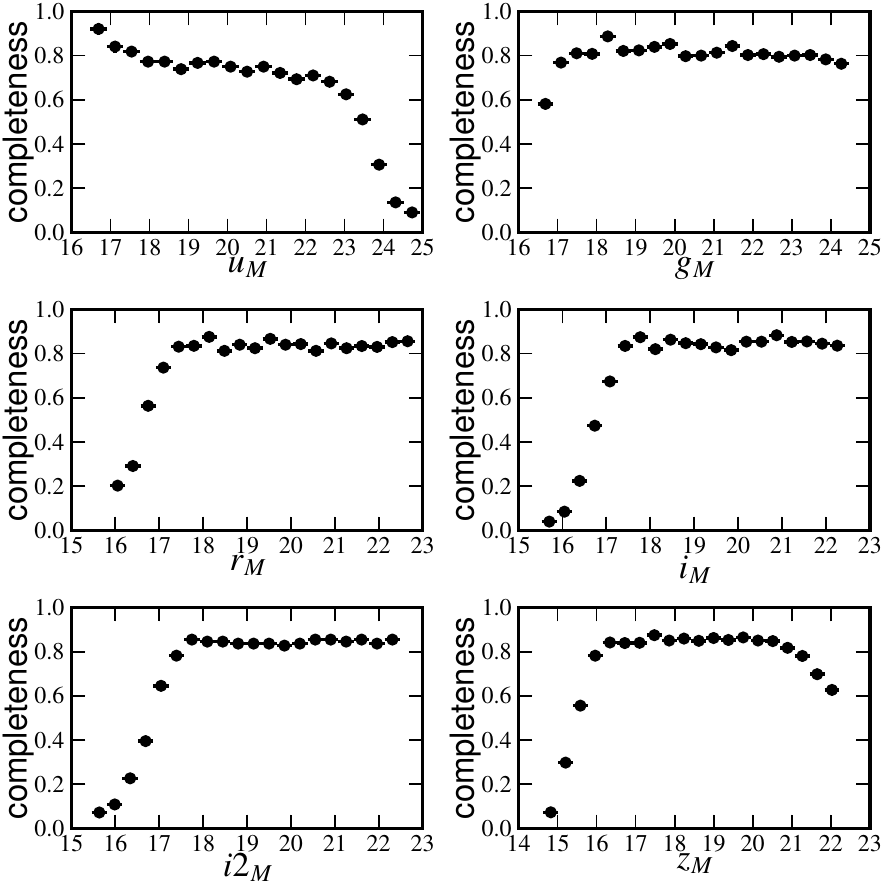}  
  \caption{Measurement completeness in the tertiary star
    catalogs. Each plot displays the ratio of selected measurements to
    the number of exposures as a function of the star magnitude in
    each band. The plateau at $\sim$85\% reflects mostly the ratio of
    the usable area in the focal plane to the total area covered by
    the catalog.}
  \label{fig:completeness}
\end{figure}

We did not attempt to correct the catalog for selection bias.
\Fixed{Biases arise for} faint \Fixed{objects} due to the
\Fixed{higher selection efficiency} of positive fluctuations, and
\Fixed{for bright objects that saturate some pixels on nights with
  good seeing,} leaving only nights with poor seeing conditions.
Completeness of the catalog, \Fixed{defined} as the fraction of nights
a given star \Fixed{is} used, is shown \Fixed{in}
Fig.~\ref{fig:completeness}.  Completeness does not reach 100\% on
average due to \Fixed{the} dithering \Fixed{needed to fill} gaps
between CCDs, contamination of apertures by cosmic rays, and
occasional defects in the instrument (part of the CCD mosaic being non
operational at a given time).  We \emph{did not} apply \Fixed{cuts} to
the catalog to discard stars potentially affected by selection bias.

\subsection{Anchoring to the Landolt system}
\label{sec:anchoring-landolt}

Nine fields containing a large subset of the Landolt equatorial
secondary standards were routinely observed by the CFHT for
calibration purposes. Those observations can be used to derive zero
points for a large number of nights and tie the tertiary catalogs to
the Landolt system.

The basic assumption underlying our \Fixed{strategy} is that, besides
the most obvious systematic differences between observations (\Fixed{such as}
airmass differences or systematic differences in the PSF),\footnote{to
  be discussed in section~\ref{sec:syst-uncert-calibr}} variations of
observing conditions are uncorrelated from one night to another,
\Fixed{and therefore these variations} will average out \Fixed{with
  many} repeated measurement of the flux ratios.  We are thus modeling
the exposure zero point as:
\begin{equation}
  \label{eq:15}
  \zp_t = \zp_n + 2.5 \logdec(t_\text{exp}) + (X_t-1)k_{\rm atm} + \delta \zp_{tn}
\end{equation}
where $\zp_t$ is the zero point of a given exposure $t$, $\zp_n$ is
the mean zero point for night $n$, $(X_t-1)k_\text{atm}$ is the
average extinction at airmass $X_t$ and $\delta \zp_{tn}$ accounts for
a variation of the photometric response in exposure $t$ with respect
to the average of the night $\zp_n$.  In the calibration process, we
treat this latter term as a noise \Fixed{that is independent for each
  night, and whose average is zero over many nights.}  Deviations from
this hypothesis, related to systematic differences between calibration
and science exposures, are discussed in
Sect~\ref{sec:syst-uncert-calibr}.

Following the scheme described in R09, we assume that the transformation
between the Landolt system and the reference MegaCam system follows an
\emph{approximate} piece-wise linear color relation $\LtoM$:
\begin{equation}
  \label{eq:21}
  \umag = L + \LtoM(C) + \delta L - \delta_\ab
\end{equation}
where $L$ is the Landolt magnitude of the star in the closest
photometric band, $C$ is the Landolt color index of the star, and
$\delta L$ accounts for small deviations of individual stars with respect to the
single parameter relation $\LtoM$. The calibration offset $\delta_\ab$
sets the magnitude system to the AB flux scale, \Fixed{and is}
determined by referring to a primary standard which is the subject of
Sect.~\ref{sec:megac-absol-calibr}.  For now we ignore
\Fixed{$\delta_\ab$ and set} the photometric calibration of the
tertiary catalogs to a common but arbitrary scale, close to a Vega
system.

\begin{figure*}
  \centering
  \igraph{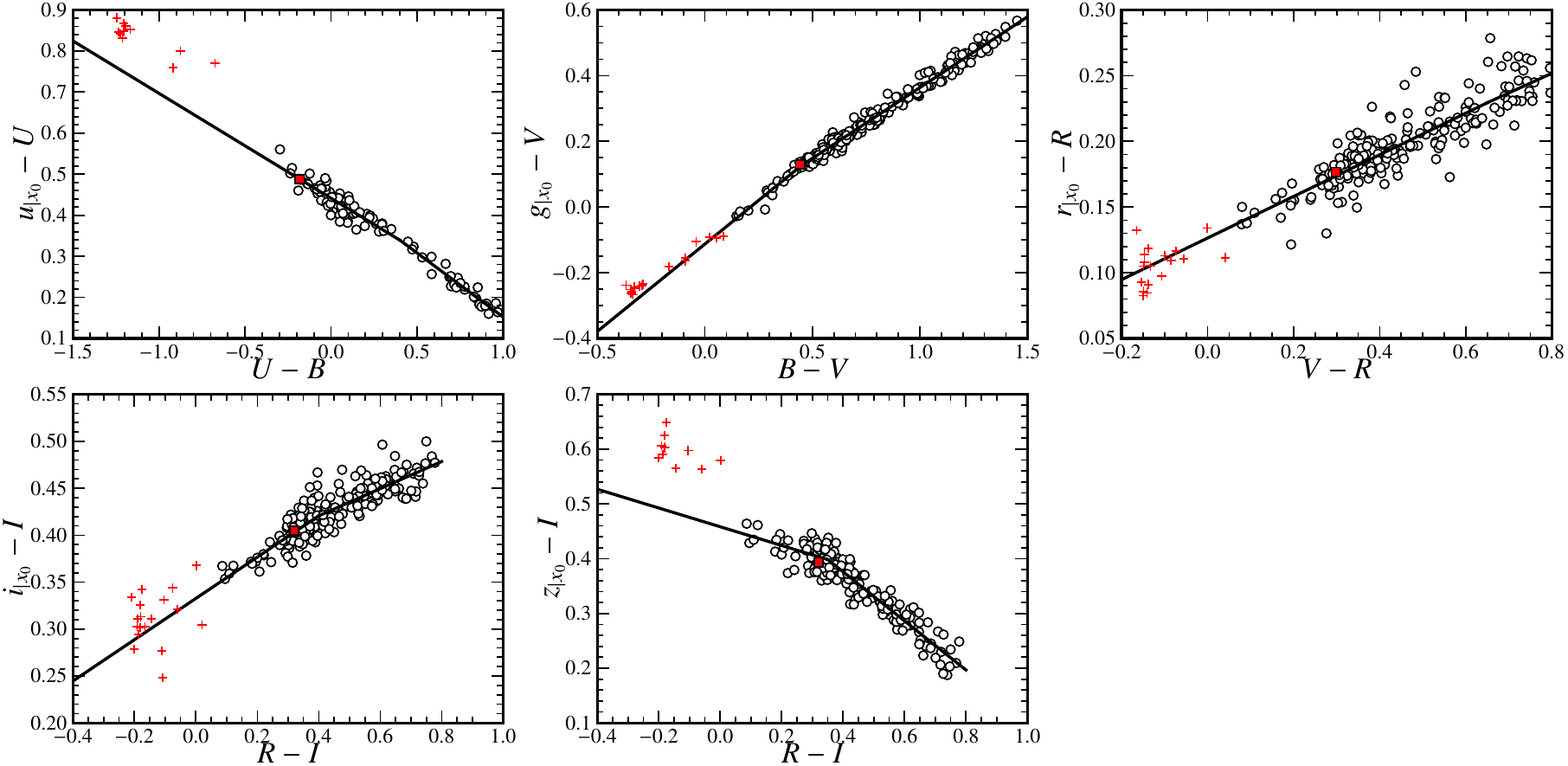}
  \caption{Color transformation between the Landolt system and the
    reference MegaCam system. 
    \Fixed{Open} black circles show the MegaCam and
    Landolt measurements of the Landolt secondary standards. The
    overall scale 
    is set by the magnitudes of the primary
    spectrophotometric standard star \bdtruc (\Fixed{solid} red square). 
    Measurements of \bdtruc magnitudes in the Landolt system
    are provided by \citet{2007AJ....133..768L}. 
    \Fixed{Since} \bdtruc cannot be
    directly observed with MegaCam, its MegaCam magnitudes are
    inferred from the average color transformation (black line)
    \Fixed{determined from}  
    the secondary standard measurements. 
    \Fixed{P}otential deviations from this average color transformation
    \Fixed{are examined}  
    in R09 (see Sect.~\ref{sec:absolute}). 
    Other \Fixed{primary standards from}
    \citet{2007AJ....133..768L} are displayed as red crosses;
    \Fixed{they are not}
    close enough to the bulk of secondary
    standards to be reliably transformed with the average
    transformation. Observations of the Landolt
    standards in \band{i2} were not numerous enough to deliver a
    robust measurement of the $\umag[i2]-I$ transformation (refer to
    Sect.~\ref{sec:interc-bandi-band} for the \band{i2} calibration).
  \label{fig:landoltcol}
}
\end{figure*}

The large number of observations allows us to solve
\Fixed{simultaneously} for the transformation and the \Fixed{exposure}
zero points.  \Fixed{We impose} the constraint that $\delta L = 0$ on
average, \Fixed{and find no} significant degeneracy between
parameters.  The complete model for \Fixed{a} photometrically
flat-fielded Landolt star \Fixed{measurement} is:
\begin{equation}
  \label{eq:22}
\begin{split}
  &\madu - \dk(x) (\umag[c] - c^0(x)) =\\ &L + \LtoM(C) + \delta L  - \zp_n -  2.5 \logdec(t_\text{exp}) - (X_t-1)k_{\rm atm}
\end{split}
\end{equation}
The relation $\LtoM$ is parameterized as:
\[\LtoM(C) = \left\lbrace
  \begin{array}{l}
    \alpha C \quad\text{if} \quad C < c_\text{break}\\
    \alpha c_\text{break} + \beta (C-c_\text{break}) \quad\text{otherwise.}\\
\end{array}
\right.
\]
The \Fixed{free parameters include the} two slopes for the piece-wise
linear color transform ($\alpha$ and $\beta$), the airmass correction
term $k_{\rm atm}$, the zero point for each night $\zp_n$, and the
deviation $\delta L$ for each star.  We impose \Fixed{the constraint}
$<\delta L > \approx 0$ by adding a prior term $\sum_s \delta
L^2/\sigma_s^2$ to the $\chi^2$ function.  We choose $\sigma_s = 0.01$
to match the observed dispersion of the MegaCam to Landolt relation.
The adjusted transformations are shown in Fig.~\ref{fig:landoltcol}.
Corresponding values for the transform parameters are given in
table~\ref{tab:ltom}.
\begin{table*}
  \centering
  \caption{Landolt to MegaCam average transformation parameters
  \Fixed{defined in Eq.~(\ref{eq:22})}.}
  \label{tab:ltom}
  
\begin{tabular}{llcrrr|rrr|r}
    \hline\hline
    band & color & color range  & $c_\text{break}$ & $\alpha$ & $\beta$ & $k_\text{atm}$\tablefootmark{a} & $k_\text{atm}$\tablefootmark{b} & $k_\text{atm}$\tablefootmark{c} & $\delta L$\tablefootmark{d} \\
    \hline

$\band{u}$ & $U-V$ & 0--1  & $0.40$   & $-0.2553 \pm 0.0082$  &  $-0.3120 \pm 0.0094$  &  $0.3075\pm 0.0036$ & $0.3052\pm 0.0028$ & $0.3161$ & -- \\
$\band{g}$ & $B-V$ & 0--2  & $0.45$   & $0.5320 \pm 0.0141$  &  $0.4309 \pm 0.0023$  &  $0.1235\pm 0.0006$ & $0.1296\pm 0.0011$ & $0.1422$ & $0.007$ \\
$\band{r}$ & $R-I$ & 0--0.8  & $0.65$   & $0.1583 \pm 0.0058$  &  $0.1493 \pm 0.0295$  &  $0.0629\pm 0.0007$ & $0.0762\pm 0.0010$ & $0.0785$ & $0.003$ \\
$\band{i}$ & $R-I$ & 0--0.8  & $0.40$   & $0.2200 \pm 0.0122$  &  $0.1446 \pm 0.0070$  &  $0.0309\pm 0.0007$ & $0.0402\pm 0.0010$ & $0.0321$ & $0.002$ \\
$\band{z}$ & $R-I$ & 0--0.8  & $0.35$   & $-0.1713 \pm 0.0267$  &  $-0.4483 \pm 0.0105$  &  $0.0154\pm 0.0014$ & $0.0367\pm 0.0015$ & $0.0206$ & $-0.010$ \\
    \hline
  \end{tabular}

  \tablefoot{\\
    \tablefoottext{a}{Average airmass term measured on data.}\\
    \tablefoottext{b}{Average airmass term measured on data for photometry in fixed aperture radii of 43 pixels. The difference between the two is attributed to correlation between airmass and aperture corrections.}\\
    \tablefoottext{c}{Airmass term according to the average extinction model at Mauna Kea from \cite{Buton2012}.}\\
    \tablefoottext{d}{Offset affecting the \bdtruc color transform derived in R09 table~7.}
  }
\end{table*}

The fit of Landolt observations delivers measurements of zero points
$\zp_n$ for a significant subset of the nights with observations of
the SNLS fields (see table~\ref{tab:landoltN}). These \Fixed{zero
  points} can be used to anchor the flux scale of the multi-epoch
catalogs of the 4 fields to a common scale (relative to the Landolt
system) according to the model given by
Eq.~(\ref{eq:15}).\footnote{Anchoring to the Landolt system also
  enables the use of spectrophotometric standards measured in the
  Landolt system to set the absolute scale of the catalogs (see
  Sect.~\ref{sec:absolute}).}  Non-photometric nights that were
discarded \Fixed{in} the \Fixed{analysis} of the multi-epoch catalog
(see Sect.~\ref{sec:catalog-constitution}) were \Fixed{also rejected}
in this averaging.  \Fixed{Since a} significant time \Fixed{period}
can occasionally separate the calibration exposures and the science
exposures, this cut does not exclude all calibration
\Fixed{observations} affected by non-photometric conditions. A more
robust assessment of the photometricity is possible when several
calibration exposures are available for a single night, as is the case
for about a third of the nights. We also discarded nights displaying a
zero point that is significantly lower (by more than 0.05 mag) than
the average zero point observed in a 10-night rolling average.
\Fixed{Finally, we discarded} nights displaying 3~$\sigma$ zero point
outliers when applying the calibration model from Eq.~(\ref{eq:15}) to
calibrate the deep field exposures.  Table~\ref{tab:landoltN}
summarizes the number of photometric nights with concomitant
observations of the Landolt and science fields, the observed
dispersion between the independent measurements of the catalog scale
provided by each night, and the resulting statistical uncertainty.
The smaller number \Fixed{of \band{z} nights} is partly related to
fewer Landolt observations in \band z and cuts.  Observations in
\band{u} were not part of the supernova survey and were not cadenced.
\begin{table}
  \centering
  \caption{Statistical uncertainty (in magnitudes) on the calibration to the Landolt system.}
  \label{tab:landoltN}
  
\begin{tabular}{c|*{3}{c}|*{3}{c}}
\hline
\hline
& & D1 & &  & D2 \\
band & \# nights & $\sigma$ & $\sigma/\sqrt{N}$  & \# nights & $\sigma$ & $\sigma/\sqrt{N}$ \\
\hline

$u_M$
 & 
7 & 0.007 & 0.003
 & 
1 & 0.000 & 0.000
\\
$g_M$
 & 
53 & 0.008 & 0.001
 & 
44 & 0.011 & 0.002
\\
$r_M$
 & 
63 & 0.009 & 0.001
 & 
64 & 0.011 & 0.001
\\
$i_M$
 & 
68 & 0.010 & 0.001
 & 
57 & 0.008 & 0.001
\\
$z_M$
 & 
30 & 0.014 & 0.002
 & 
25 & 0.012 & 0.002
\\
\hline
& & D3 & &  & D4 \\
\hline

$u_M$
 & 
8 & 0.010 & 0.004
 & 
8 & 0.015 & 0.005
\\
$g_M$
 & 
56 & 0.010 & 0.001
 & 
55 & 0.009 & 0.001
\\
$r_M$
 & 
74 & 0.011 & 0.001
 & 
64 & 0.014 & 0.002
\\
$i_M$
 & 
63 & 0.012 & 0.001
 & 
51 & 0.010 & 0.001
\\
$z_M$
 & 
25 & 0.013 & 0.003
 & 
27 & 0.013 & 0.002
\\
\hline
\end{tabular}
\end{table}

\subsection{Systematic uncertainties in the calibration to the Landolt
  system}
\label{sec:syst-uncert-calibr}
We now turn to a review of potential sources of systematic
\Fixed{uncertainty} in the calibration path through Landolt
observations.  \Fixed{Here} we only discuss uncertainties that affect
the determination of the absolute flux scale of the tertiary catalog.
\Fixed{These uncertainties} are summarized in
Table~\ref{tab:snlserrorbudget}.

\subsubsection{Systematic differences in aperture corrections}
\label{sec:aperture-corrections}

The first \Fixed{is} from potential systematic differences in the
photometry of standard and tertiary stars.  The brightness of standard
stars requires a short integration time (2--3~s) \Fixed{to avoid
  saturation}.  During the course of the survey, a small defocus
(\Fixed{smearing the PSF core by about 1 pixel}) on Landolt fields has
also been applied to \Fixed{observe} the brightest secondary standards
of the \cite{2002AJ....123.2121S} set. It is likely that this
\Fixed{defocus} introduces systematic differences between the
effective PSF of tertiary star observations and primary or secondary
standard observations, resulting in slightly different aperture
corrections.

To first order, PSF variations are approximately accounted for by the
scaling of aperture radius in our photometry method. We statistically
investigated deviations to this correction by computing aperture
corrections between aperture of radius 7.5 and 20 times the image
quality.  We found a remaining variation of aperture corrections with
image quality, and offsets between Landolt and deep field exposures
dependent on the defocusing of Landolt observations.

\begin{figure}
  \centering
  \igraph{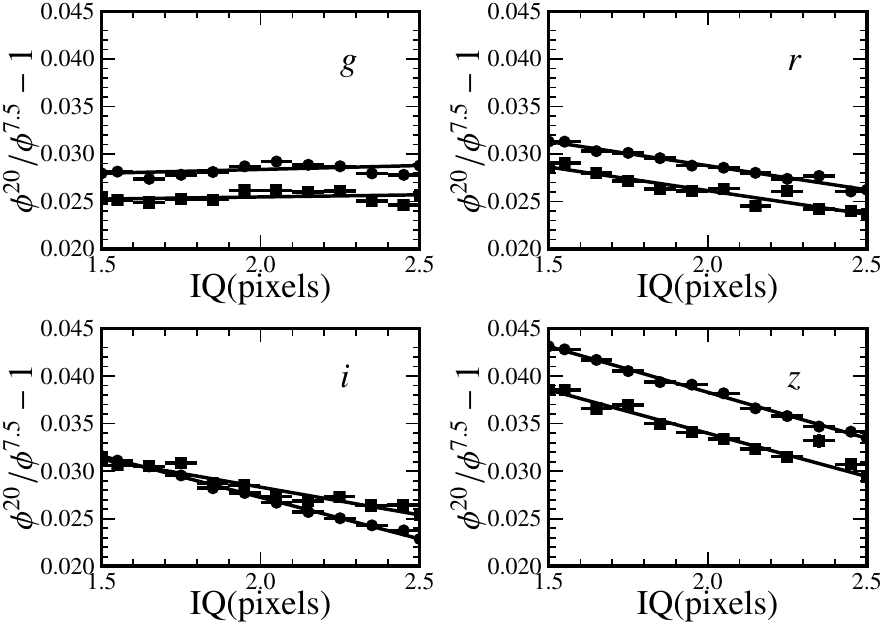}
  \caption{\Fixed{Variation} of aperture correction from small to
    large apertures as a function of the image quality, \Fixed{IQ,
      defined as} sigma of a Gaussian fit to the PSF \Fixed{(in
      pixels)}.  Apertures are scaled with the IQ: small apertures
    correspond to a radius of $7.5\times$IQ and large apertures to a
    radius of $20\times$IQ. The aperture corrections are computed as
    $\phi^{7.5}/\phi^{20} -1$ where $\phi^{7.5}$ and $\phi^{20}$
    \Fixed{denote} instrumental \Fixed{fluxes} in the small and large
    aperture, respectively.  Data points correspond to the median of
    aperture corrections in bins of IQ over all the exposures in the
    survey. Results for short (and slightly out-of-focus) exposures on
    Landolt fields (black circles) and long exposures on deep fields
    (black squares) are displayed separately.}
  \label{fig:seeing_cor}
\end{figure}

The change in aperture corrections with image quality is illustrated
in Fig.~\ref{fig:seeing_cor}.  The \Fixed{changes for tertiary
  standards and Landolt stars are} similar, \Fixed{except for} the
offset related to defocusing.  \Fixed{The IQ dependence is}
approximately linear with a small positive slope in \Fixed{the} \band
g \Fixed{band} and a negative slope in \Fixed{the} other bands.

\begin{figure}
  \centering
  \igraph{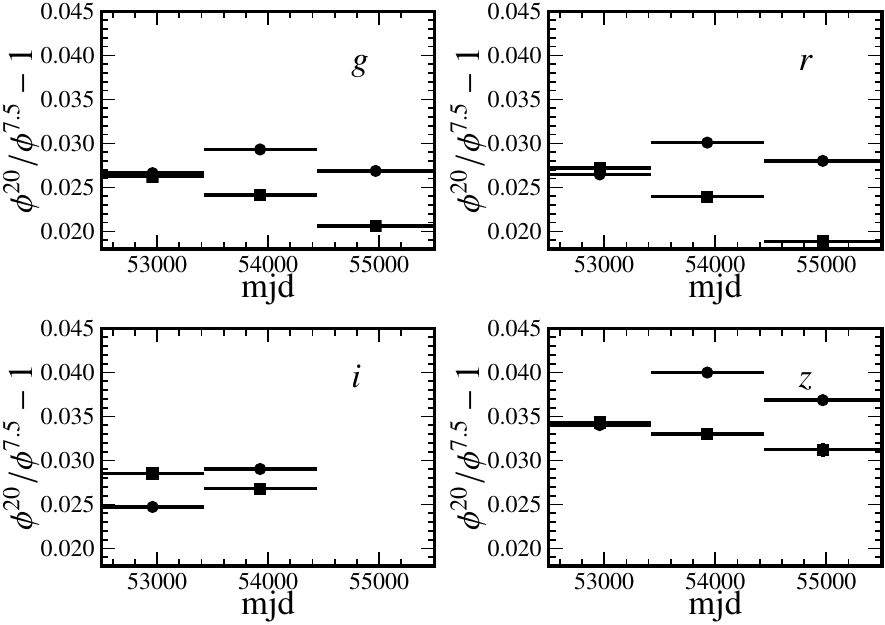}
  \caption{Average aperture correction 
    (between 7.5 and 20 IQ radius)
    \Fixed{is shown for the} 3 periods corresponding to different 
    observational setups. 
    Landolt stars (short exposures, slightly out of focus)
    \Fixed{are shown in black circles},
    and deel fields (long exposures)
    \Fixed{are shown in black squares.}
    The statistical uncertainties on the averages are negligible.}
  \label{fig:focus_cor}
\end{figure}

The observation strategy of the Landolt fields slightly evolved over
the course of the survey. Calibration stars were originally exposed
\Fixed{for} 3~s with the telescope in focus.  The exposure time was
reduced to 2~s in February 2005 and the telescope focus was offset by
0.5~mm with respect to the nominal focus to avoid saturation of the
brightest stars.  The amount of defocus was \Fixed{later} reduced to
0.3~mm in August 2007.  \Fixed{We therefore have} three distinct time
periods corresponding to different observational setups. We computed
the average aperture correction offset between Landolt and deep field
observations in bins corresponding to each time period.  The measured
offsets, shown in Fig.~\ref{fig:focus_cor}, \Fixed{were applied to}
the calibrated magnitudes to correct for the systematic differences on
aperture corrections in the determination of the zero points.

A \Fixed{reliable} estimate of the error on this correction is
difficult to obtain.  \Fixed{The correction is based} on the
assumption that the total flux loss is indeed proportional to the
aperture correction, or in other words, that the fraction of flux
captured in the larger aperture is stable. Whether this assumption is
verified or not is a tricky question.  A model independent hint on the
accuracy of this correction is \Fixed{indicated} by the agreement
between the three \Fixed{time periods} once the correction is applied,
and by the residual dispersion of the calibration as a function of
seeing difference. The different periods are statistically compatible.
\Fixed{To be} conservative, we took the difference between the two
periods with the greater statistical accuracy (2003-2005 and
2005-2007) as a reasonable upper bound on the systematic related to
aperture correction.

\subsubsection{Flat-field error}
\label{sec:flat-field-error}

\begin{figure}
  \centering
  \igraph{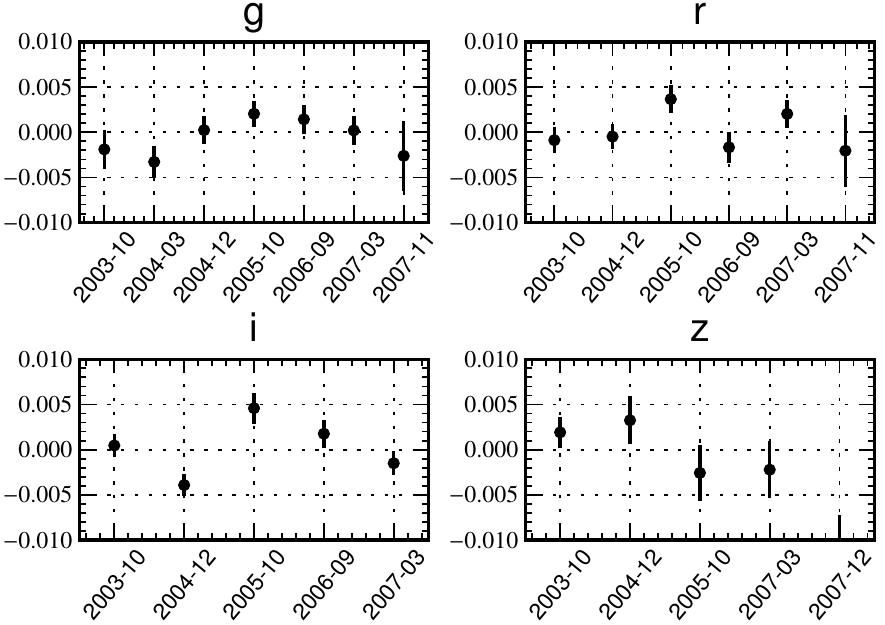}
  \caption{Catalog zero point as a function of the grid solution.  
    The error bars reflect the statistical uncertainty on the 
    mean estimated from the
    dispersion of measurements in the period.}
  \label{fig:griddisp}
\end{figure}

The first effect of the flat-field error is to \Fixed{degrade} the
uniformity of the tertiary catalogs as discussed in
Sect.~\ref{sec:catalog-constitution} and \Fixed{in}
Appendix~\ref{sec:more-megac-flatf}.  It can also result in an error
on the average calibration because secondary standards are observed
preferentially at the center of the focal plane. The Landolt
calibration is made with sufficient statistics to provide a strict
limit on \Fixed{this} effect.  The dispersion of the calibration
residuals as a function of the flat-field solution applied is shown in
Fig.~\ref{fig:griddisp}. Each point on the plot corresponds to a
mostly independent realization of the SNLS calibration, as it involves
disjoint data samples for both the determination of the flat-field and
the zero points. The measured dispersion is between 1 and 3~mmag in
all bands. This dispersion is dominated by the statistical uncertainty
on each point, \Fixed{and} can be used to set an upper bound on the
contribution from the flat-field error.  \Fixed{Since} the flat-field
error is likely to be uncorrelated between different observations of
the grid fields (\emph{cf.} Sect.~\ref{sec:instr-model-assessm}),
\Fixed{w}e divided the measured dispersion by the square-root of the
number of grid observations to estimate the flat-field error in
Table~\ref{tab:snlserrorbudget}.

\subsubsection{Shutter precision}
\label{sec:other-probl-phot}

The precision of the MegaPrime shutter could bias the photometric 
ratio between short and long exposures. 
The uncertainty on the exposure time was 
estimated in R09 to be less than $3$~ms. 
\Fixed{For photometric standards exposed for 2~s,
this uncertainty affects the flux by at most $0.15\%$.}

\subsubsection{Residual background contamination}
\label{sec:resid-backgr-cont}

The \Fixed{sky} background subtraction is a difficult problem for
aperture photometry. Our background estimate is based on the
distribution of image pixels in tiles of $256\times256$ pixels after
application of a mask discarding objects up to a given contamination
level. The final estimate is a smooth interpolated version of the
resulting low-resolution background map. Higher frequency structure
remains in the background subtracted maps, in particular, \Fixed{far
  in the} PSF tails from bright objects.  \Fixed{Since R09 showed}
significant hints for background residuals in the aperture photometry
of tertiary stars, we investigated the origin of this bias.

\Fixed{The basic idea is the following: structure in the background is
  expected at scales smaller than $256$~pixels. Important contributors
  to such structure are PSF tails from bright objects and residuals
  from fringe subtraction in \Fixed{the} $i$ and $z$ bands. These
  structures remain in the background subtracted images and
  contaminate aperture photometry. This is an intrinsic limitation of
  the aperture photometry method which assumes isolated objects on a
  background with noise expected to be zero on average, an hypothesis
  not verified in practice.}
 
\Fixed{PSF tails are particularly annoying in our case as they make
  the average level of aperture contamination dependent on the star
  location. For our tertiary stars, which are selected to be fairly
  isolated objects, one expects the average contamination of apertures
  to be lower than for randomly distributed objects.  Therefore, our
  hypothesis to explain the bias spotted in R09 is that the background
  subtraction algorithm is ill-suited to the specific locations of our
  tertiary stars (typically an overestimate as they are isolated).}

\Fixed{We seek a quantitative estimate of this effect. Let us denote
  $b(\x)$ the residual background level at a given position $\x$ in
  the image \emph{after background subtraction}, and $C(\x)$ the flux
  carried to this position by the PSF tails of neighboring
  objects. According to the discussion above, $b$ and $C$ are
  related. This relation is rather complex as the specifics of the
  background subtraction algorithm comes into play to suppress the
  large scales of the structure. In addition, the value of $C$ is
  difficult to obtain because it requires knowledge of the PSF far out
  in the tails.  Estimating the effect from first principles is thus
  not feasible.  }

\Fixed{We can characterize the residual background statistically by
  measuring an empirical relation between $b$ and a quantity crudely
  related to $C$.  We define a contamination index $\tilde C(\x)$ as:}
\begin{equation}
  \tilde C(\x)= \sum_o \Phi_o  \psi(\x-\x_o)\label{eq:5}
\end{equation}
\Fixed{where $\psi$ is an approximation of the PSF normalized to 1,
  $o$ runs across objects detected in the image and {$\Phi_o$} is the
  flux of object $o$.  The actual value of $\tilde C$ does not {have a
    clear physical interpretation because} it depends on the number of
  objects included in the sum and on the approximation used to
  describe the PSF.  We can expect, however, that positions
  corresponding to larger $\tilde C$ correspond to positions with
  greater residuals.}

\begin{figure}
  \centering
  \igraph{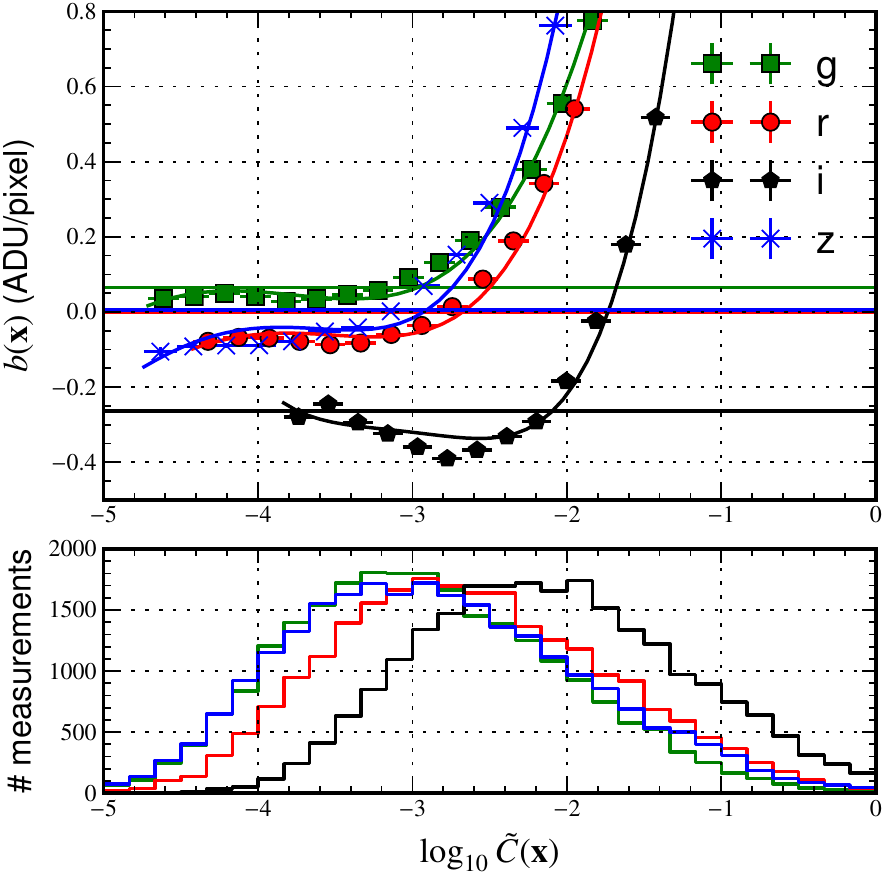}
  \caption{Background residuals as a function of the contamination
    index. The contamination index at a given location for a given
    aperture is defined as the flux in the aperture expected from
    objects detected in the field, assuming they have a Moffat PSF
    profile  (see Eq.~\ref{eq:5}). 
    The top figure displays the average background
      level (in ADU per pixel) in random apertures as a function of
      this contamination index. We concentrate on random positions
    with low contamination index (apertures expected to be mostly
    empty), sampling the background level far from detected
    objects. The shape of the curves reflect the remaining structure
    in the background due to PSF tails. The bottom plot displays the
    distribution of the contamination indexes at the position of
    tertiary stars (excluding the contribution from the tertiary star
    itself). Horizontal lines in the top plot display the average
    value of the curve weighted for tertiaries, i.e. the expected
    per-pixel background offset for tertiaries.
  }
  \label{fig:backstud}
\end{figure}

\Fixed{To determine the empirical relation between $b$ and $\tilde C$,
  we sampled the deep field images in the survey at uniformly
  \Fixed{spaced} positions. For each position $\x$ we compute
  \Fixed{two quantities.}  \Fixed{First we estimate $b(\x)$ from} the
  average residual sky level (after background subtraction) in a
  $32\times32$ pixel aperture centered at this position.
  \Fixed{Second,} the contamination index $\tilde C(\x)$ \Fixed{is
    computed} under the assumption that the PSF follows a Moffat
  function \citep{1969A&A.....3..455M} with parameters adjusted on the
  images.  The top panel of Fig.~\ref{fig:backstud} \Fixed{shows}
  clear evidence \Fixed{for} the dependence between the residual
  background level and the isolation of the location.  }

\Fixed{The average contamination in tertiary standard measurements is
  obtained as follows: we first compute the contamination index
  \Fixed{(Eq.~\ref{eq:5})} for the positions of our tertiary
  standards, excluding the contribution of the tertiary star itself.
  We then convert \Fixed{each} contamination \Fixed{index} into
  \Fixed{a} residual background level according to the relation
  \Fixed{shown} in Fig.~\ref{fig:backstud}, and compute the average
  over all tertiary star positions.  The distribution of the
  contamination index of tertiary stars is shown on the bottom panel
  of Fig.~\ref{fig:backstud}. As expected, the tertiary standards are
  well isolated objects with \Fixed{other image objects adding less
    than} a fraction of \Fixed{an} ADU per pixel.  We found an average
  of $0.06$, $0$, $-0.26$ and $0.01$ ADU/pixel, respectively, in
  \Fixed{the} \band{g}\band{r}\band{i} and \band{z} bands.}

\Fixed{Those \Fixed{contamination} numbers are in perfect agreement with 
the estimate of
  background residuals obtained in R09 Sect.~(4.2) by studying aperture
  corrections as a function of flux. We thus conclude that the
  proposed mechanism explains the observed background subtraction
  bias. To account for the resulting bias on aperture magnitudes, we
  subtracted the average contamination level to the aperture flux
  measurements. Note that this is only an average correction,
  accounting for the overall bias introduced by background
  residuals. Contamination of apertures by background structure still
  contributes noise on individual aperture measurements. The
  distribution of this noise has a positive skewness (whose
  significance depends on the magnitude range considered) and some
  caution should be exercised when using the aperture catalog for
  calibration.
}

\begin{figure}
  \centering
  \igraph{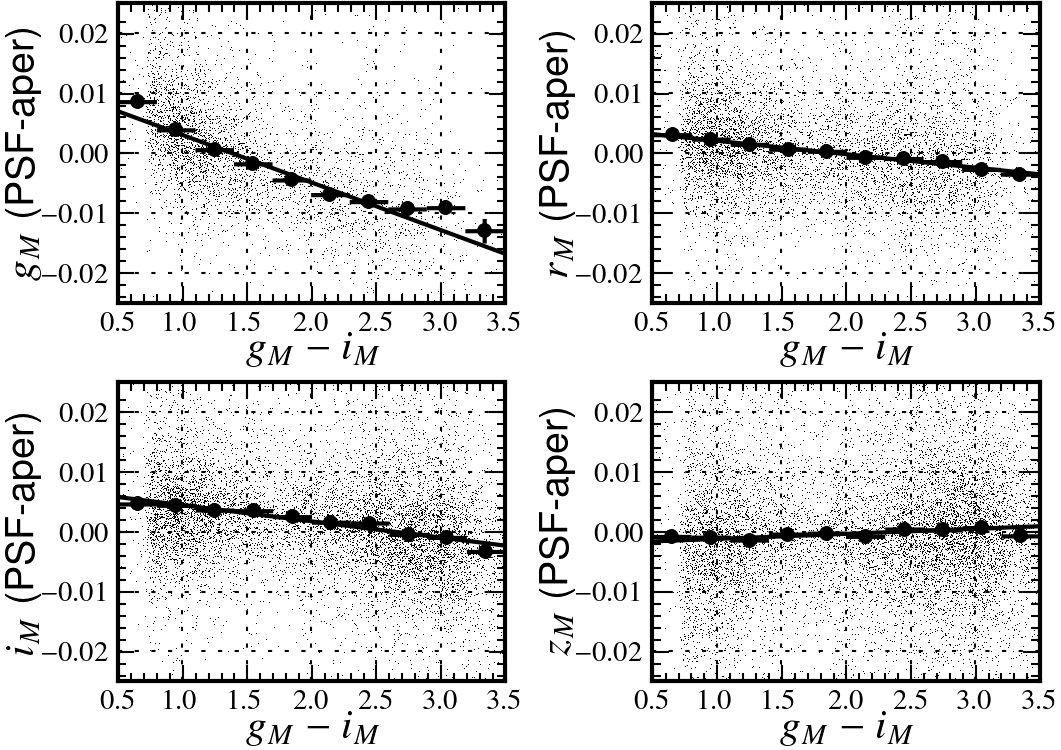}
  \\[1em]
  \caption{Difference between PSF and aperture photometry for tertiary
    stars as a function of star color. The black line displays a
    linear fit of the chromatic trend observed.}
  \label{fig:psfcomp_color}
\end{figure}

\Fixed{To illustrate the amplitude of the remaining effect, we
  \Fixed{give} a comparison of catalog magnitudes with PSF photometry.
  The PSF photometry \Fixed{software} independently adjusts and
  subtracts a background offset for each measurement, and is thus
  insensitive to residual background fluctuations.  A known chromatic
  effect (see \citealt{2010A&A...523A...7G}), attributed to
  \Fixed{wavelength-dependent} changes \Fixed{in} the PSF \Fixed{that
    were} neglected in the PSF photometry method,\footnote{This is
    handled in SNLS by interpreting the PSF magnitudes in a modified
    photometric system (see \citealt{2010A&A...523A...7G} for further
    discussion).} has to be accounted for before performing the
  comparison. The dispersion of the difference between aperture and
  PSF photometry of the tertiary stars as a function of the star color
  are displayed \Fixed{in} Fig.~\ref{fig:psfcomp_color}. The chromatic
  difference is \Fixed{easily seen} in \Fixed{the} \band{g} band.
  Smaller \Fixed{but} significant trends are also visible in
  \Fixed{the} \band{r} and \band{i} bands.  }

\begin{figure}
  \centering
  \igraph{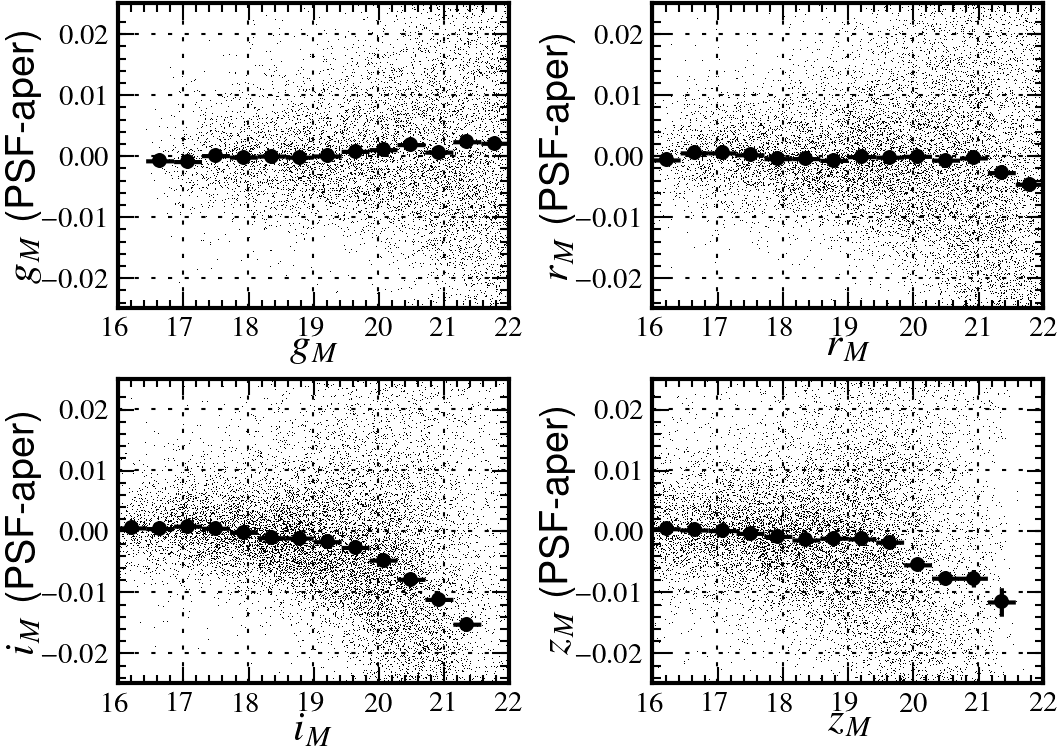}
  \caption{Difference between PSF and aperture photometry for tertiary
    stars as a function of star magnitudes (with the chromatic
      trend from Fig.~\ref{fig:psfcomp_color} and the average
      background contamination of apertures subtracted). 
      Black circles display the average of data points in bins of 
      magnitudes. Averages are computed with a clipping at 2.5$\sigma$ 
      as used in the calibration of SNLS supernovae 
      light-curves \citep{2010A&A...523A...7G}. }
  \label{fig:psfcomp_mag}
\end{figure}

\Fixed{After removing the chromatic trend and correcting the aperture
  magnitudes for the average background contamination computed above,
  \Fixed{the} PSF and aperture magnitudes \Fixed{agree} as a function
  of the star magnitude.  As illustrated in
  Fig.~\ref{fig:psfcomp_mag}, \Fixed{there is excellent} agreement in
  the magnitude range 16--21 \Fixed{for the $\band g$ and $\band r$
    bands}, indicating that the average \Fixed{aperture} contamination
  is mostly removed.  The range of agreement appears a bit smaller in
  \Fixed{the} \band i \Fixed{and} \band{z} \Fixed{bands: 16--19.5.}
  \Fixed{To obtain the} most precise calibration, we suggest
  \Fixed{using} stars \Fixed{only} in \Fixed{the above} confidence
  ranges.}

Applying the same analysis to the Landolt measurements, we found that
the residual background level in short time exposures is consistently
positive and varies around a mean value of 0.1 ADU/pixels. This is
attributed to a small biasing of our background estimator, which is
based on a combination of the mean and the median of pixel values. It
delivers biased estimates on the asymmetric distribution of background
pixels in short exposures. The impact of such a bias on the
determination of zero points is smaller than 1~mmag and is included in
the systematic error budget.


\subsubsection{Survey uniformity}
\label{sec:second-stars-unif}

The uniformity of the photometry between the 4 SN fields relies on the
uniformity of the \Fixed{photometry of}
Landolt equatorial standards.  The dispersion of
zero points between individual Landolt 
\Fixed{fields} has been estimated in
R09 to be 0.002 in $gri$ and $0.004$ in $z$. 
As the \Fixed{true} error is likely to have some spatial correlation, 
we \Fixed{use these} numbers as estimates of
the systematic \Fixed{uncertainty} induced by non uniformity in the secondary
network. Note that \Fixed{this error} potentially affects the 
average calibration as the photometric standard observations are 
calibrated against a subset of the secondary standards.

\begin{figure*}
  \centering
  \igraph{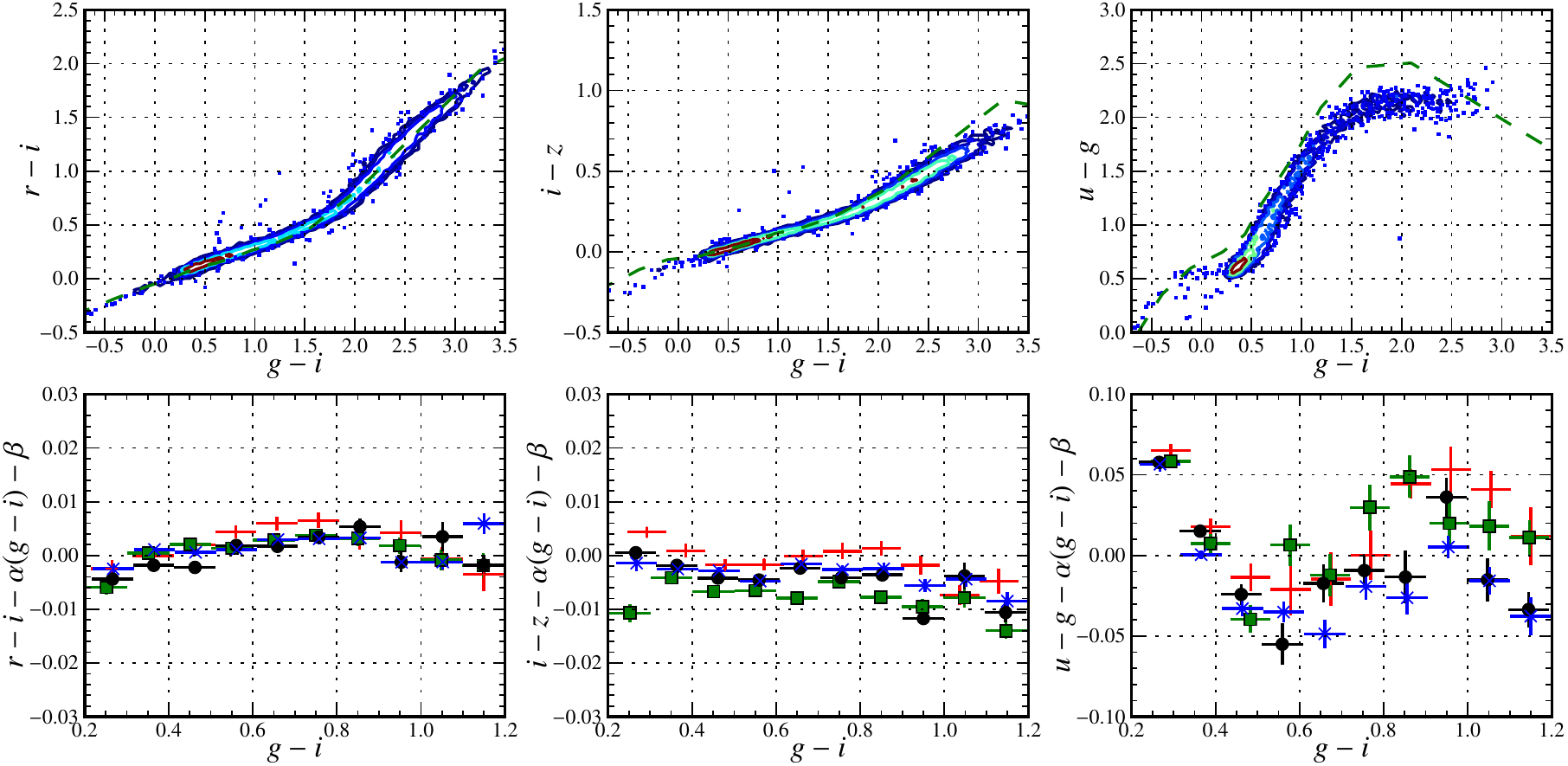}
  \caption{\emph{Top:} Color-color diagram of the 4 SNLS tertiary star
    catalogs. \emph{Bottom:} Residuals to a linear fit of the average
    stellar locus in the range $0.2<g-i<1.2$. The residuals for the 4
    deep fields are displayed separately 
    (D1: red \Fixed{crosses}, 
    D2: green squares, 
    D3: black circles, 
    D4: blue \Fixed{asterisks}). }
  \label{fig:colorcolor}
\end{figure*}

The uniformity is confirmed by the generally good agreement of stellar
locus between fields as displayed \Fixed{in} Fig.~\ref{fig:colorcolor}. 
The dispersion of the stellar locus is compatible with the statistical
uncertainty on the zero point in each field.

\section{MegaCam absolute calibration measurements.}
\label{sec:megac-absol-calibr}

The \Fixed{straightforward way to set the} physical flux scale of the star catalog
\Fixed{is to equate synthetic magnitudes of spectrophotometric
  standard stars with their corresponding magnitudes in the system
  defined by the catalog}. Unfortunately, \Fixed{no such standards are
  currently} present in the SNLS fields, although several pure
hydrogen white-dwarves have been identified
\citep{2008MNRAS.383..957L,2012arXiv1205.1306L} that might serve this
purpose in the future. For now, \Fixed{our absolute flux calibration
  relies} on external observations of spectrophotometric standard
stars.

\Fixed{Measurements from the Hubble Space Telescope imaging spectrograph 
(STIS) provide} the best available set of spectrophotometric standard stars 
in the visible part of the \Fixed{spectrum}. The calibrated spectra of HST 
fundamental standards are \Fixed{part of} the CALSPEC database. Most of 
\Fixed{these} stars are too bright \Fixed{for MegaCam to routinely and 
reliably observe}. \Fixed{Therefore, until now, observations of the Landolt 
equatorial secondary standards \citep{landolt_ubvri_1992} have} been the 
preferred \Fixed{method to set} the \Fixed{SNLS field} absolute calibration,
with anchoring to the HST flux scale \Fixed{indirectly provided by} 
\citet{2007AJ....133..768L} observations of HST standards in the Landolt system.

\Fixed{However, recent work \citep{bohlin_nicmos_2008} has extended the CALSPEC 
database to include redder and fainter standards. Direct MegaCam observation of 
these new standards opens an alternate calibration path for the SNLS fields}.

We start this section by deriving AB offsets for the tertiary catalogs
from the Landolt calibration observations. We then explore the
possibility of direct calibration through observations of three solar
analogs. \Fixed{Cross-calibration with the SDSS, presented in 
Sect.~\ref{sec:sdsssnls-direct-inte}, offers yet a third calibration 
alternative. We devote section~\ref{sec:concl} to the comparison of 
these various options}.

\subsection{Absolute calibration through Landolt observations}
\label{sec:absolute}

\subsubsection{Transformation of Landolt standard measurements}
\label{sec:transf-land-stand}

\Fixed{We have used MegaCam observations of Landolt secondary 
standards to adjust the color relation shown in Eq.~(\ref{eq:21}).
This relation} can be used to transform \Fixed{Landolt system 
measurements of stars (particularly} measurements of standard stars) 
to the \Fixed{tertiary standard-defined} SNLS system.  Calibration 
constraints can be obtained \Fixed{by applying} this relation to a 
spectrophotometric standard whose expected MegaCam AB magnitude $\umag[\tilde m]$ 
can be synthesized from the calibrated CALSPEC spectrum. 
This \Fixed{gives}:
\begin{equation}
  \label{eq:landaboffset}
  \delta_\ab =  L + \LtoM(C) + \delta L - \umag[\tilde m]\,.
\end{equation}

\Fixed{Any of the \cite{2007AJ....133..768L} Primary Spectrophotometric 
Standard Stars (hereafter PSSS) can be used to set the unknown calibration 
coefficient $\delta_\ab$, provided the corresponding $\delta L$ are known}. 
The red crosses on Fig.~\ref{fig:landoltcol} show primary standards from 
the CALSPEC database with accurate \Fixed{Landolt system} magnitudes.

The average color transformation $\LtoM$ is adjusted so that
$\langle\delta L\rangle=0$ on average for Landolt secondary stars.  As
\Fixed{shown in} Fig.~\ref{fig:landoltcol}, the only
spectrophotometric standard with Landolt magnitudes sufficiently close
in color to the average of Landolt secondary standards (so that the
corresponding $\delta L$ is expected to be small) is \bdtruc.
\Fixed{R09 has done a careful estimate of the \bdtruc $\delta L$
  values, reproduced here in table~\ref{tab:ltom}. We use these values
  to} compute estimates of the calibration offsets according to
Eq.~(\ref{eq:landaboffset}).  \Fixed{These calibration offsets} are
given in table~\ref{tab:SA}.

\subsubsection{Uncertainties on the MegaCam magnitudes of \bdtruc}
\label{sec:color-transf-bdtr}

The dominant contribution to the systematic error budget is the
uncertainty on \Fixed{$\delta L$, the term quantifying a star's departure 
from the average} transformation between the Landolt system and the MegaCam
system. Intrinsic dispersion around the average transformation is
expected due to variations in the SED of individual stars. Without any
further assumptions\Fixed{,} the dispersion of Landolt secondary standards 
around the average color transformation \Fixed{gives an estimate of the related 
uncertainty. Dispersion} ranges from $13$~mmag rms in band $\band{g}$ to
$20$~mmag in band \band{z}.

As already stated, a more precise determination of $\delta L$,
\emph{i.e.} of the exact position of \bdtruc with respect to the bulk
of the Landolt secondaries in the \Fixed{Fig.~\ref{fig:landoltcol}
  color-color diagram}, was attempted in R09. \Fixed{This work used
  synthetic photometry on stellar libraries} to model the small
deviations induced by \Fixed{specific} properties of \bdtruc
(extinction, metallicity, $\log g$, binarity) with respect to the mean
properties of the Landolt standards, leading theoretically to a
precise determination of $\delta L$ \Fixed{and} hence, to accurate
magnitudes for the standards in the MegaCam system.  Unfortunately a
few contributions, like the effect of a faint companion to \bdtruc,
\Fixed{were} difficult to estimate \Fixed{accurately} and quite large
uncertainties remain on the \Fixed{\bdtruc MegaCam magnitudes},
especially in \band z.  \Fixed{The R09 $\delta L$ values are given in
  the last column of table~\ref{tab:ltom}; R09 uncertainty estimates
  on $\delta L$ are shown in the `Landolt to Megacam transformation'
  row of table~\ref{tab:snlserrorbudget}.}

\Fixed{In band \band{u} the value of $\delta L$ has very large uncertainty, 
large enough that its determination was not even attempted by R09}. 
\Fixed{Consequently,} we quote the dispersion of the secondary standards around
the color transformation as the uncertainty on the MegaCam magnitude
of \bdtruc.\footnote{The dispersion is measured on secondary stars in
  a limited color-range enclosing the color of \bdtruc.} It amounts to
0.02 magnitudes in \band u.

The Landolt measurements of \bdtruc magnitudes are also affected by a
small statistical uncertainty (about 0.002) that we include in the
systematic error budget.  The uncertainties affecting the \Fixed{Landolt 
observations calibration method} (including relevant uncertainties
discussed in Sect.~\ref{sec:syst-uncert-calibr}) are summarized in
Table~\ref{tab:snlserrorbudget}.

\begin{table*}
  \centering
  \caption{Uncertainty budget in the cross-calibration between SNLS and the HST via the observation of Landolt secondary standards.}
  \label{tab:snlserrorbudget}
  
\begin{tabular}{l*{5}{r}}
\hline\hline
 & \band{u} & \band{g} & \band{r} & \band{i}\tablefootmark{a} & \band{z}\\
\hline
Aperture corrections & $ \pm 0.0045 $ & $ \pm 0.0023 $ & $ \pm 0.0014 $ & $ < 0.001$ & $ \pm 0.0036 $\\
Flat-field error & $ \pm 0.0012 $ & $ < 0.001$ & $ < 0.001$ & $ < 0.001$ & $ < 0.001$\\
Landolt uniformity & $ \pm 0.0040 $ & $ \pm 0.0020 $ & $ \pm 0.0020 $ & $ \pm 0.0020 $ & $ \pm 0.0040 $\\
Shutter precision & $ \pm 0.0015 $ & $ \pm 0.0015 $ & $ \pm 0.0015 $ & $ \pm 0.0015 $ & $ \pm 0.0015 $\\
Background residuals & $ \pm 0.0010 $ & $ \pm 0.0010 $ & $ \pm 0.0010 $ & $ \pm 0.0010 $ & $ \pm 0.0010 $\\
BD +17 4708 landolt measurements & $ \pm 0.0034 $ & $ \pm 0.0027 $ & $ \pm 0.0028 $ & $ \pm 0.0028 $ & $ \pm 0.0030 $\\
Landolt to MegaCam transformation & $ \pm 0.0228 $ & $ \pm 0.0023 $ & $ \pm 0.0043 $ & $ \pm 0.0026 $ & $ \pm 0.0178 $\\
\hline
Total & $ \pm 0.0239 $ & $ \pm 0.0050 $ & $ \pm 0.0060 $ & $ \pm 0.0047 $ & $ \pm 0.0189 $\\
\hline
\end{tabular}
  \tablefoot{
    \tablefoottext{a}{The \band{i2} band can be calibrated relatively to \band{i} with a precision of $2$~mmag (\emph{cf.} Sect.~\ref{sec:absolute}).}}
\end{table*}

\subsection{Direct calibration to the HST}
\label{sec:direct-hst-standard}

\subsubsection{Observations of 3 solar analogs.}
\label{sec:observations-3-solar}

The \Fixed{addition} of fainter standards \Fixed{to} the CALSPEC database 
\Fixed{raises the prospect of a more straightforward calibration method}.
We investigated the possibility of using direct observations of the faintest PSSS to set
the photometric calibration of the SNLS. Three stars\Fixed{, SNAP-2, P177D and P330E,} 
lie close enough to the \Fixed{D3} field to be easily observed \Fixed{together} in a single 
short observing block with a potentially small airmass lever arm. 

Observations were conducted at CFHT in June 2011 in short sequences\Fixed{,} 
with exposures \Fixed{of} standard stars bracketing a single exposure on
D3. \Fixed{To avoid saturation and to study the behavior of aperture corrections
on out-of-focus exposures, two exposures (one in focus, one out-of-focus)
were taken of the brighter stars P177D and P330E}. Exposures on SNAP-2 begin and 
close the sequence, \Fixed{enabling} the monitoring of \Fixed{atmospheric
extinction stability}. Standard fields are exposed for 3~s\Fixed{; D3 fields are exposed 
for 120~s}.

Images were processed by the standard Elixir pipeline and aperture
photometry was performed in several predefined radii. Images were
flat-fielded using the appropriate grid solution. Standard stars were
observed on CCD \#22. \Fixed{D3 flux measurements} are matched
with the tertiary star catalog and used to determine\Fixed{d} a zero point for
the sequence. The zero point is \Fixed{then} used to calibrate the \Fixed{standard
 star measurements}, correcting for the effect of different airmass\Fixed{es}.
\Fixed{This allows us to determine} the offset between the Landolt based calibration 
and an AB system by:
\begin{equation}
  \label{eq:32}
  \delta_\ab = -2.5\logdec(\madu[\phi]/t_\text{exp}) + \zp + k_\text{atm} (X_D - X_S) - \cmag* 
\end{equation}
where $\madu[\phi]/t_\text{exp}$ is the instrumental aperture flux
corrected for PSF variations between the current exposure and the deep
field exposure on which the sequence zero point $\zp$ has been
determined, $X_D$ and $X_s$ are the respective airmass\Fixed{es} of the deep
field and the standard star, $k_\text{atm}$ is the average linear
correction for atmospheric extinction and \cmag* is the natural AB
magnitude of the standard star at the observed position synthesized
from its CALSPEC spectrum. Again, we use the notation \cmag* in place
of \cmag to distinguish synthetic from measured magnitudes.

\subsubsection{Error model}
\label{sec:error-model}

We build an error model for \Fixed{these} measurements by considering the
following sources of uncertainty: the photon noise of the standard
star measurements, the flat-field noise, the tertiary measurement noise, 
the time variation of the zero point during the sequence and the uncertainty 
on the airmass correction. We are only interested in uncertainties affecting 
the flux measurements, for now. Any error on the calibration offset related to 
uncertainties affecting the synthetic magnitude \cmag*, such as uncertainties 
on passbands and error on the reference spectra, will be discussed later (in
Sect.~\ref{sec:concl}) and are not considered in what follows.

The photon noise can be readily determined from the sky level\Fixed{ and} the
magnitude of the star. The typical contribution of background residuals to aperture 
contamination has been measured on the numerous observations of the Landolt
fields. We use the measurements from the \Fixed{D3} exposure to estimate the size 
of the flat-field noise for the sequence. We adjust the error model~(\ref{eq:31}) 
to the dispersion of measurements with respect to the magnitude of the corresponding 
\Fixed{tertiary catalog star}. \Fixed{Typically, the} corresponding flat-fielding 
error term ($f$) is found to be \Fixed{smaller} than $0.4$\%.

The 72 tertiary stars on CCD \# 22 allow a \Fixed{zero point determination} 
with \Fixed{typical} statistical uncertaint\Fixed{ies of} $2$~mmag in bands 
$\band g \band r \band i$, \Fixed{and} $5$~mmag in bands $\band z$ and \band{u}. 
This error affects all measurements of the sequence in the same way but is
expected to be uncorrelated from \Fixed{one night to} another.

\Fixed{The photometric ratio between the two \Fixed{SNAP-2} exposures
  bracketing each sequence was used to estimate \Fixed{gray}
  atmospheric extinction variation for the corresponding set of
  observations}.  \Fixed{The extra noise contribution of this
  component} is found to \Fixed{be} smaller than $2$~mmag.

\Fixed{Due to external constraints in the observing schedule, the deep field 
observation airmasses tend to be higher, on average, than the primary standard 
observation airmasses}. This is unfortunate as \Fixed{it} makes the final measurement 
sensitive to the variations of the atmospheric extinction around its average value 
$k_\text{atm}$ (\emph{cf.} table~\ref{tab:ltom}).

\Fixed{Computations of the night-to-night dispersion of the SNLS zero point
were used to estimate the typical variation of extinction around its average 
value. Magnitude variations were found to be 0.028, 0.009, 0.01, 0.01, 0.015
in $ugriz$ bands, respectively. These} estimates \Fixed{do} incorporate contributions 
unrelated to \Fixed{atmospheric} extinction \Fixed{such as} instrumental effects, 
variation of the PSF uncorrected by the IQ scaling rule, and noise from the zero point 
determination itself. \Fixed{Therefore,} effective variation of atmospheric extinction 
\Fixed{on} clear nights may \Fixed{actually be} somewhat smaller. \Fixed{On the other 
hand, these numbers are} in good agreement with \Fixed{similar} SNFactory measurements 
\citep{Buton2012}, except
 for \band z where our estimate
is significantly larger than the SNFactory estimate (about $0.3\%$), and
\band u where they found a variability closer to $4\%$. \Fixed{We used the 
largest numbers in each band as a conservative estimate} of the uncertainty on the 
airmass correction term.

\subsubsection{Systematic errors on the solar analog measurements}
\label{sec:systematic-errors}

The \Fixed{solar analog-based} measurement of the \Fixed{SNLS tertiary catalog AB offsets} 
is also affected by systematic differences in the PSF between short and long exposures, 
uncertainty on short exposure\Fixed{ durations}, residual contamination of apertures, and, 
as the observations are coupled only with the SNLS deep field D3, \Fixed{non-}uniformity 
in the survey.

We carefully monitored the variation of aperture corrections between
exposures by computing aperture photometry in a large panel of
radii. We choose to use \Fixed{16-pixel radius aperture} measurements to limit 
the \Fixed{background noise contribution to} individual measurements. We also 
measure the median photometric ratio between \Fixed{these} apertures and a 
\Fixed{43-pixel radius aperture}, which is assumed to capture a stable fraction 
of the total flux.  \Fixed{All measurements were corrected} by this measured photometric 
ratio. The statistical uncertainty on this aperture correction for individual exposures is
typically smaller than 1~mmag and is included as \Fixed{extra} noise.

\begin{figure}
  \centering
  \igraph{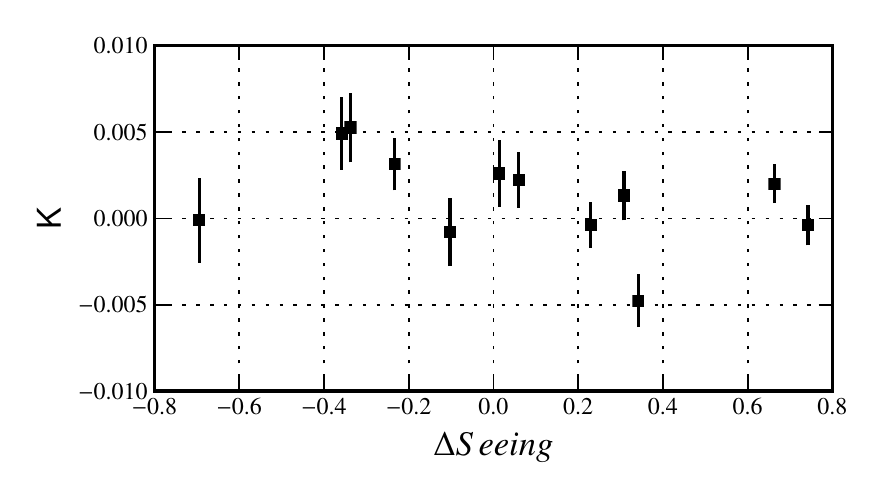}
  \caption{Stability of the flux measurements during the calibration
    sequences in $\band g \band r \band i$ bands. The difference of
    instrumental magnitudes in an aperture of 43 pixels between the
    first and the last exposure of the sequence is displayed as a
    function of the seeing evolution \Fixed{between the exposures}. }
  \label{fig:intrinsicatm}
\end{figure}

The important matter is how well the 43-\Fixed{pixel} aperture is
representative of the total flux. A hint is given by the variation of
the photometric ratio between the first and last exposures of the
sequence. Figure~\ref{fig:intrinsicatm} displays this photometric
ratio for the 43-\Fixed{pixel} aperture as a function of the difference in
image quality between the two exposures. While the seeing can vary a
lot, the photometric ratio is typically stable, with a rms of $2.3$~mmag. 
This sets an upper bound on the combined contribution of the atmospheric extinction 
variation, shutter noise and variation of the flux loss outside the 43\Fixed{-pixel} 
aperture. We conservatively use this number as a (pessimistic) estimate of the 
systematic flux difference between photometry on short and long exposures.

The background residual is measured for each exposure on a per CCD
basis and all measurements are corrected for it. The extra noise
introduced by the contamination of individual apertures is included in
the measurement covariance matrices. 

The uniformity of the survey has been discuss\Fixed{ed} in
Sect.~\ref{sec:second-stars-unif}. We use the statistical uncertainty
on the average zero points of the individual SNLS D3 field, as given
in table~\ref{tab:landoltN}, as an estimate of the departure of this
field from the average survey system.

Finally, as for the previous calibration \Fixed{method}, the shutter precision 
\Fixed{potentially biases} the flux estimate \Fixed{of} the short standard
exposures. However, no field has been exposed less than 3~s,
decreasing the impact of this systematic to less than 1~mmag.

\subsubsection{Calibration offsets from the solar analogs.}
\label{sec:offs-determ-solar}

\Fixed{Our} error estimates were used to build a covariance matrix $\tens R$
of \Fixed{measurement} errors accounting for correlation between the
different \Fixed{error sources}. The final \Fixed{AB offset} measurement is
obtained as the least-square average of the individual measurements
$\delta^j_\ab$:
\[ \delta_\ab = \sigma_\ab^2 \sum_i \sum_j \tens R^{-1}_{ij}
\delta^j_\ab \] where $\sigma_\ab^2 = 1/\sum_{ij} \tens R^{-1}_{ij}$
is the variance of the final estimate.

The results are summarized in table~\ref{tab:SA}. The offset can be
compared to the offset inferred from Landolt observations of
\bdtruc. The results are further discussed in Sect.~\ref{sec:concl}
where they are compared and combined with the other calibration
measurements.

\begin{table}
  \centering
  \caption{Determination of the calibration offsets for the SNLS catalogs.}
  \label{tab:SA}
  
\begin{tabularx}{\linewidth}{l@{\extracolsep{\fill}}*{6}{r@{\extracolsep{\fill}}}}
\hline
\hline
 & $u_M$ & $g_M$ & $r_M$ & $i_M$ & $i2_M$\tablefootmark{a} & $z_M$\\
\hline
BD+17&\\
\hline
$\delta_{\rm ab}$&$-0.439$&$0.114$&$-0.123$&$-0.332$&$-0.316$&$-0.458$\\
$\sigma_{\rm ab}$\tablefootmark{b}&$0.024$&$0.005$&$0.006$&$0.005$&$0.005$&$0.019$\\
\hline
SA&\\
\hline
$\delta_{\rm ab}$&$-0.431$&$0.111$&$-0.137$&--&$-0.335$&$-0.447$\\
$\sigma_{\rm ab}$\tablefootmark{c}&$0.017$&$0.005$&$0.008$&--&$0.007$&$0.011$\\
$\chi^2/{\rm d.o.f.}$& $8.4/8$&$15/14$&$8.1/9$&$0/0$&$10/15$&$3/8$\\
\hline
$\Delta$\tablefootmark{d}&$0.008$&$-0.003$&$-0.013$&--&$-0.019$&$0.011$\\
\hline
\end{tabularx}
  \tablefoot{\\
    \tablefoottext{a}{\Fixed{For comparison purposes, cross-calibration of the \band i and \band{i2} catalogs was used to provide a pseudo-measurement of \bdtruc in \band{i2} }(see Sect.~\ref{sec:interc-bandi-band})}\\
    \tablefoottext{b}{Errors quoted from the systematic error budget (\emph{cf.} Table~\ref{tab:snlserrorbudget}).}\\
    \tablefoottext{c}{Estimate of the uncertainty affecting the combination of solar \Fixed{analog} measurements. The model includes the (potentially correlated) contributions from measurement noise, zero point determination uncertainty, flat-field noise, atmospheric variation, airmass correction, survey uniformity and \Fixed{aperture correction uncertainty}. }\\
    \tablefoottext{d}{Difference between the two determinations of the offset.}\\
  }
\end{table}

\subsection{Intercalibration of \band{i} and \band{i2} catalogs}
\label{sec:interc-bandi-band}
The number of epochs with coincident observations of Landolt and SNLS
fields in band \band{i2} is small. \Fixed{Consequently}, no accurate
calibration of the \band{i2} \Fixed{band} measurements can be obtained \Fixed{in this
manner}. On the other hand, no direct measuremen\Fixed{ts} of the HST standards \Fixed{were}
obtained before the breaking of the \band i filter. \Fixed{Even so,} a relatively 
precise cross-calibration of the \band{i} and \band{i2} catalog\Fixed{s} is possible 
due to \Fixed{the} similarity of \Fixed{the two} filters.

We measured the average transformation between $\umag[i]$ and
$\umag[i2]$ for stars in the color range $ 0 < \umag[g] - \umag[i] <
1$ encompassing the \bdtruc color ($g-i = 0.34$). The measured color
term between the two bands is $\umag[i2] - \umag[i] = 0.025 (\pm
0.001) \times (\umag[g] - \umag[i])$. The dispersion around this
linear transformation is measured to be $4$~mmag for stars brighter
than $\band i < 18$. This is comparable to the expected
measurement error of $3.5$~mmag, dominated by the flat-field induced
error. \Fixed{Therefore, intrinsic} dispersion is expected to be \Fixed{no larger
than} $2$~mmag. Applying this transformation to BD+17 allows \Fixed{us} to infer the
\band{i2} magnitude of the primary standard from its $\band g$ and
\band{i} magnitudes with a precision better than 2~mmag, thus
anchoring together the catalog flux scales in \Fixed{the} \band{i2} and \band i
bands.

\section{SDSS tertiary star catalog}
\label{sec:SDSS}

The SDSS follows a calibration scheme similar to what was presented
for the SNLS. \Fixed{To calibrate individual supernova measurements, 
a high-accuracy photometric catalog of non-variable stars was generated 
from a coaddition of repeated observations of stripe 82.} 
The \Fixed{construction} of this catalog is described in \citet{ivezi_sloan_2007}. 
For brevity, this star catalog will be referred to as the ``coadd'' 
catalog.

\Fixed{A comparison of MegaCam Stripe 82 observations (described in 
Sect.~\ref{sec:sdsssnls-direct-inte}) and the coadd catalog revealed evidence 
for a percent-level residual non-uniformity in the final flat-fielding solution 
of the coadd catalog}. While the effect on individual SN photometry is negligible 
and the errors tend to average out \Fixed{when cosmological parameters are determined}, 
the non-uniformity triggered investigations of the origin of the 
flat-fielding error and its possible impact on \Fixed{calibration}. 
We start this section by summarizing \Fixed{the key steps of the construction}
 of the coadd catalog and then describe a revision of the coadd flat-fielding.

\subsection{\Fixed{Construction} of the SDSS coadd catalog}
\label{sec:sdss-phot-syst}

The coadd catalog was obtained by averaging SDSS measurements of $58$
photometric runs from Stripe 82 delivering on average \Fixed{10}
measurements per star. Quality cuts based on the repeatability of the
measurements were applied to discard variable stars. The version used
in the SN processing consists of 681301 stars.

The flat-fielding of the SDSS camera is simplified \Fixed{by the drift 
scanning technique used to obtain imaging data. In drift scanning, each 
point on the sky is sampled by each CCD row; therefore, the effective 
response is averaged over all rows. However, row averaging} does not
eliminate spatial and temporal variations of the atmospheric transparency. 
Long term drift is controlled by anchoring the catalog photometry to the PT 
secondary patches placed throughout the SDSS survey area.  Stripe 82 has 
an unusually high density of these secondary patches: on average, one every 
4 degrees in RA but they are not uniformly spaced and there is a gap of 29 
degrees between RA=60 and RA=89.  

\Fixed{Averaging over photometric runs improves measurement noise and several 
sources of systematics, most notably variations in atmospheric transparency.}
\Fixed{However, declination pointings remain similar from run to run, such that 
Stripe 82 stars are only observed in a narrow range of pixel columns}.
\Fixed{Therefore, cross-scan variations (including} flat-fielding errors and 
passband differences) are not smoothed out by \Fixed{run averaging}. 

The original technique for flat-fielding the images based on sky
levels proved to be problematic for reasons similar to those
preventing the direct use of twilights in MegaCam. A new procedure
that determined flat-field vectors for the SDSS camera based on the
\citet{ivezic2004} stellar locus was then implemented.\footnote{See
  \url{http://www.sdss.org/dr7/algorithms/flatfield.html} for a detailed
  description of the DR7 flat-fielding.} \Fixed{This} technique stems 
from the observation that most of the stellar population lines up in a
narrow locus in the 4D $ugriz$ color-space. The stellar colors are
projected onto 4 color-color diagrams where the bulk of the stellar
population is tightly clustered around a line in color-color space.
The requirement that the stellar locus be independent of the position
on the focal plane can be used to determine difference\Fixed{s} in flat-fields
between the bands.

To apply the stellar locus technique to the Stripe 82 observations,
average principal colors have been computed in bins of 0.01 degrees in
declination, over a region spanning Galactic coordinates. It is
assumed that all bins probe mostly the same population of stars,
thanks to the narrow range of declination of stripe 82, so that
differences in principal color are not induced by changes in the
stellar population.  The flat-field differences were adjusted to
ensure constant principal colors across declination bins.

The technique only determines the difference in flat field between
different filter bands. To determine the overall ``gray'' scale as a
function of declination, \citet{ivezi_sloan_2007} relied on the
calibration provided by the PT.  The 2.5~m photometry in band $gri$ is
compared to the PT photometry of secondary stars, and the average
difference is used as the reference flat-field vector for $r$-band. In
this procedure, any error in the PT flat-fielding is transferred to
the 2.5 meter telescope as an error common to all the passbands.

One last correction was applied to improve the uniformity of the
catalog by taking into account the differences in the filter passbands
for the different camera columns. In the published version,\footnote{\url{http://www.astro.washington.edu/users/ivezic/sdss/catalogs/stripe82calibStars_v2.6.dat.gz}} all
stellar measurements were transformed to the system defined by the
official SDSS transmission curves\footnote{available from
  \url{http://www.sdss.org/dr5/instruments/imager/index.html\#filters}}. The
present work is based on an earlier (unpublished) version of the
catalog that was used by \citep{holtzman_sloan_2008} to calibrate the
photometry of the supernova light curves.  In this earlier version, no
color transformation is attempted and all magnitudes are delivered in
the natural system.

The coadd catalog photometric scale is anchored to the PT secondary
patches by the SDSS pipeline. The PT photometry, in turn, is tied to
the USNO star network \cite{2002AJ....123.2121S}.  The calibration of
the secondary patches is based on the PT observations of 3 HST CALSPEC
standards \citep{holtzman_sloan_2008} so we do not rely on the
calibration of the USNO star network.  We do rely on \Fixed{its} uniformity
and suitability to provide an accurate measurement of atmospheric
extinction. Observations are transformed from the PT natural system to
the 2.5~m natural system using canonical color transformations. \Fixed{Standard 
star observations} have been conducted with the target object roughly at the 
center of the camera image, whereas the color transformations have been 
derived from \Fixed{measurements over} the entire image. In this observational 
setup, non-uniformity in the PT photometry may result in some calibration bias. 
In addition, the color transformations were determined quite early in the 
survey and were not monitored closely.

As variations of the PT response \Fixed{as large as} 2\% were found (\emph{cf.}
appendix~\ref{sec:pt-flat-field}), \Fixed{the following sections describe
our determination of} a flat-fielding correction to the coadd catalog 
which removes most of the sensitivity to PT photometry. In
Sect.~\ref{sec:sdss-pt-calibration}, we will use this uniform catalog
to determine corrections to the PT flat-field, update the PT to 2.5~m
transformation equation and revise the AB offsets for the corrected
catalog from PT observations of HST photometric standards.

\subsection{Corrected SDSS flat-field}
\label{sec:dr8-flat-field}

A separate effort \citep{2008ApJ...674.1217P} sought to enforce the
uniformity of the SDSS survey with naturally overlapping fields and
special crossing scans that observed the sky in directions that were
approximately perpendicular to the SDSS Stripes.  The observation of
stars in different camera columns allowed a recalibration of the SDSS
that did not rely directly on the PT flat-fielding. In addition, it
calibrated each filter separately and did not rely on the stellar
locus or other assumptions about the objects being observed.  This
technique was used for the SDSS Data Release
8\footnote{\url{http://www.sdss3.org/dr8/}.}. This catalog will be
referred to as DR8 and was retrieved from the SDSS database using the
query given in appendix~\ref{sec:dr8-catal-retr}.

The DR8 catalog is based on a single observation of each star\Fixed{. 
Consequently it is statistically less} precise than the coadd catalog.  
In addition, typical SDSS observations take place at low airmass making 
it difficult to determine the atmospheric extinction from the SDSS 2.5~m 
data alone, requiring some additional assumptions.  
In contrast, the PT routinely measures standard stars at
both low and high airmass and measures atmospheric extinction with
much better accuracy.

Since we know that the PT flat-field has significant errors we want to
use the DR8 data to correct the coadd catalog as a function of
declination. We correct the coadd catalog, rather than using DR8
directly, because the multiple star observations incorporated into the
coadd catalog greatly decrease the photometric errors and make it
possible to eliminate various types of outliers. In the following, we
will refer to the resulting catalog as the ``corrected''
catalog. Corresponding magnitudes will be denoted with a $c$
subscript.

\subsubsection{Matching the coadd and DR8 catalogs}
\label{sec:analys-diff-betw}

We analyzed the differences between the two catalogs. The results of
the analysis are restricted to the resulting catalog of matched
stars. The matching criteria were:
\begin{itemize}
\item The separation of two objects in the coadd and DR8 catalogs is
  less than 1\arcsec.
\item There is no other object in the coadd catalog that is within
  3.6\arcsec\ in both right ascension and declination.
\item The object in the coadd catalog is not within 3.6\arcsec in
  both right ascension and declination of any other object in the DR8
  catalog.
\end{itemize}
A star must have 4 or more observations in the $g$, $r$, and $i$ bands to be
included in the catalog.  In the following analysis, the measured $u$
and $z$ magnitudes are used only if there are 4 or more observations.
We have also chosen to require $14.0< r <20.4$

\subsubsection{Right ascension trend}
\label{sec:right-ascens-trend}

\begin{figure}
\igraph{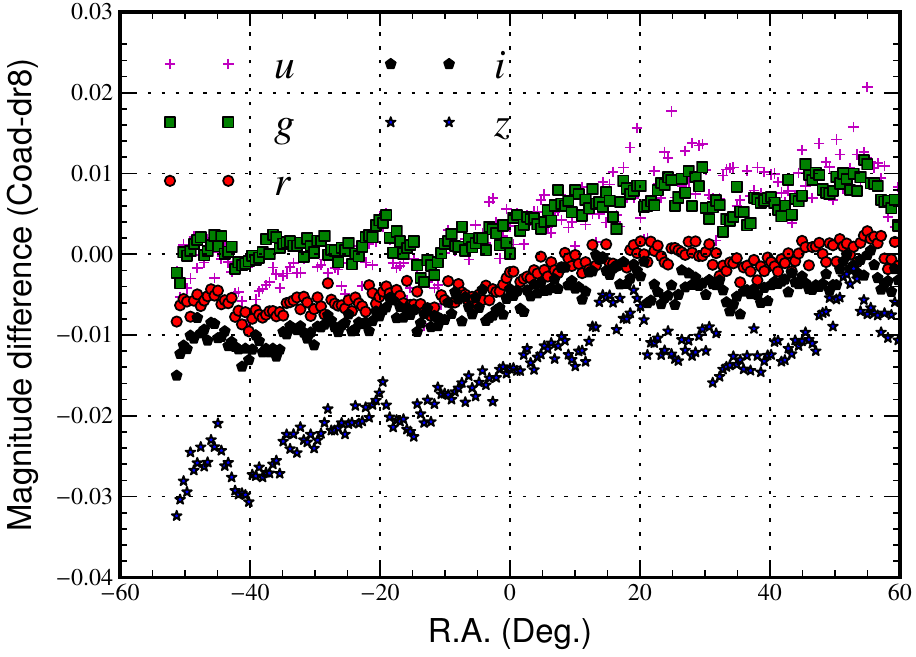}
\caption{The difference in magnitude (coadd$-$DR8) for the 5 SDSS filters is shown as a function of right ascension.  \label{fig:ra_nocorr}}
\end{figure}

Figure \ref{fig:ra_nocorr} \Fixed{shows} the difference in the two catalogs as
a function of right ascension. \Fixed{The data show a clear trend} that
is approximately linear in right ascension ($\alpha$).  We fit the
data for each filter to the form
\begin{equation}
\Delta m = a \alpha + b
\end{equation}
The results of the fits for each filter are shown in Table
\ref{tab:raline}.

Given the significant trend in the difference, the question arises as
to which catalog is more nearly correct.  The SDSS DR8 procedure
\citep{2008ApJ...674.1217P} assumes the extinction to be of the form
\begin{equation}
k(t) = k_0 + \frac{dk}{dt} (t-t_0) \label{eqn:kslope}
\end{equation}
where $k_0$ and $dk/dt$ are constants, $t$ is the time of observation
and $t_0$ a reference time (in this case 0700UT). The assumed mean
values and dispersion of ${dk}/{dt}$ are taken from
\citet{2008ApJ...674.1217P}, converted into mmag/deg and shown in the
``DR8'' column of Table~\ref{tab:raline} assuming an airmass of 1.2,
a typical airmass for a Stripe 82 run.  The slope of the difference
between the coadd and DR8 are similar to the values assumed by DR8,
but not identical.

The mtPipe calibration \citep{2006AN....327..821T} \Fixed{used} for
SDSS data production does not include any time-dependent terms
although occasionally different photometric solutions were used for
different time intervals during the night.  The PT calibration is
imprinted on the SDSS 2.5~m data through the secondary patches.  Thus,
we would expect a \Fixed{time variation-induced} calibration error 
in the SDSS catalog data with the effect being proportional, on average,
to the difference in time between the 2.5~m observations and the PT
observations.  A typical scan of Stripe 82 starts in the early evening
at approximately -60$^\circ$ and ends in the early morning at
+60$^\circ$, resulting in a tight correlation between RA and the time
of night.  In contrast, the 178 \Fixed{Stripe 82} secondary PT patches
\Fixed{used} to calibrate the coadd catalog were taken
throughout the night at more-or-less random times.

\begin{table}
\centering
\caption{Trend line fits in RA  \label{tab:raline}}
\begin{tabular}{l|cc|c|c}
  \hline\hline
  Filter & \multicolumn{2}{c|}{DR8 - coad} & DR8 & PT - coadd \\
  & $a$ &  $b$ &  $dk/dt$ &  $a$  \\
  & {\scriptsize (mmag/deg)} & {\scriptsize (mmag)} & {\scriptsize (mmag/deg)}& {\scriptsize (mmag/deg)}\\
  \hline
  u  &  0.142  &    3.1 & $0.096 \pm 0.200$ & $ 0.082  \pm 0.291$ \\
  g  &  0.082  &    3.7 & $0.056 \pm 0.136$ & $-0.006  \pm 0.027$ \\
  r   &  0.075 &   -3.4 & $0.080 \pm 0.136$ & $ 0.013  \pm 0.017$ \\
  i   &  0.087 &   -6.6 & $0.096 \pm 0.120$ & $ 0.004  \pm 0.024$ \\
  z  &  0.199  &  -16.7 & $0.176 \pm 0.136$ & $-0.022  \pm 0.059$ \\
  \hline
\end{tabular}
\tablefoot{The comparison with \citet{2008ApJ...674.1217P} assumes an airmass=1.2, which is typical for a Stripe 82 run.}
\end{table}

Given the number of patches and their distribution in time of
observation, it seems likely that the secondary patches should provide
a good average over observing conditions throughout the night.
Seasonal variations could produce a bias in calibration as a function
of right ascension since higher right-ascension patches tend to be
observed later in the season, but any variation that is slower than a
night should be removed by the nightly calibration.  We therefore
assume that the PT patches have no bias as a function of right
ascension.

In the processing of the 2.5~m data a single zeropoint is computed for
each CCD so changes in atmospheric extinction could result in a bias
in calibration as a function of right ascension.  On the other hand, a
typical photometric run in the stripe 82 catalog is often much shorter
than the maximum of 8 hours: the average run length is 3 hours.  Thus
the effect of any change in atmospheric extinction should be
substantially reduced for the shorter run.  Regardless of what we
might expect from the SDSS photometric processing, we can examine the
agreement between the coadd catalog and the PT patches, which we \Fixed{
assume} to be unbiased with respect to right ascension.  For each PT
patch we determine the average difference between coadd and PT
observations of the same stars.  We then fit a line to the data as a
function of RA.  The results are shown in table~\ref{tab:raline}.  The
individual PT patches show considerably more scatter than would be
expected from the statistical errors; presumably this is due to small
time variations in atmospheric extinction (which was assumed to be
constant over the course of the night).  We characterize the
distributions by fitting \Fixed{straight} lines whose slopes are reported in
the last column of table~\ref{tab:raline}. The quoted uncertainties
are computed from the scatter of the data, not the statistical errors
on the individual points.  Also given in table~\ref{tab:raline} are
the slopes expected if the DR8 trend were correct.  The \Fixed{fitted} slopes 
are consistent with zero and, for all \Fixed{bands} except $u$, are
significantly different from the slopes \Fixed{suggested by the DR8 catalog}.
We conclude that this comparison disfavors the trends of the DR8
catalog and we remove the slopes reported in Table~\ref{tab:raline}
from the DR8 data\Fixed{, bringing} it into agreement with the coadd.

There are remaining trends in the data that are not well fit by the
straight-line approximation.\Fixed{The origin of these variations is
likely to be fluctuations in } atmospheric transparency \Fixed{which
affect} the DR8 single epoch photometry more than the coadd catalog.
However, we are primarily interested in the flat-fielding as a
function of declination, and the major effect of remaining trends on
the flat-fielding is a slight increase of the dispersion.

\subsubsection{Flat-field correction}
\label{sec:flat-field-corr}

\begin{figure*}
  \igraph{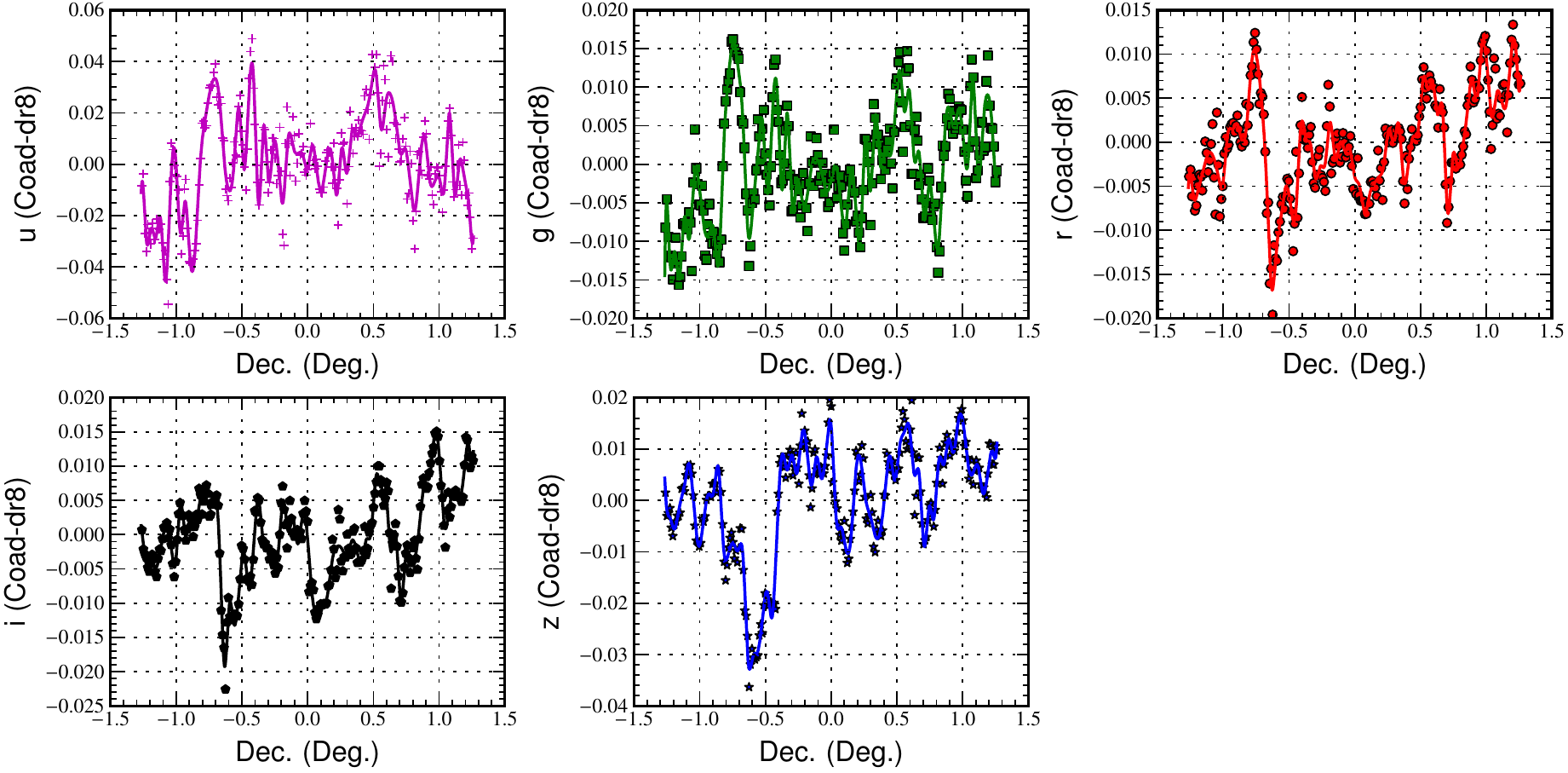}
  \caption{Difference between the coadd \Fixed{catalog} and DR8 as 
    a function of declination. The symbols represent the binned data and the 
    lines are a cubic spline function fit to the data with 71 control
    points.\label{fig:decfit}}
\end{figure*}

\Fixed{Figure~\ref{fig:decfit} shows the comparison between the DR8 
catalog and the coadd as a function of declination after the linear 
trend in RA has been removed from DR8}. The data are binned into 226
\Fixed{declination ranges} (\Fixed{each} bin is approximately 0.01\arcdeg wide).  
\Fixed{The catalog differences display similar patterns from band to band
(particularly in $r$, $i$, and $z$),} reinforcing the notion that the PT 
flat-field is a significant source of error in the coadd catalog. 
However, the bands also exhibit significant trends and the pattern of 
residuals does not show the exact 0.6 degree periodicity that \Fixed{would be} 
expected if a gray scale error in the PT were the only effect.

The solid lines on Fig.~\ref{fig:decfit} are cubic B-spline curves,
each of which has 71 control points \Fixed{determined} by a fit to
the binned data. The decision to use 71 control points was somewhat
arbitrary but puts the angular resolution at about 0.03.  The analysis
of the PT flat field leads us to expect a sharp discontinuity where
the secondary patches abut one another. This would argue for a large
number of control points in order to adjust those expected
discontinuities. However, it was suspected that the binning of data to
produce the coadd originally may have introduced some numerical noise,
so the number of control points was chosen to be smaller than the
number of data bins in order to produce some smoothing effect. The
$\chi^2$ of the fit decreases slowly as the number of bins is
increased and doesn't seem to offer any useful clues as to an
appropriate cutoff.

We correct the coadd catalog by subtracting the fitted smooth
correction. \Fixed{Because we redetermine a (small) absolute calibration 
offset for the corrected catalog (described in the next section), we do 
not require the average correction to be zero.}
The residual difference between the DR8 and the corrected coadd catalogs 
\Fixed{shows a rapid 2~mmag rms fluctuation} on small angular scales\Fixed{
but} is constant over larger scales. An assessment of the \Fixed{flat-fielding 
quality} of the corrected catalog is given by the comparison with MegaCam 
measurements in Sect.~\ref{sec:spatial-uniformity}. In particular, 
Fig.~\ref{fig:uniformity} suggests that the flat-fielding is now correct at 
the 0.5\% level in $griz$.

\section{SDSS PT calibration measurements}
\label{sec:sdss-pt-calibration}
We now turn to the determination of AB offsets for the corrected coadd
catalog using \Fixed{SDSS PT} observations of HST standards. The 
non-uniformity of the PT response has a potential impact on this calibration 
transfer and must be corrected for. The basic idea is to use the uniformity 
of the corrected coadd catalog to improve the flat-field of the PT, redetermine 
the color relation between the corrected coadd catalog and the corrected PT 
photometry, and use those corrected relations to transform PT observations of 
the primary standards to the 2.5~m photometric system.

\subsection{Correction of PT measurements}
\label{sec:pt-based-absolute}

The calibration of the 2.5~m telescope native system to the absolute
(AB) system relies on repeated \Fixed{SDSS PT} observations of the HST
CALSPEC\footnote{http://www.stsci.edu/hst/observatory/cdbs/calspec.html.}
standard stars \citep{bohlin_nicmos_2008} \Fixed{. The} PT measurements are 
calibrated against the \Fixed{set of USNO standard} stars and \Fixed{transformed} 
to the SDSS 2.5~m natural system using fixed color transformations. The PT calibration is
robustly transferred to the 2.5~m telescope through the PT secondary
patches.

With the discovery of photometric errors in the PT, it becomes
important to redetermine (or check) the transformation equations
between the two systems. We did so by looking for corrections to the
already transformed magnitudes taking the form:
\begin{eqnarray}
u_c &=& u + b_u(u-g-1.362) +a_u\label{eq:PT2SDSSa}\\
g_c &=& g + b_g(g-r-0.596)+ a_g\\
r_c &=& r + b_r(r-i-0.241) + a_r\\
i_c &=& i + b_i(r-i-0.250)+ a_i\\
z_c &=& z + b_z(i-z-0.128)+ a_z\label{eq:PT2SDSSc}\,,
\end{eqnarray}
Here we denote the magnitude coming from the corrected coadd catalog
with the subscript $m_c$, to avoid confusion with the PT measurements
transformed to the $ugriz$ using the nominal color transformations.

As we are fitting for corrections, we would recover b = a = 0 in all
filters if the nominal transformation described the data perfectly. In
order to compute these transformations the PT observations of
secondary stars are matched with the corrected catalog of 2.5~m
observations with the following criteria:
\begin{itemize}
\item Match within 0.001 degrees 
\item Matches must be unique.
\end{itemize}

\begin{figure*}
  \igraph{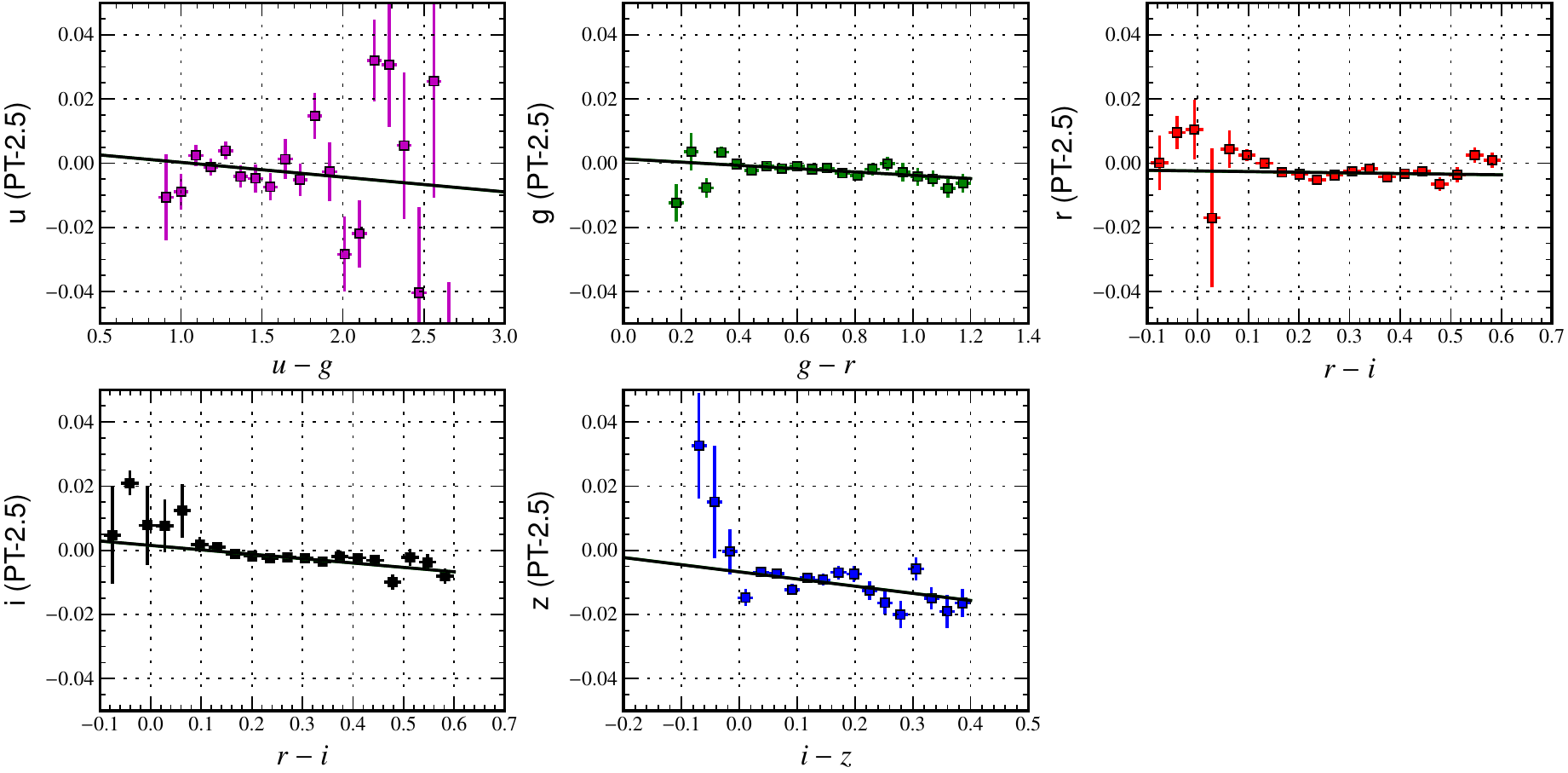}
  \caption{SDSS 2.5~m to PT measured color transformations. The
    individual stellar measurements are averaged in color bins of
    0.05 width for clarity. The error bars display the 1$\sigma$
    uncertainty on the average estimated from the dispersion of
    individual measurements. The straight black line is the
    linear fit to the data.\label{fig:PTcolor} }
\end{figure*}

Since we know that the PT is subject to flat-fielding errors we
want to allow for the possibility that $a$ varies with position on the
focal plane.  The color term $b$, however, is likely to be nearly
constant. So, we fit our data to equations
(\ref{eq:PT2SDSSa}--\ref{eq:PT2SDSSc}) in two steps.  First, we
determine the color term using the central portion of the PT CCD and
the following criteria:
\begin{itemize}
\item Valid PT magnitude and SDSS magnitudes for color
\item At least 4 observations in the corrected catalog
\item 2.5~m $r$-band magnitude in the range $14.0<r<19.0$
\item The star must be in the center of the PT focal plane ($400<x<1448$
  and $400<y<1448$), where x and y are the pixel coordinates of the
  2048$\times$2048 pixel CCD.
\item Color in the range given by Table~\ref{tab:crange}.
\end{itemize}
The fit is iterated and any magnitude that deviates by more than
$5\sigma$ is dropped from the fit.  The data are shown in Figure
\ref{fig:PTcolor} and the results of the fits are given in Table
\ref{tab:PTcolor}. In general, the color terms and offsets between the
two systems are quite small.  While some of the $b$ terms are
statistically significant their magnitudes are small.

\begin{table}
\centering
\caption{Color range used to compute the color transformations.  \label{tab:crange}}
\begin{tabular}{crr}
\hline\hline
Color & Minimum & Maximum\\
\hline
$u-g$  &  0.7   &  2.7   \\
$g-r$  &  0.15  &  1.2   \\
$r-i$   &  -0.1  &  0.6   \\
$i-z$  &  -0.2  &  0.4   \\
\hline
\end{tabular}
\end{table}

\begin{table}
\caption{PT Color Transforms \label{tab:PTcolor}}
\begin{tabular}{crrrrr}
\hline\hline
Parameter & $u$ & $g$ & $r$ & $i$ & $z$\\
\hline
$a$        & -0.0022 & -0.0021 & -0.0022 & -0.0014 & -0.0078 \\
$\sigma_a$ & 0.0011  & 0.0003  & 0.0003  & 0.0003  & 0.0006 \\
$b$        & -0.0016 & -0.0040 & -0.0011 & -0.0145 & -0.0255 \\
$\sigma_b$ & 0.0041  & 0.0018  & 0.0027  & 0.0030  & 0.0067 \\
$\chi^2$   & 3739    & 10003   & 11427   & 10820   & 6215 \\
d.o.f.     & 3012    & 10100   & 12215   & 11391   & 6636 \\
\hline
\end{tabular}
\end{table}

\begin{figure}
  \igraph{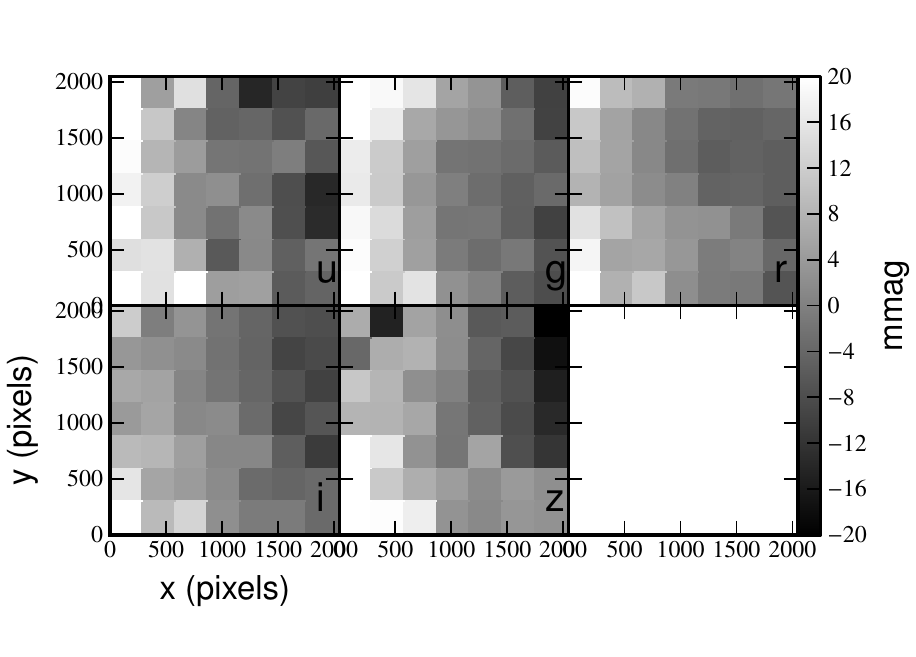}
  \caption{PT flat-field corrections.\label{fig:PTfocal} }
\end{figure}

Next, we determine a position dependent $a$ term by fitting the data in
bins of a uniformly spaced $7\times7$ grid across the CCD focal plane.
Although the $b$ terms are not significant, the data is corrected for color
prior to determining position dependent $a$ terms. The resulting
pattern of $a$ terms is shown in Fig.~\ref{fig:PTfocal}. The main feature is a 
$\sim4\%$ gradient between the 2
opposite corner of the CCD and has a similar shape in all the bands.

\subsection{Absolute calibration}
\label{sec:absolute-calibration}

We can now use the PT measurements to \Fixed{update the} SDSS supernova 
survey calibration. The original HST standard \Fixed{star} PT
measurements are displayed in Table~\ref{tab:STstar}. They are
reported, as usual, in the 2.5~m photometric system by the application
of the nominal (uncorrected) color transformations. 
\Fixed{Table~\ref{tab:SDSSstar} lists the same PT measurements in the 
corrected catalog system, obtained by applying equations (\ref{eq:PT2SDSSa}--\ref{eq:PT2SDSSc})
to the original (Table~\ref{tab:STstar}) magnitudes. }
\Fixed{As described earlier in the text, these corrections are PT focal plane 
position-dependent, and compensate for the non-uniformity of the PT response. }
\Fixed{AB offsets calculated from the corrected HST standard measurements are
given in Table~\ref{tab:PTOff}.}

\begin{table*}
\caption{PT measurements of HST standard star magnitudes  \label{tab:STstar}}
\centering
\begin{tabular}{lrrrrrrrrrrrrrrr}
  \hline\hline
  Star & $m_u$ & $\delta m_u$ & $n_u$ & $m_g$ & $\delta m_g$ & $n_g$ & $m_r$ & $\delta m_r$ & $n_r$ & $m_i$ & $\delta m_i$ & $n_i$ & $m_z$ & $\delta m_z$ & $n_z$
  \\
  \hline
  G191B2B	& 11.045 & 0.031 & 7 &11.461 & 0.011 & 7 & 12.014 & 0.011 & 7 &12.397	&0.010&	7&	12.743&	0.005&	7 \\
  GD153 &		12.693	&0.020	&6&	13.055	&0.018&	6&	13.570&	0.006&	6	&13.943&	0.010&	6	&14.296&	0.021&	6 \\
  GD71 & 12.439 &0.016	&6	&12.753	&0.011	&7	&13.242&	0.012&	7&	13.610&	0.005&7 &13.967&	0.023&	7 \\
  P041C & 		13.569&	0.015&	11&	12.260&	0.010&	11&	11.842&	0.009&	11	&11.718&	0.024&	11	&11.702&	0.011&	11 \\
  P177D &	15.120	&0.017&	13	&13.746&	0.010&	13	&13.299&	0.009	& 13	&13.158&	0.008&	13	&13.126&	0.009&	13 \\
  P330E &14.551	&0.018	&17	&13.281&	0.009&	17	&12.842	&0.007&	17	&12.701&	0.007&	17	&12.672	&0.013	& 17\\
  BD+17\arcdeg4708 & 10.559 & 0.020 & 169 & 9.630 & 0.029 & 169 & 9.350 & 0.027 & 169	&9.248	&0.030 & 169	& 9.235 &	0.032 & 169 \\
  \hline
\end{tabular}
\tablefoot{The HST CALSPEC stars are measured in the PT native system but reported in the SDSS 2.5~m (unprimed) system using canonical color \Fixed{transformations. 
For each filter $x$, we report the average magnitude $m_x$, the standard deviation (not the error on the mean) $\delta m_x$ and the number of observations  $n_x$.}}
\end{table*}

\begin{table*}
  \caption{PT Focal Plane Offsets.\tablefootmark{a} \label{tab:PTOff}}
  \centering
  \begin{tabular}{lrrrrrrr}
    \hline\hline
    Filter & $x$ & $y$ & $a_u$ & $a_g$ &$a_r$ & $a_i$ &$a_z$\\
    \hline
    G191B2B          &   816 & 1013 &   0.0018 & 0.0026 & 0.0015 &  0.0016 & 0.0039 \\   
    GD71             &   261 & 1251 &   0.0153 & 0.0149 & 0.0080 &  0.0058 & 0.0100 \\   
    GD153            &   914 & 1426 &  -0.0006 & 0.0021 & -0.0010 & -0.0008 &0.0022 \\   
    P041C            &   977 &   898 &   0.0021 & 0.0004 &  0.0003 &  0.0015 &-0.0003 \\ 
    P177D            &   712 & 1239 &   0.0040 & 0.0050 &  0.0018 &  0.0013 & 0.0038 \\  
    P330E            &   876 &   633 &   0.0001 & 0.0018 &  0.0046 &  0.0031 & 0.0024 \\ 
    BD+17\arcdeg4708 & 1024 & 1024 &  0.0024 & -0.0001 &  0.0002 &  0.0018 &-0.0015 \\   
    Error            & 1024 & 1024 &  0.0028 &  0.0009 &  0.0009 &   0.0019 & 0.0018 \\  
    \hline
  \end{tabular}
  \tablefoot{
    \tablefoottext{a}{Values of the $a$ term from Eq.~(\ref{eq:PT2SDSSa}--\ref{eq:PT2SDSSc}) at the position of the PT focal plane where the star is measured. See Fig.~\ref{fig:PTfocal} for a map of the correction.}}
\end{table*}

\begin{table}
\caption{Standard star magnitudes in the SDSS system after all corrections. \label{tab:SDSSstar}}
  \begin{tabular}{p{50 pt}rrrrr}
 \hline
 \hline
 Star & $m_u$ & $m_g$ & $m_r$ & $m_i$ & $m_z$\\
 \hline
 G191B2B              & 11.048 & 11.456 & 12.014 & 12.388 & 12.735 \\    
 GD153                & 12.699 & 13.051 & 13.573 & 13.936 & 14.289 \\  
 GD71                 & 12.429 & 12.736 & 13.236 & 13.597 & 13.946 \\ 
 P041C                & 13.569 & 12.261 & 11.844 & 11.716 & 11.707 \\   
 P177D                & 15.118 & 13.743 & 13.299 & 13.157 & 13.128 \\  
 P330E                & 14.553 & 13.280 & 12.839 & 12.697 & 12.675 \\  
 BD17\arcdeg4708      & 10.560 &  9.631 &  9.352 &  9.245 &  9.241 \\   
 \hline
 \end{tabular}
\tablefoot{
  The color transformation and position dependent offsets described in the text have been applied to the measurement listed in Table~\ref{tab:STstar}.
}
\end{table}

\Fixed{HST spectra and SDSS filter responses \citep{2010arXiv1002.3701D}
are used to compute expected magnitudes for the HST standard stars via
the synthetic magnitude formula given in Eq.~(\ref{eq:}).}
\Fixed{Results} of this calculation are given in Table~\ref{tab:SYNstar}.
\Fixed{As the CALSPEC version somewhat affects the resulting magnitudes,
Table~\ref{tab:SYNstar} also notes the specific files used.}
\Fixed{Given in Table~\ref{tab:ABoff}, the} differences between the 
synthetic \Fixed{and the measured} magnitudes are the AB offsets of the 
SDSS photometric system. The standard calibration uses the solar analogs 
because their colors are well within the range of validity of the 
PT$\rightarrow$SDSS system color transformation.  
The dispersion in the results of the 3 solar analogs gives an estimate of 
the error in the AB offset -- about 0.002 to 0.003 \Fixed{magnitudes --} although 
this does not include any systematic effects that might result in common 
errors to the solar analogs.  We also show \bdtruc, which serves as the 
primary reference for the SDSS photometry \Fixed{ as well as for several other} 
experiments.  The difference between \bdtruc and the solar analogs is another
indication of \Fixed{AB offset} errors. The WD result is also shown\Fixed{.
However, WD colors} are far beyond the region of validity for the color
transformation \Fixed{thus} their computed AB offsets should not be regarded as
reliable.
\begin{table*}
  \centering
  \caption{Synthetic Magnitudes of HST Standard Stars.\tablefootmark{a} \label{tab:SYNstar}}

\begin{tabular}{l*{5}{r}l}
\hline\hline
Star & $u$ & $g$ & $r$ & $i$ & $z$ & File\\
\hline
G191B2B&11.008&11.471&12.011&12.398&12.746 & g191b2b\_stisnic\_003.ascii\\
GD153&12.677&13.067&13.580&13.959&14.300 & gd153\_stisnic\_003.ascii\\
GD71&12.433&12.770&13.258&13.627&13.970 & gd71\_stisnic\_003.ascii\\
P041C&13.507&12.281&11.852&11.747&11.734 & p041c\_stisnic\_003.ascii\\
P177D&15.053&13.771&13.308&13.178&13.143 & p177d\_stisnic\_003.ascii\\
P330E&14.471&13.295&12.840&12.708&12.675 & p330e\_stisnic\_003.ascii\\
BD 17\degree4708&10.499&9.647&9.351&9.255&9.239 & bd\_17d4708\_stisnic\_003.ascii\\
\hline
\end{tabular}
  \tablefoot{\\
    \tablefoottext{a}{The SDSS filters are from \citet{2010arXiv1002.3701D}}.
  }
\end{table*}

\begin{table}
  \caption{AB Offsets for SDSS.\label{tab:ABoff}\tablefootmark{a}}
  
  \begin{tabular}{crrrrr}
    \hline\hline
Filter & $u$ & $g$ & $r$ & $i$ & $z$\\
\hline
Solar Average & 0.069 & -0.021 & -0.006 & -0.021 & -0.014\\
Solar Error & 0.006 & 0.004 & 0.003 & 0.006 & 0.008\\
\hline
 BD+17\arcdeg4708 & 0.061 & -0.016 & 0.001 & -0.010 & 0.002\\
\hline
WD Average & 0.020 & -0.021 & -0.009 & -0.021 & -0.015\\
WD Error & 0.013 & 0.006 & 0.007 & 0.006 & 0.004\\
\hline
\end{tabular}

  \tablefoot{\\
    \tablefoottext{a}{\Fixed{Offsets} should be subtracted from the SDSS native magnitudes to \Fixed{obtain} calibrated AB magnitudes.}}
\end{table}

\subsection{Discussion}

Besides the hints given by the variation of the calibration as a
function of the selected standard, we can try to review and quantify
the sources of uncertainty affecting the calibration of the corrected
catalog. As for the SNLS, our concern here are the uncertainties
affecting the measured magnitudes of the HST standards. The discussion
of \Fixed{the spectra uncertainties} is common to both surveys and 
delayed to section~\ref{sec:concl}.

\subsubsection{Survey uniformity}
\label{sec:survey-uniformity-1}

The PT calibration can be affected by a spatial non-uniformity in the
USNO star network.  In particular, non-uniformity in the network could
result in a systematic difference between the calibration of the solar
analog measurements ($ \alpha \approx 15h$) and the calibration of the
secondary patches ($20h < \alpha < 4h$). The question is thus how much
error there might be in transferring the calibration 12 hours ahead
(or behind) in RA. Possible non-uniformities are not directly
discussed in the literature but the USNO network is expected to be
especially robust because of nearly continuous measurements of the
primary standards over a two year period of observation.  From the
uncertainties quoted in \cite{2002AJ....123.2121S} for the primary
standard star measurements in the global fit, we estimate that the
relative uncertainty in uniformity is less than 1.5 mmag in $u$ band
and 1 mmag in the other bands.

The PT observation program was designed so that the USNO standard
stars were observed at the center of the focal plane. In contrast to
the measurements of the HST standards, the non\Fixed{-}uniformity of the
response thus has no impact on the PT calibration to the USNO standard
star network.

\subsubsection{Color transformation}
\label{sec:color-transformation}

The color transformation from the PT to the 2.5~m system introduces
some uncertainty on the precision to which the 2.5~m magnitudes of the
standard stars are known. Two effects must be considered: the fact that
individual stars \Fixed{do not fall exactly} on the average color
transformation, and the fact that in the SDSS camera each CCD has its
own filter.

We found the intrinsic dispersion around the average transformation by
looking at the dispersion of magnitudes synthesized in both systems
from a stellar spectral atlas (Gunn-Stryker-Bruzal).  We use this
dispersion to estimate how much spectral variations of the HST
standards might vary from the average stellar population which
determines the color transformations.  This dispersion is probably an
overestimate because we ignore the effects of noise in the spectral
atlas.  In addition, the spectral diversity is likely over represented
by the spectral atlas.

An alternative estimate of the magnitude of this error can be obtained
from the variation of the AB-offset with the spectral type and color
of the standard considered (available from table~\ref{tab:ABoff}). We
\Fixed{obtain} similar numbers, except for $u$ band where the variation
is inflated by the fact that the blue WDs clearly lie outside the
validity range of the linear color transformation, and $g$ band where
the dispersion in the atlas seems artificially high. However, we do
not attempt to correct the dispersion measured from the atlas and
accept this dispersion as \Fixed{a} conservative error
estimate.  This error dominates the uncertainty budget for the
PT-based calibration as shown in table~\ref{tab:sdsssys}.  This is
similar to the situation in SNLS, where the transformation of the
\bdtruc Landolt magnitudes dominates the uncertainty on the
Landolt-based calibration.

The color transformation is adjusted using data covering the entire
width of the SDSS camera. As a consequence it translates PT
measurements to the ``average'' SDSS photometric system.  Our
corrected tertiary catalog stems from a version of the coadd catalog
with star magnitudes reported in the native SDSS photometric
system. The flat-fielding correction applied to match the DR8
photometry does not use any color-dependent term, and leaves
magnitudes in the natural system. This calibration procedure should
adjust the zeropoint of each camera column so that a star of average
color (as is nearly the case \Fixed{for} solar analogs) will have the same
magnitude in each camera column in the native system. \footnote{Of
  course, there will be differences in magnitude for stars of
  different colors.  }

The filters have been measured independently. Table~\ref{tab:SYN6}
shows the synthetic magnitudes for the 6 camera columns for one of the
HST white dwarfs and one of the solar analogs.  The differences in
magnitudes are small (~1\%) but not completely negligible. The
variation of the AB magnitude of the solar analog around its nominal
value illustrates the variation of the calibration offset effectively
applicable to individual columns. Neglecting the filter changes adds a
calibration \emph{noise} (centered on zero) of 6, 4, 1, 0.2 and
0.6~mmag rms in $ugri$ and $z$ bands, perfectly negligible for SN
measurements.

As far as we are only interested in accurate calibration on average
(which is the case for SN-Ia science), the global AB offsets are
adequate, and we do not account for any related effect in our
systematic budget.  When considering only a subset of the camera
columns however, some attention should be paid to the change of
photometric system, and calibration offsets for individual columns may
have to be determined from table~\ref{tab:SYN6}.

\begin{table*}
  \caption{Synthetic and natural SDSS magnitudes of HST standard stars by camera column\label{tab:SYN6}}
  \centering
  \begin{tabular}{lcccccc}
    \hline
    \hline
    Star & column & $u$ & $g$ & $r$ & $i$ & $z$\\
    \hline
    G191B2B & 1 & 11.0087 & 11.4721 & 12.0166 & 12.4006 & 12.7460 \\
    G191B2B & 2 & 11.0061 & 11.4686 & 12.0152 & 12.4036 & 12.7595 \\
    G191B2B & 3 & 11.0133 & 11.4752 & 12.0087 & 12.4039 &12.7520 \\
    G191B2B & 4 & 11.0110 & 11.4742 & 12.0148 & 12.4020 & 12.7542 \\
    G191B2B & 5 & 11.0058 & 11.4749 & 12.0141 & 12.4018 & 12.7470 \\
    G191B2B & 6 & 11.0112 & 11.4750 & 12.0147 & 12.4006 & 12.7430 \\
    G191B2B & nom & 11.0094 &11.4733 & 12.0140 & 12.4021 &12.7504 \\
    P041C & 1 & 13.5112 & 12.2836 & 11.8512 &	11.7471 & 11.7339 \\ 
    P041C & 2 & 13.5185 & 12.2888 & 11.8518 &	11.7467 & 11.7327 \\
    P041C & 3 & 13.5025 & 12.2767 & 11.8547 &	11.7466 & 11.7333 \\
    P041C & 4 & 13.5067 & 12.2786 & 11.8519 &	11.7468 & 11.7330 \\
    P041C & 5 & 13.5183 & 12.2776 & 11.8522 &	11.7469 & 11.7339 \\
    P041C & 6 & 13.5073 & 12.2774 & 11.8520 &	11.7470 & 11.7346 \\
    P041C & nom &13.5107 & 12.2805 & 11.8523 & 11.7469 & 11.7336 \\
    \hline
  \end{tabular}
\end{table*}

\begin{table*}
  \centering
  \caption{Systematic uncertainty budget for SDSS}\label{tab:sdsssys}
  \begin{tabular}{l*{5}{c}}
    \hline
    \hline
    Source & u& g& r&i&z\\
    \hline
    PT measurement error\tablefootmark{a} & 0.0026 & 0.0015 & 0.0013 & 0.0025 & 0.0018 \\
    PT transformation to 2.5\tablefootmark{b} & 0.0029 & 0.0009 & 0.0008 & 0.0009 & 0.0017\\
    Dispersion around the transformation & 0.008 & 0.007 & 0.003 & 0.004 &0.004 \\
    USNO uniformity & 0.0015 & 0.0010 & 0.0010 & 0.0010 & 0.0010 \\
    \hline
  \end{tabular}
  \tablefoot{
    \tablefoottext{a}{This is based on the variance of the PT observations.}\\
    \tablefoottext{b}{This is based on the computed statistical error. The error is dominated by the zeropoint offset between the PT and 2.5~m.}}
\end{table*}
\section{SDSS/SNLS direct cross-calibration}
\label{sec:sdsssnls-direct-inte}

\subsection{Cross-calibration dataset}
\label{sec:interc-datas}
\Fixed{In theory,} several overlapping areas between the CFHTLS and 
the SDSS (in particular the CFHTLS deep fields D2 and D3) could be
used to tighten \Fixed{both surveys' photometric calibration}. 
\Fixed{However, the utility of existing published data sets is limited 
by two factors: the precision of the SDSS} zero point determination 
for single epoch photometry and the \Fixed{difficult-to-evaluate precision
of the inter-calibration between the D2-D3 DR8 and the Stripe 82 
observations}. In this work we rely instead on a \Fixed{specially} 
designed observation program of \Fixed{Stripe 82} with MegaCam.

The original program, called \mapc (for MegaCam absolute calibration
program), \Fixed{imaged} the CFHTLS deep fields D1 and D4 \Fixed{and} 
2 fields \Fixed{in Stripe} 82 (hereafter SDSS36 and SDSS326) in short 
``observing blocks''. Each observing block (OB) \Fixed{was} observed in 
about half an hour in two bands. Since switching filters with \megacam takes 
about 2 minutes, all the OB targets were first observed in one band, 
then another filter \Fixed{was} selected to re-observe all the targets.  
\Fixed{To enable the monitoring of extinction stability, each} observing block 
begins and ends \Fixed{with} an exposure \Fixed{of} the same field.
Observations \Fixed{extended} from December 2006 to February 2008.
The cross-calibration data set was complemented \Fixed{by an early-2011}
program of joint observations of D1 and SDSS36 alone (hereafter
OB\_D1\_ext) with a reduced scope. 

In the original program, 5 successive exposures \Fixed{of Stripe} 82
were taken with a coarse dithering applied. This \Fixed{allowed} us to
map a slightly larger field ($-0.8^\circ$ to $0.8^\circ$ in
declination\Fixed{, covering} two thirds of the width of Stripe 82)
while providing a cross check of the MegaCam flat-fielding. The
position offsets applied to each exposure are given in
table~\ref{tab:DitheringPattern} and roughly correspond to half of the
focal plane size. Short exposures on standard fields were also
included in the original program to provide an alternative
\Fixed{determination of the MegaCam absolute calibration}, but
\Fixed{this method proved more sensitive to observation systematics
  than the one} presented in Sect.~\ref{sec:direct-hst-standard}. Both
\Fixed{of these} complications were dropped from the complementary
program to shorten the observing block as much as possible. \Fixed{The
  definition and specifics of observing blocks for the two programs
  can be found in table~\ref{tab:OB_desc}.}

\begin{table}
  \centering
  \caption{Field definition in the MAPC program.}
  \begin{tabular}{ccc}
    \hline
    \hline
    Field & RA & DEC \\
    \hline
    D1 &2:26:00.0 &-04:30:00 \\
    D4 &22:15:31.0 &-17:44:05 \\
    SDSS36 &2:25:59.9 &-0:00:00 \\
    SDSS326 &21:46:00.0 &-0:00:00 \\
    SA95 &3:54:13.9 &+0:02:41 \\
    HZ4 &3:55:14.2 &+9:39:30 \\
    GD71 &5:52:22.2 &15:45:24 \\
    SA 113 &21:41:57.0 &0:19:32 \\
    Feige110 &23:19:53.1 &-5:17:55 \\
    BD+284211 &21:51:04.6 &28:44:01 \\
    \hline
  \end{tabular}

\end{table}

\begin{table*}
  \centering
  \caption{Observing group description in the \mapc program and its extension.}\label{tab:OB_desc}
  \begin{tabular}{l*{6}{c}lcl}
\hline
\hline
Name  & \band{u}\tablefootmark{a} & \band{g} & \band{r} & \band{i} & \band{i2} & \band{z} & Field & $t_\text{exp}$ (s) & Comment \\
\hline 
\multirow{6}{*}{OB\_D1} & \multirow{6}{*}{2} & \multirow{6}{*}{2} & \multirow{6}{*}{5} & \multirow{6}{*}{2} & \multirow{6}{*}{2} & \multirow{6}{*}{2} &
       D1 & 120\\
&&&&&&&SDSS36& 120 & 5 dithered exposures taken in a row \\
&&&&&&&SA95 & 3 & out of focus 1.5~mm\\
&&&&&&&HZ4 & 2& out of focus 1.5~mm\\
&&&&&&&GD71 & 2& out of focus 1.5~mm\\
&&&&&&&D1 & 120 \\ 
\hline
\multirow{6}{*}{OB\_D4} & \multirow{6}{*}{2} & \multirow{6}{*}{2} & \multirow{6}{*}{4} & \multirow{6}{*}{2} & \multirow{6}{*}{2} & \multirow{6}{*}{3} & 
      D4& 120\\
&&&&&&&SDSS326 & 120 & 5 dithered exposures taken in a row \\
&&&&&&&SA113& 3 &out of focus 1.5~mm\\
&&&&&&&Feige110& 2 & out of focus 1.5~mm\\
&&&&&&&BD+284211& 2 &  out of focus 1.5~mm\\
&&&&&&&D4 & 120 \\
\hline
\multirow{3}{*}{OB\_D1\_ext} & \multirow{3}{*}{2} & \multirow{3}{*}{7} & \multirow{3}{*}{3} & \multirow{3}{*}{0} & \multirow{3}{*}{7} & \multirow{3}{*}{2} & SDSS36 & 120\\
&&&&&&&D1 & 120\\
&&&&&&&SDSS36 & 120\\ 
\hline\end{tabular}
\tablefoot{ 
\tablefoottext{a}{This column and the following give the number of times each observing block was completed in each band. Measures were subsequently selected.}
}
\end{table*}

\begin{table}
  \centering
\caption{The dithering pattern in \mapc.}
\label{tab:DitheringPattern}
\begin{tabular}{ccc}
\hline
\hline
obs & $\delta_\text{ra}$ & $\delta_\text{dec}$\\
\hline
1 &0.00' &0.00' \\
2 &-24’ &+18’\\
3 &-24’ &-18’\\
4 &+24’ &-18’\\
5 &+24’ &+18’\\
\hline
\end{tabular}
\end{table}

From this data sample we derive \Fixed{Stripe 82 star catalogs} in
MegaCam natural magnitudes, calibrated on the SNLS deep fields that
can be readily compared to the SDSS catalog.

\subsection{Calibrated MegaCam measurements of the Stripe 82 fields}
\label{sec:calibr-megac-meas}

Images from the cross-calibration data are processed using
the standard SNLS photometric pipeline. Each resulting catalog is then
matched (regarding astrometry) to the tertiary star catalog of the
corresponding field, namely the SNLS tertiary star catalog described
in Sect.~\ref{sec:tert-catal-constr} for SNLS deep fields and the
corrected coadd SDSS catalog described in Sect.~\ref{sec:SDSS} for
Stripe 82 fields. A matching radius of 1 arcsec was used and a few other
selection criteria were applied to the SDSS corrected catalog:
\begin{itemize}
\item The number of \Fixed{$z$ band} observation\Fixed{s} of the star must be greater than 4
\item The \Fixed{rms of the $gri$ band star measurements} must be less than 30~mmag.
\end{itemize}

\Fixed{From studying MegaCam photometric flat-field corrections, we know} 
that the accuracy of single epoch flat-field\Fixed{s are} limited by
errors that are expected to average out \Fixed{in the construction of the
deep field tertiary star catalogs}. A simple way to propagate this
improvement to single epoch photometry is to align the photometry of
deep field exposures on the corresponding tertiary star catalog. For
each observing group we compute a rough flattening correction to the
photometry by computing a zero point deviation per chip in addition to
the global zero point. The following model is adjusted to the deep
field observations available in the observing group.
\[\cmag - \madu - \zp_t = \dzp(ccd)\,\] 
where \cmag designates the magnitude reported in the tertiary
catalogs, \madu the flat-fielded instrumental magnitudes and $\zp_t$
has been adjusted for each exposure in a first pass neglecting the
$\dzp$ term. The fit is \Fixed{done in flux} and measurements are
weighted according to the error model introduced in
Sect.~\ref{sec:catalog-constitution}. The $\dzp(ccd)$ corrections are
then applied to all images of the observing group.

The procedure described in section~\ref{sec:catalog-constitution} is
applied to build averaged MegaCam catalog\Fixed{s} of the 2 Southern Stripe
fields. Separate catalogs are built for different pointings.
The catalogs are then calibrated according to the zero points computed
on deep field exposures. The uncertainty on the catalog zero point is
estimated as the night-to-night dispersion of the fit residuals, divided by the
square root of the number of nights.

Direct comparison with the SDSS is enabled by the transformation of
the measurements to the SNLS uniform photometric system. This is done
using the transformation derived in Sect.~\ref{sec:transl-meas} for
stars lying in the valid color range of the transformation.

This procedure results in two sets of roughly 2000 and 6000 stars with
natural SDSS magnitudes (denoted $\smag$) and uniform SNLS magnitude
(denoted $\umag$). Depending on the pointing and photometric band 
\Fixed{between 2 and 9 nights are averaged} (see Table~\ref{tab:OB_desc}). 
\Fixed{For such short observing blocks, the rms of the nightly zero points
is typically below 5~mmag}.

\subsection{Passband consistency}
\label{sec:passband-consistency}

The first output\Fixed{s obtained} from the comparison of SNLS and
SDSS measurements of the Stripe 82 stars are precise color 
transformations between the two systems. We determined linear 
color-transformation\Fixed{s} between \Fixed{SDSS} and SNLS measurements
 for stars in the color range $0.5<g-i<1.5$ according to the model:
\begin{equation}
\umag - \smag = \alpha (g-i-0.8) + \beta\label{eq:27}
\end{equation}

\begin{figure}
  \centering
  \igraph{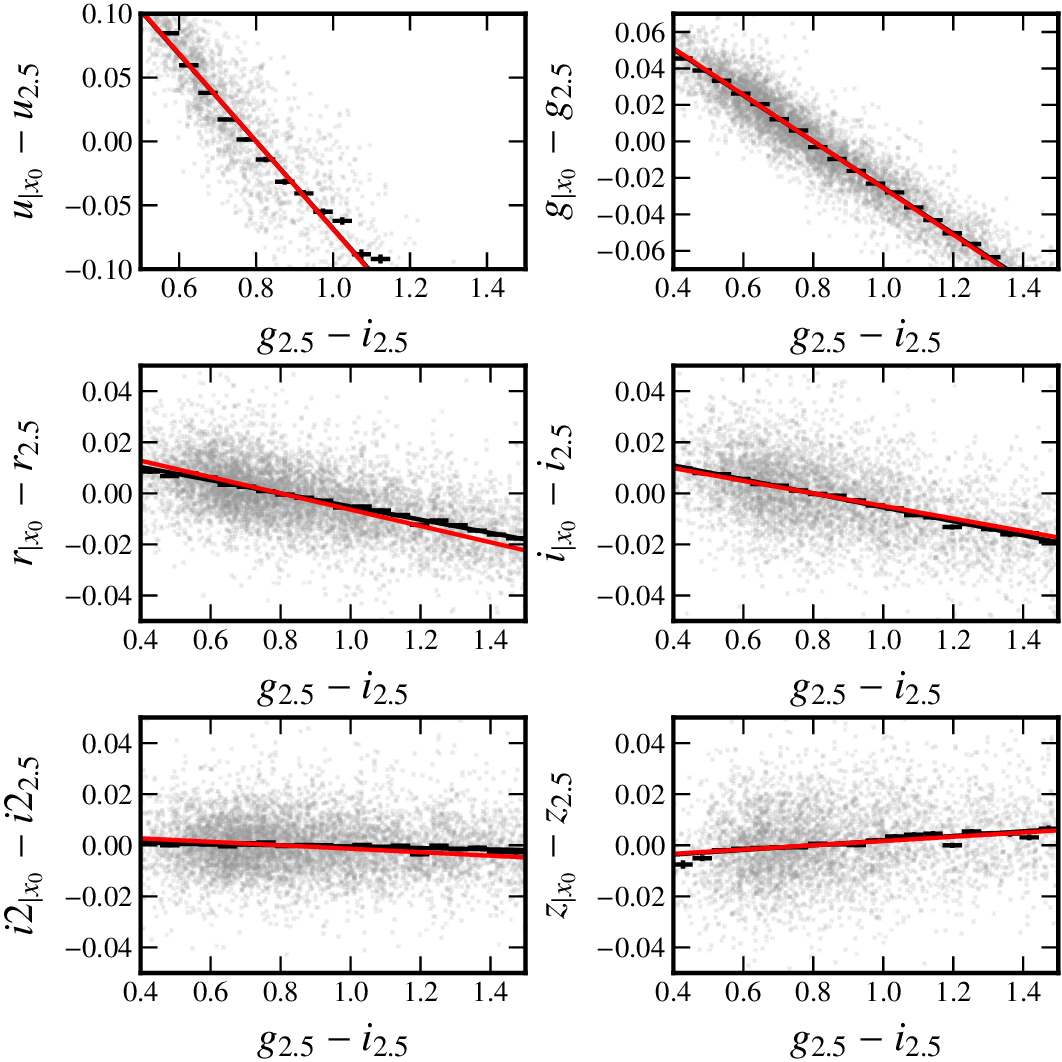}
  \caption{Linear color transformation\Fixed{s} between the SDSS and SNLS
    uniform magnitude system. The black lines represent the best
    linear fit of the data, whereas the red lines are based on
    synthetic magnitudes of the Gunn-Stryker-Bruzal stellar library
    and the SNLS and SDSS transmission models described in
    Sect.~\ref{sec:effective-passbands}.}
  \label{fig:sdsssnlsaveragetrans}
\end{figure}

Measurements are weighted according to the errors reported in both
catalogs added in quadrature. \Fixed{To account for residual variation
in the catalog zero point, i}ndependent $\beta$ are determined for
each MegaCam pointing. The transformations are illustrated in
Fig.~\ref{fig:sdsssnlsaveragetrans} and the measured slopes are given
in table~\ref{tab:trans}. We emphasize that
\Fixed{these equations describe transformations of the reference
MegaCam system (filters at the center of the focal plane) to the SDSS
system. The reference MegaCam system is not equivalent to and should 
not be confused with the average natural MegaCam system.}

\begin{table*}
  \centering
  \caption{Transformation equation between SNLS and SDSS systems.}
  \label{tab:trans}
    
\begin{tabular}{l*{6}{r}}
\hline\hline
 & $u$ & $g$ & $r$ & $i$ & $i2$ & $z$ \\
\hline
Expected color term &$-0.343 \pm 0.081$&$-0.128 \pm 0.004$&$-0.032 \pm 0.001$&$-0.025 \pm 0.002$&$-0.007 \pm 0.002$&$0.008 \pm 0.001$\\
Measured color term &$-0.341 \pm 0.002$&$-0.126 \pm 0.000$&$-0.026 \pm 0.000$&$-0.027 \pm 0.001$&$-0.003 \pm 0.000$&$0.009 \pm 0.001$\\
Apparent shift (nm) &$-0.1 \pm 4.9$&$-0.2 \pm 0.5$&$-2.2 \pm 0.5$&$1.8 \pm 1.7$&$-3.0 \pm 1.2$&$-0.5 \pm 0.7$\\
\hline
\end{tabular}
\end{table*}

\Fixed{These transformations} can be compared to \Fixed{synthetic 
photometry-based} predictions as a relative assessment of the accuracy 
of \Fixed{both instruments' effective passbands}. We synthesized stellar 
magnitudes for the 2 photometric systems from 2 stellar libraries, 
and determined a linear relationship for the resulting color-color plots. 
The slopes are given in table~\ref{tab:trans} and over-plotted in red on
Fig.~\ref{fig:sdsssnlsaveragetrans}. They are to be compared with the
fitted slope in black. We report the statistical uncertainty on the
slope determined on the Gunn-Stryker-Bruzal library, and the
difference of slope obtained with respect to the Pickles library as an
estimate of the systematic uncertainty on this quantity.

The discrepancy between the expected and measured slopes can be used
to estimate the uncertainty on the passband knowledge. We provide in
table~\ref{tab:trans} the wavelength shift to be applied to the SNLS
passbands to explain the whole effect. As can be seen the agreement is
good in all bands, except in \band r and \band{i2} \Fixed{where} a 
\Fixed{ $3\sigma$ shift} is detected. Note that the \band u measurement 
does not \Fixed{provide useful} constraints on the filter model.

\begin{figure}
  \centering
  \igraph{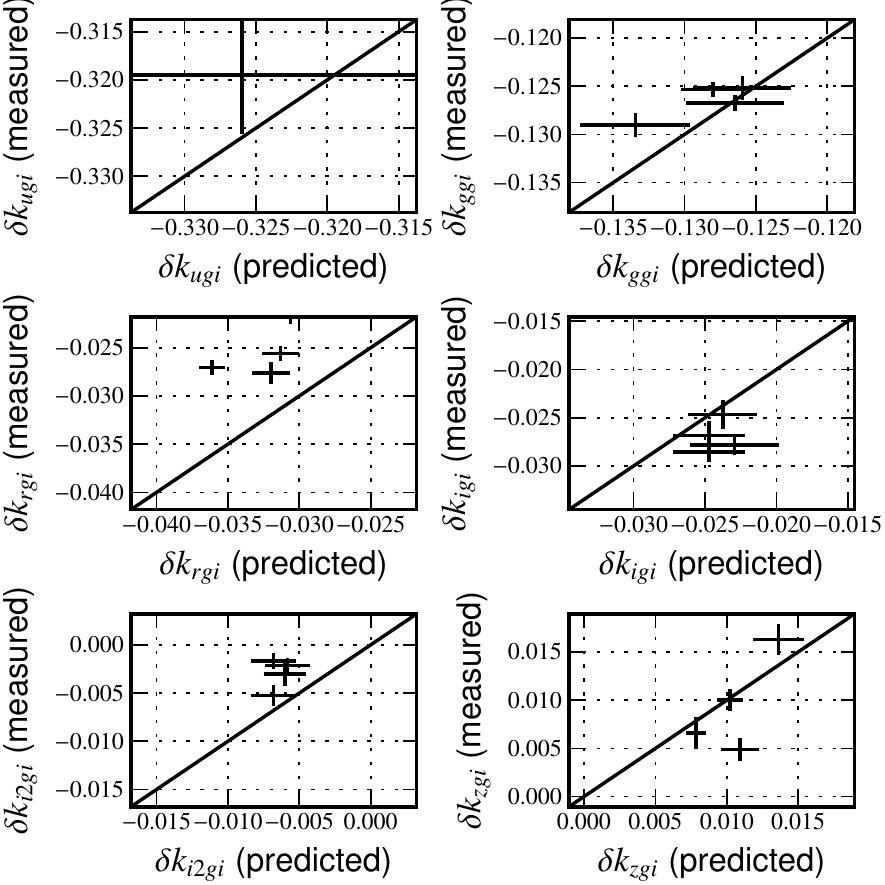}
  \caption{Comparison between synthetic and measured stellar locus
    slopes in the SNLS-SDSS color diagram for the 4 SDSS camcols
    covered in the observation program.}
  \label{fig:camcols}
\end{figure}

The effective SDSS passbands actually vary slightly from one chip to
another. In order to investigate whether \Fixed{passband differences
may account for some of the observed discrepancies}, we \Fixed{subdivided}
 the sample \Fixed{into} declination slices roughly corresponding to the 4 SDSS 
camera columns covered by our measurements. We compare \Fixed{these} results to 
slopes determined separately for the camera columns using the independent 
passband\Fixed{ determinations} from \cite{2010arXiv1002.3701D} in 
Fig.~\ref{fig:camcols}. There may be some correlation between the small 
observed and predicted \Fixed{slope variations} from one camera column to 
another, but the \Fixed{flat-fielding} uncertainty and other systematics make 
this comparison problematic. \Fixed{This is especially the case in \band u, where
both} the predicted and measured color terms are affected by large uncertainties, 
making the comparison \Fixed{inconclusive}. However, the measurements 
consistently disagree with the models in \band r and \band{i2} 
bands\Fixed{, indicating} that \Fixed{SDSS camera column transmission variations} 
cannot explain the discrepancies noticed in Table~\ref{tab:trans}. 
It is more likely that these \Fixed{discrepancies originate from} real 
\Fixed{deficiencies} in the MegaPrime passband\Fixed{ models}. We further 
discuss \Fixed{passband uncertainties} and their impact on the calibration 
in Sect.~\ref{sec:concl}.

\subsection{Uniformity of the photometry}
\label{sec:spatial-uniformity}

As they are subject to different systematics, a thorough assessment of
the photometry uniformity can be delivered by the comparison of the
two measurement sets. Due to the drift-scanning of the SDSS
observations, the uniformity of the SDSS photometry is expected to be
excellent at small scales in right ascension. In contrast, no specific
direction is expected to be preferred in MegaCam observations.

\begin{figure*}[!h]
  \centering
  \igraph{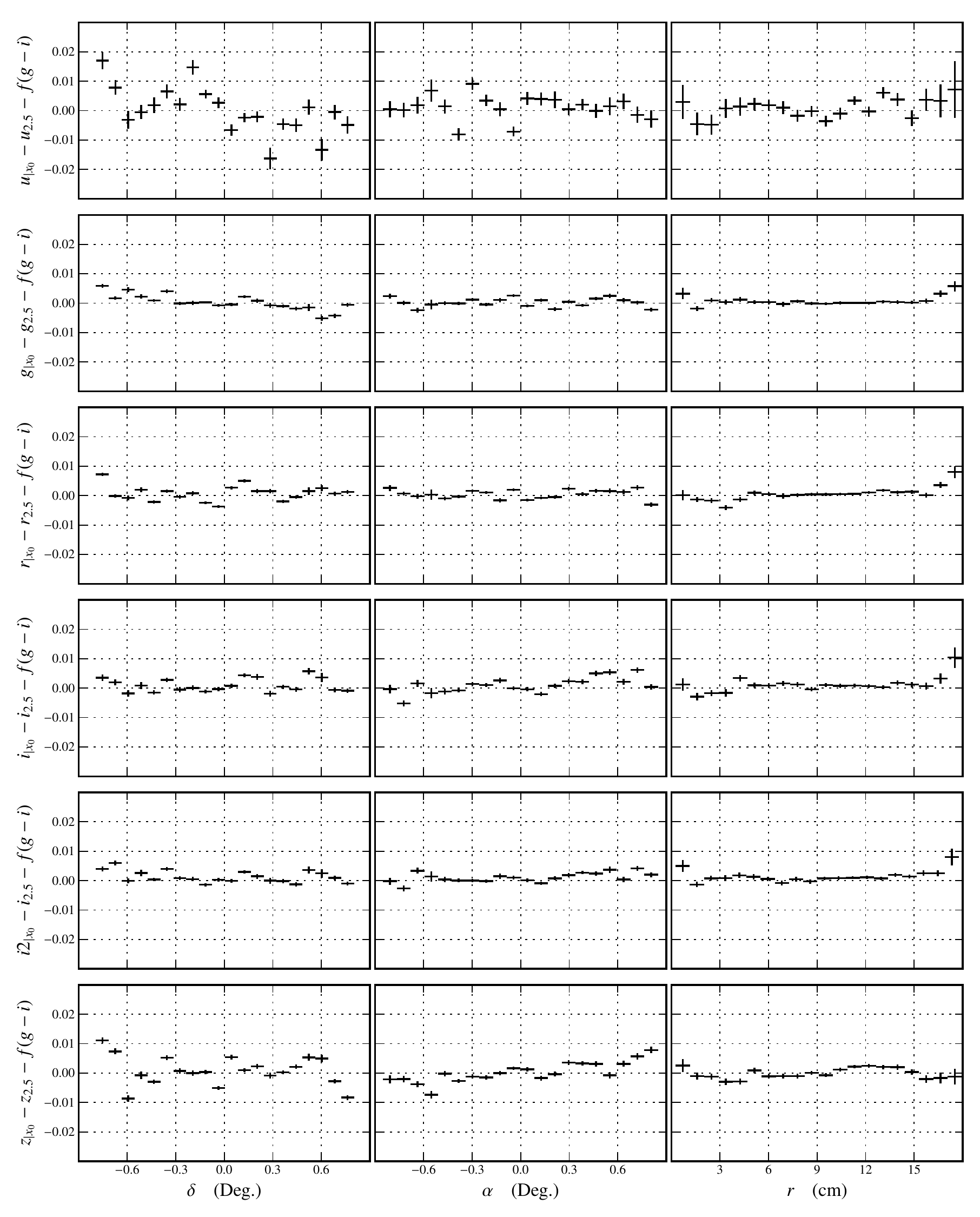}
  \caption{Relative uniformity of SNLS and SDSS photometry.}
  \label{fig:uniformity}
\end{figure*}

This study led to the discovery of the flat-field problem in the PT
and triggered the recalibration of the SDSS coadd catalog. The
photometry residuals (after correction of the Stripe 82 catalog) are
shown as a function of right ascension, declination and MegaCam radius
in Fig.~\ref{fig:uniformity}. Errors in SDSS flat-fielding would be
expected to be most prominently seen when plotted versus declination
while SNLS flat-fielding errors could appear in any of the plots.  The
radial plots correspond to an approximate \Fixed{MegaCam symmetry} and
exhibit some residual structure in $r$ and $z$ bands. The rms
of residuals in bins of right ascension are 4, 1, 2, 3, 2 and 3~mmag
in \band u \band g\band r\band i\band {i2} and \band z. Those
numbers agree with the expected \Fixed{large-scale flat-fielding} accuracy 
of \Fixed{the} SNLS tertiary catalogs.

\subsection{Average offsets}
\label{sec:cross-calibration}

We can now tie the SNLS and SDSS calibrations together. \Fixed{Ultimately}, 
the quantity of interest \Fixed{is the} flux interpretation of a calibrated
measurement. As filters \Fixed{themselves have uncertainties}, the error on
this quantity depends on the SED of the object itself (to first order \Fixed{in} 
its color). We discuss this question in the next section. 
We display in table~\ref{tab:caliboffset} the offset between
the two surveys for a star of color $g-i=0.8$ (defined as the $\beta$
term in Eq.~(\ref{eq:27})).

The offset is computed separately on each field. The statistical
uncertainty is estimated from the night-to-night dispersion of
measurements divided by the square root of the number of observation
nights. We then combine the two numbers and use the difference between
the two fields as an estimate of the uncertainty on the mean related
to the residual non-uniformity of the surveys.

\begin{table}
  \caption{$\beta$ values as defined by Eq.~\ref{eq:27} between SDSS and SNLS surveys.
  \label{tab:caliboffset}}
  \centering
  
  \begin{tabular}{rccc}
\hline\hline
 & SDSS36 & SDSS326 & Combined\\
\hline
u&
$-0.432 \pm 0.002 $&
$-0.447 \pm 0.002 $&
$-0.438 \pm 0.001 \pm 0.014$\\
g&
$-0.082 \pm 0.002 $&
$-0.074 \pm 0.001 $&
$-0.076 \pm 0.001 \pm 0.008$\\
r&
$-0.017 \pm 0.001 $&
$-0.016 \pm 0.001 $&
$-0.017 \pm 0.001 \pm 0.001$\\
i&
$-0.001 \pm 0.001 $&
$-0.003 \pm 0.001 $&
$-0.002 \pm 0.000 \pm 0.002$\\
i2&
$0.010 \pm 0.003 $&
$0.012 \pm 0.002 $&
$0.012 \pm 0.001 \pm 0.003$\\
z&
$0.030 \pm 0.001 $&
$0.035 \pm 0.002 $&
$0.031 \pm 0.001 \pm 0.005$\\
\hline
\end{tabular}
  \tablefoot{All numbers are expressed in magnitudes. Those numbers should not be interpreted as calibration offsets, part of the difference is expected from the definition of the magnitude systems. A discussion about AB offsets can be found in Section~\ref{sec:concl}.}
 
\end{table}
\section{Combined calibration}
\label{sec:concl}

The calibration products for both surveys consist of:
\begin{itemize}
\item measured natural magnitudes for a set of tertiary standards $m^3$,
  in an arbitrary magnitude system.
\item magnitudes of primary spectrophotometric standards $ m^s$
  in the same magnitude system, affected by some measurement and
  systematic errors $n_s$.
\end{itemize}
The primary spectrophotometric standards are to be used to set the
actual flux scale (equivalently determining a magnitude offset to the
AB system $\delta_\ab$) of the first by equating the measured
magnitude of the standard $m^s$, with its \Fixed{expected magnitude} 
$\tilde m^s$ synthesized from the spectrophotometric reference 
measurement:
\begin{equation}
  \label{eq:33}
  m^s = \tilde m^s + \delta_\ab\,.
\end{equation}
with:
\[\tilde m^s = -2.5\logdec\frac{\int_\lambda \lambda T(\lambda) S^s(\lambda) d\lambda}{\int_\lambda \lambda T(\lambda) S_\ab(\lambda) d\lambda}\]
The result is affected by uncertainties on the effective instrument
passbands $T$ and on the SED $S^s$ of the reference, as well as the
error on the measured magnitude $n_s$.  

In our case we have at our disposal:
\begin{itemize}
\item 59 direct measurements of 3 solar-analogs (P330E, P177D and SNAP2) performed with MegaCam and calibrated to the SNLS system (see Sect.~\ref{sec:direct-hst-standard}).
\item \citet{2007AJ....133..768L} measurements of the star \bdtruc,
  color transformed to the SNLS system in 5 bands (see Sect.~\ref{sec:absolute}).
\item PT measurements of 4 photometric standards (\bdtruc, P330E,
  P177D, P041C) color-transformed to the 2.5~m system in the 5
  photometric bands (see Sect.~\ref{sec:sdss-pt-calibration}). We do not
  use the PT observations of bluer standards as they \Fixed{are outside 
  the valid range of the PT$\rightarrow$SDSS system color transformation.}
\end{itemize}
In addition, the cross-calibration data described in 
Sect.~\ref{sec:sdsssnls-direct-inte} \Fixed{yields 6 constraints on the 
relation between the two  surveys' tertiary catalog flux scales, and 
a} straightforward constraint on the \band{i} and \band{i2} magnitudes
is \Fixed{given} by the comparison of tertiary star  magnitudes 
in these two bands (see Sect.~\ref{sec:interc-bandi-band}). \Fixed{Altogether, 
we have} a total of 91 measurements (or pseudo-measurements) 
\Fixed{and }associated uncertainty estimates \Fixed{with which to} 
constrain the AB offsets of the tertiary catalogs in the 11 photometric bands 
considered.

Our purpose here is to \Fixed{optimally combine the entire set} of 
calibration data, to check its consistency, and to derive a reliable 
estimate of the uncertainty on the resulting calibration. As we want to combine
several standard stars measured in different photometric systems
related by linear transformations, we start by discussing and
parametrizing the uncertainty on the instrument passbands and primary
standard SEDs.

\subsection{Uncertainties on passbands}
\label{sec:filter-uncertainty}

The uncertainty on the effective instrument passband \Fixed{comes from various sources} 
and \Fixed{may} take some arbitrarily complex form. 
\Fixed{For instance, the portion of passband uncertainty from} 
optical elements \Fixed{whose transmissions can be measured} should \Fixed{approximately} 
reduce to the uncertainty \Fixed{of} the wavelength calibration of 
the monochromator (\emph{i.e.} a global wavelength shift).
\Fixed{On the other hand, the portion of passband uncertainty due to}
other elements, like \Fixed{the} CCD QE, can be harder to characterize. 
The determination of the average atmospheric transmission is also affected by 
uncertainties, which \Fixed{primarily impact} the $u$ band calibration. 
In what follows, we assume that \Fixed{we can parametrize} the error affecting the 
effective passband defining the photometric band $b$ \Fixed{by} a shift in the 
\Fixed{passband's} mean wavelength \wshift[b]. 
\Fixed{In other words, when} performing first 
order computation\Fixed{s} in the next section, we will assume that \Fixed{passband
uncertainties} result from a global shift of the filter. We now attempt to derive 
\Fixed{an} estimate of the amplitude of the uncertainty on the mean wavelength $\wshift$.

\subsubsection{SDSS passbands}
\label{sec:sdss-filters}
Measurement errors on the SDSS passbands are not completely
characterized by \citet{2010arXiv1002.3701D}, but we will estimate the
uncertainty from data in that paper. The accuracy of the monochromator
calibration is given as $\pm3\angstrom$, which we assume is a constant error
for all wavelengths in a given filter band. The ability to reproduce
an independent quantum efficiency measurement (Figure 2 in
\citealt{2010arXiv1002.3701D}) suggests that the amplitude error is
less than 2\%. Assuming a $\pm2\%$ error that is linear in wavelength
across the band, we calculate an error in the mean filter wavelength that is
0.3\% of the filter bandwidth. Using $1500\angstrom$ for the filter width
and combining that uncertainty with the monochromator uncertainty,
results in a $6\angstrom$ uncertainty in mean filter response. The $1500\angstrom$
width is slightly too large for g, r, and i bands and a factor of 3
too large for u and z. However, we don’t decrease the error estimate
for these bands because there are probably other (unquantified)
uncertainties for these bands (\Fixed{e.g.} the difficulty of determining the 
UV transmission and the red cutoff in z).

\subsubsection{SNLS passbands}
\label{sec:snls-passbands}

The lack of monitoring for the MegaCam passbands proved to be a weak
point in the SNLS survey. \Fixed{It was only recently realized} that two of
the interference filters may have undergone significant alteration in
their early days.

The REOSC filters were measured twice (in 2002 and 2006) and both
measurements \Fixed{appear} to be in rough agreement in all bands but \band i
and \band r. Two arguments support the hypothesis that the 2006
measurements yield a better representation of the survey passbands:
first, the stability of the color locus of tertiary stars during the 5
years of the survey, presented in Sect.~\ref{sec:passband-variations},
exclude\Fixed{s} the hypothesis of a slow aging of the filters between 2003 and
2008. Second, recent lab measurements of pieces of the broken \band i
filter were found \Fixed{to be} in agreement with the 2006 measurement 
while incompatible with the 2002 ones. 
\Fixed{Because the} 2006 measurements \Fixed{only cover} the outer \Fixed{edge}
of the filters (12-15 cm away from the center), it was not possible to use them 
directly. \Fixed{Instead,} we opt for applying a correction to
the 2002 \band i and \band r measurements (see
appendix~\ref{sec:megacam-band-i}). The unaltered 2002 measurements
are used \Fixed{as-is} for the \band u\Fixed{, }\band g\Fixed{,} and \band z 
filters. We now try to quantify the size of the uncertainty on the mean passbands
resulting from these choices.

The mean wavelength shift between the 2002 and 2006 measurements
provides an estimate of the error at a specific position (about 15cm
away from the center).  The measured differences amount to $5$, $3$,
$-37$, $-31$ and $6$ ~\angstrom\ in \band{u}, \band{g}, \band{r},
\band i and \band{z} bands respectively. The spatial consistency of
the filter model is then ensured by the agreement of modeled and
measured color terms between different positions (as illustrated on
Fig.~\ref{fig:measvssynthdk}). A noticeable exception is the \band r
filter for which the measured and modeled color terms start to
disagree at radii greater than 10cm. Without proper measurements of
the \band r filter at radius smaller than 10 cm, it is thus not
possible to trust the corrected filter model more than the uncorrected
model. \Fixed{Therefore,} we assign an uncertainty of $37\angstrom$ to 
\Fixed{the} \band r filter passband. The situation appears a 
bit better for the \band i filter, since the accuracy of the 2006 measurement 
has been checked (\emph{cf.} appendix \ref{sec:megacam-band-i}), and the
agreement between predicted and measured color-terms is quite
good. Nevertheless, it seems premature to assign an uncertainty
smaller than 31~\angstrom\ to the \band i passband without better
\Fixed{understanding the source of the filter change}. For the \band 
u\Fixed{, }\band g\Fixed{,} and \band z filters, we also use the quoted 
differences between the two available measurements, namely $5$, $3$ and $6$
\angstrom, as an estimate of the measurement uncertainty.

The \band{i2} BARR filter was never completely measured. In particular
its effective refraction index is unknown and assumed to be similar to
\Fixed{that} of the \band{i} filter. We base our estimate of the
\band{i2} \Fixed{passband uncertainty} on the stellar locus in the
plane $\band{i2}-i$ vs $g-i$ \Fixed{which has been well-constrained by
  observations of Stripe 82 stars and synthetic photometry predictions
  (see Sect.~\ref{sec:passband-consistency})}. \Fixed{The comparison
  between synthetic predictions and measurements suggests} an
\Fixed{\band{i2}} shift of $-30$~\angstrom (Table~\ref{tab:trans}).

The uncertainty on the average atmospheric extinction curve
is estimated in \citet{Buton2012} to be smaller than $0.02$
mag/airmass. This has a negligible impact on the definition of any of
the passbands. A noteworthy exception is the \band z filter, which
covers a region of strong H$_2$O absorption
(8916-9929~\angstrom). Errors in the modeling of the average
extinction could cause the mean wavelength of the effective filter to
shift by a few angstrom\Fixed{s}. \Fixed{However,} the error is \Fixed{unlikely} 
to exceed the uncertainty quoted for this filter. Given their high throughput 
and the smoothness of their transmission, we do not expect significant 
contribution\Fixed{s} to a wavelength shift from any of the instrument elements 
besides the interference filters.

\subsection{SED of the primary standard stars}
\label{sec:sed-primary-standard}

\Fixed{Five} spectrophotometric standard stars from the CALSPEC
database \Fixed{are being considered in our final calibration data:
the} F sub-dwarf \bdtruc and the 4 G-type stars P041C, P177D, P330E and SNAP2. 
All \Fixed{these} stars have STIS spectra calibrated by the model spectra 
of \Fixed{the} white dwarfs G191B2B, GD153, and GD71. Any error in the spectral modeling 
of those three fundamental standard stars would equally affect our 5 standard stars.

\Fixed{The calibration reference is affected by both the} white dwarf atmosphere 
model error \Fixed{and} uncertainties in the determination of the primary standard 
surface gravity and effective temperature \Fixed{from observations of their Balmer line profiles}. 
\Fixed{Following R09, we} base our estimate on considerations from \citet{2002acs..rept....4B}
\Fixed{and adopt} a 0.5\% slope uncertainty (1~$\sigma$) over the range 3000\angstrom-
10000\angstrom. \Fixed{Relative to $g$ band\footnote{We are not interested in the overall 
flux scale. Otherwise the uncertainty would be dominated by the determination of the Vega 
absolute flux, which \Fixed{does} not play any role here.}, the} resulting (correlated) 
uncertainties are $0.7$, $1.1$, $2.2$ and 3.2~mmag in $u$, $r$, $i$ and $z$
bands respectively. 

In addition, some measurement error is expected to affect the
individual measurements. Similarly to what was done in R09, our
estimate of the covariance of broadband magnitudes synthesized from
STIS spectra is based on the repeatability of the monitoring spectra
of AGK +81 266. The individually measured spectra were integrated in the MegaCam
and SDSS passbands and an empirical covariance matrix of the
measurements was built. We found a repeatability of 5, 4, 5, 6 and 11 mmag in
broadband magnitudes $ugriz$ for both instruments, with significant
correlation between neighboring bands, and (as expected) nearly
perfect correlation between overlapping bands. The full covariance
matrix is delivered in table~\ref{tab:stiscov}. Some of the standard
spectra were established with several independent
measurements. When \Fixed{determining} the expected broadband noise of the
star measurement the covariance matrix is divided by the number
of \Fixed{observations}.

\subsection{Mixing calibration data}
\label{sec:mixing-calibrations}

For each available measurement (or pseudo-measurement in the case of a
color-transformed magnitude) of a photometric standard $s$ in a
photometric band $b$, denoted $m^s_b$,  \Fixed{the corresponding expected 
standard AB magnitude $\tilde m^s_b$ } can be calculated from its CALSPEC 
spectrum. The model for the calibration data is:
\begin{equation}
  \label{eq:37}
  m^s_b - \tilde m^s_b = \delta_\ab^b + e_{s,b}\,,
\end{equation}
where $e$ is the error term affecting the estimate of $m^s_b -
\tilde m^s_b$. It is further decomposed as:
\begin{equation}
  e_{s,b} = n_{s,b} + c_{s,b} + \frac {\partial \tilde m^s_b}{\partial \weff[b]} \wshift[b]\,, \label{eq:35}
\end{equation}
where $n_{s,b}$ is the measurement error affecting $m^s_b$
(including systematic uncertainties on the measurement), $c_{s,b}$ is
the uncertainty affecting the broadband CALSPEC magnitudes, and $\frac
{\partial \tilde m^s_b}{\partial \weff[b]} \wshift[b]$ describe the first
order impact of the filter uncertainty on $\tilde m^s_b$.

The cross-calibration data (between SNLS and SDSS as well as between
bands \band{i} and \band{i2}) \Fixed{adds} further constraints. 
\Fixed{These} take the generic form of approximate linear relations between 
very similar filters, applicable on average to the tertiary stars:
\begin{equation}
  \label{eq:34}
  m^s_b - m^s_{b'} \approx \alpha c + \beta
\end{equation}
with some uncertainty on $\alpha$ and $\beta$.\footnote{A complete
  interpretation of these relations is not trivial as it requires some
  spectral model of the stellar population composing the tertiary
  standard stars. The estimation of resulting uncertainties would be
  somewhat complex. We prefer to rely on a star with available and
  reliable spectrophotometry and model how it deviates from the
  average stellar population.}  As before, we translate these color
relations into cross-calibration constraints by assuming that those
relations also apply to a spectrophotometric standard lying in the
color range of the transformation. The extra error coming from the
dispersion around the color transformation has to be estimated from
stellar libraries. We thus add to our model the following \Fixed{terms}:
\begin{equation}
 \beta_{bb'} - \alpha_{bb'} c^s - \tilde m_b^s +  \tilde m_{b'}^s= \delta_\ab^b  - \delta_\ab^{b'} + e_{bb'}
\end{equation}
where $(b,b')$ ranges \Fixed{over} $\lbrace(\band{u}, u),$ $(\band{g}, g),$
$(\band{r}, r),$ $(\band{i}, i),$ $(\band{i2}, i),$ $(\band{z}, z),$
$(\band{i2}, \band{i})\rbrace$.  Again the error term is compound:
\begin{equation*}
e_{bb'} = c_{s,b} - c_{s,b'} + \frac {\partial m^s_b}{\partial
    \weff[b]} \wshift[b] - \frac {\partial m^s_{b'}}{\partial
    \weff[b']} \wshift[b'] + e_\alpha c + e_\beta + d\,,
\end{equation*}
where $d$ is an uncertainty term accounting for the dispersion of
stars around the color relation. We obtain the expected rms of the $d$
term by measuring the dispersion around the average transformation of
magnitudes synthesized from a stellar spectrum library. We find $48$,
$6$, $2$, $4$, $3$ and $2$ mmag of dispersion around the relations
between the SNLS and corresponding SDSS band for
\band{u},\band{g},\band{r},\band{i},\band{i2},\band{z} respectively.

To ease the handling of correlated uncertainties between all the
measurements, we can gather the 91 calibration measurements into a
single vector $\vec y$  and write the linear model for the calibration data in
matrix form:
\begin{equation}
\vec{y} = \tens{A} \vec{x} + \vec{e}
\end{equation}
We want to solve for the vector of calibration offsets $\vec x =
(\delta_\ab^{\band{u}}, \cdots, \delta_\ab^{\band{z}}, \delta_\ab^u,
\cdots, \delta_\ab^z)$. The Jacobian matrix $\tens A$ relates each
measurement to the corresponding offset (or a difference between two
offset\Fixed{s} in the case of cross-calibration constraints), and $\vec e$ is
the vector of error terms.
\subsection{Covariance matrix of calibration data }
\label{sec:over-corr-matr}

Most of the measurement systematic uncertainties \Fixed{coherently} affect
different measurements (even across different bands). Neglecting the
correlations would significantly underestimate the true error. We thus
formed the full covariance matrix $\tens R= \cov(\vec e)$ of the
calibration measurements accounting for all the error terms in the
model above.

We already described the computation of the covariance of broadband
CALSPEC magnitudes in section~\ref{sec:sed-primary-standard}. The
computation of the resulting contribution to $\tens R$ accounts for
the STIS measurement error and the uncertainty on white \Fixed{dwarf} primary
calibration differently. The STIS measurement errors are assumed to be
uncorrelated from one star to another and divided by the number of
\Fixed{observations} for each star. The calibration error is assumed to be a
systematic error that produces the same error for all the measurements
of all the stars.

The error affecting the determination of the \Fixed{filter} effective
wavelength is assumed unrelated from one filter to another. Its
contribution $\tens R^F$ to $\tens R$ is obtained as:
\[\tens R^F = \tens H^\dag \diag (\wshift[b])^2_b \tens H
\]
where $\tens H$ is the matrix holding the terms $\frac {\partial
  m^s_b}{\partial \weff[b]}$ corresponding to each measurement. 

\Fixed{Due to the large number of systematic uncertainties to be 
accounted for, the construction} of the covariance matrix of the 
measurement error terms is less straightforward. In \Fixed{brief, we have
assumed} that \Fixed{all} statistical uncertainties arising from measurement
noise (photon noise or flat-field noise) \Fixed{are} uncorrelated from one
measurement to another, \Fixed{whereas part of the} systematic \Fixed{uncertainties} 
affect measurements coherently and
\Fixed{are therefore taken to be fully correlated}.
\Fixed{Full} details of the assumptions made \Fixed{in the covariance matrix 
computation} are given in appendix~\ref{sec:covariance-matrix}. 

\subsection{Least-square estimate of the calibration}
\label{sec:least-square-estim}

We derived the combined estimate of the calibration offsets through
weighted least-square minimization. The resulting offsets are given in
table~\ref{tab:combcalib}, along with the diagonal component of their
covariance matrix. The corresponding correlation matrix is given in
table~\ref{tab:corr}. The offsets are applied to the catalog of
tertiary standard stars to provide our best estimate of \Fixed{their} 
AB magnitudes.

The global $\chi^2$ of the fit is 79.8 for 80 degree\Fixed{s} of freedom.  
Note that we are mainly manipulating actual systematic uncertainties rather
than Gaussian-distributed random variables. A better assessment of the
agreement between the different set\Fixed{s} of primary standard measurements
is provided by Fig.~\ref{fig:calibcomp}. We defer the discussion of
these results to the next section of the paper.

\begin{table}
  \centering
  \caption{Correlation matrix of the final calibration.}
  \label{tab:corr}
  
\begin{equation*} \left(\begin{array}{*{11}{l}}
1.0 & 0.1 & 0.0 & -0.0 & -0.0 & -0.0 & 0.0 & 0.0 & 0.0 & -0.0 & -0.1
\\
 & 1.0 & 0.1 & 0.1 & 0.1 & -0.0 & 0.1 & 0.2 & 0.1 & 0.0 & -0.1
\\
 &  & 1.0 & 0.3 & 0.3 & 0.2 & 0.0 & 0.0 & 0.6 & 0.3 & 0.2
\\
 &  &  & 1.0 & 0.9 & 0.5 & -0.1 & -0.0 & 0.4 & 0.8 & 0.6
\\
 &  &  &  & 1.0 & 0.5 & -0.1 & -0.0 & 0.4 & 0.8 & 0.6
\\
 &  &  &  &  & 1.0 & -0.1 & -0.1 & 0.3 & 0.5 & 0.8
\\
 &  &  &  &  &  & 1.0 & 0.1 & 0.0 & -0.1 & -0.1
\\
 &  &  &  &  &  &  & 1.0 & 0.0 & -0.0 & -0.1
\\
 &  &  &  &  &  &  &  & 1.0 & 0.4 & 0.4
\\
 &  &  &  &  &  &  &  &  & 1.0 & 0.6
\\
 &  &  &  &  &  &  &  &  &  & 1.0
\\
\end{array}\right)
\begin{array}{*{1}{r}}
u_M\\g_M\\r_M\\i_M\\i2_M\\z_M\\
u\\g\\r\\i\\z\\
\end{array}\end{equation*}
\end{table}

\section{Results}
\label{sec:calibration-products}

\subsection{Calibration products}
\label{sec:calibration-products-1}

We deliver natural broadband AB magnitudes for a large set of selected
stars (the tertiary standards) in the SNLS and SDSS science
fields. \Fixed{These} tertiary standards will be used to calibrate the
science images of the surveys and may be useful for other, similar
purposes. 

The calibration of the catalogs to the AB system results from the
combination of all the calibration data gathered in this work. The
details of the combination are described in
Sect.~\ref{sec:concl}. Briefly, we combined calibration constraints
coming from observations of 5 different spectrophotometric standard
stars from the CALSPEC database. Those observations were conducted
following 3 mostly independent paths presented in Sect.~\ref{sec:project}
and described respectively in
Sect.~\ref{sec:absolute}, Sect.~\ref{sec:direct-hst-standard} and
Sect.~\ref{sec:sdss-pt-calibration}. In addition, the SNLS and SDSS
tertiary catalogs are tied together by a cross-calibration measurement
described in Sect.~\ref{sec:sdsssnls-direct-inte}.

We emphasize that the magnitudes are delivered in the natural
photometric systems of the corresponding instrument, as defined in
Sect.~\ref{sec:effective-passbands}. As a consequence, their correct
interpretation is given by Eq.~(\ref{eq:24}) using the effective
filters corresponding to the average position of the star observation
in the field of view. We deliver a model of MegaCam passbands along
with this paper and refer the reader to the work from
\citet{2010arXiv1002.3701D} for the SDSS effective passbands.

Another consequence \Fixed{of} the variation of the natural photometric
system with the position in the field of view is that, for \Fixed{calibration
purposes}, the reference catalogs are \Fixed{best} matched with exposures
sharing a compatible pointing (same pointing for MegaCam or same
declination for SDSS). In other cases, small color terms accounting
for the slight differences in the effective filters may have to be
considered to \Fixed{obtain} accurate zero points.

One may also note that the catalogs have been established using
IQ-scaled aperture photometry in the case of SNLS and the usual PSF
photometry algorithm in the case of SDSS. While the result does not
depend on the photometry method \Fixed{to} first order, small spatial or
chromatic effects may need to be accounted for when comparing to
measurements obtained using different photometry methods (see
\emph{e.g.} Sect.~\ref{sec:other-probl-phot}).

For most comparison purposes, it is also convenient to report MegaCam
measurements in a single photometric system, close to the natural
system. In this paper, we follow the convention adopted in R09 and
define the MegaCam uniform magnitudes as the natural magnitudes that
would have been observed for objects \emph{at the center} of the focal
plane\footnote{Note that this choice is not representative of the
  ``average'' MegaCam natural system. Passbands at the center are
  redder than the spatial average.}. We have derived color transformations
between the natural and uniform system (see Sect. ~\ref{sec:transl-meas}) 
\Fixed{and include in the catalog MegaCam uniform} magnitudes for 
all stars that can be accurately transformed.

Finally, we determine measured color terms between the uniform MegaCam
photometric system and the average SDSS system in
Sect.~\ref{sec:passband-consistency}. \Fixed{These facilitate} comparison of
other data specifically obtained in these photometric systems.

Catalogs can be found in tables~\ref{tab:S82_catalog}
and~\ref{tab:D1_catalog}.\footnote{Full catalogs can be retrieved from
  the CDS or the dedicated webpage
  \url{http://supernovae.in2p3.fr/snls_sdss/}.} AB offsets
given in table~\ref{tab:combcalib} are \emph{already} applied to the
released catalogs. The covariance matrix associated with this calibration
 is given in tables~\ref{tab:combcalib}
and~\ref{tab:corr}. Finally the \Fixed{Megacam effective transmission
models} to be used in Eq~(\ref{eq:2}) are given in table
\ref{tab:u_band_table}-\ref{tab:z_band_table}.\footnote{Higher
  resolution versions can be found on the webpage.} What follows is a
general discussion of these results.

\subsection{Agreement between the available calibration paths}
\label{sec:agre-betw-avail}

\begin{figure}
  \centering
  \igraph{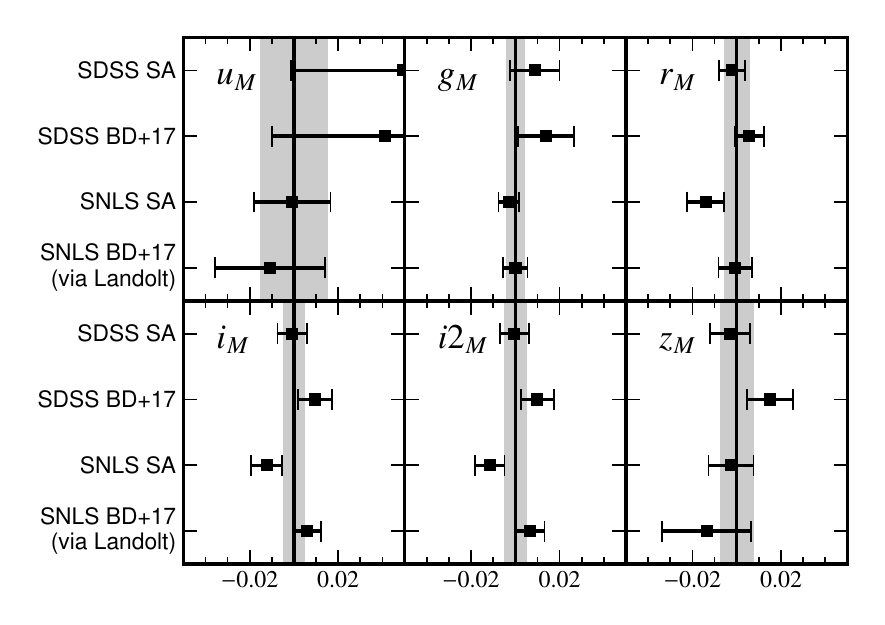}
  \caption{Agreement between \Fixed{complementary} determinations of the 
    SNLS AB offsets. SNLS BD+17 and SNLS SA lines refer to
    the SNLS calibration on Landolt standards and direct solar \Fixed{analog}
    measurements respectively
    (\emph{cf.} Sect.~\ref{sec:tert-catal-constr}). The SDSS SA and SDSS
    BD+17 lines refer to the calibration obtained by translating the
    PT measurements of the solar analogs and \bdtruc respectively, to
    the SNLS system via the relations obtained in
    Sect.~\ref{sec:sdsssnls-direct-inte}. The data points \Fixed{correspond}
    to the calibration offset relative to the combined calibration
    including all the measurements and are not independent. The shaded
    region \Fixed{illustrates} the uncertainty on the combined calibration.}
  \label{fig:calibcomp}
\end{figure}

A key feature of this work is the redundancy between several
calibration paths established independently and subject to different
systematics. An illustration of the agreement between the different
paths is provided by Fig.~\ref{fig:calibcomp}.

The agreement between the SNLS and SDSS calibration appears better
than one percent in the $gri$ and $z$ bands. There is an apparent
offset of about 4\% in $u$ which is still compatible with the low
precision of the cross-calibration data in this band. Internally,
there is a small tension between the Landolt and solar analog\Fixed{ 
calibrations} of the SNLS in \band{i}/\band{i2} and \band{r} ($\sim 1.9$
and $1.3$\%). To a lesser extent, a similar trend is visible in the
SDSS which suggests that at least a part of the tension may be
explained by an actual calibration difference between the \bdtruc and
the solar \Fixed{analog} CALSPEC reference spectra.

Overall, we reckon that the whole data sample is compatible with the
modeled uncertainties. Consequently we opted for combining the whole
data sample to set the flux scale of the tertiary catalogs and we
believe that the uncertainties resulting from the combination are
reliable.

\begin{table*}\centering
  \caption{SDSS and SNLS combined AB offsets.}
  \label{tab:combcalib}
  
\begin{tabular}{l|*{6}{l}|*{5}{l}}
\hline\hline
Band & $u_M$ & $g_M$ & $r_M$ & $i_M$ & $i2_M$ & $z_M$ & $u$ &  $g$ &  $r$ &  $i$ &  $z$ \\
\hline
$\delta_\ab$ &0.011 & -0.000 & 0.001 & -0.006 & -0.007 & 0.014 & 0.068 & -0.020 & -0.005 & -0.018 & -0.010\\
$\sigma$ &0.0145 & 0.0035 & 0.0051 & 0.0042 & 0.0043 & 0.0069 & 0.0089 & 0.0050 & 0.0031 & 0.0039 & 0.0060\\
\hline
With perfect passbands\tablefootmark{a} &0.0122 & 0.0033 & 0.0033 & 0.0041 & 0.0041 & 0.0069 & 0.0059 & 0.0040 & 0.0030 & 0.0039 & 0.0060\\
With perfect reference\tablefootmark{b} &0.0144 & 0.0031 & 0.0046 & 0.0029 & 0.0030 & 0.0044 & 0.0086 & 0.0047 & 0.0022 & 0.0024 & 0.0029\\
\hline
\end{tabular}
  \tablefoot{Figures are given in magnitudes.\\
    \tablefoottext{a}{Remaining uncertainty on the AB calibration when assuming that the passband \Fixed{measurements} are perfect.}\\
    \tablefoottext{b}{Remaining uncertainty on the AB calibration when assuming that the STIS spectra are perfect.}
  }
\end{table*}

\subsection{Uncertainties on the combined calibration}
\label{sec:uncert-comb-calibr}

A few points pertaining to the calibration uncertainties ought to be
discussed.

First, one can observe from the correlation matrix in
table~\ref{tab:corr} that the cross-calibration data \Fixed{brings}
significant correlations in the calibration of \Fixed{all} the SNLS and the SDSS overlapping bands \Fixed{but} $u$ and $g$, where the accuracy of the 
cross-calibration is limited by \Fixed{the less-similar} filters and the 
\Fixed{higher dispersion of the} stellar population about the color-tranformations. 
\Fixed{Consequently}, flux measurements in the two surveys \Fixed{can now} be compared
with low cross-calibration uncertainty.

It is also interesting to compare the relative importance of the
various contributions to the global calibration uncertainty. The last
row of table~\ref{tab:combcalib} shows that the uncertainty on the
primary standard spectra amounts for about half the uncertainty (in
variance) for the redder bands ($riz$). The imperfect knowledge of
filters impacts mostly the $u$ and $\band u$ band\Fixed{ calibration} 
due to the large variation of the standard star spectra in this \Fixed{wavelength 
region} and the \band r band due to the larger uncertainty remaining 
on the transmission of this filter. 

Evidence \Fixed{for} potential evolution of the \band r and \band i filters
\Fixed{has} been found and the resulting uncertainty on the filter passbands
is discussed in Sect.~\ref{sec:filter-uncertainty}. \Fixed{We emphasize 
that the impact of filter passband uncertainty depends on the color 
of the object whose flux is being determined. Specifically, this} error cancels 
for objects whose spectrum is close to the spectrum of our standard stars (main 
sequence stars of color $g-i \sim 0.55$). \Fixed{This is why the} poor 
knowledge of MegaCam \band i and \band r passbands \Fixed{only minimally 
affects AB calibration accuracy: the primary standards we observe
have SEDs close to the AB spectrum in those bands (stars with $g-i \sim
0.55$ have $r-i\sim0.1$)}. \Fixed{This is not the case for type Ia supernovae;
therefore, the impact of filter passband uncertainty on their flux measurements} 
is bound to be much larger.

\subsection{Comparison with SNLS3}
\label{sec:snls3}

\begin{figure}
  \centering
  \igraph{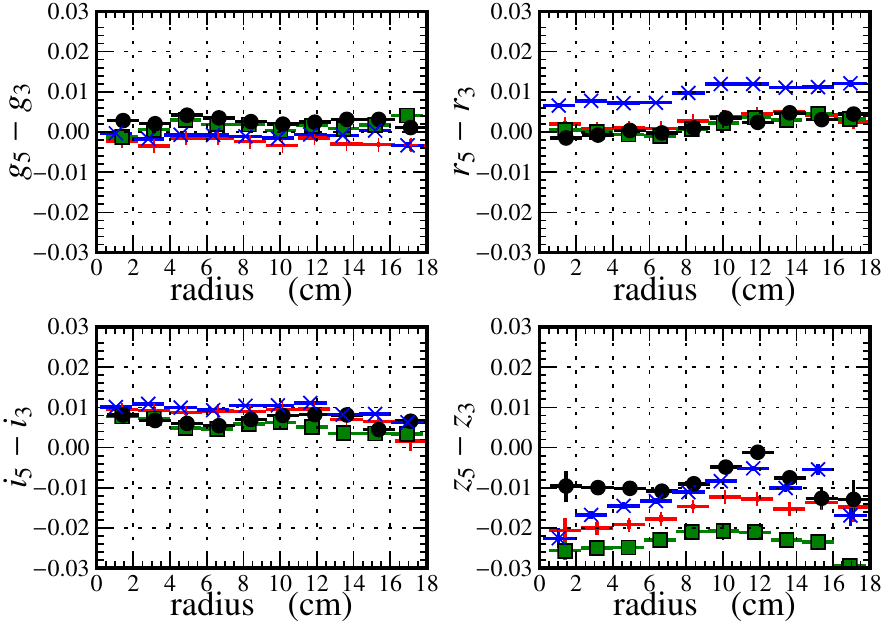}
  \caption{Difference between SNLS-calibrated tertiary star
    magnitudes and previous R09 release as a function of the distance
    to the focal plane center. The 4 deep fields are compared
    separately (D1: red plus, D2: green squares, D3: black circles,
    D4: blue crosses.)}
  \label{fig:r09comp}
\end{figure}

For the four photometric bands already calibrated in R09, we \Fixed{compared 
uniformity and absolute calibrations between the new and prior tertiary catalogs.}
A sign mistake affecting the R09 catalog was discovered in this comparison. 
It contributes to a slight deterioration of the catalog uniformity and shifts 
the average flux scale of the catalog by 8, 5, 6, 3~mmag on average for band
\band{g},\band{r},\band{i} and \band{z} respectively. \Fixed{Once the error sign
is corrected, the remaining offsets} are given in table~\ref{tab:snls3off}.  The
comparison of calibrated \Fixed{magnitude} differences as a function of the
distance to the center of the focal plane is given in Fig.~\ref{fig:r09comp}.

\begin{table}
  \centering
  \caption{Calibration offset between R09 and the present release of the SNLS calibration.}
  \label{tab:snls3off}

\begin{tabular}{l*{4}{r}}
\hline
\hline
& \band g&\band r&\band i&\band z\\
\hline
offset\tablefootmark{a} & -0.000&0.006&0.008&-0.013\\
rms\tablefootmark{b} & 0.004&0.006&0.004&0.012\\
\hline
\end{tabular}
\tablefoot{\\
  \tablefoottext{a}{The offset should be added to the previous release to match the present result, and comes in addition to a sign error correction in the previous release (see text).}\\
  \tablefoottext{b}{Rms of the differences between the two catalogs for stars brighter than $\band r<18$.}}
\end{table}

\subsection{Comparison with the previous calibration of the SDSS SN
  survey.}
\label{sec:comp-with-prev}
\Fixed{Following the discovery of a non-uniformity in the SDSS PT flat-field,
this work introduces two changes to the SDSS tertiary catalogs.} The first 
is a correction of the tertiary catalog flat-fielding based on the 
\Fixed{Data Release 8 SDSS catalog \citep{2008ApJ...674.1217P} flat-fielding 
solution}. The second is the new determination of the corresponding AB offsets. 
The previous calibration of the SDSS SN survey was based on the uncorrected 
version of the coadd catalog scaled to the AB system by the application of AB
offsets from \citet[table 1]{holtzman_sloan_2008}. As supernovae are
uniformly distributed in declination, the calibration offset to be
expected from this release is readily given by the weighted average of
magnitude differences between the two calibrated catalogs. This offset
is given in table~\ref{tab:sdss_13_offset}.

\begin{table}
  \centering
  \caption{Calibration offsets between first and third year release of the SDSS-II SN survey. \label{tab:sdss_13_offset}}

\begin{tabular}{l*{5}{r}}
\hline
\hline
& u&g&r&i&z\\
\hline
rms\tablefootmark{b} & 0.018&0.007&0.006&0.006&0.011\\
ptp\tablefootmark{c} & 0.085&0.031&0.029&0.035&0.050\\
\hline
offset\tablefootmark{a} & -0.031&-0.004&-0.000&-0.000&-0.006\\
\hline
\end{tabular}
  \tablefoot{\\
    \tablefoottext{a}{Calibration offset in mag (average difference between the present work and \citet{holtzman_sloan_2008}).}\\
    \tablefoottext{b}{Standard deviation of the photometry difference between the two catalogs (in mag).}\\
    \tablefoottext{c}{Peak to peak difference between the two catalogs (in mag).}\\
    }
  
\end{table}

The difference \Fixed{in} the average photometry \Fixed{is} fairly
small compared to the size of the PT flat-field correction. This is a
fortunate consequence of the fact that the effect is close to symmetric 
with respect to the center of the PT focal plane, where standard stars 
were observed. The row labeled rms in table~\ref{tab:sdss_13_offset} 
gives an estimate of the gain in photometric precision that is to be 
expected from the improved flat-fielding.

\subsection{Comparison with other work}
\label{sec:comp-with-other}

While we have achieved excellent agreement with the SNLS and SDSS
supernova surveys we did not attempt to compare our photometry with any
other survey.  The SDSS SN photometry has been compared with CSP
\citep{2012AJ....144...17M}, where good agreement was found.  The
Pan-STARRS survey \citep{2004AN....325..636H} has recently reported a
small discrepancy with the SDSS stellar photometry
\citep{2012ApJ...750...99T}, but larger than might be expected on the
basis of this work.  Their analysis used the publicly available SDSS
catalogs which lack the corrections applied in this paper and they did
not use the filter transmission measurements of Doi, et al.  However,
the discrepancy in photometry might very well persist even with our
revised catalog.

\section{Conclusion}
\label{sec:conclusion}

We have delivered updates to the tertiary standard star catalogs that
define the photometric calibration of the SNLS and the SDSS supernovae
surveys. \Fixed{The catalogs provide} \Fixed{standard star calibrated
  AB magnitudes}, in the \emph{natural} photometric system of both
instruments. The products are fully described in
Sect.~\ref{sec:calibration-products-1}.

The flux calibration is based on the observation of 5 different HST primary
spectrophotometric standard stars from the CALSPEC database. In
addition, the SNLS and SDSS tertiary standard stars are now tied
together by specifically designed cross-calibration observations. 
\Fixed{Three independent routes, each subject to different observational
systematics, were used to transfer the HST primary standard star calibration
 to the tertiary standard stars}.
Good agreement was found between these 3 routes, which were \Fixed{then} combined 
to provide the final products. In particular, the offsets derived in 
Sect.~\ref{sec:concl} from simultaneously fitting all available information 
have been applied to the delivered catalogs. \Fixed{These offsets have been 
determined with CALSPEC version 3 spectra, and may be updated as new CALSPEC 
versions are released. See Appendix~\ref{sec:ancalibration-products} for more details.}

With respect to \Fixed{previous} releases, the photometric calibration
evolved by less than 1\% in all photometric bands but $u$. Major
improvements arise from better flat-fielding processes (that also
improve the uniformity of the resulting catalogs), and the
introduction of more data \Fixed{to constrain} the calibration transfer.

The calibration transfer from HST standard stars to stars in
supernovae fields is thought to be accurate to about $3$~mmag in the
$riz$ photometric bands. This is comparable, or better than the error
related to the calibration uncertainty of the primary standard stars
themselves. The error in band $g$ is about $4$~mmag and remains
dominated by the error in the calibration transfer. The uncertainty on
the $u$ band calibration amounts to about 1\% in the SDSS and 2\% in
the SNLS, dominated by the uncertainty on the instrument effective
passbands.

The photometric calibration was complicated by the need to
overcome several instrumental effects as well as weaknesses in \Fixed{
instrument} or survey designs. \Fixed{The most important complications
are summarized here:}
\begin{itemize}
\item \Fixed{Uniform illumination of wide-field instruments is not enough
  to achieve accurate flat-fielding; internal reflections in the optical
  path must} be accounted for. At least two elements, CCDs and filters, 
  are reflective enough to feed \Fixed{significant amounts of light back into} 
  the wide-field corrector optics. \Fixed{Reflected by convex
  lens surfaces, this light causes} ghosting as well as diffuse and 
  structured contamination in the flat-fields. For this reason, \Fixed{photometric 
  uniformity of both the SDSS and SNLS} instruments ultimately relies on the
  analysis of dithered stellar observations. The variation of
  atmospheric conditions during the observation sequences ultimately
  limits \Fixed{the flat-field solution accuracy}.
\item \Fixed{Similarly}, small-angle scattering of light caused by
  dust or defects in the optics \Fixed{can} affect the
  photometry of point sources while leaving the flat-field twilight
  images unchanged. A remarkable illustration of such a phenomenon
  occurred during the SNLS survey: the progressive deposit of metal
  dust from the filter jukebox on one corner of the top corrector
  lens caused a local \Fixed{ 6\% deterioration of the effective 
  throughput of the photometry, and was only detected and measured 
  through dithered stellar observations}. Such effects can also be 
  detected by monitoring apparent star brightness in often-visited fields.
\item For SNLS, spatial variations of the instrument effective
  passbands had to be accounted for in order to accurately conduct the
  calibration transfer and deliver relevant flat-fielding solutions.
  An extensive survey of the filters' transmission would make the
  calibration more accurate.
\item The \Fixed{characterization} of the SNLS photometric system
  would have been \Fixed{simplified by the implementation of a
    well-planned filter passband monitoring program}. Nonetheless, no
  significant chromatic differences \Fixed{have been observed in
    repeat stellar photometry over the course of the survey, providing
    direct evidence of the photometric system's long-term stability}.
\item \Fixed{Slight readout electronics instabilities cause percent-level 
  noise in the MegaCam flat-fielding. For the SNLS, this effect is mitigated
  by repeated measurements over the long duration of the survey; for shorter
  surveys, this effect would have been difficult to overcome. Reproducible
  illumination provided by artificial light sources} might provide \Fixed{more
   reliable} monitoring of the instrument throughput.
\item The SDSS calibration relied on a separate telescope (the PT),
  but neither the flat-field nor the filter throughput was \Fixed{
  well-characterized enough} for the precision photometry goals of this
  paper. \Fixed{Instead} we had to rely on empirical color-transformations and
  correct the flat-field from external data.
\end{itemize}

\Fixed{Most of these} effects were \Fixed{overcome} by 
introducing redundancy in the measurements, or by the analysis of
specifically designed additional observations. While \Fixed{such strategies
allowed us to reach an accuracy of 0.3\%}, it \Fixed{would be prohibitively time-consuming 
to reduce this accuracy further}. Future photometric surveys such as 
SkyMapper \citep{2007PASA...24....1K}, DES \citep{2012ApJ...753..152B} 
and LSST \citep{2009arXiv0912.0201L} might take advantage of current developments 
regarding instrumental calibration with artificial calibration sources. 
\Fixed{Information} provided by the SNDICE illumination prototype
\citep{barrelet_direct_2008,sndice} installed at the CFHT \Fixed{has} already
proved immensely valuable \Fixed{in understanding} MegaPrime instrumental effects 
such as flat-field pollution with ghosts and \Fixed{readout electronics} gain variations. 
\Fixed{Ongoing analysis of SNDICE data may also give us a final understanding of the 
MegaCam filter passbands. A comparable device \citep{2010ApJS..191..376S} has aided in 
the calibration of the Pan-STARRS instrument; the Dark Energy Survey is 
taking a similar approach with  DECal \citep{2012AAS...21924802R}.}

\begin{acknowledgements}
  The authors heartily acknowledge the fruitful collaboration of the
  Canada France Hawaï telescope team, in gathering part of the data
  required \Fixed{for} the present effort. We warmly thank the SNDICE
  collaboration for their help in the measurement of the SNLS
  passbands as well as for enlightening discussions on instrumental effects. 
  We thank C. Buton and the SNF Collaboration for providing us early 
  versions of their atmospheric transmission curve prior to publication. 
  We are grateful to D. Tucker for giving us details about the SDSS 
  calibration. This work was supported in part by the U.S. Department of Energy 
  under contract number DE-AC0276SF00515.  Part of the data reduction was 
  carried out at the Centre de Calcul de l'IN2P3 (CCIN2P3, Lyon, France).

\end{acknowledgements}

\bibliography{calib5}
\bibliographystyle{aa}
\appendix

\section{Monitoring of the MegaCam flat-field variability from deep
  fields}
\label{sec:more-megac-flatf}

\Fixed{Variations of the instrumental response are determined by the time distribution of the }
photometric coefficient $\delta \zp_{t,ccd}$ (where $t$ indexes the
exposures, and $ccd$ indexes the 36 \Fixed{MegaCam chips.
We recall that
$\delta \zp_{t,ccd}$ is computed for each deep field 
by building} an average photometric catalog of non-variable stars as
described in Sect.~\ref{sec:tert-catal-constr}. We then compute for
each available exposure a zero point relative to the average
catalog. \Fixed{Once the exposure $t$ is scaled by this zero point,
we recompute a zero point offset for each CCD ($\delta
\zp_{t,ccd}$) relative to the average catalog. Non-zero offsets
indicate that the throughput of the instrument has varied non
uniformly. This can be due either to 
clouds or to a variation of the instrument response that is not
corrected by the twilight flat-fielding.}

\begin{figure*}
  \centering
  \igraph{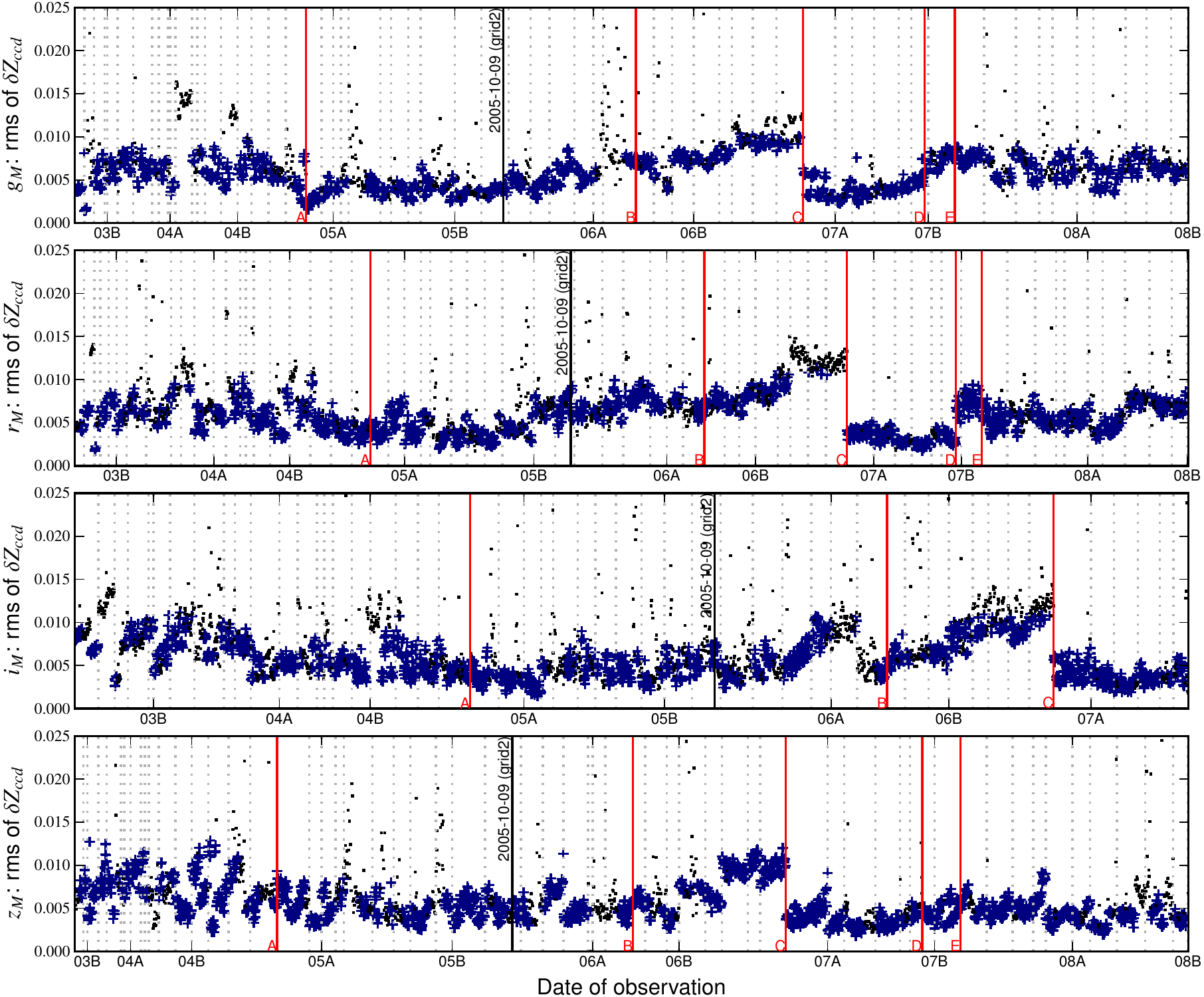}
  \caption{The standard deviation of the relative CCD zero points for each exposure in the 
  survey. Black dots denote exposures that
    were excluded from the average because of cuts on the stability of
    the observing conditions or on the uniformity of the effective
    instrument response. Exposures are ordered by increasing
    observation date. \Fixed{The} vertical gray dotted lines indicate the separation
    between different MegaCam runs (roughly one run per lunation). The
    twilight flat-field is \Fixed{common} to all observations in a run.
    The vertical red lines \Fixed{indicate significant} changes to the
    telescope. The letters refer to the events described in
    table~\ref{tab:grid_obs}. In this figure, the same photometric
    flat-field correction, determined from observations on 2005-10-09
    (black vertical line), is applied to all the observations.}
  \label{fig:zpperexpalt}
\end{figure*}

For each exposure $t$ we compute the rms of $\delta \zp_{t,ccd}$ \Fixed{of}
the 36 CCDs. The statistical uncertainty on the determination of
$\delta \zp_{t,ccd}$ is typically  \Fixed{2~mmag, so that sets the minimum expected value. 
The rms values observed throughout the survey are}
displayed in Fig.~\ref{fig:zpperexpalt}. The measured rms is typically
about 5~mmag. \Fixed{The observed excess relative to the minimum value of 2~mmag  is to}
be expected for at least 3 reasons: 1) structured variations of the
atmospheric transparency, such as clouds, affect the
observations, 2) \Fixed{the gains of some CCDs amplifiers vary on
time scales shorter than a single run, and therefore can't be calibrated by a single twilight
flat-field}, and 3) \Fixed{telescope modifications can} have an
impact on the photometric response, while leaving the twilight
observations virtually unchanged. The effects of 1) and 3) can be
observed directly in Fig.~\ref{fig:zpperexpalt}, while 2) is a noise
that contributes to a global increase of the measured rms.

Clouds are occasionally responsible for a large increase in the
measured rms. \Fixed{These} non-photometric conditions are easily detected and
excluded from the averaging of the tertiary catalogs. Exposures
detected as non photometric are displayed as black dots on
Fig.~\ref{fig:zpperexpalt}.

We \Fixed{have} displayed telescope modifications that had a noticeable impact on the
uniformity of the photometry as red vertical lines. The beginning of
the survey was marked by continuous adjustment of the optical
setup. We retained only the main modification (that had the greatest
impact on the image quality): the flip of the third lens in the wide
field corrector. It is denoted as event A on the graph. The rms of
exposures obtained prior to this event can be seen to be 
\Fixed{higher}. The second and most significant
phenomenon was the accumulation of metal dust on a corner of the wide
field corrector optics. Its effect was the scattering at small angles
of a growing part of the light, decreasing progressively the flux
fraction contained in the \Fixed{core} of the PSF (and hence the throughput
of the photometry), while leaving the twilight images essentially
unchanged. The effect is clearly visible in Fig.~\ref{fig:zpperexpalt}
as a progressive increase of the rms during the year 2006. Its
detection and cleaning occurred at the end of 2006 and is denoted as
event C. The beginning of the accumulation is harder to determine. We
approximately determined that the effect becomes sufficiently
important at the middle of 2006 (event B). The last events are the
destruction (D) and replacement (E) of the \band i filter in
2007. Exposures in \band g and \band r \Fixed{were apparently also affected by the same event (D), 
but we do not understand why they changed}.

\begin{figure*}
  \centering
  \igraph{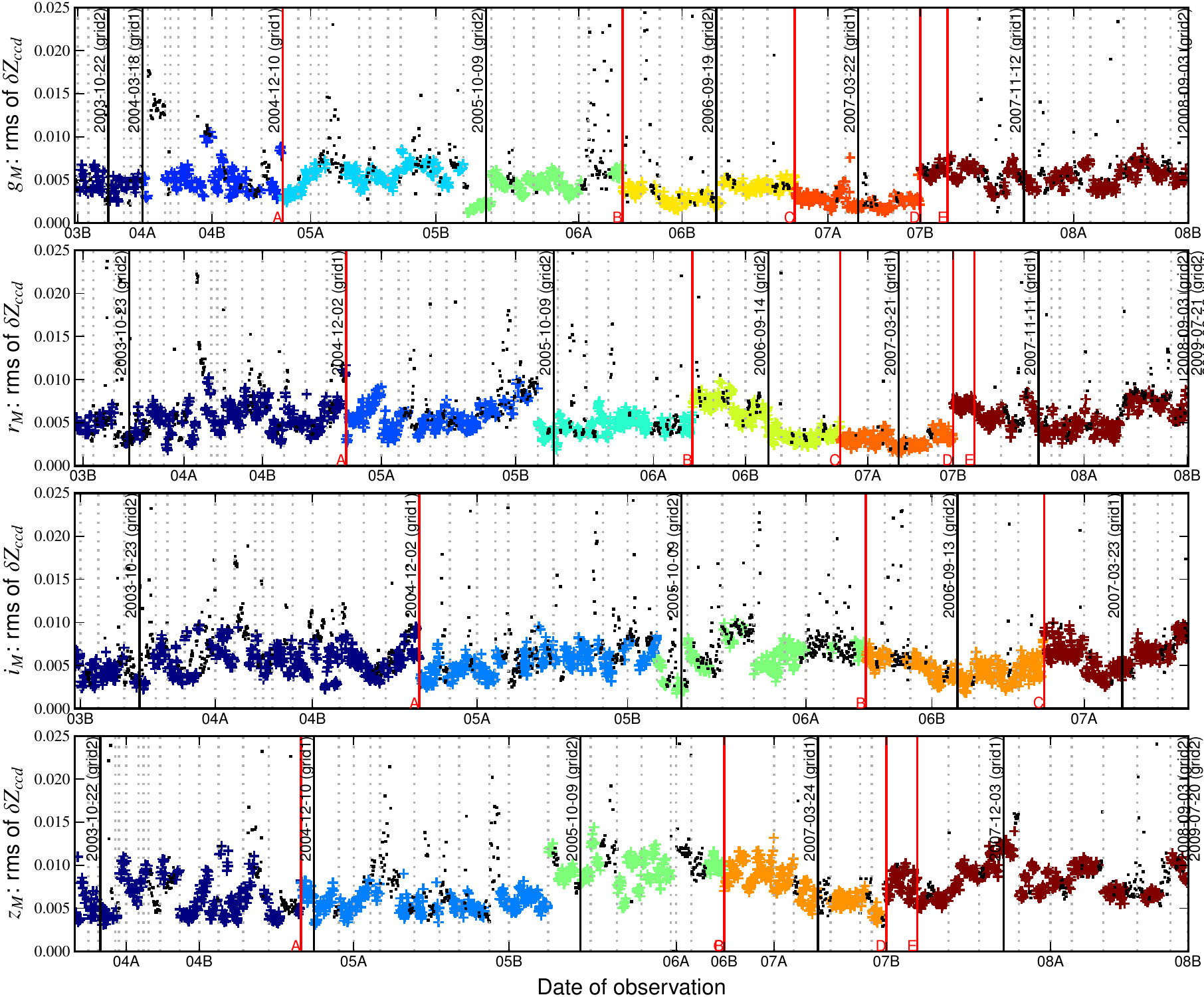}
  \caption{This figure is similar to Fig.~\ref{fig:zpperexpalt}, but different photometric 
  flat-field corrections were used for the 
    different periods of the survey. The vertical black lines
    indicate the date of grid observations used to compute the
    photometric corrections. The changes of data point color indicate
    the different flat-field corrections applied.  These corrections were used to
    produce the multi-epoch tertiary catalogs described in
    Sect.~\ref{sec:tert-catal-constr}.}
  \label{fig:zpperexp}
\end{figure*}

The events are summarized in table~\ref{tab:grid_obs}. They define 5
periods for which independent photometric corrections can be
determined. The application of our set of photometric corrections to
the survey flat-fielding is illustrated in
Fig.~\ref{fig:zpperexp}. This figure is similar to
Fig.~\ref{fig:zpperexpalt}, but made after the application of
photometric corrections. \Fixed{After correction,  the
measurements are consistent within an error
$\sigma_{ff}$, which is estimated as the average rms of $\delta
\zp_{ccd}$.  Values of $\sigma_{ff}$ were given in
table~\ref{tab:fferror} in the main text.}

\begin{table}
  \centering
  \caption{Grid observations and associated telescope events.}
  \label{tab:grid_obs}
  
\begin{tabular*}{\linewidth}{@{\extracolsep{\fill}}lllcp{2.5cm}}
\hline
\hline
 Run\tablefootmark{a} & Date & Bands\tablefootmark{b} & Field & Events \\
\hline
03Bm02
& 2003-10-02 & i & 2 &   \\
& 2003-10-03 & r & 2 &   \\
\hline
03Bm03
& 2003-10-22 & \textbf{g},\textbf{u},\textbf{z} & 2 &   \\
& 2003-10-23 & \textbf{i},\textbf{r} & 2 &   \\
\hline
04Am02
& 2004-03-18 & u,\textbf{g},r,i,z & 1 &   \\
\hline
04Bm05
& 2004-12-02 & g,\textbf{r},\textbf{i} & 1 &   \\
& 2004-12-02 &  &  &  (A) L3 lens flipped upside down. \\
& 2004-12-10 & \textbf{g},\textbf{u},\textbf{z} & 1 &   \\
\hline
05Bm02
& 2005-10-09 & \textbf{u},\textbf{g},\textbf{r},\textbf{i},\textbf{z} & 2 &   \\
\hline
06Am04
& 2006-05-19 &  &  &  (B) Growing metal dust deposit. \\
\hline
06Bm02
& 2006-09-13 & \textbf{i} & 2 &   \\
& 2006-09-14 & \textbf{r} & 2 &   \\
& 2006-09-15 & z & 2 &   \\
& 2006-09-16 & \textbf{u} & 2 &   \\
& 2006-09-19 & \textbf{g} & 2 &   \\
\hline
06Bm05
& 2006-12-27 &  &  &  (C) Top lens cleaning. \\
\hline
07Am02
& 2007-03-21 & \textbf{r} & 1 &   \\
& 2007-03-22 & \textbf{g},\textbf{u} & 1 &   \\
& 2007-03-23 & \textbf{i} & 1 &   \\
& 2007-03-24 & \textbf{z} & 1 &   \\
\hline
07Am06
& 2007-07-20 &  &  &  (D) \band i filter broken. \\
\hline
07Bm01
& 2007-08-24 &  &  &  (E) Recoating. \\
\hline
07Bm03
& 2007-10-18 &  &  &  (F) New \band{i2} filter installed. \\
\hline
07Bm04
& 2007-11-08 & i2 & 1 &   \\
& 2007-11-11 & \textbf{r} & 1 &   \\
& 2007-11-12 & \textbf{g} & 1 &   \\
\hline
07Bm05
& 2007-12-03 & \textbf{u},\textbf{z} & 1 &   \\
\hline
08Bm02
& 2008-09-03 & \textbf{u},\textbf{g},\textbf{r},\textbf{i2},\textbf{z} & 2 &   \\
\hline
09Am06
& 2009-07-20 & g,\textbf{z} & 2 &   \\
& 2009-07-21 & \textbf{r},\textbf{u},\textbf{i2} & 2 &   \\
\hline
10Am01
& 2010-02-06 & \textbf{u} & 1 &   \\
& 2010-02-18 & \textbf{r},\textbf{z} & 1 &   \\
& 2010-02-19 & i2 & 1 &   \\
& 2010-02-20 & g & 1 &   \\
\hline
10Bm02
& 2010-10-05 & r & 1 &   \\
& 2010-10-09 & \textbf{g},\textbf{i2} & 2 &   \\
& 2010-10-13 & r & 2 &   \\
\hline
10Bm03
& 2010-11-08 & \textbf{g},u & 2 &   \\
\hline
11Am06
& 2011-07-30 & z & 2 &   \\
& 2011-08-02 & g,r,u,y & 2 &   \\
\hline
11Bm01
& 2011-08-28 & g,r,z & 2 &   \\
& 2011-08-29 & g,u,y & 2 &   \\
\hline
\end{tabular*}
  \tablefoot{
    \tablefoottext{a}{Run id in the elixir pipeline nomenclature.}\\
    \tablefoottext{b}{Observations selected as valid are bold-faced in this column.}
  }
\end{table}
\begin{table}
  \centering
  \caption{Dithering pattern of the grid exposures.}\label{tab:grid_dp}
  \begin{tabular}{ccc}
    \hline
    \hline
    Exposure & $\alpha$ (deg.)\tablefootmark{a} & $\delta$ (deg.)\tablefootmark{a}\\ 
    \hline
    1&0.000 &           0.000 \\
    2&0.028 &           0.000   \\
    3&0.056 &           0.000   \\
    4&0.112 &           0.000   \\
    5&0.225 &           0.000   \\
    6&0.451 &           0.000   \\
    7&0.902 &           0.000   \\
    8&0.000 &           0.027   \\
    9&0.000 &           0.055   \\
    10&0.000 &           0.111   \\
    11&0.000 &           0.222   \\
    12&0.000 &           0.444   \\
    13&0.000 &           0.888   \\
    14\tablefootmark{b}&0.000 &           0.000   \\
    \hline
  \end{tabular}
  \tablefoot{\tablefoottext{a}{Pointing deviation relative to the reference pointing of the field.}
    \tablefoottext{b}{A concluding 14th exposure,
that comes back to the first pointing has been added to grid sequences
since September 2008. This redundancy provides a way to monitor the
the stability of the observation conditions during the sequence.}
  }
\end{table}

\section{MegaCam filters measurements}
\label{sec:megacam-band-i}
\begin{figure}
  \centering
  \igraph{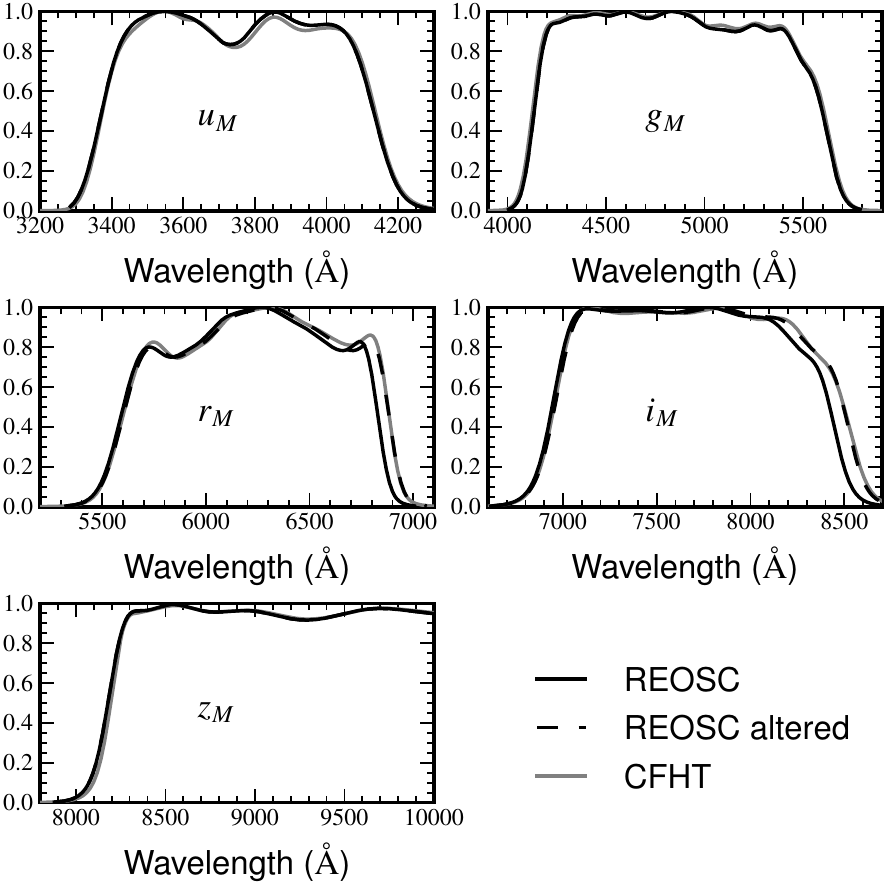}
  \caption{Comparison of available measurements of the REOSC filter
    transmission curves at a distance of 15cm from the center. }
  \label{fig:benedictvsreosc}
\end{figure}

The transmission curves of the original Sagem/REOSC MegaCam filters
were measured twice: the first time by the manufacturer before their
installation in MegaCam in 2002, and the second time in 2006 by the
CFHT operational team. The 2006 measurements are available only at
positions close to the border of the filter due to limitations in the
bench clearance, and their spectral resolution is coarser. Comparison
between the two measurements at 15cm from the center (after
convolution of the REOSC measurements to match resolutions) is
given in Fig.~\ref{fig:benedictvsreosc}. The position of the \band{r}
and \band{i} filter cut-offs can be seen to be shifted $\sim 8$~nm
redder in the 2006 measurements. The origin of these discrepancies and
the reason why only two filters are affected are unclear.

A confirmation of the 2006 measurements was recently given by the
measurement of the transmission of one of the broken pieces of the
\band{i} filter. The measurement was conducted on the calibration
bench of the SNDICE illumination device \citep{barrelet_direct_2008},
and display a perfect agreement with the 2006 transmission curves 12~cm
away from the center (the appropriate curve for the broken piece).

The possibility of a continuous (slow) aging of the filters is
excluded by the examination of color-terms between the beginning and
the end of the survey (\emph{cf.}
Sect.~\ref{sec:passband-variations}). It is thus likely that the 2006
measurements are the most representative of the state of MegaCam
filters during the survey. The lack for measurements at the center of
the filters in the 2006 data \Fixed{requires additional data or assumptions}. We opt for
correcting the \band{i} and \band{r} REOSC measurements to make them
match the 2006 transmission curves. The application of a linear
stretch in wavelength of the form:
\[T^\star(\lambda) = T\left(\lambda_1 + \frac{\lambda_3
    -\lambda_1}{\lambda_2 - \lambda_1}(\lambda - \lambda_1)\right)\]
was found to reconcile both measurement sets. We used $(\lambda_1,
\lambda_2, \lambda_3) = (5300 \angstrom, 6837\angstrom,
6891\angstrom)$ for \band{r} and $(6600\angstrom, 8464\angstrom,
8539\angstrom)$ for \band i. The resulting transmission model at 15cm
is shown as a dashed black line in Fig.~\ref{fig:benedictvsreosc}; \Fixed{similar agreement is obtained with the 12cm data}.
 
\section{SDSS Photometric Telescope (PT) flat-field}
\label{sec:pt-flat-field}
There are some overlapping secondary patches in stripe 82 that can be
used to check for deviations from a uniform response \Fixed{of} the PT. This
section provides an analysis of that data.

The data sample consists of the 13 secondary patches listed in Table
\ref{tab:PTdata}.  The fields labeled A, B, C, and D are the standard
locations in declination for secondary patches in stripe 82.  The
field of view is 40 arc-minutes (2/3 degree), so there is no overlap
between the A, B, C, and D secondary patches.  The fields labeled E,
F, and G are special patches that were placed approximately half-way
in declination between the standard locations, providing a 50\%
overlap with the standard patches.  The camera pointings are all in
the same relative orientation with East-West being in the CCD column
direction.

 \begin{table}
   \caption{The PT data with secondary patch overlaps in SDSS stripe 82.  \label{tab:PTdata}}
   \centering
   \begin{tabular}{lrrrr}
\hline\hline
Field & Date (MJD) & RA (deg) &  Dec (deg) 
& Frame \\
\hline
82A12 &  52551  & 315.0  & -1.00 & 161156  \\
82A16 &  52551  &   25.0  & -1.00 & 161196  \\
82B12 &  52577  & 315.0  & -0.33 & 165233  \\
82B16 &  52551  &   25.0  & -0.33 & 161246 \\
82C16 &  52577  &   25.0  & 0.33  & 165263 \\
82D12 &  52577  & 315.0  & 1.00  & 162033 \\
82D16 &  52584  &   25.0  & 1.00  & 166668 \\
82E12 &  52557  & 315.0  & -0.66 & 162058 \\
2E16 &  52553  &   25.0  & -0.66 & 161813 \\
82F12 &  52557  & 315.0  &  0.00 & 162068 \\
82F16 &  52553  &   25.0  &  0.00 & 161823 \\
82G12 &  52578 & 315.0  &  0.66 & 165524 \\
82G16 &  52557 &   25.0  &  0.66 & 162098 \\
\hline
\end{tabular}
\end{table}

The analysis involved matching stars in the overlapping pointings and
computing the difference in magnitude.  The data come from the SDSS
data processing (mtPipe) \citep{2006AN....327..821T}.  The differences
in magnitude measure the difference in response of the focal plane
between a given column and another one that is displaced by 1/2 the
size of the CCD.  If the response is uniform all the differences
would be zero.  It is convenient to measure the distance along the
focal plane in terms of degrees of declination. We determine this
distance as the declination of a matched star minus the reported
declination of the telescope pointing. The matching star is nominally
at a declination lower by 20
arc-minutes. 

The results are summarized in Table~\ref{tab:PTfits}.  The table gives
the total number of comparison stars that were found and a fit to the
equation
\begin{equation}
\Delta m=a \delta + b\label{eqn:line}
\end{equation}
where $\Delta m$ is the magnitude difference, $\delta$ is the position
on the focal plane in units of degrees of declination, and $a$ and $b$
are fit parameters.  A $\chi^2$ is given for each fit. The fit is made
to 6 binned data points so there are 4 degrees of freedom for each
fit, except for the combined fit to \textit{g}, \textit{r}, and
\textit{i} (last line in Table~\ref{tab:PTfits}).  This fit is
performed on 18 binned data points from \textit{g}, \textit{r}, and
\textit{i} so there are 16 degrees of freedom for that fit.  The
chi-squared's are generally good although they could probably be
improved by the addition of a quadratic term.  However, it was felt
that a quadratic term was unnecessary and possibly sensitive to
unknown systematic errors.  Every filter shows a consistent trend of
positive residuals and a negative slope with declination.  However,
the fits are quite precise and there are statistically significant
differences between the filters.

\begin{table}
  \centering
  \caption{The results of PT fits.\label{tab:PTfits}}
  \begin{tabular}{lrrrrrr}
    \hline
    \hline
    Band & Stars & $\chi^2$ &  $a$ 
    & $\sigma_a$ & $b$ & $\sigma_b$ \\
    \hline
    $u$   &   699      & 5.6 & -0.129  & 0.022 & 0.041 & 0.004  \\
    $g $  &   2514    & 2.2 & -0.039  & 0.009 &0.027 & 0.002    \\
    $r$    &   2971    & 6.5 & -0.037  & 0.008 &0.022 & 0.001   \\
    $i$    &   2581    & 8.0 & -0.019  & 0.009 & 0.023 & 0.002  \\
    $z$   &  1127     & 3.0 & -0.053  & 0.016 & 0.038 & 0.003   \\
    $gri$ &  8066   & 39.5 & -0.032 & 0.005 & 0.024 & 0.001     \\
    \hline
  \end{tabular}
\end{table}

The results above only give the difference in response between the two
halves of the PT. Determining the actual response requires additional
data or assumptions. There seem to be two obvious models that could
explain some features of the data:
\begin{itemize}
\item The CCD readout uses different amplifiers for the two halves of
  the chip we are considering.  It is possible that the gain
  difference between the two amplifiers has not been properly
  calibrated.  Such a mis-calibration could occur, for example, if the
  light levels in the flat-field exposures resulted in a non-linear
  amplifier response that was different for the two amplifiers while
  normal exposure levels remained in the linear regime. This scenario
  could explain the offset between the two halves of the CCD but
  doesn't seem very plausible and would not explain the observed
  slopes.
\item The illumination of the flat-field exposures may be non-uniform.
  This would lead to an error in the calibration of the CCD response.
  The non-uniformity is most likely to be a slowly varying function of
  position.
\end{itemize}
In any case the data are not compatible with a uniform photometric
response in the PT flat-fielded exposures. The PT flat-field error has
two consequences. First it is transferred to the 2.5~m as a
flat-fielding error.  It should imprint a periodic pattern 0.6 degrees
wide in declination on the photometry. Second, as primary standard
stars are observed preferentially at the center of the focal plane, it
can bias the calibration relative to the tertiary stars that are
observed throughout the focal plane. It is thus necessary to provide a
correction of the 2.5~m flat-field independent of the PT \emph{and} to
characterize the PT flat-field error to derive accurate calibration
from its measurement.

\section{DR8 catalog retrieval}
\label{sec:dr8-catal-retr}

The Stripe 82 DR8 catalog has been retrieved from the DR8 database using the following query:
\begin{lstlisting}
select psfmag_u,psfmag_g,psfmag_r,psfmag_i,
  psfmag_z,ra,dec,run,camcol, psfmagerr_u,
  psfmagerr_g,psfmagerr_r,psfmagerr_i,
  psfmagerr_z into mydb.MyTable from
  phototag where dec between -1.3
  and 1.3 and ((ra between 300 and 360) 
  or (ra between 0 and 60)) and type = 6
  and mode = 1 and psfmag r < 22.5
\end{lstlisting}

\section{SNLS and SDSS calibration products}
\label{sec:ancalibration-products}

\Fixed{Our main delivery are catalogs of natural magnitudes for a
  large set of tertiary standard stars calibrated to the AB flux
  scale.  We also provide the material required for the interpretation
  of those magnitudes in term of physical fluxes (instrument
  transmission functions) as well as the material required to update
  the AB calibration as new CALSPEC spectra for our primary standards
  are released.}

\Fixed{All the material described below is available from the
  SNLS and SDSS joint calibration webpage:
  \url{http://supernovae.in2p3.fr/snls_sdss/}.  }

\subsection{\Fixed{Tertiary stars catalogs}}
\label{sec:tert-stars-catal}

\Fixed{
We provide catalogs of natural AB magnitudes in each of the MegaCam
filters for selected tertiary standard stars in the four CFHTLS deep
fields in the natural MegaCam photometric system. This system is
described in Sect.~\ref{sec:instrument-bandpass} and the corresponding
 transmission curves are given below
(Sect.~\ref{sec:megac-transm-funct}). Note that the transmission
depends on the location on the focal plane. A subsample of the D1
catalog is given in table~\ref{tab:D1_catalog}.
}

\Fixed{
Similarly, we deliver natural SDSS AB magnitudes for selected
tertiaries in the Stripe 82. The magnitudes are given for the average
filter response as given by \citet{2010arXiv1002.3701D}, i.e., they
should not be adjusted for the small differences in the filter
responses of the different CCD columns.  An excerpt from the catalog
is given in Table~\ref{tab:S82_catalog}. 
}

\Fixed{
In both cases the released magnitudes are \emph{already} placed on the
AB system by the addition of the zero points given in table
\ref{tab:ABoff}. Full catalogs in electronic format can be retrieved from
the CDS or from the SNLS and SDSS joint calibration webpage.
}

\longtab{1}{

\begin{landscape}
\begin{longtable}{*{8}{c}}
\caption{Catalog of natural AB SDSS magnitudes for selected tertiary standards in the Stripe 82. \label{tab:S82_catalog}} \\
\hline
\hline 
    $\rm{R.A.}$ 
              & $\rm{Dec.}$ 
              & $u$ 
              & $g$ 
              & $r$ 
              & $i$ 
              & $z$ \\

  $(\rm{Deg.})$    &  $(\rm{Deg.})$
              & $(\rm{mag})$
              & $(\rm{mag})$
& $(\rm{mag})$
& $(\rm{mag})$
& $(\rm{mag})$ \\
\hline 
\endfirsthead
\caption{continued.} \\
\hline 
\hline 
    $\rm{R.A.}$ 
              & $\rm{Dec.}$ 
              & $u$ 
              & $g$ 
              & $r$ 
              & $i$ 
              & $z$ \\

  $(\rm{Deg.})$    & $ (\rm{Deg.})$
              & $(\rm{mag})$
              & $(\rm{mag})$
& $(\rm{mag})$
& $(\rm{mag})$
& $(\rm{mag})$ \\

\endhead
\hline 
\endfoot

$-51.50000$ & $-0.60412$ & $-0.089\pm0.000$ & $20.386\pm0.009$ & $18.795\pm0.005$ & $17.927\pm0.004$ & $17.413\pm0.007$\\
$-51.50000$ & $1.21455$ & $-0.066\pm0.000$ & $19.443\pm0.005$ & $17.980\pm0.003$ & $17.178\pm0.003$ & $16.735\pm0.005$\\
$-51.49960$ & $-0.59146$ & $-0.092\pm0.000$ & $21.264\pm0.016$ & $20.219\pm0.010$ & $19.823\pm0.009$ & $-0.001\pm0.000$\\
$-51.49960$ & $0.25187$ & $-0.075\pm0.000$ & $19.973\pm0.007$ & $19.395\pm0.006$ & $19.163\pm0.006$ & $0.001\pm0.000$\\
$-51.49951$ & $0.63092$ & $-0.109\pm0.000$ & $20.665\pm0.008$ & $19.948\pm0.006$ & $19.690\pm0.007$ & $0.007\pm0.000$\\
$-51.49951$ & $0.70459$ & $-0.057\pm0.000$ & $19.446\pm0.005$ & $18.796\pm0.004$ & $18.555\pm0.005$ & $18.406\pm0.012$\\
$-51.49951$ & $0.69510$ & $19.719\pm0.014$ & $18.417\pm0.004$ & $17.859\pm0.003$ & $17.659\pm0.003$ & $17.568\pm0.007$\\
$-51.49930$ & $0.70751$ & $-0.057\pm0.000$ & $21.059\pm0.013$ & $20.043\pm0.008$ & $19.668\pm0.009$ & $-0.006\pm0.000$\\
$-51.49930$ & $-0.43479$ & $-0.114\pm0.000$ & $21.322\pm0.016$ & $19.707\pm0.007$ & $18.857\pm0.005$ & $18.353\pm0.011$\\
$-51.49930$ & $-0.11023$ & $-0.055\pm0.000$ & $20.765\pm0.011$ & $19.228\pm0.005$ & $18.408\pm0.004$ & $17.935\pm0.009$\\
$-51.49921$ & $0.33849$ & $-0.051\pm0.000$ & $20.151\pm0.007$ & $19.591\pm0.006$ & $19.409\pm0.007$ & $0.009\pm0.000$\\
$-51.49890$ & $0.77932$ & $-0.077\pm0.000$ & $19.379\pm0.005$ & $18.159\pm0.004$ & $17.640\pm0.003$ & $17.352\pm0.006$\\
$-51.49890$ & $0.75056$ & $-0.057\pm0.000$ & $20.494\pm0.009$ & $20.118\pm0.009$ & $19.971\pm0.011$ & $-0.012\pm0.000$\\
$-51.49890$ & $0.30273$ & $-0.051\pm0.000$ & $20.948\pm0.011$ & $19.425\pm0.006$ & $18.553\pm0.005$ & $18.068\pm0.010$\\
$-51.49869$ & $-0.98086$ & $-0.066\pm0.000$ & $19.407\pm0.006$ & $18.430\pm0.004$ & $18.080\pm0.004$ & $17.844\pm0.008$\\
$-51.49860$ & $-0.17120$ & $-0.057\pm0.000$ & $21.328\pm0.016$ & $19.899\pm0.007$ & $19.319\pm0.007$ & $18.981\pm0.019$\\
$-51.49860$ & $-0.98718$ & $-0.062\pm0.000$ & $20.157\pm0.008$ & $18.835\pm0.004$ & $18.319\pm0.004$ & $17.968\pm0.009$\\
$-51.49851$ & $1.19985$ & $18.017\pm0.008$ & $16.588\pm0.002$ & $16.010\pm0.002$ & $15.813\pm0.003$ & $15.723\pm0.005$\\
$-51.49838$ & $1.08040$ & $-0.091\pm0.000$ & $20.493\pm0.009$ & $19.620\pm0.007$ & $19.258\pm0.008$ & $0.005\pm0.000$\\
$-51.49838$ & $-0.10840$ & $20.161\pm0.021$ & $18.967\pm0.005$ & $18.407\pm0.004$ & $18.230\pm0.004$ & $18.118\pm0.009$\\
$-51.49829$ & $0.22556$ & $-0.068\pm0.000$ & $20.316\pm0.008$ & $19.720\pm0.007$ & $19.528\pm0.008$ & $-0.009\pm0.000$\\
$-51.49829$ & $1.04860$ & $-0.076\pm0.000$ & $20.535\pm0.007$ & $19.268\pm0.004$ & $18.757\pm0.004$ & $18.449\pm0.010$\\
$-51.49820$ & $0.33468$ & $-0.074\pm0.000$ & $18.492\pm0.005$ & $17.645\pm0.007$ & $17.362\pm0.011$ & $17.197\pm0.011$\\
$-51.49820$ & $0.70993$ & $-0.058\pm0.000$ & $20.563\pm0.009$ & $19.099\pm0.005$ & $18.413\pm0.004$ & $18.029\pm0.008$\\
$-51.49789$ & $-0.57415$ & $19.323\pm0.014$ & $18.103\pm0.004$ & $17.498\pm0.004$ & $17.261\pm0.003$ & $17.131\pm0.007$\\
$-51.49768$ & $1.19547$ & $-0.060\pm0.000$ & $20.814\pm0.011$ & $20.401\pm0.013$ & $20.242\pm0.017$ & $0.000\pm0.000$\\
$-51.49750$ & $0.21928$ & $-0.076\pm0.000$ & $21.277\pm0.015$ & $20.227\pm0.009$ & $19.838\pm0.009$ & $-0.003\pm0.000$\\
$-51.49750$ & $0.73620$ & $-0.075\pm0.000$ & $20.000\pm0.009$ & $18.555\pm0.005$ & $17.961\pm0.005$ & $17.622\pm0.009$\\
$-51.49735$ & $1.10949$ & $-0.044\pm0.000$ & $21.276\pm0.015$ & $19.847\pm0.007$ & $18.292\pm0.004$ & $17.452\pm0.007$\\
$-51.49735$ & $-0.89879$ & $-0.062\pm0.000$ & $21.218\pm0.019$ & $19.712\pm0.007$ & $18.974\pm0.006$ & $18.518\pm0.012$\\
$-51.49735$ & $1.23392$ & $19.262\pm0.012$ & $18.128\pm0.003$ & $17.643\pm0.003$ & $17.475\pm0.003$ & $17.407\pm0.007$\\
$-51.49728$ & $0.72424$ & $-0.060\pm0.000$ & $20.440\pm0.009$ & $19.457\pm0.006$ & $19.127\pm0.006$ & $18.916\pm0.015$\\
$-51.49719$ & $-0.03586$ & $-0.073\pm0.000$ & $21.276\pm0.015$ & $20.061\pm0.008$ & $19.649\pm0.008$ & $0.005\pm0.000$\\
$-51.49719$ & $-1.01131$ & $-0.047\pm0.000$ & $20.443\pm0.010$ & $19.173\pm0.005$ & $18.649\pm0.005$ & $18.303\pm0.011$\\
$-51.49719$ & $0.74200$ & $-0.075\pm0.000$ & $20.551\pm0.009$ & $19.213\pm0.005$ & $18.690\pm0.005$ & $18.406\pm0.010$\\
$-51.49701$ & $-0.93142$ & $-0.075\pm0.000$ & $19.961\pm0.008$ & $19.298\pm0.005$ & $19.045\pm0.006$ & $18.895\pm0.017$\\
$-51.49701$ & $1.07940$ & $19.788\pm0.016$ & $18.390\pm0.003$ & $17.788\pm0.003$ & $17.578\pm0.003$ & $17.463\pm0.007$\\
$-51.49692$ & $-0.54050$ & $-0.079\pm0.000$ & $19.940\pm0.007$ & $18.949\pm0.005$ & $18.540\pm0.005$ & $18.270\pm0.010$\\
$-51.49692$ & $-0.61986$ & $-0.098\pm0.000$ & $21.225\pm0.015$ & $20.040\pm0.009$ & $19.578\pm0.008$ & $0.002\pm0.000$\\
$-51.49692$ & $-0.14299$ & $-0.031\pm0.000$ & $20.106\pm0.007$ & $19.590\pm0.006$ & $19.447\pm0.007$ & $0.006\pm0.000$\\
$-51.49680$ & $-0.00895$ & $-0.061\pm0.000$ & $20.092\pm0.007$ & $19.409\pm0.005$ & $19.171\pm0.006$ & $19.021\pm0.021$\\
$-51.49670$ & $-0.95457$ & $-0.052\pm0.000$ & $20.108\pm0.008$ & $19.480\pm0.006$ & $19.227\pm0.007$ & $19.056\pm0.034$\\
$-51.49655$ & $1.12279$ & $17.199\pm0.007$ & $16.071\pm0.002$ & $15.631\pm0.002$ & $15.486\pm0.003$ & $15.435\pm0.004$\\
$-51.49655$ & $-0.50931$ & $-0.091\pm0.000$ & $21.206\pm0.017$ & $20.087\pm0.009$ & $19.634\pm0.008$ & $0.009\pm0.000$\\
$-51.49616$ & $-0.83636$ & $-0.071\pm0.000$ & $20.087\pm0.006$ & $19.479\pm0.005$ & $19.220\pm0.005$ & $19.034\pm0.017$\\
\hline\end{longtable}\end{landscape}
}
\longtab{2}{

\begin{landscape}
\begin{longtable}{*{8}{c}}
\caption{Catalog of natural AB MegaCam magnitudes for selected tertiary standards in the D1 field. \label{tab:D1_catalog}} \\
\hline
\hline 
    $\rm{R.A.}$ 
              & $\rm{Dec.}$ 
              & $u_M$ 
              & $g_M$ 
              & $r_M$ 
              & $i_M$ 
              & $i2_M$ 
              & $z_M$ \\

  $(\rm{Deg.})$    &  $(\rm{Deg.})$
              & $(\rm{mag})$
              & $(\rm{mag})$
& $(\rm{mag})$
& $(\rm{mag})$
& $(\rm{mag})$
& $(\rm{mag})$ \\
\hline 
\endfirsthead
\caption{continued.} \\
\hline 
\hline 
    $\rm{R.A.}$ 
              & $\rm{Dec.}$ 
              & $u_M$ 
              & $g_M$ 
              & $r_M$ 
              & $i_M$ 
              & $i2_M$ 
              & $z_M$ \\

  $(\rm{Deg.})$    & $ (\rm{Deg.})$
              & $(\rm{mag})$
              & $(\rm{mag})$
& $(\rm{mag})$
& $(\rm{mag})$
& $(\rm{mag})$
& $(\rm{mag})$ \\

\endhead
\hline 
\endfoot

$36.38510$ & $-4.99688$ & $21.273\pm0.009$ & $19.131\pm0.001$ & $17.887\pm0.001$ & $16.971\pm0.001$ & $17.044\pm0.001$ & $16.599\pm0.001$\\
$36.53275$ & $-4.99662$ & $19.306\pm0.002$ & $18.510\pm0.001$ & $18.143\pm0.001$ & $17.988\pm0.001$ & $17.997\pm0.001$ & $17.952\pm0.001$\\
$36.09325$ & $-4.99260$ & $19.705\pm0.003$ & $17.795\pm0.000$ & $16.834\pm0.000$ & $16.412\pm0.001$ & $16.436\pm0.002$ & $16.236\pm0.001$\\
$36.47386$ & $-4.99281$ & $18.428\pm0.002$ & $17.413\pm0.000$ & $16.940\pm0.000$ & $16.749\pm0.001$ & $16.763\pm0.001$ & $16.691\pm0.001$\\
$36.54650$ & $-4.99460$ & $23.000\pm0.046$ & $20.823\pm0.002$ & $19.560\pm0.001$ & $18.022\pm0.001$ & $18.131\pm0.001$ & $17.401\pm0.001$\\
$36.58690$ & $-4.99623$ & $21.494\pm0.015$ & $20.979\pm0.002$ & $20.839\pm0.002$ & $20.843\pm0.004$ & $21.162\pm0.009$ & $20.833\pm0.018$\\
$36.26135$ & $-4.99125$ & $21.188\pm0.008$ & $19.048\pm0.001$ & $17.794\pm0.000$ & $16.895\pm0.001$ & $16.960\pm0.001$ & $16.524\pm0.001$\\
$36.30102$ & $-4.99325$ & $20.781\pm0.006$ & $19.402\pm0.001$ & $18.794\pm0.001$ & $18.558\pm0.001$ & $18.573\pm0.001$ & $18.481\pm0.002$\\
$36.84051$ & $-4.99460$ & $21.328\pm0.011$ & $20.712\pm0.002$ & $20.464\pm0.002$ & $20.386\pm0.003$ & $20.381\pm0.004$ & $20.420\pm0.012$\\
$36.59163$ & $-4.99306$ & $22.907\pm0.061$ & $20.994\pm0.002$ & $19.902\pm0.001$ & $19.303\pm0.001$ & $19.355\pm0.002$ & $19.051\pm0.003$\\
$36.25865$ & $-4.99148$ & $23.322\pm0.057$ & $21.129\pm0.002$ & $19.846\pm0.001$ & $18.797\pm0.001$ & $18.890\pm0.001$ & $18.363\pm0.002$\\
$36.04552$ & $-4.99251$ & $23.224\pm0.056$ & $21.252\pm0.002$ & $20.076\pm0.001$ & $19.351\pm0.001$ & $19.388\pm0.002$ & $19.026\pm0.003$\\
$36.89012$ & $-4.99279$ & $21.641\pm0.017$ & $21.056\pm0.002$ & $20.830\pm0.002$ & $20.743\pm0.003$ & $20.737\pm0.006$ & $20.816\pm0.017$\\
$36.65816$ & $-4.99284$ & $22.096\pm0.019$ & $21.496\pm0.003$ & $21.236\pm0.003$ & $21.155\pm0.005$ & $21.170\pm0.010$ & $21.192\pm0.024$\\
$36.14973$ & $-4.98829$ & $19.426\pm0.002$ & $18.532\pm0.001$ & $18.156\pm0.001$ & $18.024\pm0.000$ & $18.030\pm0.001$ & $18.001\pm0.001$\\
$36.25174$ & $-4.98940$ & $21.277\pm0.009$ & $19.248\pm0.001$ & $18.138\pm0.000$ & $17.563\pm0.000$ & $17.601\pm0.001$ & $17.321\pm0.001$\\
$36.54013$ & $-4.99079$ & $21.751\pm0.013$ & $21.055\pm0.002$ & $20.795\pm0.002$ & $20.710\pm0.003$ & $20.720\pm0.005$ & $20.764\pm0.014$\\
$36.95613$ & $-4.99045$ & $21.979\pm0.020$ & $21.798\pm0.004$ & $21.570\pm0.004$ & $21.592\pm0.007$ & $21.572\pm0.012$ & $21.403\pm0.026$\\
$36.15851$ & $-4.98740$ & $19.821\pm0.003$ & $18.594\pm0.000$ & $18.046\pm0.000$ & $17.833\pm0.000$ & $17.843\pm0.001$ & $17.760\pm0.001$\\
$36.16302$ & $-4.98960$ & $23.390\pm0.064$ & $22.019\pm0.004$ & $21.221\pm0.003$ & $20.833\pm0.003$ & $20.843\pm0.006$ & $20.702\pm0.013$\\
$36.27337$ & $-4.98826$ & $23.671\pm0.262$ & $22.113\pm0.007$ & $20.755\pm0.002$ & $18.805\pm0.001$ & $18.914\pm0.001$ & $17.997\pm0.001$\\
$36.26092$ & $-4.98886$ & $24.089\pm0.103$ & $21.933\pm0.004$ & $20.657\pm0.002$ & $19.645\pm0.001$ & $19.714\pm0.002$ & $19.203\pm0.003$\\
$36.83998$ & $-4.98953$ & $23.237\pm0.053$ & $22.078\pm0.005$ & $21.438\pm0.003$ & $21.145\pm0.005$ & $21.144\pm0.008$ & $21.038\pm0.018$\\
$36.37521$ & $-4.98974$ & $22.810\pm0.036$ & $22.120\pm0.005$ & $21.802\pm0.005$ & $21.621\pm0.006$ & $21.632\pm0.011$ & $21.555\pm0.027$\\
$36.52906$ & $-4.98853$ & $23.932\pm0.093$ & $21.838\pm0.003$ & $20.629\pm0.002$ & $19.625\pm0.001$ & $19.710\pm0.002$ & $19.231\pm0.003$\\
$36.60251$ & $-4.98351$ & $18.040\pm0.002$ & $17.153\pm0.000$ & $16.771\pm0.000$ & $16.621\pm0.001$ & $16.630\pm0.001$ & $16.590\pm0.001$\\
$36.50657$ & $-4.98856$ & $21.359\pm0.009$ & $21.190\pm0.002$ & $21.141\pm0.003$ & $20.941\pm0.004$ & $21.029\pm0.007$ & $20.993\pm0.016$\\
$36.98082$ & $-4.98706$ & $22.351\pm0.025$ & $21.096\pm0.002$ & $20.481\pm0.001$ & $20.189\pm0.002$ & $20.199\pm0.004$ & $20.086\pm0.008$\\
$36.04928$ & $-4.98642$ & $23.804\pm0.080$ & $21.710\pm0.003$ & $20.438\pm0.001$ & $19.208\pm0.001$ & $19.268\pm0.002$ & $18.671\pm0.002$\\
$36.31037$ & $-4.98424$ & $19.416\pm0.002$ & $18.453\pm0.000$ & $17.951\pm0.000$ & $17.718\pm0.000$ & $17.732\pm0.001$ & $17.638\pm0.001$\\
$36.58115$ & $-4.98026$ & $18.937\pm0.002$ & $17.374\pm0.000$ & $16.640\pm0.001$ & $16.307\pm0.001$ & $16.336\pm0.002$ & $16.175\pm0.001$\\
$36.98122$ & $-4.98534$ & $24.936\pm0.418$ & $22.913\pm0.010$ & $21.643\pm0.004$ & $20.711\pm0.003$ & $20.748\pm0.006$ & $20.296\pm0.009$\\
$36.52662$ & $-4.98517$ & $21.308\pm0.008$ & $20.659\pm0.001$ & $20.400\pm0.001$ & $20.321\pm0.002$ & $20.316\pm0.003$ & $20.369\pm0.008$\\
$36.87478$ & $-4.98405$ & $23.535\pm0.073$ & $21.339\pm0.002$ & $20.073\pm0.001$ & $19.352\pm0.001$ & $19.397\pm0.002$ & $19.040\pm0.003$\\
$36.42190$ & $-4.98455$ & $22.233\pm0.018$ & $21.885\pm0.003$ & $21.975\pm0.005$ & $21.523\pm0.006$ & $21.295\pm0.008$ & $21.485\pm0.023$\\
$36.31201$ & $-4.97952$ & $21.392\pm0.008$ & $19.281\pm0.000$ & $18.067\pm0.000$ & $16.687\pm0.001$ & $16.859\pm0.001$ & $16.127\pm0.001$\\
$36.97860$ & $-4.98196$ & $24.141\pm0.114$ & $21.709\pm0.003$ & $20.400\pm0.001$ & $19.028\pm0.001$ & $19.096\pm0.001$ & $18.441\pm0.002$\\
$36.48674$ & $-4.98136$ & $21.088\pm0.007$ & $20.911\pm0.001$ & $20.642\pm0.001$ & $20.675\pm0.003$ & $20.660\pm0.004$ & $20.572\pm0.010$\\
$36.98416$ & $-4.98185$ & $20.459\pm0.005$ & $19.759\pm0.001$ & $19.443\pm0.001$ & $19.315\pm0.001$ & $19.314\pm0.002$ & $19.309\pm0.004$\\
$36.15043$ & $-4.98003$ & $23.940\pm0.088$ & $21.840\pm0.003$ & $20.574\pm0.001$ & $19.050\pm0.001$ & $19.116\pm0.001$ & $18.419\pm0.002$\\
$36.81017$ & $-4.97627$ & $18.645\pm0.001$ & $17.567\pm0.000$ & $17.101\pm0.000$ & $16.922\pm0.000$ & $16.935\pm0.001$ & $16.882\pm0.001$\\
$36.85117$ & $-4.98075$ & $22.211\pm0.017$ & $21.595\pm0.003$ & $21.349\pm0.003$ & $21.223\pm0.004$ & $21.230\pm0.007$ & $21.269\pm0.020$\\
$36.70455$ & $-4.97555$ & $18.696\pm0.001$ & $17.362\pm0.000$ & $16.800\pm0.000$ & $16.594\pm0.001$ & $16.607\pm0.001$ & $16.525\pm0.001$\\
$36.45521$ & $-4.97674$ & $22.788\pm0.029$ & $20.812\pm0.001$ & $19.617\pm0.001$ & $18.076\pm0.000$ & $18.181\pm0.001$ & $17.452\pm0.001$\\
$36.30635$ & $-4.97895$ & $21.902\pm0.011$ & $21.249\pm0.002$ & $20.936\pm0.002$ & $20.818\pm0.003$ & $20.811\pm0.005$ & $20.804\pm0.012$\\
\hline\end{longtable}\end{landscape}
}

\subsection{MegaCam transmission functions}
\label{sec:megac-transm-funct}

Low resolution version of the Megacam effective transmission curves
are tabulated in
tables~\ref{tab:u_band_table}-\ref{tab:z_band_table}. Higher
resolution versions are available on the joint calibration webpage.

\Fixed{As described in Sect.~\ref{sec:instrument-bandpass}, the
  transmission functions depends on the location in the MegaCam focal
  plane. Assuming that the transmission vary continuously with the
  radius from the focal plane center provides an effective
  description. We deliver tabulated measurements at regularly spaced
  radii. Those measurements can be linearly interpolated to other
  positions.  }

\longtab{3}{

\begin{landscape}
\begin{longtable}{*{11}{c}}
\caption{$\band{u}$ transmission curve at an airmass of 1.25. \label{tab:u_band_table}} \\
\hline
\hline 
$\lambda$    & $T_{u_M}(\lambda; r)$ 
              & $T_{u_M}(\lambda; r)$ 
              & $T_{u_M}(\lambda; r)$ 
              & $T_{u_M}(\lambda; r)$ 
              & $T_{u_M}(\lambda; r)$ 
              & $T_{u_M}(\lambda; r)$ 
              & $T_{u_M}(\lambda; r)$ 
              & $T_{u_M}(\lambda; r)$ 
              & $T_{u_M}(\lambda; r)$ 
              & $T_{u_M}(\lambda; r)$ \\

  $(\angstrom)$    & $r = 0\ \rm{mm}$
              & $r = 23\ \rm{mm}$
              & $r = 47\ \rm{mm}$
              & $r = 70\ \rm{mm}$
              & $r = 93\ \rm{mm}$
              & $r = 117\ \rm{mm}$
              & $r = 140\ \rm{mm}$
              & $r = 163\ \rm{mm}$
              & $r = 186\ \rm{mm}$
              & $r = 210\ \rm{mm}$ \\
\hline 
\endfirsthead
\caption{continued.} \\
\hline 
\hline 
$\lambda$    & $T_{u_M}(\lambda; r)$ 
              & $T_{u_M}(\lambda; r)$ 
              & $T_{u_M}(\lambda; r)$ 
              & $T_{u_M}(\lambda; r)$ 
              & $T_{u_M}(\lambda; r)$ 
              & $T_{u_M}(\lambda; r)$ 
              & $T_{u_M}(\lambda; r)$ 
              & $T_{u_M}(\lambda; r)$ 
              & $T_{u_M}(\lambda; r)$ 
              & $T_{u_M}(\lambda; r)$ \\

  $(\angstrom)$    & $r =  0\ \rm{mm}$
              & $r = 23\ \rm{mm}$
              & $r = 47\ \rm{mm}$
              & $r = 70\ \rm{mm}$
              & $r = 93\ \rm{mm}$
              & $r = 117\ \rm{mm}$
              & $r = 140\ \rm{mm}$
              & $r = 163\ \rm{mm}$
              & $r = 186\ \rm{mm}$
              & $r = 210\ \rm{mm}$ \\

\endhead
\hline 
\endfoot

3200 & 0.0000 & 0.0000 & 0.0000 & 0.0000 & 0.0000 & 0.0000 & 0.0000 & 0.0000 & 0.0000 & 0.0000\\
3250 & 0.0001 & 0.0002 & 0.0003 & 0.0003 & 0.0002 & 0.0002 & 0.0001 & 0.0001 & 0.0002 & 0.0002\\
3300 & 0.0047 & 0.0065 & 0.0098 & 0.0090 & 0.0075 & 0.0062 & 0.0054 & 0.0054 & 0.0065 & 0.0068\\
3350 & 0.0427 & 0.0489 & 0.0576 & 0.0562 & 0.0526 & 0.0492 & 0.0464 & 0.0471 & 0.0511 & 0.0524\\
3400 & 0.1015 & 0.1059 & 0.1106 & 0.1117 & 0.1124 & 0.1128 & 0.1132 & 0.1163 & 0.1222 & 0.1242\\
3450 & 0.1357 & 0.1371 & 0.1381 & 0.1400 & 0.1429 & 0.1463 & 0.1501 & 0.1553 & 0.1613 & 0.1633\\
3500 & 0.1687 & 0.1689 & 0.1684 & 0.1688 & 0.1698 & 0.1717 & 0.1740 & 0.1775 & 0.1818 & 0.1831\\
3550 & 0.1860 & 0.1862 & 0.1862 & 0.1876 & 0.1890 & 0.1902 & 0.1912 & 0.1926 & 0.1946 & 0.1948\\
3600 & 0.1883 & 0.1893 & 0.1910 & 0.1938 & 0.1970 & 0.1995 & 0.2013 & 0.2023 & 0.2024 & 0.2010\\
3650 & 0.2055 & 0.2052 & 0.2050 & 0.2069 & 0.2093 & 0.2108 & 0.2111 & 0.2099 & 0.2074 & 0.2047\\
3700 & 0.1916 & 0.1899 & 0.1886 & 0.1913 & 0.1955 & 0.1996 & 0.2038 & 0.2085 & 0.2155 & 0.2160\\
3750 & 0.1888 & 0.1890 & 0.1907 & 0.1946 & 0.2006 & 0.2079 & 0.2179 & 0.2295 & 0.2438 & 0.2465\\
3800 & 0.2373 & 0.2393 & 0.2427 & 0.2462 & 0.2512 & 0.2574 & 0.2659 & 0.2742 & 0.2825 & 0.2840\\
3850 & 0.2934 & 0.2932 & 0.2922 & 0.2939 & 0.2962 & 0.2987 & 0.3014 & 0.3021 & 0.3020 & 0.3021\\
3900 & 0.2977 & 0.2952 & 0.2924 & 0.2930 & 0.2947 & 0.2961 & 0.2979 & 0.2981 & 0.2984 & 0.2983\\
3950 & 0.3074 & 0.3063 & 0.3054 & 0.3046 & 0.3036 & 0.3030 & 0.3034 & 0.3025 & 0.3019 & 0.3019\\
4000 & 0.3163 & 0.3168 & 0.3181 & 0.3187 & 0.3189 & 0.3185 & 0.3180 & 0.3167 & 0.3159 & 0.3155\\
4050 & 0.3135 & 0.3077 & 0.2952 & 0.2985 & 0.3049 & 0.3093 & 0.3120 & 0.3127 & 0.3115 & 0.3098\\
4100 & 0.2514 & 0.2298 & 0.1925 & 0.2000 & 0.2169 & 0.2317 & 0.2424 & 0.2426 & 0.2344 & 0.2312\\
4150 & 0.1133 & 0.0940 & 0.0669 & 0.0723 & 0.0854 & 0.0990 & 0.1111 & 0.1124 & 0.1049 & 0.1028\\
4200 & 0.0289 & 0.0228 & 0.0155 & 0.0170 & 0.0207 & 0.0251 & 0.0297 & 0.0309 & 0.0285 & 0.0277\\
4250 & 0.0071 & 0.0058 & 0.0041 & 0.0045 & 0.0053 & 0.0063 & 0.0073 & 0.0076 & 0.0070 & 0.0067\\
4300 & 0.0020 & 0.0015 & 0.0009 & 0.0011 & 0.0013 & 0.0015 & 0.0018 & 0.0020 & 0.0018 & 0.0017\\
4350 & 0.0003 & 0.0002 & 0.0000 & 0.0001 & 0.0001 & 0.0002 & 0.0002 & 0.0003 & 0.0002 & 0.0002\\
4400 & 0.0000 & 0.0000 & 0.0000 & 0.0000 & 0.0000 & 0.0000 & 0.0000 & 0.0000 & 0.0000 & 0.0000\\
4450 & 0.0000 & 0.0000 & 0.0000 & 0.0000 & 0.0000 & 0.0000 & 0.0000 & 0.0000 & 0.0000 & 0.0000\\
4500 & 0.0000 & 0.0000 & 0.0000 & 0.0000 & 0.0000 & 0.0000 & 0.0000 & 0.0000 & 0.0000 & 0.0000\\
4550 & 0.0000 & 0.0000 & 0.0000 & 0.0000 & 0.0000 & 0.0000 & 0.0000 & 0.0000 & 0.0000 & 0.0000\\
4600 & 0.0000 & 0.0000 & 0.0000 & 0.0000 & 0.0000 & 0.0000 & 0.0000 & 0.0000 & 0.0000 & 0.0000\\
4650 & 0.0000 & 0.0000 & 0.0000 & 0.0000 & 0.0000 & 0.0000 & 0.0000 & 0.0000 & 0.0000 & 0.0000\\
4700 & 0.0000 & 0.0000 & 0.0000 & 0.0000 & 0.0000 & 0.0000 & 0.0000 & 0.0000 & 0.0000 & 0.0000\\
4750 & 0.0000 & 0.0000 & 0.0000 & 0.0000 & 0.0000 & 0.0000 & 0.0000 & 0.0000 & 0.0000 & 0.0000\\
4800 & 0.0000 & 0.0000 & 0.0000 & 0.0000 & 0.0000 & 0.0000 & 0.0000 & 0.0000 & 0.0000 & 0.0000\\
4850 & 0.0000 & 0.0000 & 0.0000 & 0.0000 & 0.0000 & 0.0000 & 0.0000 & 0.0000 & 0.0000 & 0.0000\\
4900 & 0.0000 & 0.0000 & 0.0000 & 0.0000 & 0.0000 & 0.0000 & 0.0000 & 0.0000 & 0.0000 & 0.0000\\
4950 & 0.0000 & 0.0000 & 0.0000 & 0.0000 & 0.0000 & 0.0000 & 0.0000 & 0.0000 & 0.0000 & 0.0000\\
5000 & 0.0000 & 0.0000 & 0.0003 & 0.0004 & 0.0004 & 0.0004 & 0.0004 & 0.0006 & 0.0009 & 0.0010\\
5050 & 0.0000 & 0.0002 & 0.0016 & 0.0019 & 0.0020 & 0.0020 & 0.0020 & 0.0021 & 0.0022 & 0.0022\\
5100 & 0.0000 & 0.0004 & 0.0015 & 0.0018 & 0.0021 & 0.0021 & 0.0020 & 0.0018 & 0.0012 & 0.0010\\
5150 & 0.0000 & 0.0001 & 0.0003 & 0.0004 & 0.0005 & 0.0005 & 0.0004 & 0.0003 & 0.0001 & 0.0000\\
5200 & 0.0000 & 0.0000 & 0.0000 & 0.0000 & 0.0000 & 0.0000 & 0.0000 & 0.0000 & 0.0000 & 0.0000\\
5250 & 0.0000 & 0.0000 & 0.0000 & 0.0000 & 0.0000 & 0.0000 & 0.0000 & 0.0000 & 0.0000 & 0.0000\\
5300 & 0.0000 & 0.0000 & 0.0000 & 0.0000 & 0.0000 & 0.0000 & 0.0000 & 0.0000 & 0.0000 & 0.0000\\
\hline\end{longtable}\end{landscape}
}
\longtab{4}{

\begin{landscape}
\begin{longtable}{*{11}{c}}
\caption{$\band{g}$ transmission curve at an airmass of 1.25. \label{tab:g_band_table}} \\
\hline
\hline 
$\lambda$    & $T_{g_M}(\lambda; r)$ 
              & $T_{g_M}(\lambda; r)$ 
              & $T_{g_M}(\lambda; r)$ 
              & $T_{g_M}(\lambda; r)$ 
              & $T_{g_M}(\lambda; r)$ 
              & $T_{g_M}(\lambda; r)$ 
              & $T_{g_M}(\lambda; r)$ 
              & $T_{g_M}(\lambda; r)$ 
              & $T_{g_M}(\lambda; r)$ 
              & $T_{g_M}(\lambda; r)$ \\

  $(\angstrom)$    & $r = 0\ \rm{mm}$
              & $r = 23\ \rm{mm}$
              & $r = 47\ \rm{mm}$
              & $r = 70\ \rm{mm}$
              & $r = 93\ \rm{mm}$
              & $r = 117\ \rm{mm}$
              & $r = 140\ \rm{mm}$
              & $r = 163\ \rm{mm}$
              & $r = 186\ \rm{mm}$
              & $r = 210\ \rm{mm}$ \\
\hline 
\endfirsthead
\caption{continued.} \\
\hline 
\hline 
$\lambda$    & $T_{g_M}(\lambda; r)$ 
              & $T_{g_M}(\lambda; r)$ 
              & $T_{g_M}(\lambda; r)$ 
              & $T_{g_M}(\lambda; r)$ 
              & $T_{g_M}(\lambda; r)$ 
              & $T_{g_M}(\lambda; r)$ 
              & $T_{g_M}(\lambda; r)$ 
              & $T_{g_M}(\lambda; r)$ 
              & $T_{g_M}(\lambda; r)$ 
              & $T_{g_M}(\lambda; r)$ \\

  $(\angstrom)$    & $r =  0\ \rm{mm}$
              & $r = 23\ \rm{mm}$
              & $r = 47\ \rm{mm}$
              & $r = 70\ \rm{mm}$
              & $r = 93\ \rm{mm}$
              & $r = 117\ \rm{mm}$
              & $r = 140\ \rm{mm}$
              & $r = 163\ \rm{mm}$
              & $r = 186\ \rm{mm}$
              & $r = 210\ \rm{mm}$ \\

\endhead
\hline 
\endfoot

4000 & 0.0006 & 0.0006 & 0.0017 & 0.0017 & 0.0017 & 0.0018 & 0.0020 & 0.0023 & 0.0043 & 0.0052\\
4050 & 0.0054 & 0.0073 & 0.0148 & 0.0151 & 0.0152 & 0.0160 & 0.0188 & 0.0217 & 0.0394 & 0.0463\\
4100 & 0.0376 & 0.0585 & 0.0890 & 0.0901 & 0.0908 & 0.0952 & 0.1082 & 0.1201 & 0.1789 & 0.1976\\
4150 & 0.1751 & 0.2334 & 0.2798 & 0.2810 & 0.2816 & 0.2872 & 0.3035 & 0.3167 & 0.3662 & 0.3784\\
4200 & 0.3377 & 0.3707 & 0.4067 & 0.4075 & 0.4089 & 0.4109 & 0.4147 & 0.4168 & 0.4215 & 0.4234\\
4250 & 0.3634 & 0.3796 & 0.4163 & 0.4168 & 0.4186 & 0.4205 & 0.4234 & 0.4252 & 0.4341 & 0.4386\\
4300 & 0.3839 & 0.4166 & 0.4395 & 0.4397 & 0.4415 & 0.4436 & 0.4468 & 0.4487 & 0.4551 & 0.4582\\
4350 & 0.4174 & 0.4348 & 0.4411 & 0.4423 & 0.4458 & 0.4493 & 0.4535 & 0.4554 & 0.4618 & 0.4648\\
4400 & 0.4271 & 0.4224 & 0.4489 & 0.4506 & 0.4544 & 0.4589 & 0.4648 & 0.4678 & 0.4753 & 0.4776\\
4450 & 0.4324 & 0.4428 & 0.4674 & 0.4682 & 0.4709 & 0.4736 & 0.4770 & 0.4785 & 0.4800 & 0.4812\\
4500 & 0.4505 & 0.4576 & 0.4700 & 0.4701 & 0.4719 & 0.4738 & 0.4764 & 0.4779 & 0.4838 & 0.4870\\
4550 & 0.4494 & 0.4504 & 0.4838 & 0.4845 & 0.4857 & 0.4878 & 0.4910 & 0.4927 & 0.4982 & 0.5001\\
4600 & 0.4524 & 0.4579 & 0.5034 & 0.5035 & 0.5037 & 0.5037 & 0.5034 & 0.5029 & 0.5001 & 0.5000\\
4650 & 0.4525 & 0.4633 & 0.5025 & 0.5024 & 0.5023 & 0.5011 & 0.4994 & 0.4984 & 0.4954 & 0.4963\\
4700 & 0.4455 & 0.4644 & 0.4965 & 0.4960 & 0.4956 & 0.4950 & 0.4950 & 0.4954 & 0.4996 & 0.5033\\
4750 & 0.4448 & 0.4793 & 0.5048 & 0.5038 & 0.5033 & 0.5036 & 0.5053 & 0.5065 & 0.5133 & 0.5168\\
4800 & 0.4680 & 0.4973 & 0.5184 & 0.5181 & 0.5183 & 0.5189 & 0.5199 & 0.5203 & 0.5230 & 0.5248\\
4850 & 0.4907 & 0.4973 & 0.5207 & 0.5212 & 0.5222 & 0.5228 & 0.5232 & 0.5229 & 0.5215 & 0.5220\\
4900 & 0.4811 & 0.4817 & 0.5232 & 0.5228 & 0.5226 & 0.5219 & 0.5209 & 0.5202 & 0.5168 & 0.5164\\
4950 & 0.4639 & 0.4805 & 0.5148 & 0.5142 & 0.5142 & 0.5128 & 0.5105 & 0.5090 & 0.5028 & 0.5030\\
5000 & 0.4331 & 0.4546 & 0.4849 & 0.4844 & 0.4857 & 0.4868 & 0.4885 & 0.4894 & 0.4953 & 0.4999\\
5050 & 0.4193 & 0.4639 & 0.4785 & 0.4770 & 0.4784 & 0.4810 & 0.4848 & 0.4872 & 0.4978 & 0.5032\\
5100 & 0.4535 & 0.4651 & 0.4671 & 0.4658 & 0.4686 & 0.4727 & 0.4787 & 0.4828 & 0.4976 & 0.5029\\
5150 & 0.4359 & 0.4503 & 0.4541 & 0.4541 & 0.4586 & 0.4643 & 0.4732 & 0.4793 & 0.4984 & 0.5042\\
5200 & 0.4444 & 0.4526 & 0.4579 & 0.4593 & 0.4643 & 0.4707 & 0.4812 & 0.4880 & 0.5070 & 0.5115\\
5250 & 0.4354 & 0.4481 & 0.4746 & 0.4765 & 0.4813 & 0.4864 & 0.4928 & 0.4956 & 0.4979 & 0.4988\\
5300 & 0.4268 & 0.4597 & 0.4590 & 0.4606 & 0.4652 & 0.4684 & 0.4717 & 0.4734 & 0.4803 & 0.4853\\
5350 & 0.4461 & 0.4425 & 0.4469 & 0.4499 & 0.4551 & 0.4608 & 0.4694 & 0.4744 & 0.4893 & 0.4930\\
5400 & 0.4284 & 0.4385 & 0.4569 & 0.4603 & 0.4656 & 0.4696 & 0.4732 & 0.4741 & 0.4641 & 0.4604\\
5450 & 0.4325 & 0.4275 & 0.4108 & 0.4142 & 0.4195 & 0.4206 & 0.4199 & 0.4191 & 0.4099 & 0.4091\\
5500 & 0.3879 & 0.3476 & 0.3516 & 0.3556 & 0.3620 & 0.3669 & 0.3725 & 0.3755 & 0.3747 & 0.3717\\
5550 & 0.3050 & 0.3017 & 0.3160 & 0.3211 & 0.3277 & 0.3303 & 0.3308 & 0.3308 & 0.3022 & 0.2880\\
5600 & 0.2608 & 0.2455 & 0.2252 & 0.2310 & 0.2376 & 0.2353 & 0.2272 & 0.2227 & 0.1705 & 0.1508\\
5650 & 0.1635 & 0.1283 & 0.1043 & 0.1085 & 0.1131 & 0.1102 & 0.1016 & 0.0965 & 0.0589 & 0.0480\\
5700 & 0.0596 & 0.0440 & 0.0343 & 0.0357 & 0.0370 & 0.0350 & 0.0305 & 0.0280 & 0.0146 & 0.0115\\
5750 & 0.0174 & 0.0129 & 0.0094 & 0.0096 & 0.0098 & 0.0089 & 0.0074 & 0.0067 & 0.0034 & 0.0027\\
5800 & 0.0042 & 0.0030 & 0.0019 & 0.0019 & 0.0021 & 0.0018 & 0.0015 & 0.0012 & 0.0003 & 0.0002\\
5850 & 0.0006 & 0.0003 & 0.0001 & 0.0001 & 0.0001 & 0.0001 & 0.0000 & 0.0000 & 0.0000 & 0.0000\\
\hline\end{longtable}\end{landscape}
}
\longtab{5}{

\begin{landscape}
\begin{longtable}{*{11}{c}}
\caption{$\band{r}$ transmission curve at an airmass of 1.25. \label{tab:r_band_table}} \\
\hline
\hline 
$\lambda$    & $T_{r_M}(\lambda; r)$ 
              & $T_{r_M}(\lambda; r)$ 
              & $T_{r_M}(\lambda; r)$ 
              & $T_{r_M}(\lambda; r)$ 
              & $T_{r_M}(\lambda; r)$ 
              & $T_{r_M}(\lambda; r)$ 
              & $T_{r_M}(\lambda; r)$ 
              & $T_{r_M}(\lambda; r)$ 
              & $T_{r_M}(\lambda; r)$ 
              & $T_{r_M}(\lambda; r)$ \\

  $(\angstrom)$    & $r = 0\ \rm{mm}$
              & $r = 23\ \rm{mm}$
              & $r = 47\ \rm{mm}$
              & $r = 70\ \rm{mm}$
              & $r = 93\ \rm{mm}$
              & $r = 117\ \rm{mm}$
              & $r = 140\ \rm{mm}$
              & $r = 163\ \rm{mm}$
              & $r = 186\ \rm{mm}$
              & $r = 210\ \rm{mm}$ \\
\hline 
\endfirsthead
\caption{continued.} \\
\hline 
\hline 
$\lambda$    & $T_{r_M}(\lambda; r)$ 
              & $T_{r_M}(\lambda; r)$ 
              & $T_{r_M}(\lambda; r)$ 
              & $T_{r_M}(\lambda; r)$ 
              & $T_{r_M}(\lambda; r)$ 
              & $T_{r_M}(\lambda; r)$ 
              & $T_{r_M}(\lambda; r)$ 
              & $T_{r_M}(\lambda; r)$ 
              & $T_{r_M}(\lambda; r)$ 
              & $T_{r_M}(\lambda; r)$ \\

  $(\angstrom)$    & $r =  0\ \rm{mm}$
              & $r = 23\ \rm{mm}$
              & $r = 47\ \rm{mm}$
              & $r = 70\ \rm{mm}$
              & $r = 93\ \rm{mm}$
              & $r = 117\ \rm{mm}$
              & $r = 140\ \rm{mm}$
              & $r = 163\ \rm{mm}$
              & $r = 186\ \rm{mm}$
              & $r = 210\ \rm{mm}$ \\

\endhead
\hline 
\endfoot

5300 & 0.0000 & 0.0000 & 0.0000 & 0.0001 & 0.0002 & 0.0003 & 0.0008 & 0.0013 & 0.0011 & 0.0007\\
5350 & 0.0003 & 0.0003 & 0.0007 & 0.0012 & 0.0017 & 0.0022 & 0.0034 & 0.0045 & 0.0043 & 0.0033\\
5400 & 0.0021 & 0.0022 & 0.0031 & 0.0041 & 0.0052 & 0.0062 & 0.0092 & 0.0119 & 0.0116 & 0.0090\\
5450 & 0.0059 & 0.0061 & 0.0080 & 0.0103 & 0.0130 & 0.0158 & 0.0240 & 0.0316 & 0.0313 & 0.0244\\
5500 & 0.0144 & 0.0149 & 0.0197 & 0.0258 & 0.0331 & 0.0404 & 0.0611 & 0.0800 & 0.0808 & 0.0649\\
5550 & 0.0354 & 0.0368 & 0.0483 & 0.0627 & 0.0795 & 0.0954 & 0.1360 & 0.1691 & 0.1728 & 0.1480\\
5600 & 0.0825 & 0.0857 & 0.1088 & 0.1356 & 0.1640 & 0.1888 & 0.2423 & 0.2783 & 0.2808 & 0.2566\\
5650 & 0.1677 & 0.1731 & 0.2068 & 0.2415 & 0.2735 & 0.2984 & 0.3431 & 0.3662 & 0.3620 & 0.3431\\
5700 & 0.2787 & 0.2845 & 0.3171 & 0.3452 & 0.3668 & 0.3822 & 0.4036 & 0.4107 & 0.4050 & 0.3933\\
5750 & 0.3705 & 0.3740 & 0.3925 & 0.4019 & 0.4077 & 0.4122 & 0.4117 & 0.4076 & 0.4071 & 0.4075\\
5800 & 0.4068 & 0.4071 & 0.4108 & 0.4051 & 0.4007 & 0.3984 & 0.3891 & 0.3810 & 0.3791 & 0.3851\\
5850 & 0.4030 & 0.4020 & 0.3997 & 0.3936 & 0.3880 & 0.3838 & 0.3792 & 0.3731 & 0.3622 & 0.3594\\
5900 & 0.3990 & 0.3983 & 0.3958 & 0.3941 & 0.3913 & 0.3884 & 0.3886 & 0.3850 & 0.3723 & 0.3626\\
5950 & 0.4021 & 0.4015 & 0.4002 & 0.3997 & 0.3993 & 0.3983 & 0.4003 & 0.3988 & 0.3891 & 0.3808\\
6000 & 0.4068 & 0.4066 & 0.4078 & 0.4105 & 0.4120 & 0.4121 & 0.4177 & 0.4198 & 0.4111 & 0.4012\\
6050 & 0.4211 & 0.4214 & 0.4246 & 0.4305 & 0.4332 & 0.4345 & 0.4433 & 0.4491 & 0.4432 & 0.4316\\
6100 & 0.4421 & 0.4425 & 0.4461 & 0.4514 & 0.4551 & 0.4578 & 0.4666 & 0.4732 & 0.4721 & 0.4645\\
6150 & 0.4591 & 0.4596 & 0.4623 & 0.4658 & 0.4700 & 0.4730 & 0.4793 & 0.4845 & 0.4863 & 0.4847\\
6200 & 0.4696 & 0.4701 & 0.4720 & 0.4756 & 0.4798 & 0.4820 & 0.4875 & 0.4914 & 0.4909 & 0.4907\\
6250 & 0.4767 & 0.4769 & 0.4782 & 0.4825 & 0.4866 & 0.4888 & 0.4938 & 0.4959 & 0.4922 & 0.4904\\
6300 & 0.4817 & 0.4818 & 0.4826 & 0.4864 & 0.4902 & 0.4923 & 0.4947 & 0.4938 & 0.4894 & 0.4879\\
6350 & 0.4893 & 0.4893 & 0.4900 & 0.4933 & 0.4950 & 0.4953 & 0.4926 & 0.4872 & 0.4832 & 0.4837\\
6400 & 0.4957 & 0.4953 & 0.4950 & 0.4948 & 0.4922 & 0.4894 & 0.4805 & 0.4707 & 0.4662 & 0.4686\\
6450 & 0.4958 & 0.4947 & 0.4930 & 0.4890 & 0.4832 & 0.4776 & 0.4659 & 0.4544 & 0.4469 & 0.4477\\
6500 & 0.4909 & 0.4892 & 0.4864 & 0.4794 & 0.4716 & 0.4652 & 0.4526 & 0.4410 & 0.4328 & 0.4317\\
6550 & 0.4799 & 0.4775 & 0.4729 & 0.4606 & 0.4514 & 0.4468 & 0.4348 & 0.4256 & 0.4194 & 0.4193\\
6600 & 0.4541 & 0.4515 & 0.4460 & 0.4310 & 0.4237 & 0.4226 & 0.4145 & 0.4110 & 0.4088 & 0.4096\\
6650 & 0.4172 & 0.4152 & 0.4113 & 0.3982 & 0.3940 & 0.3970 & 0.3946 & 0.3993 & 0.4046 & 0.4071\\
6700 & 0.3795 & 0.3786 & 0.3774 & 0.3697 & 0.3697 & 0.3763 & 0.3832 & 0.3976 & 0.4088 & 0.4112\\
6750 & 0.3525 & 0.3522 & 0.3535 & 0.3524 & 0.3594 & 0.3707 & 0.3920 & 0.4157 & 0.4258 & 0.4248\\
6800 & 0.3401 & 0.3407 & 0.3460 & 0.3599 & 0.3797 & 0.3942 & 0.4171 & 0.4184 & 0.4216 & 0.4305\\
6850 & 0.3490 & 0.3503 & 0.3579 & 0.3728 & 0.3717 & 0.3718 & 0.3258 & 0.2731 & 0.2811 & 0.3209\\
6900 & 0.3414 & 0.3370 & 0.3217 & 0.2594 & 0.2100 & 0.1965 & 0.1233 & 0.0839 & 0.0931 & 0.1264\\
6950 & 0.1990 & 0.1909 & 0.1670 & 0.0916 & 0.0598 & 0.0533 & 0.0275 & 0.0185 & 0.0216 & 0.0313\\
7000 & 0.0536 & 0.0507 & 0.0431 & 0.0208 & 0.0140 & 0.0127 & 0.0076 & 0.0057 & 0.0066 & 0.0089\\
7050 & 0.0123 & 0.0118 & 0.0103 & 0.0057 & 0.0042 & 0.0039 & 0.0024 & 0.0017 & 0.0022 & 0.0031\\
7100 & 0.0036 & 0.0034 & 0.0030 & 0.0015 & 0.0010 & 0.0009 & 0.0004 & 0.0002 & 0.0004 & 0.0007\\
7150 & 0.0007 & 0.0007 & 0.0005 & 0.0001 & 0.0000 & 0.0000 & 0.0000 & 0.0000 & 0.0000 & 0.0000\\
\hline\end{longtable}\end{landscape}
}
\longtab{6}{

\begin{landscape}
\begin{longtable}{*{11}{c}}
\caption{$\band{i}$ transmission curve at an airmass of 1.25. \label{tab:i_band_table}} \\
\hline
\hline 
$\lambda$    & $T_{i_M}(\lambda; r)$ 
              & $T_{i_M}(\lambda; r)$ 
              & $T_{i_M}(\lambda; r)$ 
              & $T_{i_M}(\lambda; r)$ 
              & $T_{i_M}(\lambda; r)$ 
              & $T_{i_M}(\lambda; r)$ 
              & $T_{i_M}(\lambda; r)$ 
              & $T_{i_M}(\lambda; r)$ 
              & $T_{i_M}(\lambda; r)$ 
              & $T_{i_M}(\lambda; r)$ \\

  $(\angstrom)$    & $r = 0\ \rm{mm}$
              & $r = 23\ \rm{mm}$
              & $r = 47\ \rm{mm}$
              & $r = 70\ \rm{mm}$
              & $r = 93\ \rm{mm}$
              & $r = 117\ \rm{mm}$
              & $r = 140\ \rm{mm}$
              & $r = 163\ \rm{mm}$
              & $r = 186\ \rm{mm}$
              & $r = 210\ \rm{mm}$ \\
\hline 
\endfirsthead
\caption{continued.} \\
\hline 
\hline 
$\lambda$    & $T_{i_M}(\lambda; r)$ 
              & $T_{i_M}(\lambda; r)$ 
              & $T_{i_M}(\lambda; r)$ 
              & $T_{i_M}(\lambda; r)$ 
              & $T_{i_M}(\lambda; r)$ 
              & $T_{i_M}(\lambda; r)$ 
              & $T_{i_M}(\lambda; r)$ 
              & $T_{i_M}(\lambda; r)$ 
              & $T_{i_M}(\lambda; r)$ 
              & $T_{i_M}(\lambda; r)$ \\

  $(\angstrom)$    & $r =  0\ \rm{mm}$
              & $r = 23\ \rm{mm}$
              & $r = 47\ \rm{mm}$
              & $r = 70\ \rm{mm}$
              & $r = 93\ \rm{mm}$
              & $r = 117\ \rm{mm}$
              & $r = 140\ \rm{mm}$
              & $r = 163\ \rm{mm}$
              & $r = 186\ \rm{mm}$
              & $r = 210\ \rm{mm}$ \\

\endhead
\hline 
\endfoot

6650 & 0.0008 & 0.0008 & 0.0006 & 0.0013 & 0.0016 & 0.0016 & 0.0021 & 0.0030 & 0.0032 & 0.0030\\
6700 & 0.0027 & 0.0026 & 0.0023 & 0.0035 & 0.0044 & 0.0046 & 0.0062 & 0.0090 & 0.0096 & 0.0091\\
6750 & 0.0054 & 0.0053 & 0.0047 & 0.0070 & 0.0089 & 0.0094 & 0.0130 & 0.0194 & 0.0212 & 0.0201\\
6800 & 0.0105 & 0.0103 & 0.0093 & 0.0141 & 0.0182 & 0.0194 & 0.0276 & 0.0424 & 0.0473 & 0.0451\\
6850 & 0.0202 & 0.0198 & 0.0176 & 0.0275 & 0.0360 & 0.0387 & 0.0558 & 0.0863 & 0.0971 & 0.0935\\
6900 & 0.0429 & 0.0420 & 0.0374 & 0.0588 & 0.0769 & 0.0827 & 0.1173 & 0.1726 & 0.1922 & 0.1874\\
6950 & 0.0953 & 0.0935 & 0.0837 & 0.1279 & 0.1620 & 0.1730 & 0.2298 & 0.3043 & 0.3285 & 0.3239\\
7000 & 0.1872 & 0.1843 & 0.1683 & 0.2365 & 0.2813 & 0.2950 & 0.3552 & 0.4162 & 0.4334 & 0.4306\\
7050 & 0.3073 & 0.3043 & 0.2866 & 0.3564 & 0.3934 & 0.4043 & 0.4430 & 0.4721 & 0.4797 & 0.4788\\
7100 & 0.4093 & 0.4077 & 0.3959 & 0.4395 & 0.4575 & 0.4632 & 0.4780 & 0.4856 & 0.4883 & 0.4884\\
7150 & 0.4484 & 0.4482 & 0.4440 & 0.4593 & 0.4640 & 0.4660 & 0.4689 & 0.4687 & 0.4696 & 0.4701\\
7200 & 0.4445 & 0.4450 & 0.4445 & 0.4467 & 0.4465 & 0.4469 & 0.4464 & 0.4446 & 0.4442 & 0.4442\\
7250 & 0.4445 & 0.4449 & 0.4450 & 0.4445 & 0.4435 & 0.4430 & 0.4421 & 0.4403 & 0.4384 & 0.4374\\
7300 & 0.4507 & 0.4509 & 0.4508 & 0.4506 & 0.4495 & 0.4488 & 0.4480 & 0.4463 & 0.4432 & 0.4412\\
7350 & 0.4635 & 0.4637 & 0.4636 & 0.4636 & 0.4624 & 0.4618 & 0.4612 & 0.4596 & 0.4561 & 0.4535\\
7400 & 0.4671 & 0.4674 & 0.4677 & 0.4671 & 0.4659 & 0.4658 & 0.4656 & 0.4641 & 0.4612 & 0.4587\\
7450 & 0.4624 & 0.4629 & 0.4637 & 0.4623 & 0.4610 & 0.4617 & 0.4618 & 0.4607 & 0.4591 & 0.4573\\
7500 & 0.4532 & 0.4539 & 0.4550 & 0.4530 & 0.4520 & 0.4533 & 0.4537 & 0.4533 & 0.4531 & 0.4520\\
7550 & 0.4216 & 0.4222 & 0.4232 & 0.4216 & 0.4210 & 0.4223 & 0.4232 & 0.4235 & 0.4242 & 0.4237\\
7600 & 0.3147 & 0.3150 & 0.3153 & 0.3148 & 0.3148 & 0.3156 & 0.3167 & 0.3175 & 0.3182 & 0.3180\\
7650 & 0.3099 & 0.3099 & 0.3093 & 0.3104 & 0.3107 & 0.3111 & 0.3126 & 0.3141 & 0.3146 & 0.3140\\
7700 & 0.3997 & 0.3993 & 0.3978 & 0.4004 & 0.4009 & 0.4010 & 0.4032 & 0.4052 & 0.4052 & 0.4041\\
7750 & 0.4256 & 0.4250 & 0.4230 & 0.4263 & 0.4265 & 0.4263 & 0.4282 & 0.4294 & 0.4288 & 0.4274\\
7800 & 0.4248 & 0.4244 & 0.4226 & 0.4249 & 0.4244 & 0.4240 & 0.4247 & 0.4240 & 0.4228 & 0.4218\\
7850 & 0.4178 & 0.4177 & 0.4167 & 0.4171 & 0.4154 & 0.4149 & 0.4138 & 0.4106 & 0.4090 & 0.4085\\
7900 & 0.4057 & 0.4060 & 0.4058 & 0.4041 & 0.4015 & 0.4006 & 0.3975 & 0.3924 & 0.3901 & 0.3899\\
7950 & 0.3929 & 0.3933 & 0.3934 & 0.3906 & 0.3873 & 0.3857 & 0.3813 & 0.3753 & 0.3720 & 0.3716\\
8000 & 0.3812 & 0.3813 & 0.3811 & 0.3785 & 0.3750 & 0.3724 & 0.3677 & 0.3618 & 0.3575 & 0.3563\\
8050 & 0.3733 & 0.3732 & 0.3725 & 0.3705 & 0.3670 & 0.3639 & 0.3594 & 0.3540 & 0.3491 & 0.3471\\
8100 & 0.3600 & 0.3600 & 0.3596 & 0.3571 & 0.3534 & 0.3509 & 0.3467 & 0.3414 & 0.3372 & 0.3353\\
8150 & 0.3311 & 0.3316 & 0.3327 & 0.3280 & 0.3241 & 0.3231 & 0.3189 & 0.3131 & 0.3111 & 0.3104\\
8200 & 0.3062 & 0.3076 & 0.3110 & 0.3030 & 0.2986 & 0.2993 & 0.2945 & 0.2878 & 0.2884 & 0.2897\\
8250 & 0.2862 & 0.2882 & 0.2934 & 0.2829 & 0.2784 & 0.2799 & 0.2744 & 0.2673 & 0.2694 & 0.2723\\
8300 & 0.2648 & 0.2666 & 0.2715 & 0.2619 & 0.2579 & 0.2585 & 0.2530 & 0.2467 & 0.2486 & 0.2516\\
8350 & 0.2484 & 0.2496 & 0.2529 & 0.2459 & 0.2423 & 0.2418 & 0.2367 & 0.2314 & 0.2326 & 0.2347\\
8400 & 0.2311 & 0.2332 & 0.2375 & 0.2284 & 0.2235 & 0.2252 & 0.2197 & 0.2126 & 0.2165 & 0.2196\\
8450 & 0.1945 & 0.2000 & 0.2127 & 0.1907 & 0.1831 & 0.1916 & 0.1841 & 0.1719 & 0.1833 & 0.1915\\
8500 & 0.1316 & 0.1399 & 0.1619 & 0.1274 & 0.1190 & 0.1325 & 0.1239 & 0.1096 & 0.1253 & 0.1383\\
8550 & 0.0676 & 0.0744 & 0.0947 & 0.0647 & 0.0590 & 0.0697 & 0.0634 & 0.0533 & 0.0650 & 0.0762\\
8600 & 0.0282 & 0.0316 & 0.0428 & 0.0268 & 0.0243 & 0.0295 & 0.0264 & 0.0217 & 0.0273 & 0.0332\\
8650 & 0.0112 & 0.0125 & 0.0170 & 0.0106 & 0.0097 & 0.0117 & 0.0105 & 0.0087 & 0.0109 & 0.0132\\
8700 & 0.0048 & 0.0053 & 0.0070 & 0.0046 & 0.0043 & 0.0050 & 0.0045 & 0.0038 & 0.0047 & 0.0056\\
8750 & 0.0022 & 0.0024 & 0.0031 & 0.0022 & 0.0020 & 0.0023 & 0.0021 & 0.0018 & 0.0022 & 0.0026\\
8800 & 0.0008 & 0.0008 & 0.0011 & 0.0008 & 0.0007 & 0.0008 & 0.0008 & 0.0007 & 0.0008 & 0.0009\\
\hline\end{longtable}\end{landscape}
}
\longtab{7}{

\begin{landscape}
\begin{longtable}{*{11}{c}}
\caption{$\band{i2}$ transmission curve at an airmass of 1.25. \label{tab:i2_band_table}} \\
\hline
\hline 
$\lambda$    & $T_{i2_M}(\lambda; r)$ 
              & $T_{i2_M}(\lambda; r)$ 
              & $T_{i2_M}(\lambda; r)$ 
              & $T_{i2_M}(\lambda; r)$ 
              & $T_{i2_M}(\lambda; r)$ 
              & $T_{i2_M}(\lambda; r)$ 
              & $T_{i2_M}(\lambda; r)$ 
              & $T_{i2_M}(\lambda; r)$ 
              & $T_{i2_M}(\lambda; r)$ 
              & $T_{i2_M}(\lambda; r)$ \\

  $(\angstrom)$    & $r = 0\ \rm{mm}$
              & $r = 20\ \rm{mm}$
              & $r = 40\ \rm{mm}$
              & $r = 60\ \rm{mm}$
              & $r = 80\ \rm{mm}$
              & $r = 100\ \rm{mm}$
              & $r = 120\ \rm{mm}$
              & $r = 140\ \rm{mm}$
              & $r = 160\ \rm{mm}$
              & $r = 180\ \rm{mm}$ \\
\hline 
\endfirsthead
\caption{continued.} \\
\hline 
\hline 
$\lambda$    & $T_{i2_M}(\lambda; r)$ 
              & $T_{i2_M}(\lambda; r)$ 
              & $T_{i2_M}(\lambda; r)$ 
              & $T_{i2_M}(\lambda; r)$ 
              & $T_{i2_M}(\lambda; r)$ 
              & $T_{i2_M}(\lambda; r)$ 
              & $T_{i2_M}(\lambda; r)$ 
              & $T_{i2_M}(\lambda; r)$ 
              & $T_{i2_M}(\lambda; r)$ 
              & $T_{i2_M}(\lambda; r)$ \\

  $(\angstrom)$    & $r =  0\ \rm{mm}$
              & $r = 20\ \rm{mm}$
              & $r = 40\ \rm{mm}$
              & $r = 60\ \rm{mm}$
              & $r = 80\ \rm{mm}$
              & $r = 100\ \rm{mm}$
              & $r = 120\ \rm{mm}$
              & $r = 140\ \rm{mm}$
              & $r = 160\ \rm{mm}$
              & $r = 180\ \rm{mm}$ \\

\endhead
\hline 
\endfoot

6600 & 0.0007 & 0.0007 & 0.0007 & 0.0009 & 0.0008 & 0.0010 & 0.0016 & 0.0015 & 0.0012 & 0.0014\\
6650 & 0.0028 & 0.0028 & 0.0028 & 0.0037 & 0.0035 & 0.0043 & 0.0078 & 0.0068 & 0.0053 & 0.0065\\
6700 & 0.0103 & 0.0100 & 0.0100 & 0.0135 & 0.0129 & 0.0160 & 0.0279 & 0.0249 & 0.0194 & 0.0237\\
6750 & 0.0356 & 0.0348 & 0.0346 & 0.0461 & 0.0440 & 0.0541 & 0.0922 & 0.0824 & 0.0648 & 0.0784\\
6800 & 0.1158 & 0.1133 & 0.1129 & 0.1448 & 0.1389 & 0.1643 & 0.2433 & 0.2253 & 0.1887 & 0.2174\\
6850 & 0.2602 & 0.2564 & 0.2561 & 0.2963 & 0.2899 & 0.3177 & 0.3818 & 0.3703 & 0.3422 & 0.3651\\
6900 & 0.3841 & 0.3819 & 0.3822 & 0.4002 & 0.3985 & 0.4093 & 0.4273 & 0.4253 & 0.4190 & 0.4247\\
6950 & 0.4510 & 0.4500 & 0.4506 & 0.4545 & 0.4545 & 0.4570 & 0.4609 & 0.4604 & 0.4598 & 0.4606\\
7000 & 0.4702 & 0.4695 & 0.4700 & 0.4714 & 0.4713 & 0.4723 & 0.4744 & 0.4736 & 0.4729 & 0.4731\\
7050 & 0.4756 & 0.4751 & 0.4756 & 0.4760 & 0.4760 & 0.4764 & 0.4769 & 0.4765 & 0.4763 & 0.4761\\
7100 & 0.4760 & 0.4755 & 0.4760 & 0.4759 & 0.4761 & 0.4762 & 0.4766 & 0.4762 & 0.4763 & 0.4759\\
7150 & 0.4617 & 0.4613 & 0.4617 & 0.4620 & 0.4621 & 0.4624 & 0.4628 & 0.4624 & 0.4623 & 0.4617\\
7200 & 0.4425 & 0.4422 & 0.4426 & 0.4424 & 0.4426 & 0.4424 & 0.4416 & 0.4414 & 0.4420 & 0.4408\\
7250 & 0.4382 & 0.4379 & 0.4383 & 0.4378 & 0.4380 & 0.4378 & 0.4375 & 0.4373 & 0.4377 & 0.4370\\
7300 & 0.4421 & 0.4418 & 0.4421 & 0.4422 & 0.4423 & 0.4426 & 0.4433 & 0.4429 & 0.4431 & 0.4429\\
7350 & 0.4541 & 0.4537 & 0.4541 & 0.4545 & 0.4545 & 0.4549 & 0.4558 & 0.4553 & 0.4554 & 0.4550\\
7400 & 0.4586 & 0.4583 & 0.4585 & 0.4589 & 0.4589 & 0.4592 & 0.4592 & 0.4589 & 0.4592 & 0.4581\\
7450 & 0.4553 & 0.4551 & 0.4553 & 0.4549 & 0.4552 & 0.4549 & 0.4533 & 0.4535 & 0.4547 & 0.4532\\
7500 & 0.4459 & 0.4458 & 0.4460 & 0.4452 & 0.4456 & 0.4452 & 0.4443 & 0.4443 & 0.4454 & 0.4445\\
7550 & 0.4155 & 0.4153 & 0.4154 & 0.4157 & 0.4157 & 0.4161 & 0.4177 & 0.4171 & 0.4167 & 0.4167\\
7600 & 0.3128 & 0.3126 & 0.3127 & 0.3134 & 0.3133 & 0.3139 & 0.3152 & 0.3146 & 0.3142 & 0.3140\\
7650 & 0.3096 & 0.3097 & 0.3096 & 0.3098 & 0.3099 & 0.3098 & 0.3095 & 0.3092 & 0.3096 & 0.3086\\
7700 & 0.3947 & 0.3949 & 0.3948 & 0.3942 & 0.3945 & 0.3940 & 0.3925 & 0.3927 & 0.3938 & 0.3926\\
7750 & 0.4106 & 0.4109 & 0.4107 & 0.4095 & 0.4101 & 0.4094 & 0.4078 & 0.4085 & 0.4103 & 0.4094\\
7800 & 0.3991 & 0.3995 & 0.3992 & 0.3985 & 0.3991 & 0.3989 & 0.3991 & 0.3992 & 0.4005 & 0.4001\\
7850 & 0.3880 & 0.3883 & 0.3879 & 0.3886 & 0.3886 & 0.3893 & 0.3914 & 0.3907 & 0.3905 & 0.3904\\
7900 & 0.3799 & 0.3802 & 0.3798 & 0.3810 & 0.3809 & 0.3815 & 0.3834 & 0.3826 & 0.3821 & 0.3818\\
7950 & 0.3727 & 0.3732 & 0.3727 & 0.3732 & 0.3733 & 0.3734 & 0.3738 & 0.3735 & 0.3738 & 0.3731\\
8000 & 0.3631 & 0.3637 & 0.3631 & 0.3631 & 0.3635 & 0.3633 & 0.3632 & 0.3633 & 0.3640 & 0.3635\\
8050 & 0.3543 & 0.3549 & 0.3543 & 0.3544 & 0.3548 & 0.3548 & 0.3547 & 0.3549 & 0.3558 & 0.3553\\
8100 & 0.3406 & 0.3413 & 0.3406 & 0.3407 & 0.3412 & 0.3412 & 0.3408 & 0.3410 & 0.3421 & 0.3413\\
8150 & 0.3152 & 0.3160 & 0.3153 & 0.3155 & 0.3159 & 0.3159 & 0.3155 & 0.3155 & 0.3162 & 0.3153\\
8200 & 0.3003 & 0.3010 & 0.3004 & 0.3006 & 0.3010 & 0.3008 & 0.3001 & 0.3003 & 0.3009 & 0.3003\\
8250 & 0.2946 & 0.2954 & 0.2947 & 0.2941 & 0.2948 & 0.2941 & 0.2919 & 0.2930 & 0.2949 & 0.2941\\
8300 & 0.2839 & 0.2847 & 0.2841 & 0.2828 & 0.2838 & 0.2831 & 0.2813 & 0.2825 & 0.2848 & 0.2840\\
8350 & 0.2717 & 0.2726 & 0.2722 & 0.2694 & 0.2721 & 0.2704 & 0.2574 & 0.2659 & 0.2740 & 0.2718\\
8400 & 0.2292 & 0.2319 & 0.2326 & 0.2147 & 0.2267 & 0.2167 & 0.1701 & 0.1948 & 0.2299 & 0.2169\\
8450 & 0.1245 & 0.1283 & 0.1299 & 0.1044 & 0.1200 & 0.1069 & 0.0638 & 0.0841 & 0.1243 & 0.1073\\
8500 & 0.0378 & 0.0397 & 0.0405 & 0.0291 & 0.0360 & 0.0306 & 0.0160 & 0.0225 & 0.0388 & 0.0314\\
8550 & 0.0090 & 0.0094 & 0.0096 & 0.0070 & 0.0087 & 0.0075 & 0.0042 & 0.0058 & 0.0097 & 0.0079\\
8600 & 0.0025 & 0.0026 & 0.0026 & 0.0020 & 0.0024 & 0.0021 & 0.0013 & 0.0017 & 0.0027 & 0.0023\\
8650 & 0.0008 & 0.0008 & 0.0008 & 0.0007 & 0.0008 & 0.0007 & 0.0004 & 0.0006 & 0.0009 & 0.0007\\
8700 & 0.0002 & 0.0002 & 0.0002 & 0.0002 & 0.0002 & 0.0002 & 0.0001 & 0.0002 & 0.0002 & 0.0002\\
\hline\end{longtable}\end{landscape}
}
\longtab{8}{

\begin{landscape}
\begin{longtable}{*{11}{c}}
\caption{$\band{z}$ transmission curve at an airmass of 1.25. \label{tab:z_band_table}} \\
\hline
\hline 
$\lambda$    & $T_{z_M}(\lambda; r)$ 
              & $T_{z_M}(\lambda; r)$ 
              & $T_{z_M}(\lambda; r)$ 
              & $T_{z_M}(\lambda; r)$ 
              & $T_{z_M}(\lambda; r)$ 
              & $T_{z_M}(\lambda; r)$ 
              & $T_{z_M}(\lambda; r)$ 
              & $T_{z_M}(\lambda; r)$ 
              & $T_{z_M}(\lambda; r)$ 
              & $T_{z_M}(\lambda; r)$ \\

  $(\angstrom)$    & $r = 0\ \rm{mm}$
              & $r = 23\ \rm{mm}$
              & $r = 47\ \rm{mm}$
              & $r = 70\ \rm{mm}$
              & $r = 93\ \rm{mm}$
              & $r = 117\ \rm{mm}$
              & $r = 140\ \rm{mm}$
              & $r = 163\ \rm{mm}$
              & $r = 186\ \rm{mm}$
              & $r = 210\ \rm{mm}$ \\
\hline 
\endfirsthead
\caption{continued.} \\
\hline 
\hline 
$\lambda$    & $T_{z_M}(\lambda; r)$ 
              & $T_{z_M}(\lambda; r)$ 
              & $T_{z_M}(\lambda; r)$ 
              & $T_{z_M}(\lambda; r)$ 
              & $T_{z_M}(\lambda; r)$ 
              & $T_{z_M}(\lambda; r)$ 
              & $T_{z_M}(\lambda; r)$ 
              & $T_{z_M}(\lambda; r)$ 
              & $T_{z_M}(\lambda; r)$ 
              & $T_{z_M}(\lambda; r)$ \\

  $(\angstrom)$    & $r =  0\ \rm{mm}$
              & $r = 23\ \rm{mm}$
              & $r = 47\ \rm{mm}$
              & $r = 70\ \rm{mm}$
              & $r = 93\ \rm{mm}$
              & $r = 117\ \rm{mm}$
              & $r = 140\ \rm{mm}$
              & $r = 163\ \rm{mm}$
              & $r = 186\ \rm{mm}$
              & $r = 210\ \rm{mm}$ \\

\endhead
\hline 
\endfoot

7850 & 0.0000 & 0.0000 & 0.0000 & 0.0000 & 0.0001 & 0.0002 & 0.0006 & 0.0008 & 0.0003 & 0.0001\\
7900 & 0.0000 & 0.0000 & 0.0003 & 0.0006 & 0.0009 & 0.0013 & 0.0022 & 0.0025 & 0.0016 & 0.0009\\
7950 & 0.0008 & 0.0007 & 0.0016 & 0.0023 & 0.0028 & 0.0034 & 0.0053 & 0.0057 & 0.0039 & 0.0026\\
8000 & 0.0026 & 0.0024 & 0.0040 & 0.0052 & 0.0064 & 0.0078 & 0.0127 & 0.0140 & 0.0092 & 0.0058\\
8050 & 0.0054 & 0.0051 & 0.0085 & 0.0114 & 0.0142 & 0.0178 & 0.0303 & 0.0340 & 0.0219 & 0.0130\\
8100 & 0.0114 & 0.0107 & 0.0186 & 0.0257 & 0.0326 & 0.0411 & 0.0697 & 0.0782 & 0.0524 & 0.0305\\
8150 & 0.0251 & 0.0233 & 0.0415 & 0.0572 & 0.0718 & 0.0885 & 0.1371 & 0.1510 & 0.1127 & 0.0707\\
8200 & 0.0579 & 0.0540 & 0.0913 & 0.1192 & 0.1426 & 0.1660 & 0.2203 & 0.2340 & 0.1994 & 0.1462\\
8250 & 0.1214 & 0.1150 & 0.1694 & 0.2021 & 0.2257 & 0.2451 & 0.2787 & 0.2850 & 0.2693 & 0.2331\\
8300 & 0.2025 & 0.1959 & 0.2436 & 0.2649 & 0.2775 & 0.2853 & 0.2936 & 0.2934 & 0.2904 & 0.2810\\
8350 & 0.2625 & 0.2589 & 0.2806 & 0.2871 & 0.2897 & 0.2903 & 0.2895 & 0.2878 & 0.2854 & 0.2854\\
8400 & 0.2808 & 0.2799 & 0.2837 & 0.2843 & 0.2842 & 0.2840 & 0.2844 & 0.2837 & 0.2793 & 0.2764\\
8450 & 0.2757 & 0.2755 & 0.2762 & 0.2770 & 0.2775 & 0.2781 & 0.2801 & 0.2802 & 0.2763 & 0.2711\\
8500 & 0.2678 & 0.2675 & 0.2694 & 0.2708 & 0.2718 & 0.2726 & 0.2740 & 0.2742 & 0.2726 & 0.2687\\
8550 & 0.2615 & 0.2612 & 0.2631 & 0.2641 & 0.2646 & 0.2649 & 0.2647 & 0.2644 & 0.2650 & 0.2641\\
8600 & 0.2546 & 0.2546 & 0.2549 & 0.2549 & 0.2548 & 0.2546 & 0.2530 & 0.2523 & 0.2537 & 0.2552\\
8650 & 0.2447 & 0.2450 & 0.2438 & 0.2431 & 0.2427 & 0.2422 & 0.2403 & 0.2394 & 0.2404 & 0.2425\\
8700 & 0.2329 & 0.2333 & 0.2315 & 0.2307 & 0.2302 & 0.2298 & 0.2285 & 0.2277 & 0.2277 & 0.2290\\
8750 & 0.2207 & 0.2210 & 0.2195 & 0.2190 & 0.2188 & 0.2187 & 0.2181 & 0.2177 & 0.2168 & 0.2168\\
8800 & 0.2090 & 0.2091 & 0.2085 & 0.2084 & 0.2084 & 0.2085 & 0.2087 & 0.2086 & 0.2073 & 0.2062\\
8850 & 0.1985 & 0.1984 & 0.1986 & 0.1988 & 0.1990 & 0.1992 & 0.1997 & 0.1998 & 0.1987 & 0.1972\\
8900 & 0.1877 & 0.1875 & 0.1881 & 0.1884 & 0.1886 & 0.1888 & 0.1891 & 0.1893 & 0.1888 & 0.1875\\
8950 & 0.1698 & 0.1696 & 0.1702 & 0.1703 & 0.1703 & 0.1704 & 0.1703 & 0.1704 & 0.1706 & 0.1700\\
9000 & 0.1534 & 0.1531 & 0.1534 & 0.1533 & 0.1530 & 0.1528 & 0.1522 & 0.1522 & 0.1528 & 0.1530\\
9050 & 0.1469 & 0.1467 & 0.1466 & 0.1461 & 0.1456 & 0.1453 & 0.1442 & 0.1440 & 0.1447 & 0.1455\\
9100 & 0.1363 & 0.1362 & 0.1356 & 0.1349 & 0.1343 & 0.1338 & 0.1326 & 0.1321 & 0.1327 & 0.1337\\
9150 & 0.1281 & 0.1281 & 0.1272 & 0.1265 & 0.1258 & 0.1253 & 0.1241 & 0.1233 & 0.1235 & 0.1245\\
9200 & 0.1249 & 0.1249 & 0.1239 & 0.1232 & 0.1225 & 0.1221 & 0.1210 & 0.1201 & 0.1197 & 0.1203\\
9250 & 0.1151 & 0.1151 & 0.1143 & 0.1137 & 0.1131 & 0.1128 & 0.1119 & 0.1111 & 0.1102 & 0.1103\\
9300 & 0.0878 & 0.0877 & 0.0872 & 0.0869 & 0.0866 & 0.0864 & 0.0859 & 0.0852 & 0.0842 & 0.0839\\
9350 & 0.0639 & 0.0638 & 0.0637 & 0.0636 & 0.0634 & 0.0634 & 0.0632 & 0.0627 & 0.0617 & 0.0611\\
9400 & 0.0647 & 0.0646 & 0.0646 & 0.0646 & 0.0646 & 0.0646 & 0.0646 & 0.0642 & 0.0630 & 0.0620\\
9450 & 0.0626 & 0.0625 & 0.0628 & 0.0628 & 0.0628 & 0.0629 & 0.0630 & 0.0626 & 0.0614 & 0.0604\\
9500 & 0.0593 & 0.0592 & 0.0595 & 0.0596 & 0.0597 & 0.0597 & 0.0599 & 0.0596 & 0.0586 & 0.0575\\
9550 & 0.0580 & 0.0578 & 0.0582 & 0.0584 & 0.0584 & 0.0585 & 0.0586 & 0.0584 & 0.0576 & 0.0566\\
9600 & 0.0568 & 0.0567 & 0.0571 & 0.0572 & 0.0572 & 0.0572 & 0.0573 & 0.0572 & 0.0566 & 0.0558\\
9650 & 0.0574 & 0.0572 & 0.0575 & 0.0576 & 0.0576 & 0.0576 & 0.0575 & 0.0575 & 0.0571 & 0.0565\\
9700 & 0.0570 & 0.0569 & 0.0571 & 0.0571 & 0.0570 & 0.0569 & 0.0568 & 0.0568 & 0.0566 & 0.0562\\
9750 & 0.0536 & 0.0536 & 0.0536 & 0.0535 & 0.0534 & 0.0533 & 0.0531 & 0.0530 & 0.0530 & 0.0529\\
9800 & 0.0505 & 0.0505 & 0.0504 & 0.0503 & 0.0501 & 0.0500 & 0.0497 & 0.0496 & 0.0497 & 0.0497\\
9850 & 0.0463 & 0.0462 & 0.0461 & 0.0459 & 0.0458 & 0.0456 & 0.0453 & 0.0452 & 0.0453 & 0.0454\\
9900 & 0.0409 & 0.0409 & 0.0407 & 0.0405 & 0.0404 & 0.0402 & 0.0400 & 0.0399 & 0.0399 & 0.0400\\
9950 & 0.0349 & 0.0349 & 0.0347 & 0.0346 & 0.0344 & 0.0343 & 0.0341 & 0.0340 & 0.0339 & 0.0340\\
10000 & 0.0178 & 0.0178 & 0.0177 & 0.0176 & 0.0175 & 0.0175 & 0.0174 & 0.0173 & 0.0173 & 0.0173\\
\hline\end{longtable}\end{landscape}
}

\subsection{HST standard measured magnitudes}
\label{sec:hst-stand-meas}

\Fixed{The tertiary catalogs define a magnitude system. This system
  has been anchored to the AB flux scale according to a calibration
  reference. We choose to use, as our reference, 5 stars from the HST
  CALSPEC database for which direct or indirect magnitudes in the
  system defined by the tertiary catalogs were available. Therefore,
  our calibration reference is the current release (stinic\_003) of
  the CALSPEC spectra for those 5 stars.
}

\Fixed{ The natural magnitudes of the measured HST standards are given
  in Table~\ref{tab:mmaghst}. Those measurements can be used to update
  the AB calibration of the catalogs, given revised synthetic
  magnitudes for these stars. Measurements in table~\ref{tab:mmaghst}
  are delivered in the same photometric system than the published
  tertiary standards catalog, \emph{i.e.} with the AB offsets derived
  in this work (Table~\ref{tab:combcalib}) \emph{already} applied. Therefore,
  any offset computed, according to Eq.~(\ref{eq:37}), from the
  difference between a measurement given in Table~\ref{tab:mmaghst}
  and the synthetic AB magnitude for the corresponding star readily
  applies to the published catalog.  }

\Fixed{When computing synthetic AB magnitudes for a star in the
  MegaCam focal plane, the peculiar position of the star in the focal
  plane must be taken into account. Also, we describe in
  Sect.~\ref{sec:covariance-matrix} the constitution of the covariance
  matrix for measurements in Table~\ref{tab:mmaghst}. This matrix can
  be used to combine all measurements similarly to what has been done
  in Sect.~\ref{sec:concl}.  }
\begin{center}
  \tablecaption{Measured magnitudes for CALSPEC
    standards.\label{tab:mmaghst}}

\tablehead{\hline\hline
Instrument & Band & Star & $x_{\rm fp}$ & $y_{\rm fp}$\footnote{$x_{\rm fp}$ and $y_{\rm fp}$ are the coordinates of measurements in the MegaCam focalplane (in cm, with the origin taken at the center of the focalplane).} & Mag\\
\hline}
\tabletail{\hline}
\begin{mpsupertabular}{lllrrr}

MegaCam & \band{u} & BD17 & 0.00 & 0.00 & 10.199\\
MegaCam & \band{g} & BD17 & 0.00 & 0.00 & 9.593\\
MegaCam & \band{r} & BD17 & 0.00 & 0.00 & 9.339\\
MegaCam & \band{i} & BD17 & 0.00 & 0.00 & 9.256\\
MegaCam & \band{z} & BD17 & 0.00 & 0.00 & 9.226\\
MegaCam & \band{u} & P330E & -0.79 & -3.12 & 14.179\\
MegaCam & \band{u} & P177D & -0.52 & -3.45 & 14.714\\
MegaCam & \band{u} & SNAP2 & -1.04 & -2.92 & 17.507\\
MegaCam & \band{u} & SNAP2 & -1.04 & -2.92 & 17.442\\
MegaCam & \band{u} & P330E & -0.79 & -3.12 & 14.163\\
MegaCam & \band{u} & SNAP2 & -1.04 & -2.92 & 17.480\\
MegaCam & \band{u} & P330E & -0.79 & -3.12 & 14.185\\
MegaCam & \band{u} & P177D & -0.52 & -3.45 & 14.731\\
MegaCam & \band{u} & SNAP2 & -1.04 & -2.92 & 17.685\\
MegaCam & \band{g} & P330E & -0.79 & -3.11 & 13.209\\
MegaCam & \band{g} & P177D & -0.53 & -3.45 & 13.693\\
MegaCam & \band{g} & SNAP2 & -1.05 & -2.91 & 16.458\\
MegaCam & \band{g} & SNAP2 & -1.05 & -2.91 & 16.443\\
MegaCam & \band{g} & P330E & -0.79 & -3.11 & 13.206\\
MegaCam & \band{g} & P177D & -0.55 & -3.46 & 13.681\\
MegaCam & \band{g} & SNAP2 & -1.04 & -2.91 & 16.448\\
MegaCam & \band{g} & SNAP2 & -1.04 & -2.91 & 16.433\\
MegaCam & \band{g} & P330E & 0.66 & -3.11 & 13.213\\
MegaCam & \band{g} & P330E & -0.79 & -3.11 & 13.208\\
MegaCam & \band{g} & P177D & -0.56 & -3.46 & 13.693\\
MegaCam & \band{g} & SNAP2 & -1.04 & -2.91 & 16.443\\
MegaCam & \band{g} & SNAP2 & -1.04 & -2.91 & 16.466\\
MegaCam & \band{g} & P330E & -0.79 & -3.11 & 13.226\\
MegaCam & \band{g} & SNAP2 & -1.04 & -2.91 & 16.445\\
MegaCam & \band{r} & P330E & -0.78 & -3.11 & 12.808\\
MegaCam & \band{r} & P177D & -0.52 & -3.44 & 13.271\\
MegaCam & \band{r} & SNAP2 & -1.04 & -2.91 & 16.035\\
MegaCam & \band{r} & SNAP2 & -1.04 & -2.91 & 16.018\\
MegaCam & \band{r} & P330E & -0.78 & -3.11 & 12.805\\
MegaCam & \band{r} & P177D & -0.52 & -3.45 & 13.267\\
MegaCam & \band{r} & SNAP2 & -1.03 & -2.91 & 15.989\\
MegaCam & \band{r} & SNAP2 & -1.03 & -2.91 & 15.999\\
MegaCam & \band{r} & P330E & -0.78 & -3.11 & 12.815\\
MegaCam & \band{r} & P177D & -0.51 & -3.45 & 13.279\\
MegaCam & \band{i2} & P330E & -0.78 & -3.11 & 12.688\\
MegaCam & \band{i2} & P177D & -0.52 & -3.45 & 13.160\\
MegaCam & \band{i2} & SNAP2 & -1.04 & -2.91 & 15.874\\
MegaCam & \band{i2} & SNAP2 & -1.04 & -2.91 & 15.890\\
MegaCam & \band{i2} & P330E & -0.78 & -3.11 & 12.686\\
MegaCam & \band{i2} & P177D & -0.52 & -3.45 & 13.153\\
MegaCam & \band{i2} & SNAP2 & -1.03 & -2.91 & 15.889\\
MegaCam & \band{i2} & SNAP2 & -1.03 & -2.91 & 15.846\\
MegaCam & \band{i2} & P330E & -0.78 & -3.11 & 12.703\\
MegaCam & \band{i2} & P177D & -0.52 & -3.45 & 13.161\\
MegaCam & \band{i2} & SNAP2 & -1.03 & -2.91 & 15.895\\
MegaCam & \band{i2} & SNAP2 & -1.03 & -2.91 & 15.903\\
MegaCam & \band{i2} & P330E & -0.79 & -3.11 & 12.699\\
MegaCam & \band{i2} & P177D & -0.55 & -3.46 & 13.169\\
MegaCam & \band{i2} & SNAP2 & -1.04 & -2.91 & 15.899\\
MegaCam & \band{i2} & SNAP2 & -1.04 & -2.91 & 15.868\\
MegaCam & \band{z} & P330E & -0.78 & -3.11 & 12.695\\
MegaCam & \band{z} & P177D & -0.51 & -3.44 & 13.140\\
MegaCam & \band{z} & SNAP2 & -1.03 & -2.91 & 15.874\\
MegaCam & \band{z} & P330E & -0.78 & -3.11 & 12.676\\
MegaCam & \band{z} & P177D & -0.52 & -3.44 & 13.130\\
MegaCam & \band{z} & SNAP2 & -1.03 & -2.91 & 15.827\\
MegaCam & \band{z} & P330E & -0.79 & -3.11 & 12.679\\
MegaCam & \band{z} & P177D & -0.52 & -3.44 & 13.136\\
MegaCam & \band{z} & SNAP2 & -1.04 & -2.91 & 15.900\\
SDSS2.5 & $u$ & P041C & -- & -- & 13.501\\
SDSS2.5 & $g$ & P041C & -- & -- & 12.281\\
SDSS2.5 & $r$ & P041C & -- & -- & 11.849\\
SDSS2.5 & $i$ & P041C & -- & -- & 11.734\\
SDSS2.5 & $z$ & P041C & -- & -- & 11.717\\
SDSS2.5 & $u$ & P330E & -- & -- & 14.485\\
SDSS2.5 & $g$ & P330E & -- & -- & 13.300\\
SDSS2.5 & $r$ & P330E & -- & -- & 12.844\\
SDSS2.5 & $i$ & P330E & -- & -- & 12.715\\
SDSS2.5 & $z$ & P330E & -- & -- & 12.685\\
SDSS2.5 & $u$ & P177D & -- & -- & 15.050\\
SDSS2.5 & $g$ & P177D & -- & -- & 13.763\\
SDSS2.5 & $r$ & P177D & -- & -- & 13.304\\
SDSS2.5 & $i$ & P177D & -- & -- & 13.175\\
SDSS2.5 & $z$ & P177D & -- & -- & 13.138\\
SDSS2.5 & $u$ & BD17 & -- & -- & 10.492\\
SDSS2.5 & $g$ & BD17 & -- & -- & 9.651\\
SDSS2.5 & $r$ & BD17 & -- & -- & 9.357\\
SDSS2.5 & $i$ & BD17 & -- & -- & 9.263\\
SDSS2.5 & $z$ & BD17 & -- & -- & 9.251\\
\end{mpsupertabular}
\end{center}

\subsection{Covariance matrices}
\label{sec:covariance-matrix}

The assumptions made to build the covariance matrix of systematic
measurement errors in the calibration data follows.  

In SNLS, we assume that the shutter bias affects identically all
measurements of flux ratios between short and long exposures. The
error resulting from aperture corrections is assumed to affect
coherently the measurements of the solar analogs in a given band, and
without correlation from one band to another. We account for the
aperture correction residuals in the Landolt-based \bdtruc
measurements independently as the observation strategy was different
and the derivation\Fixed{s} of the corrections were independent. The survey
non-uniformity affects coherently the measurement of solar analogs
that are related to the survey through the SNLS D3 field.  The Landolt
non-uniformity is included to account for a potential departure of the
\bdtruc measurements from the average Landolt system. The errors on
the color transformation of \bdtruc are assumed to be independent from
one band to another. This may not be fully justified, \Fixed{but} this
error matters only in band $z$ and potential correlations with other
bands are negligible.

In SDSS, the statistical uncertainty on the color transformation
between the PT and 2.5~m system affects coherently all the measurements
in a given band. Star-to-star dispersion around the color
transformation is assumed independent from one star to another. The
non uniformity of the Smith catalog \Fixed{coherently affects}  the measurement
of solar analogs as they \Fixed{lie} at approximately the same hour angle.

\Fixed{ The resulting covariance matrix of measurements in
  Table~\ref{tab:mmaghst} (including measurement noise and
  systematics) is available from the SNLS/SDSS calibration webpage.  }

\Fixed{ As described in Sect.~\ref{sec:concl}, when combining the
  available measurements to provide the final calibration, we account
  for other contributions to the covariance of the AB offsets: the
  HST-STIS spectrum measurement error, the STIS calibration error, and
  the uncertainty on instrument filters. A corresponding covariance
  matrix for each of these contributions is also available from the
  SNLS/SDSS calibration webpage.  }

\Fixed{
The estimation of the STIS measurement error is discussed in
Sect.~\ref{sec:sed-primary-standard}. The corresponding covariance
matrix for a single epoch STIS measurement is displayed in
Table~\ref{tab:stiscov}.
}

\begin{table*}
\centering
\caption{Covariance matrix of synthetic STIS magnitudes.}
\label{tab:stiscov}
\begin{equation} {\rm cov}(m, m')= 10^{-6}\left(\begin{array}{*{11}{l}}
25.8 & 17.2 & 9.8 & -2.9 & -2.1 & -16.4 & 26.3 & 18.7 & 10.2 & -1.4 & -16.6
\\
 & 15.5 & 3.9 & -4.4 & -3.7 & -15.6 & 16.8 & 16.3 & 4.2 & -3.2 & -15.9
\\
 &  & 26.0 & 21.0 & 20.8 & 26.0 & 9.9 & 3.8 & 26.6 & 20.8 & 26.9
\\
 &  &  & 35.5 & 33.5 & 58.6 & -3.6 & -5.0 & 21.1 & 32.9 & 60.2
\\
 &  &  &  & 31.8 & 54.7 & -2.7 & -4.4 & 20.9 & 31.2 & 56.2
\\
 &  &  &  &  & 119.1 & -17.6 & -17.2 & 26.1 & 52.6 & 122.5
\\
 &  &  &  &  &  & 27.5 & 18.3 & 10.3 & -2.0 & -17.9
\\
 &  &  &  &  &  &  & 17.3 & 4.0 & -3.8 & -17.5
\\
 &  &  &  &  &  &  &  & 27.3 & 20.9 & 26.9
\\
 &  &  &  &  &  &  &  &  & 30.7 & 54.1
\\
 &  &  &  &  &  &  &  &  &  & 125.9
\\
\end{array}\right)
\begin{array}{*{1}{r}}
u_M\\g_M\\r_M\\i_M\\i2_M\\z_M\\
u\\g\\r\\i\\z\\
\end{array}\end{equation}
\end{table*}

\end{document}